\documentclass[12pt]{report}
\usepackage[francais]{babel}
\usepackage{amsmath}
\usepackage{fancyhdr}
\usepackage{a4wide}
\usepackage{color}
\usepackage[ansinew]{inputenc}
\usepackage{amsfonts}
\usepackage{amssymb}
\usepackage{amstext}
\usepackage{amsthm}
\usepackage{newlfont}
\usepackage{graphicx}
\usepackage{titlesec}
\usepackage{times}
\usepackage{color}
\usepackage[dvipdfm]{hyperref}
\usepackage{hyperref}
\hypersetup{
backref=true, 
pagebackref=true,
hyperindex=true, 
colorlinks=true, 
breaklinks=true, 
urlcolor=blue, 
linkcolor=blue, 
citecolor=blue,
bookmarks=true, 
bookmarksopen=true, 
pdftitle={Mon fabuleux livre}, 
pdfauthor={Pejvan BEIGUI}, 
pdfsubject={Mac OS X} 
}
\definecolor{webred}{rgb}{0.5,0,0}
\definecolor{linkcol}{rgb}{0,0,0.4}
\definecolor{citecol}{rgb}{0.5,0,0}
\definecolor{citecol}{rgb}{0,0,0.4}
\makeatletter
\def\ps@headings{\def\@oddfoot{}%
\def\@oddhead{\makebox[\textwidth][l]{\underline{\hbox to \textwidth{\bf
\firstmark\hfill\thepage}}}}%
\def\@evenfoot{}%
\def\@evenhead{\makebox[\textwidth][r]{\underline{\hbox to \textwidth{\bf
\thepage\hfill\@lhead}}}}%
\def\chaptermark##1{\mark{}\def\@lhead{##1}}}%
\titleformat{\chapter}[display]
 {\normalfont\Large\filcenter\bf}
 {
   \vspace{1pt}%
   \vspace{1pc}%
   \raggedright\LARGE{\chaptername} \thechapter}
 {1pc}
 {\titlerule
    \vspace{1pc}%
   \Huge}[{\vspace{1pc}\titlerule}]

\begin{document}
\begin{titlepage}
\begin{center}
\vspace*{-2cm}{\bf République Algérienne Démocratique et Populaire}\vspace{0.1cm}
\\{\bf Ministère de l'Enseignement Supérieur et de la Recherche Scientifique}\vspace{0.1cm}
\\{\bf Université \textsc{a. mira} de Béja\"{\i}a}\vspace{0.1cm}
\\{\bf Faculté des Sciences Exactes}\vspace{0.1cm}
\\{\bf Département de Physique}\vspace{2cm}
\end{center}
\begin{figure}[\here]
   \centering
  \includegraphics[width=3cm]{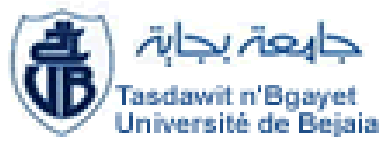}
\end{figure}

\vspace{0.5cm}
\begin{center}
{\Huge\bf Thèse de Doctorat en Physique}\vspace{1cm}
\\{\large\bf Présentée par Mr. BELABBAS Abdelmoumene}\vspace{0.3cm}
\\{\large\bf En vue de l'obtention du grade de Docteur en Sciences}
\vspace{1cm}
\\{\large\bf Option: Physique Théorique}\\
\vspace{1.5cm}
{\large\bf Thème}\\
\vspace{0.5cm}

\hrule\hrule\hrule\hrule\hrule
\vspace{0.3cm}
{\Huge \textsc{Les Interactions Fondamentales}}\\
\vspace{0.3cm} {\Huge \textsc{et la Structure de l'espace-temps}}\\
\vspace{0.3cm}
\hrule\hrule\hrule\hrule\hrule
\vspace{1.5cm}
{\bf Soutenue publiquement le $17/11/2011$ devant le jury suivant:}\vspace{0.5cm}
\begin{tabular}{lllll}
\hspace{-0.5cm}{\bf Président} & Mr.~~\textsc{BOUDRAHEM Sma\"{\i}l} & Professeur & U. A. Mira&
Béja\"{\i}a\\
\hspace{-0.5cm}{\bf Examinateur} & Mr.~~\textsc{AIT MOUSSA Karim}& Professeur & U. Mentouri &
Constantine\\
\hspace{-0.5cm}{\bf Examinateur} & Mr.~~\textsc{BELALOUI Nadir} & Professeur &
U. Mentouri &
Constantine\\
\hspace{-0.5cm}{\bf Rapporteur} & Mr.~~\textsc{BOUDA Ahmed} & Professeur & U. A. Mira &
Béja\"{\i}a\\
\end{tabular}\vspace{0.5cm}
\end{center}
\end{titlepage}
\pagestyle{empty}

\noindent\textbf{Résumé}\\

\vspace{0.25cm}
\noindent A la lumière des résultats surprenants de C.C.Barros, nous explorons dans cette thèse les possibilités d'interprétation géométrique de toutes les interactions fondamentales dans l'espoir de les unifier. Plus précisément nous essayons de fournir une description géométrique unifiée de la gravitation et de l'électromagnétisme.

\noindent L'analyse de l'approche standard de Huei de la Gravité Linéaire, dans laquelle les équations d'Einstein prennent la forme d'équations de type Maxwell, a révélé l'existence de quelques imperfections. En effet,  l'étude est restreinte au régime stationnaire, la relation entre le potentiel scalaire et le champ type électrique n'est valable que dans le cas particulier de la jauge harmonique et un facteur 4 indésirable apparait dans la partie magnétique de la force de Type Lorentz.  L'adoption d'un choix subtile de conditions de jauge nous a permis d'éliminer ces insuffisances et de revisiter la Gravité Linéaire de telle sorte à aboutir à une analogie parfaite avec l'électromagnétisme. Dans le cas linéaire, nous avons pu montrer que les Equations de Maxwell pouvaient être dérivées à partir d'une version électromagnétique d'Equations d'Einstein et que les  termes d'ordres supérieurs sont complètement négligés dans le domaine actuel de l'application de l'électromagnétisme.

\vspace{2cm}

\noindent\textbf{Abstract}\\

\vspace{0.25cm}
\noindent In the light of intriguing results of C.C.Barros, we investigate in this thesis the possibilities of geometrical interpretation of all the fundamental interactions in order to unify them. More exactly we try to supply a unified geometrical description for gravitation and electromagnetism.

\noindent The analysis of Huei's standard approach of Linear Gravity, in which the Einstein equations can be written in the same form of the Maxwell ones, revealed the existence of some imperfections. In fact, the relation between the scalar potential and the electric-type field is not valid except in the harmonic gauge, the Lorentz-type force is obtained with a time independence restriction and an undesired factor 4 appears in the magnetic-type part. A subtle gauge conditions allows us to eliminate these imperfections and to revisit the Linear Gravity in a way to get a strong similarity with electromagnetism. In the linear case, we showed that Maxwell's equations could be derived from an electromagnetic version of the Einstein ones, and that the higher order terms are negligible in the current domain of application of electromagnetism.

\newpage
\begin{flushright}
\emph{A mes très chers parents et à mon frère} $\cdots$
\end{flushright}
\newpage
\pagestyle{empty} {\Huge \textbf{Remerciements}}

\vspace{1cm}Ce travail a été réalisé au Laboratoire de Physique Théorique de l'Université \textsc{a. mira} de Béja\"{\i}a, sous la direction de Monsieur le Professeur \textsc{Bouda ahmed}. Je tiens à lui exprimer ici toute ma gratitude et ma sincère reconnaissance d'avoir assuré une grande partie de ma formation et d'avoir accepté de diriger ce travail de Thèse. En effet, Mr.~~\textsc{bouda} a non seulement le mérite de m'avoir initié aux théories de la Relativité Restreinte et Générale, à l'Electromagnétisme et à la Mécanique Quantique, mais aussi celui d'être à l'origine de mes premiers pas dans le domaine de l'enseignement. Il m'a ensuite initié au monde passionnant de la recherche scientifique. Les années de formation, passées sous sa direction, ont été d'un apport considérable, à la fois, sur le plan scientifique et humain. Je ne le remercierais jamais assez de m'avoir guidé et conseillé pendant ces dix dernières années. \\
\\
\indent C'est pour moi un honneur que Mr.~~\textsc{boudrahem s.}, Professeur à l'université \textsc{a.mira} de Béja\"{\i}a, accepte de présider le jury de ma soutenance. Qu'il veuille agréer l'expression de ma très haute considération.\\
\\
\indent Que Messieurs \textsc{ait moussa k.} et \textsc{belaloui n.}, Professeurs à l'université \textsc{Mentouri} de \textsc{Constantine}, puissent trouver ici l'expression de ma profonde gratitude pour avoir accepté de juger ce travail.\\
\\
\indent Un grand merci à tous les enseignants qui ont assuré ma
formation tout au long de mon parcours. Je tiens à rendre un hommage particulier à trois enseignants pour lesquels j'ai énormément de respect:\\
\\
\noindent  Mr. \textsc{adel kassa}, enseignant de Mathématiques et de Théorie des Champs à l'Université de Béja\"{\i}a; le physicien qui a su ouvrir nos yeux sur les horizons infinis du savoir.\\
\\
\noindent  Mr. \textsc{tatah mustapha}, enseignant de Mathématiques au Lycée Ibn Sina; très fin pédagogue et l'un des enseignants les plus charismatiques que j'ai connu.\\
\\
\noindent Mr. \textsc{amaouche abdelaziz}, enseignant de Mathématiques au \textsc{cem} Ibn Toumert; l'enseignant qui a su communiquer l'amour des Mathématiques à des générations d'élèves.\\
\\
\indent Je n'oublierais pas mon cher frère Imad, physicien chercheur, à qui je dois ma passion pour la physique. \\
\\
\indent Toute tentative pour rendre hommage à mes très chers parents est vaine $\cdots$ les mots ne pourraient être à la hauteur de leur dévouement ni de ce qu'ils représentent pour mon frère et moi $\cdots$ Je leurs rend un vibrant hommage du fond de mon c{\oe}ur !\\


\tableofcontents

\pagestyle{fancy} \lhead{chapitre\;1}\rhead{Introduction}
\chapter{Introduction}

\noindent Tous les phénomènes naturels, dans leur grande diversité, sont décrits à l'aide de quatre interactions fondamentales. L'interaction Gravitationnelle, responsable de l'attraction universelle entre corps massifs et l'interaction électromagnétique, responsable de la stabilité des atomes, se manifestent à notre échelle macroscopique à cause de leurs portées infinies; alors que les deux autres interactions ne se manifestent qu'à l'échelle nucléaire: l'interaction faible, responsable de la désintegration de certaines particules et l'interaction nucléaire forte, responsable de la stabilité des noyaux atomiques.

\vspace{0.5cm}\noindent Un des objectifs les plus ambitieux de la physique moderne consiste à vouloir unifier toutes ces interactions, i.e. arriver à les voir comme des manifestations, à des échelles d'énergie différentes, d'une seule interaction plus fondamentale.

\vspace{0.5cm}\noindent A l'heure actuelle, nous disposons de deux grandes théories "complémentaires" pour rendre compte des quatre interactions fondamentales : La théorie de la Relativité Générale (RG) qui décrit la gravité dans un cadre géométrique en reliant le champ gravitationnel à la courbure de l'espace-temps et la Mécanique Quantique (MQ) qui est le cadre conceptuel fondamental de toutes les théories visant à décrire les interactions non gravitationnelles (électromagnétique, nucléaire forte et faible).

\vspace{0.5cm}\noindent Au niveau fondamental, la description actuelle des interactions non gravitationnelles se fait dans le cadre de la Théorie des Champs; formalisme qui découle d'une description, à la fois, quantique, relativiste et causale. Une description similaire, de toutes ces interactions, est rendue possible en ayant recours à un principe de jauge, au moyen duquel on exprime la volonté d'avoir une théorie invariante par rapport à certaines transformations de champs, effectuées indépendamment en différents points de l'espace. Bien que ces interactions non gravitationnelles soient ainsi mathématiquement jumelées, néanmoins la gravité reste celle qui pose le plus de problèmes et demeure, jusqu'à présent, l'interaction qui se refuse à toute tentative d'unification avec les autres. L'obstacle majeur qui bloque le grand projet d'unification est la réconciliation de la RG et de la MQ, autrement dit, le problème de réconcilier nos deux visions "découplées" pour décrire l'univers aux échelles macroscopique et microscopique. Nous pensons que l'obstacle majeur à la réconciliation de ces deux théories fondamentales est dû au fait qu'elles s'appuient sur des conceptions philosophiques très différentes. En effet, alors que la RG est une théorie déterministe, la MQ, dans le cadre de l'interprétation standard de l'école de Copenhague, est une théorie probabiliste.

\vspace{0.5cm}\noindent Les tentatives de réconciliation entre ces deux théories peuvent être classées en deux grandes familles
\begin{itemize}
  \item Approches visant à établir une théorie probabiliste de la gravitation, autrement dit, approches où il est question de formuler une version probabiliste de RG. Toutes les tentatives de quantification de la gravité n'ont pas encore été couronnées de succès du fait de l'apparition de problèmes, jusque là,
      insurmontables, notamment l'impossibilité de réduire les infinités qui apparaissent dans les calculs (théories non renormalisables). Parmi elles, citons par exemple: la Quantification Perturbative, la Gravité Quantique à Boucles, la Géométrie Non Commutative, La Théorie des Cordes.
  \item Approches visant à formuler une version déterministe de la MQ. Alors que les prévisions expérimentales de la MQ standard, pleinement vérifiées, sont admises par toutes la communauté scientifique, néanmoins, les partisans du déterminisme contestent ses fondement philosophiques, notamment le fait de ne définir l'état quantique qu'à partir d'informations accessibles à l'expérience et le fait de nier l'existence de la valeur d'une grandeur physique avant sa mesure.
      Contrairement à ce qui se fait en MQ probabiliste, où il n'est plus question de décrire les processus physiques mais seulement de les prévoir, les déterministes tentent plutôt à rétablir l'existence d'une réalité objective indépendante de nos connaissances, et ce à travers une reformulation des postulats de base.  Parmi les tentatives les plus significatives dans ce sens, nous pouvons citer: l'approche de De Broglie \cite{broglie1}-\cite{broglie14}, l'approche de Bohm \cite{Bohm1}-\cite{Bohm8}, l'approche de Floyd \cite{floyd}-\cite{mathdeuxfloyd}, l'approche de Farragi et Matone \cite{farggi}-\cite{resumefarggi}, l'approche de Bouda et al. \cite{boudaprob}-\cite{boudareplydjama}, $\cdots$
\end{itemize}

\vspace{0.5cm}\noindent La Géométrisation des interactions fondamentales a suscité l'intérêt de plusieurs physiciens comme Einstein \cite{einstein1, einstein2}, Weyl \cite{Weyl1}-\cite{Weyl5}, Schr\"{o}dinger \cite{Schrodinger1}, Eddington \cite{Eddington}, Kaluza et Klein \cite{Kaluza, Klein1, Klein2} et bien d'autres \cite{randers}-\cite{Borzou}. Récemment, C.C.Barros \cite{Barros1, Barros2, Barros3} est arrivé à décrire l'atome d'hydrogène de façon tout à fait inédite et ce en partant de l'hypothèse que les interactions non gravitationnelles peuvent affecter la structure de l'espace-temps. Dans le contexte de la solution de Schwarzschild, de façon similaire à ce qui se fait en RG avec la gravité, en incorporant l'interaction "proton-électron" dans la métrique de l'Espace-temps, il arrive à dériver le spectre de l'atome d'hydrogène, prévu par la théorie de Dirac, et ce dans le cadre de l'approximation du champ faible.

\vspace{0.5cm}\noindent A la lumière de ce résultat "spectaculaire", nous explorons dans cette Thèse les possibilités d'une interprétation géométrique de toutes les interactions fondamentales, en vue d'offrir un cadre et une description unifiée de celles-ci. Plus précisément, nous voulons donner une porté plus considérable aux résultats de C.C.Barros en les resituant dans le contexte plus général d'une éventuelle unification des interactions fondamentales.

\vspace{0.5cm}\noindent La forte analogie entre l'électromagnétisme et la gravité fait de l'interaction électromagnétique le meilleur candidat potentiel, parmi toutes les interactions non gravitationnelles déjà jumelées, pour être décrite en termes géométriques. Nous avons opté, par conséquent, pour une stratégie qui consiste à se focaliser, dans un premier temps, sur la gravité et l'électromagnétisme, tout en espérant que cette éventuelle description unifiée soit étendue, plus tard, aux interactions faible et nucléaire forte.

\vspace{0.5cm}\noindent L'essentiel de nos investigations a été réalisé en deux périodes distinctes. La première étape, dédiée à une critique de l'approche de Barros, a révélé deux points intéressants \cite{Belabbas}
      \begin{enumerate}
        \item Depuis l'apparition de la RG, il est établi que contrairement aux autres interactions la gravité se manifeste à travers la structure de l'Espace-temps. Cette description est rendue possible grâce au principe d'Equivalence, présenté par Einstein comme un puissant argument en faveur de l'adoption du postulat de la RG, i.e. covariance des lois physiques sous l'action de transformation arbitraires de coordonnées. L'argument crucial qui justifie l'adoption de ce principe est l'égalité des masses pesante et inertielle d'un même corps; égalité qui se traduit par une propriété caractérisant exclusivement le champ gravitationnel : Tous les corps se déplaçant uniquement sous l'influence d'un champ de gravitation sont soumis à la même accélération, indépendamment de leurs masses, de leurs substances et de leurs états physiques. Contrairement à la gravité, il n'y a pas d'argument similaire en faveur d'une interprétation géométrique des interactions non gravitationnelles.
        \item Une adoption d'une métrique de Schwarzschild n'est pas également justifiée. En effet, une telle solution découle d'une utilisation des Equations d'Einstein, écrites exclusivement pour le champ de gravitation. Il faut savoir que les Equations d'Einstein sont les équations fondamentales permettant de décrire le champ de gravitation; elles ont la même importance que les Equations de Maxwell pour le champ électromagnétique.
      \end{enumerate}

\vspace{0.5cm}\noindent Ces deux remarques vont être à l'origine de la seconde étape d'investigations, représentée par le travail de Thèse, proprement dit. En effet, le premier point nous a incité à nous poser des questions sur les fondements de la RG, précisément sur le rôle du principe d'Equivalence en RG: Est-il indispensable pour formuler la Théorie de la RG ? Est-ce que l'exigence de covariance des lois physiques, à elle seule, est suffisante pour servir de base à la RG ? Faut-il penser à reformuler le Postulat d'Equivalence pour l'étendre aux interactions non gravitationnelles ? Le second point nous a poussé à émettre l'hypothèse de l'existence d'une nouvelle version des Equations d'Einstein et des géodésiques pour l'interaction coulombienne.

\vspace{0.5cm}\noindent Dans le but de traiter le champ électromagnétique au même pied d'égalité que le champ de gravitation, nous nous sommes penchés sur le domaine de la Gravité Linéaire, où des phénomènes très intéressants ont lieu. En effet, en plus du champ de gravitation radial crée par une masse (champ gravitoéléctrique), il y a apparition d'un champ orthoradial qui jouerait le rôle d'un analogue au champ magnétique pour la gravitation (champ gravitomagnétique). Cette analogie nous a permis de développer un nouveau point de vue pour décrire le champ électromagnétique de manière géométrique. Le point de départ était l'analyse de l'approche de Huei \cite{Huei}, où il est montré que les Equations d'Einstein linéarisées peuvent se mettre sous la forme d'Equations de type Maxwell pour la gravité. Dans cette formulation, et aussi dans l'approche de Wald \cite{Wald}, nous avons souligné quelques imperfections
\begin{enumerate}
  \item La relation entre les potentiels et le champ gravitoéléctrique, telle qu'elle est définie en Electromagnétisme, n'est obtenue que dans le cas particulier de la jauge harmonique.
  \item Dans les équations de géodésique, au niveau de la partie magnétique de la force de type Lorentz, il y a apparition d'un facteur 4 indésirable comparé à la force de Lorentz usuelle de l'Electromagnétisme.
  \item La force de type Lorentz n'est obtenue que dans le cas où les champs sont stationnaires.
\end{enumerate}
\noindent Soulignons qu'en redéfinissant les champs, Carroll \cite{Carroll1} est parvenu à éliminer le facteur 4 indésirable, mais sans satisfaire pour autant les Equations de Type Maxwell.

\vspace{0.5cm}\noindent A travers l'adoption d'un choix subtile des conditions de jauge, nous sommes parvenus à éliminer les inperfections précédentes, ce qui nous a permis de revisiter la Gravité Linéaire de telle sorte à aboutir à une très grande similarité avec l'Electromagnétisme. Les résultats de Barros et la gravité linéaire revisitée nous ont incité à croire à l'extension du principe d'Equivalence aux interactions non gravitationnelles. Dans le cas linéaire, nous avons pu montré que les Equations de Maxwell de l'électromagnétisme pouvaient être dérivées à partir d'une nouvelle version d'Equations d'Einstein \cite{Belabbas2}. Autrement dit, le champ électromagnétique est décrit par des équations non linéaires de type Einstein qui se réduisent, à l'ordre 1 de perturbation, aux équations de Maxwell. De plus, en poussant les calculs jusqu'à l'ordre 2 de perturbation, nous sommes parvenus à apporter des corrections aux équations de Maxwell. Finalement, nous avons pu montré que les termes d'ordre supérieur sont négligeables dans le domaine usuel de l'Electromagnétisme, de plus, une analyse qualitative de ces termes supérieurs nous a permis de souligner une différence fondamentale entre la gravité et l'électromagnétisme. En effet, alors que l'interaction gravitationnelle est essentiellement à caractère non linéaire pour laquelle la présence de la source $M$ est suffisante pour affecter la structure de l'espace-temps, l'interaction électromagnétique est essentiellement linéaire pour laquelle la présence de la source $Q$ n'affecte la métrique de l'espace-temps qu'après avoir interagit avec une charge test $q$.

\vspace{0.5cm}\noindent Le manuscrit est organisée comme suit
\begin{itemize}
  \item Le Chapitre 2 est dédié, à la fois, au rappel de quelques notions fondamentales de RG et à l'exposé de l'essentiel de l'approche de Barros.
  \item Le Chapiter 3 sera consacré à la Gravité Linéaire. Il sera question, d'abord, de rappeler la version "standard" dans laquelle les équations du champ de gravitation sont de Type Maxwell, dans le but de proposer, ensuite, une nouvelle version dans laquelle les imperfections de l'approche standard vont être surmontées grâce à un choix judicieux des conditions jauge, ce qui va nous permettre de revisiter la Gravité linéaire de telle sorte à aboutir à une plus grande analogie entre la gravité et l'électromagnétisme.
  \item Au Chapitre 4, il sera question de développer une nouvelle façon de décrire l'Electromagnétisme. En effet, nous allons montrer que le champ électromagnétique est décrit par des Equations de type Einstein qui se réduisent, dans le cas linéaire, aux équations de Maxwell d'Electromagnétisme.
  \item Le Chapitre 5 est réservé, à la fois, à la discussion et critique des résultats obtenus.
\end{itemize}

\newpage
\pagestyle{fancy} \lhead{chapitre\;2}\rhead{Relativité Générale et Extension aux Systèmes Subatomiques}
\chapter{Relativité Générale et Extension aux Systèmes Subatomiques}

\section{Introduction}

La Mécanique Classique (MC) de Newton est applicable dans des référentiels d'inertie, dans lesquelles tout corps libre a tendance à conserver son état de mouvement en se déplaçant à vitesse constante. Toute variation de cette vitesse est due à l'action d'une force extérieure $\overrightarrow{f}=m_{i}\overrightarrow{a}$. De plus, si un corps 1 agit sur un autre corps 2 avec une force $\overrightarrow{f_{12}}$, alors ce dernier agit aussi sur le premier avec une force opposée $\overrightarrow{f_{21}}=-\overrightarrow{f_{12}}$. Outre les trois lois précédentes de Newton, la MC s'appuie aussi sur une loi de gravitation.

Le référentiel fondamental de Newton est l'espace absolu muni d'un système de coordonnées cartésien. En plus des propriétés géométriques euclidiennes de cet espace, le temps s'écoule de la même manière pour tous ses points.

Alors que le principe fondamental de la dynamique est covariant sous une transformation de Galilée\footnote{On admet implicitement que $t^{'}=t$}
$\overrightarrow{r}^{'}=\overrightarrow{r}-\overrightarrow{u}t$, les équations de Maxwell de l'électromagnétisme sont plutôt covariantes sous la transformation de Lorentz.

En postulant l'invariance de la vitesse de la lumière $c$ dans tous les référentiels inertiels et la covariance de toutes les lois de la nature au passage d'un référentiel d'inertie à un autre, Einstein (1905) arrive à jeter les bases la théorie de la Relativité Restreinte (RR) dans laquelle les interactions n'agissent plus à distance et l'espace et le temps perdent leur caractère absolu pour former une entité plus fondamentale appelée espace-temps à quatre dimensions de Minkowski. Les éléments de l'espace-temps sont les quadrivecteurs $(x^{0}=c\,t,x^{1}=x,x^{2}=y,x^{3}=z)$ alors que sa métrique est $\eta_{\mu\nu}=(+1,-1,-1,-1)$ telle que le carré de l'intervalle séparant deux événements infiniment voisins
\begin{equation}\label{carre interval inertiel}
    ds^{2}=\sum_{\mu=0}^{3}\sum_{\nu=0}^{3}\eta_{\mu\nu}dx^{\mu}dx^{\mu}=(dx^{0})^{2}-(dx^{1})^{2}-(dx^{2})^{2}-(dx^{3})^{2},
\end{equation}
soit un invariant relativiste. Il faut savoir que la RR se réduit à la MC dans le cas particulier où les vitesses mises en jeux sont négligeables devant la vitesse de la lumière $v\ll c$.

La transmission instantanée de la force gravitationnelle de Newton est en désaccord avec le principe fondamental de la RR selon lequel, la vitesse de la lumière est la vitesse limite de propagation d'une information. Au lieu d'essayer de rendre conforme la gravitation de Newton à la RR, par l'introduction d'un champ qui jouerait un rôle analogue au champ magnétique pour la gravitation, Einstein a compris qu'il fallait plutôt généraliser la relativité à tous les référentiels,  quels que soient leurs états de mouvements. Ainsi la RR serait un cas particulier de cette nouvelle  théorie, applicable   dans   le cas  où  le  champ gravitationnel est très faible, voir inexistant.

En se basant sur le postulat selon lequel "Toutes les lois de la nature prennent la même forme dans tous les référentiels, quels que soient leurs états de mouvement" Einstein parvient à décrire la gravité dans un cadre géométrique, en reliant le champ gravitationnel à la courbure de l'espace-temps. Cette relation est présentée par Einstein comme une conséquence du Principe d'Equivalence en interprétant convenablement l'égalité des masses gravitationnelle et inertielle.

Pour garantir une covariance des lois de la nature vis-à-vis de transformations arbitraires de coordonnées, elles sont exprimées sous forme tensorielle. Les équations d'Einstein sont la généralisation relativiste de la loi de Newton de la gravitation; elle permettent de décrire comment le champ gravitationnel est généré par une source matérielle. De plus, l'équation qui décrit comment la matière "répond" à ce champ, qui se manifeste à travers la courbure de l'espace-temps, est l'équation des géodésiques.

Dans ce chapitre il sera question de rappeler, dans un premier temps, quelques notions de la Théorie de la Relativité Générale indispensables à la compréhension du contenu de cette thèse. Pour un exposé plus détaillé, nous recommandons la consultation des ouvrages spécialisés \cite{landau},\cite{Weinberg}, \cite{Khriplovitch}-\cite{BoudaRG} qui ont servi à l'élaboration de ces rappels. Dans un deuxième temps, l'essentiel de l'approche de Barros \cite{Barros1}-\cite{Barros3} sera présenté. Une analyse critique de cette approche va révéler que le résultat le plus significatif, i.e. description de l'atome d'hydrogène en incorporant l'interaction électron-proton dans la métrique de l'espace-temps, est du en fait à une utilisation implicite d'un nouveau Principe d'Equivalence, étendu à l'interaction coulombienne. On termine par une présentation de quelques arguments en faveur d'une interprétation géométrique de l'interaction électromagnétique.

\section{Rappels d'Analyse Tensorielle}
Les lois de la nature doivent répondre à deux exigences \cite{Weinberg}:
\begin{enumerate}
    \item Une fois formulées dans un système de coordonnées particulier, elles doivent avoir la même forme quand on passe à un système de coordonnée quelconque. Autrement dit, les lois de la nature doivent être covariantes sous n'importe quelle changement de coordonnées.
    \item En l'absence de gravitation, le tenseur métrique $g_{\mu\nu}$ tend à être égal au tenseur de Minkowski $\eta_{\mu\nu}$ et on retombe sur les lois déjà établies dans le cadre de la théorie de la relativité restreinte.
\end{enumerate}

Le formalisme tensoriel est un outil indispensable pour garantir l'adoption du Postulat de la Relativité Générale. En effet, la formulation tensorielle des lois de la nature leurs confèrent une covariance manifeste vis-à-vis des transformations arbitraires de coordonnées; autrement dit, une fois écrite sous forme tensorielle, une lois physique possède nécessairement une forme indépendante du système de coordonnées.

Le contenu de cette section est dédié au rappel de quelques notions d'analyse tensorielle; d'abord dans un espace euclidien et ensuite dans un espace riemannien.

\subsection{Composantes contravariantes et covariantes}
Soit une base quelconque $\{\overrightarrow{e_{i}}\}_{i=1,\cdots,n}$ d'un espace vectoriel euclidien
$E$ de dimension $n$. On appelle composantes contravariantes d'un
vecteur $\overrightarrow{A}\in E$, les quantités $\{A^{i}\}_{i=1,\cdots,n}$, telles que
\begin{equation}\label{composcontavdef}
    \overrightarrow{A}=\displaystyle\sum_{i=1}^{n}A^{i}\overrightarrow{e_{i}}.
\end{equation}

On appelle composantes covariantes d'un vecteur $\overrightarrow{A}\in E$, les
quantités $\{A_{i}\}_{i=1,\cdots,n}$, tel que
\begin{eqnarray}\label{composcovdef}
    \left\{%
\begin{array}{ll}
    A_{i}=\overrightarrow{A}.\overrightarrow{e_{i}},\\
    i=1,\cdots,n.\\
\end{array}%
\right.
\end{eqnarray}

Pour déterminer, comment ces grandeurs se transforment sous l'action d'une
transformation de coordonnées quelconque: $\{x^{i}\}\longrightarrow\{x^{'i}\}$, il
faut d'abord définir les coordonnées curvilignes.

\subsection{Coordonnées curvilignes} Considérons un point $M$ d'un espace vectoriel
euclidien $E$ et un système de coordonnées $\{x_{i}\}$, où $i=1,\cdots,n$. On associe au point
$M$ un repère naturel qui admet $M$ pour origine et
$\{\overrightarrow{e_{i}}\}_{i=1,\cdots,n}$ pour base
\begin{eqnarray}\label{deei}
\left\{%
\begin{array}{ll}
    \overrightarrow{e_{i}}=\displaystyle\frac{\partial \,\overrightarrow{OM}}
    {\partial x^{i}}\\
     i=1\rightarrow n.\\
\end{array}%
\right.
\end{eqnarray}
de sorte que
\begin{eqnarray}
    d\,\overrightarrow{OM}&=&\displaystyle\sum_{i=1}^{n}\displaystyle\frac{\partial \,\overrightarrow{OM}}
    {\partial x^{i}}\;dx^{i}=\displaystyle\sum_{i=1}^{n}\overrightarrow{e_{i}}\;dx^{i}.\nonumber
\end{eqnarray}

Pour pouvoir faire le passage d'un système de coordonnées $\{x^{i}\}$ à un autre
système de coordonnées $\{x^{'i}\}$, déterminons comment se transforment les vecteurs
de la base $\{\overrightarrow{e_{i}}\}\rightarrow\{\overrightarrow{e_{i}}^{'}\}$. On
suppose que les deux systèmes de coordonnées sont reliés par une transformation
inversible\footnote{On peut montrer que dans ce cas $\displaystyle\sum^{n}_{\ell=1}\displaystyle\frac{\partial x^{m}}{\partial x^{'\,\ell}}\;\displaystyle\frac{\partial x^{'\,\ell}}{\partial x^{p}}=\delta^{m}_{p}$ et que $\displaystyle\sum^{n}_{\ell=1}\displaystyle\frac{\partial x^{'\,m}}{\partial x^{\ell}}\;\displaystyle\frac{\partial x^{\ell}}{\partial x^{'\,p}}=\delta^{m}_{p}$} de la forme
\begin{eqnarray}
\left\{%
\begin{array}{ll}
    x^{'1}=x^{'1}(x^{1},x^{2},...,x^{n})\\
    x^{'2}=x^{'2}(x^{1},x^{2},...,x^{n}) \\
    \vdots\\
    x^{'n}=x^{'n}(x^{1},x^{2},...,x^{n}) \\
\end{array}%
\right.\hspace{0.5cm}\Leftrightarrow\hspace{0.5cm}\left\{%
\begin{array}{ll}
    x^{'i}=x^{'i}(x^{j})  \\
    i,j=1,\cdots,n \\
\end{array}%
\right.
\end{eqnarray}
En utilisant la définition (\ref{deei}), nous avons
\begin{eqnarray}\label{ejprimei}
    \overrightarrow{e_{j}}^{'}&=&\displaystyle\frac{\partial \,\overrightarrow{OM}}
    {\partial x^{'j}}\nonumber\\
    &=&\displaystyle\sum_{i=1}^{n}\displaystyle\frac{\partial \,\overrightarrow{OM}}{\partial x^{i}}
    \;\displaystyle\frac{\partial x^{i}}{\partial
    x^{'j}}\nonumber\\
    &=&\displaystyle\sum_{i=1}^{n}\displaystyle\frac{\partial x^{i}}{\partial
    x^{'j}}\;\overrightarrow{e_{i}},
\end{eqnarray}
et de manière analogue, il est aussi possible de montrer que
\begin{equation}\label{eiejprim}
     \overrightarrow{e_{i}}=\displaystyle\sum_{j=1}^{n}\displaystyle\frac{\partial x^{'j}}{\partial
    x^{i}}\;\overrightarrow{e_{j}}^{'}.
\end{equation}
Les équations (\ref{ejprimei}) et (\ref{eiejprim}) sont les formules de passage
entre les deux bases $\{\overrightarrow{e_{i}}\}$ et $\{\overrightarrow{e_{i}}^{'}\}$

\subsection{Transformation des composantes d'un vecteur}
\subsubsection{Composantes contravariantes} Déterminons comment se transforment les
composantes contravariantes d'un vecteur $\overrightarrow{A}$, lors d'une
transformation de coordonnées $\{x^{i}\}\rightarrow\{x^{'i}\}$ et vis versa:
\begin{eqnarray}
    \overrightarrow{A}&=&\displaystyle\sum_{i=1}^{n}A^{i}\;\overrightarrow{e_{i}},\hspace{1.25cm}
    \textrm{dans la base} \{\overrightarrow{e_{i}}\},\nonumber\\
    &=&\displaystyle\sum_{j=1}^{n}A^{'j}\;\overrightarrow{e_{j}}^{'},\hspace{1cm}
    \textrm{dans la base} \{\overrightarrow{e_{j}}^{'}\}.\nonumber
\end{eqnarray}
En utilisant (\ref{ejprimei}), nous avons, d'une part,
\begin{eqnarray}
    \overrightarrow{A}&=&\displaystyle\sum_{j=1}^{n}A^{'j}\;\Bigg(\displaystyle\sum_{i=1}^{n}
    \displaystyle\frac{\partial x^{i}}{\partial
    x^{'j}}\;\overrightarrow{e_{i}}\Bigg)\nonumber\\
    &=&\displaystyle\sum_{i=1}^{n}\Bigg(\displaystyle\sum_{j=1}^{n}A^{'j}\;
    \displaystyle\frac{\partial x^{i}}{\partial
    x^{'j}}\Bigg)\;\overrightarrow{e_{i}}\nonumber\\
    &=&\displaystyle\sum_{i=1}^{n}A^{i}\;\overrightarrow{e_{i}}.\nonumber
\end{eqnarray}
D'autre part, une identification membre à membre nous permet d'écrire
\begin{equation}\label{aiaprimj}
       A^{i}=\displaystyle\sum_{j=1}^{n}A^{'j}\;
    \displaystyle\frac{\partial x^{i}}{\partial
    x^{'j}}.
\end{equation}
De même, en utilisant (\ref{eiejprim}), il est possible de montrer de manière analogue que
\begin{equation}\label{aprimjai}
       A^{'j}=\displaystyle\sum_{i=1}^{n}A^{i}\;
    \displaystyle\frac{\partial x^{'j}}{\partial
    x^{i}}.
\end{equation}
Les formules (\ref{aiaprimj}) et (\ref{aprimjai}) montrent comment se transforment les
composantes contravariantes d'un vecteur $\overrightarrow{A}$.

\subsubsection{Composantes covariantes} Pour déterminer comment se transforment les
composantes covariantes, il faut écrire la définition (\ref{composcovdef}) dans les
deux bases tout en utilisant (\ref{eiejprim}) et (\ref{ejprimei}) pour avoir
\begin{eqnarray}
    \left\{%
\begin{array}{ll}
    A_{i}=\overrightarrow{A}.\overrightarrow{e_{i}}=\overrightarrow{A}.\displaystyle\sum_{j=1}^{n}\displaystyle\frac{\partial x^{'j}}{\partial
    x^{i}}\;\overrightarrow{e_{j}}^{'}=\displaystyle\sum_{j=1}^{n}\displaystyle\frac{\partial x^{'j}}{\partial
    x^{i}}\;A_{j}^{'}, \\
    A_{j}^{'}=\overrightarrow{A}.\overrightarrow{e_{j}}^{'}=\overrightarrow{A}.\displaystyle\sum_{i=1}^{n}\displaystyle\frac{\partial x^{i}}{\partial
    x^{'j}}\;\overrightarrow{e_{i}}=\displaystyle\sum_{i=1}^{n}\displaystyle\frac{\partial x^{i}}{\partial
    x^{'j}}\;A_{i}.\\
\end{array}%
\right.
\end{eqnarray}

\subsection{Définition d'un tenseur}
On appelle composante «$p$ fois contravariante» et «$q$ fois covariante» d'un tenseur $A$
mixte d'ordre $p+q$, toute quantité: $A_{\ell_{1}...\ell_{q}}^{k_{1}...k_{p}}$ se
transformant comme le produit de $p$ composantes contravariantes et $q$ composantes
covariantes d'un vecteur, lors d'un changement de coordonnées:
$\{x^{i}\}\rightarrow\{x^{'i}=x^{'i}(x^{j})\}$. La transformation se fait conformément à la loi
\begin{equation}\label{tranformtanseur}
    A_{j_{1}...j_{q}}^{'i_{1}...i_{p}}=\displaystyle\sum_{k_{1}=1}^{n}...
    \displaystyle\sum_{k_{p}=1}^{n}\displaystyle\sum_{\ell_{1}=1}^{n}
    ...\displaystyle\sum_{\ell_{p}=1}^{n}\;\displaystyle\frac{\partial x^{'i_{1}}}
    {\partial x^{k_{1}}}...\displaystyle\frac{\partial x^{'i_{p}}}{\partial
    x^{k_{p}}}\;
    \displaystyle\frac{\partial x^{\ell_{1}}}{\partial x^{'j_{1}}}
    ...\displaystyle\frac{\partial x^{\ell_{q}}}{\partial
    x^{'j_{q}}}\;A_{\ell_{1}...\ell_{q}}^{k_{1}...k_{p}},
\end{equation}
alors que la transformation inverse est donnée plutôt par l'expression
\begin{equation}\label{invtranformtanseur}
    A_{j_{1}...j_{q}}^{i_{1}...i_{p}}=\displaystyle\sum_{k_{1}=1}^{n}...
    \displaystyle\sum_{k_{p}=1}^{n}\displaystyle\sum_{\ell_{1}=1}^{n}
    ...\displaystyle\sum_{\ell_{p}=1}^{n}\;\displaystyle\frac{\partial x^{i_{1}}}
    {\partial x^{'k_{1}}}...\displaystyle\frac{\partial x^{i_{p}}}{\partial
    x^{'k_{p}}}\;
    \displaystyle\frac{\partial x^{'\ell_{1}}}{\partial x^{j_{1}}}
    ...\displaystyle\frac{\partial x^{'\ell_{q}}}{\partial
    x^{j_{q}}}\;A_{\ell_{1}...\ell_{q}}^{'k_{1}...k_{p}}.
\end{equation}
L'ensemble des tenseurs $p$ fois contravariants et $q$ fois covariants est noté $\textbf{T}^{p}_{q}$.

Pour alléger les écritures, la convention d'Einstein sera adopté dans tout ce qui suit; elle consiste à faire une somme de $1$ à $n$ sur chaque indice répété. Avec cette convention les formules précédentes s'écrivent sous la forme
\begin{eqnarray}\label{tranformtanseur sommation}
    A_{j_{1}...j_{q}}^{'i_{1}...i_{p}}=\displaystyle\frac{\partial x^{'i_{1}}}
    {\partial x^{k_{1}}}...\displaystyle\frac{\partial x^{'i_{p}}}{\partial
    x^{k_{p}}}\;
    \displaystyle\frac{\partial x^{\ell_{1}}}{\partial x^{'j_{1}}}
    ...\displaystyle\frac{\partial x^{\ell_{q}}}{\partial
    x^{'j_{q}}}\;A_{\ell_{1}...\ell_{q}}^{k_{1}...k_{p}},\\\nonumber\\
    A_{j_{1}...j_{q}}^{i_{1}...i_{p}}=\displaystyle\frac{\partial x^{i_{1}}}
    {\partial x^{'k_{1}}}...\displaystyle\frac{\partial x^{i_{p}}}{\partial
    x^{'k_{p}}}\;
    \displaystyle\frac{\partial x^{'\ell_{1}}}{\partial x^{j_{1}}}
    ...\displaystyle\frac{\partial x^{'\ell_{q}}}{\partial
    x^{j_{q}}}\;A_{\ell_{1}...\ell_{q}}^{'k_{1}...k_{p}}.\label{invtranformtanseur sommation}
\end{eqnarray}

\subsection{Tenseur métrique}
Considérons un système orthonormé cartésien $\{x^{(0)i}\}$ et un
système curviligne $\{x^{i}\}$. Soit $A^{(0)i}$ et $B^{(0)i}$ les composantes
contravariantes respectives dans $\{x^{(0)i}\}$ de deux vecteurs $\overrightarrow{A}$
et $\overrightarrow{B}$, et $A^{i}$ et $B^{i}$ leurs composantes contravariantes dans
$\{x^{i}\}$. Conformément à (\ref{aprimjai}), nous avons
\begin{eqnarray}\label{aokai}
    \left\{%
\begin{array}{ll}
    A^{(0)k}=\displaystyle\frac{\partial x^{(0)k}}{\partial x^{i}}
    \;A^{i},\\\nonumber\\
    B^{(0)k}=\displaystyle\frac{\partial x^{(0)k}}{\partial x^{i}}
    \;B^{i}.\\
\end{array}%
\right.
\end{eqnarray}
Calculons le produit scalaire
\begin{eqnarray}
    \overrightarrow{A}.\overrightarrow{B}&=&\left(
    A^{(0)k}\;\overrightarrow{e_{k}}^{0}\right).\left(
    A^{(0)\ell}\;\overrightarrow{e_{\ell}}^{0}\right)\nonumber\\
    &=&A^{(0)k}\;A^{(0)\ell}
    \underbrace{(\overrightarrow{e_{k}}^{0}.\;\overrightarrow{e_{\ell}}^{0})}_{\delta_{k\ell}}\nonumber\\
    &=&\delta_{k\ell}\;
    \left(\displaystyle\frac{\partial x^{(0)k}}{\partial
    x^{i}}\;A^{i}\right)\;\left(\displaystyle\frac{\partial x^{(0)k}}{\partial x^{j}}\;B^{j}\right)\nonumber\\
    &=&\left(\delta_{k\ell}\;\displaystyle\frac{\partial x^{(0)k}}{\partial
    x^{i}}\;\displaystyle\frac{\partial x^{(0)k}}{\partial x^{j}}\right)\;A^{i}\,B^{j},
\end{eqnarray}
et posons
\begin{equation}\label{gijdelta}
    g_{ij}=\delta_{k\ell}\;\displaystyle\frac{\partial x^{(0)k}}{\partial
    x^{i}}\;\displaystyle\frac{\partial x^{(0)k}}{\partial x^{j}}.
\end{equation}
Le produit scalaire s'écrit alors, d'une part,
\begin{equation}\label{produiscalair1}
   \overrightarrow{A}.\overrightarrow{B}=g_{ij}\;A^{i}\,B^{j},
\end{equation}
et il s'exprime, d'autre part, dans le système $\{x^{i}\}$ par l'expression
\begin{eqnarray}\label{produiscalair2}
   \overrightarrow{A}.\overrightarrow{B}&=&\left(A^{i}\;
   \overrightarrow{e_{i}}\right).\left(B^{j}\;
   \overrightarrow{e_{j}}\right)\nonumber\\
   \overrightarrow{A}.\overrightarrow{B}&=&
   (\overrightarrow{e_{i}}\,.\overrightarrow{e_{j}})A^{i}\;B^{j}.
\end{eqnarray}
La comparaison entre (\ref{produiscalair1}) et (\ref{produiscalair2}) conduit à poser
\begin{equation}\label{tensmetriqueiej}
    g_{ij}=\overrightarrow{e_{i}}\,.\overrightarrow{e_{j}}.
\end{equation}

D'après (\ref{gijdelta}) et (\ref{tensmetriqueiej}), il est clair que les $g_{ij}$
forment les composantes d'un tenseur symétrique d'ordre deux, dit: tenseur métrique.
Einstein a eu l'intuition de décrire le champ de gravitation par le tenseur métrique.

\subsection{Passage entre composantes covariantes et contravariantes} Soit un vecteur $\overrightarrow{A}\in E$. Il est possible de montrer que ses composantes covariantes données par (\ref{tensmetriqueiej}) peuvent s'exprimer en fonction des composantes contravariantes, et ce en utilisant (\ref{tensmetriqueiej}). En effet,
\begin{eqnarray}\label{passage cov vers contrav}
    A_{i}&=&\overrightarrow{A}.\overrightarrow{e_{i}}
         =\left(A^{j}\;\overrightarrow{e_{j}}\right).\;
         \overrightarrow{e_{i}}\nonumber\\
         &=&(\underbrace{\overrightarrow{e_{j}}.
         \overrightarrow{e_{i}}}_{g_{ji}})
         \;A^{j}
         =(\underbrace{\overrightarrow{e_{i}}.
         \overrightarrow{e_{j}}}_{g_{ii}})
         \;A^{j}\nonumber\\
    A_{i}&=&g_{ij}\;A^{j}.
\end{eqnarray}
Il est clair que le tenseur métrique permet d'élever l'indice covariant $i$ .\\

Soit $G$ la matrice dont les éléments sont les $g_{ij}$
$$G=(g_{ij}),$$
et notons $g^{ij}$ les éléments de la matrice inverse
$$G^{-1}=(g^{ij}),$$
telle que $$G\;G^{-1}=1,$$ ou encore
\begin{equation}
    g_{ik}\;g^{kj}=\delta_{i}^{j}.
\end{equation}
En calculant l'expression
\begin{eqnarray}
    g^{\ell\,i}\;A_{i}&=&g^{\ell\,i}\;
    \left(g_{ij}\;A^{j}\right)
    =\left(g^{\ell\,i}\;
    g_{ij}\right)\;A^{j}=\delta_{j}^{\ell}\;A^{j}
    =A^{\ell},\nonumber
\end{eqnarray}
nous déduisons finalement que
\begin{equation}
    A^{i}=g^{ij}\;A_{j}.
\end{equation}
Les éléments $g^{ij}$ du tenseur inverse permettent d'abaisser l'indice contravariant $i$.

D'une manière générale, pour élever ou abaisser les indices des composantes d'un
tenseur, on utilise autant de fois, respectivement, les $g_{ij}$ ou $g^{ij}$. En effet, nous avons d'une part
\begin{eqnarray}\label{passage cov vers contrav gle}
    A_{i_{1}...i_{p}}&=&g_{i_{1}j_{1}}\;
    A_{i_{2}...i_{p}}^{j_{1}}=
    g_{i_{1}j_{1}}\;g_{i_{2}j_{2}}\;A_{i_{3}...i_{p}}^{j_{1}j_{2}}=
    g_{i_{1}j_{1}}...g_{i_{p}j_{p}}\;A^{j_{1}...j_{p}},
\end{eqnarray}
et d'autre part
\begin{eqnarray}
    A^{i_{1}...i_{p}}&=&g^{i_{1}j_{1}}\;
    A^{i_{2}...i_{p}}_{j_{1}}=
    g^{i_{1}j_{1}}\;g^{i_{2}j_{2}}\;A^{i_{3}...i_{p}}_{j_{1}j_{2}}=
    g^{i_{1}j_{1}}...g^{i_{p}j_{p}}\;A_{j_{1}...j_{p}}.
\end{eqnarray}
C'est aussi valable même si les indices sont mélangés, dans le cas d'un tenseur mixte
\begin{eqnarray}
    A^{i_{1}i_{2}...i_{p}}_{j_{1}j_{2}...j_{q}}&=&
    g^{i_{1}\ell_{1}}\; A^{i_{2}...i_{p}}_{\ell_{1}j_{1}j_{2}...j_{q}}.
\end{eqnarray}

\subsection{Les symboles de Christoffel}
\subsubsection{Définition}
 Considérons un point $M$ de l'espace et soit
$(M,\{\overrightarrow{e_{i}}\}_{i=1,\cdots,n})$ le repère naturel attaché à un tel point. En
un point infiniment voisin $M^{'}$, repéré par:
$\overrightarrow{OM}^{'}=\overrightarrow{OM}+d\,\overrightarrow{OM}$, le repère
naturel associé est par définition
$$(M^{'},\{\overrightarrow{e_{i}}^{'}\}_{i=1,\cdots,n})=(M^{'},\{\overrightarrow{e_{i}}+
d\overrightarrow{e_{i}}\}_{i=1,\cdots,n}),$$ tel que
\begin{equation}
     d\,\overrightarrow{e_{i}}=
     \frac{\partial \overrightarrow{e_{i}}}{\partial x^{k}}\;dx^{k}.
\end{equation}
Le développement de $\partial \overrightarrow{e_{i}}/\partial x^{k}$ dans la
base $\{\overrightarrow{e_{i}}\}_{i=1,\cdots,n}$ est donné par
\begin{equation}
    \displaystyle\frac{\partial \overrightarrow{e_{i}}}{\partial x^{k}}\equiv
    \partial_{k}\overrightarrow{e_{i}}=
     \Gamma_{i\,k}^{j}\;\overrightarrow{e_{j}},
\end{equation}
où les coefficients du développement $\Gamma_{i\,k}^{j}$ sont les symboles de Christoffel
de seconde espèce. Dans ce cas, la différentielle des vecteurs de base s'exprime sous forme
\begin{equation}\label{etoile filante}
     d\overrightarrow{e_{i}}=\Gamma_{i\,k}^{j}\;dx^{k}\;\overrightarrow{e_{j}}.
\end{equation}

Pour exprimer les symboles $\Gamma_{i\,k}^{j}$ en fonction de la métrique, calculons la différentielle des composantes du tenseur métrique. Nous avons
d'une part,
\begin{eqnarray}\label{dgij1}
    dg_{ij}&=&d(\overrightarrow{e_{i}}.\overrightarrow{e_{j}})\nonumber\\
    &=&d\overrightarrow{e_{i}}.\overrightarrow{e_{j}}+\overrightarrow{e_{i}}.
    d\overrightarrow{e_{j}}\nonumber\\
    &=&\Gamma_{i\,k}^{\ell}\;dx^{k}\;\overrightarrow{e_{\ell}}
     .\overrightarrow{e_{j}}+\overrightarrow{e_{i}}.\Gamma_{j\,k}^{\ell}\;dx^{k}\;\overrightarrow{e_{\ell}}
     \nonumber\\
     &=&\Big(\Gamma_{i\,k}^{\ell}\;g_{\ell\,j}+
     \Gamma_{j\,k}^{\ell}\;g_{i\,\ell}\Big)\;dx^{k}.
\end{eqnarray}
et d'autre part,
\begin{equation}\label{dgij2}
    dg_{ij}=\displaystyle\frac{\partial g_{ij}}{\partial
    x^{k}}\;dx^{k}=\partial_{k}g_{ij}\;dx^{k}.
\end{equation}
D'après (\ref{dgij1}) et (\ref{dgij2}), nous déduisons l'expression des dérivées partielles des composantes du tenseur métrique
\begin{equation}\label{dkgij}
    \partial_{k}g_{ij}=\Big(\Gamma_{i\,k}^{\ell}\;g_{\ell\,j}
    +\Gamma_{j\,k}^{\ell}\;g_{i\,\ell}\Big).
\end{equation}
En introduisant les symboles de Christoffel de première espèce, définis par
\begin{equation}\label{christoffel1}
    \Gamma_{i\,k\,,j}\equiv g_{mj}\;\Gamma_{i\,k}^{m},
\end{equation}
alors (\ref{dkgij}) se réécrit da la façon suivante
\begin{equation}\label{dkgij bis}
    \partial_{k}g_{ij}=\Gamma_{i\,k\,,j}+
     \Gamma_{j\,k\,,i}.
\end{equation}

En utilisant la propriété $\Gamma_{i\,k\,,j}=\Gamma_{k\,i\,,j}$,
il est possible de montrer que
\begin{equation}\label{importantegij}
   \Gamma_{i\,j}^{k}=\frac{1}{2}\,g^{km}
   \Big(\partial_{i}g_{mj}+\partial_{j}g_{im}-
   \partial_{m}g_{ij}\Big).
\end{equation}
\subsubsection{Une autre Définition}
Une autre façon de définir les symboles de Christoffel consiste à passer du système de coordonnées cartésien $\{x^{(0)i}\}$ au
système curviligne $\{x^{i}\}$. Alors que la transformation des composantes du tenseur métrique $g_{ij}$ se fait conformément à (\ref{gijdelta}), les éléments du tenseur inverse se transforment plutôt comme suit

\begin{equation}\label{gij inverse delta}
    g^{ij}=\delta^{k\ell}\;\displaystyle\frac{\partial
    x^{i}}{\partial x^{(0)k}}\;\displaystyle\frac{\partial x^{j}}{\partial x^{(0)k}}.
\end{equation}
En remplaçant les expressions (\ref{gijdelta}) et (\ref{gij inverse delta}) dans (\ref{importantegij}), il est possible d'aboutir à une définition équivalente à (\ref{importantegij}) des symboles de Christoffel
\begin{equation}\label{nouvelle christoffel}
    \Gamma_{i\,j}^{k}=\displaystyle\frac{\partial x^{k}}{\partial x^{(0)\ell}}\;\displaystyle\frac{\partial^{\,2} x^{(0)\ell}}{\partial x^{i}\,\partial x^{j}}.
\end{equation}

\subsubsection{Les symboles de Christoffel ne sont pas les composantes d'un tenseur}
Lors du passage du système de coordonnées $\{x^{i}\}$ vers $\{x^{'\,i}\}$, les symboles de Christoffel se transforment, conformément à (\ref{nouvelle christoffel}), comme suit
\begin{eqnarray*}
    \Gamma_{i\,j}^{'\,k}&=&\displaystyle\frac{\partial x^{'\,k}}{\partial x^{(0)\ell}}\;\displaystyle\frac{\partial^{\,2} x^{(0)\ell}}{\partial x^{'\,i}\,\partial x^{'\,j}}\\
    &=&\displaystyle\frac{\partial x^{'\,k}}{\partial x^{m}}\displaystyle\frac{\partial x^{m}}{\partial x^{(0)\ell}}\;\displaystyle\frac{\partial}{\partial x^{'\,i}}\left(\displaystyle\frac{\partial x^{(0)\ell}}{\partial x^{'\,j}}\right)\\
    &=&\displaystyle\frac{\partial x^{'\,k}}{\partial x^{m}}\displaystyle\frac{\partial x^{m}}{\partial x^{(0)\ell}}\;\displaystyle\frac{\partial}{\partial x^{'\,i}}\left(\displaystyle\frac{\partial x^{(0)\ell}}{\partial x^{n}}\,\displaystyle\frac{\partial x^{n}}{\partial x^{'\,j}}\right)\\
    &=&\displaystyle\frac{\partial x^{'\,k}}{\partial x^{m}}\displaystyle\frac{\partial x^{m}}{\partial x^{(0)\ell}}\;\left[\displaystyle\frac{\partial}{\partial x^{'\,i}}\left(\displaystyle\frac{\partial x^{(0)\ell}}{\partial x^{n}}\right)\,\displaystyle\frac{\partial x^{n}}{\partial x^{'\,j}}+\displaystyle\frac{\partial x^{(0)\ell}}{\partial x^{n}}\,\displaystyle\frac{\partial}{\partial x^{'\,i}}\left(\displaystyle\frac{\partial x^{n}}{\partial x^{'\,j}}\right)\right]\\
    &=&\displaystyle\frac{\partial x^{'\,k}}{\partial x^{m}}\displaystyle\frac{\partial x^{m}}{\partial x^{(0)\ell}}\;\left[\displaystyle\frac{\partial}{\partial x^{p}}\left(\displaystyle\frac{\partial x^{(0)\ell}}{\partial x^{n}}\right)\displaystyle\frac{\partial x^{p}}{\partial x^{'\,i}}\,\displaystyle\frac{\partial x^{n}}{\partial x^{'\,j}}+\displaystyle\frac{\partial x^{(0)\ell}}{\partial x^{n}}\,\left(\displaystyle\frac{\partial^{\,2} x^{n}}{\partial x^{'\,i}\,\partial x^{'\,j}}\right)\right]\\
    &=&\displaystyle\frac{\partial x^{'\,k}}{\partial x^{m}}\displaystyle\frac{\partial x^{m}}{\partial x^{(0)\ell}}\;\left[\displaystyle\frac{\partial^{\,2} x^{(0)\ell}}{\partial x^{p}\,\partial x^{n}}\,\displaystyle\frac{\partial x^{p}}{\partial x^{'\,i}}\,\displaystyle\frac{\partial x^{n}}{\partial x^{'\,j}}+\displaystyle\frac{\partial x^{(0)\ell}}{\partial x^{n}}\,\left(\displaystyle\frac{\partial^{\,2} x^{n}}{\partial x^{'\,i}\,\partial x^{'\,j}}\right)\right]\\
    &=&\displaystyle\frac{\partial x^{'\,k}}{\partial x^{m}}\;\displaystyle\frac{\partial x^{p}}{\partial x^{'\,i}}\;\displaystyle\frac{\partial x^{n}}{\partial x^{'\,j}}\left(\displaystyle\frac{\partial x^{m}}{\partial x^{(0)\ell}}\;\displaystyle\frac{\partial^{\,2} x^{(0)\ell}}{\partial x^{p}\,\partial x^{n}}\right)
    +\displaystyle\frac{\partial x^{'\,k}}{\partial x^{m}}\underbrace{\displaystyle\frac{\partial x^{m}}{\partial x^{(0)\ell}}\;\displaystyle\frac{\partial x^{(0)\ell}}{\partial x^{n}}}_{\delta^{m}_{n}}\,\left(\displaystyle\frac{\partial^{\,2} x^{n}}{\partial x^{'\,i}\,\partial x^{'\,j}}\right).\\
\end{eqnarray*}
Nous aboutissons finalement à la loi de transformation
\begin{equation}\label{christoffel non tenseur}
    \Gamma_{i\,j}^{'\,k}=\displaystyle\frac{\partial x^{'\,k}}{\partial x^{m}}\;\displaystyle\frac{\partial x^{p}}{\partial x^{'\,i}}\;\displaystyle\frac{\partial x^{n}}{\partial x^{'\,j}}\;\Gamma_{pn}^{m}
    +\displaystyle\frac{\partial x^{'\,k}}{\partial x^{n}}\;\displaystyle\frac{\partial^{\,2} x^{n}}{\partial x^{'\,i}\,\partial x^{'\,j}}.
\end{equation}
La présence du deuxième terme dans (\ref{christoffel non tenseur}) fait que les symboles de Christoffel ne se transforment pas comme les composantes d'un tenseur.

\subsubsection{Annulation des $\Gamma^{i}_{kl}$ en un point arbitraire}
La transformation inverse des symboles de Christoffel, lors d'un changement de coordonnée arbitraire $x^{i}\rightarrow x^{'\,i}$ est donnée par
\begin{equation}\label{un 1}
\Gamma^{i}_{kl}=\frac{\partial x^{i}}{\partial x^{'m}}\;\frac{\partial x^{'n}}{\partial x^{k}}\;\frac{\partial x^{'p}}{\partial x^{l}}\;\Gamma^{'\,m}_{np}+\frac{\partial^{2} x^{'m}}{\partial x^{k} \partial x^{l}}\;\frac{\partial x^{i}}{\partial x^{'m}}.
\end{equation}
Soit un point arbitraire $O$ et soient $(\Gamma^{i}_{kl})_{0}$ les valeurs des $\Gamma^{i}_{kl}$ en ce point. Effectuons au voisinage de l'origine, la transformation de coordonnées suivante
\begin{equation}\label{deux 1}
x^{'i}=x^{i}+\frac{1}{2}\,(\Gamma^{i}_{kl})_{0}\,x^{k}\,x^{l}.
\end{equation}
Dans le but de déterminer l'expression des $(\Gamma^{i}_{kl})_{0}$ dans le nouveau système de coordonnées, procédons par les étapes suivantes:

Commençons, dans un premier temps, par la dérivation de (\ref{deux 1})
\begin{eqnarray}
  \frac{\partial x^{'i}}{\partial x^{m}} &=& \frac{\partial x^{i}}{\partial x^{m}}+\frac{1}{2}\,\frac{\partial (\Gamma^{i}_{kl})_{0}}{\partial x^{m}}\,x^{k}\,x^{l}+\frac{1}{2}\,(\Gamma^{i}_{kl})_{0}\left[\frac{\partial x^{k}}{\partial x^{m}}\,x^{l}+x^{k}\,\frac{\partial x^{l}}{\partial x^{m}}\right] \nonumber\\
  &=& \delta^{i}_{m}+\frac{1}{2}\,\frac{\partial (\Gamma^{i}_{kl})_{0}}{\partial x^{m}}\,x^{k}\,x^{l}+\frac{1}{2}\,(\Gamma^{i}_{kl})_{0}\left(\delta^{k}_{m}\,x^{l}+x^{k}\,\delta^{l}_{m}\right),\nonumber
\end{eqnarray}
tel que
\begin{eqnarray}
(\Gamma^{i}_{kl})_{0}\left(\delta^{k}_{m}\,x^{l}+x^{k}\,\delta^{l}_{m}\right)&=& (\Gamma^{i}_{ml})_{0}\,x^{l}+(\Gamma^{i}_{km})_{0}\,x^{k}\nonumber\\
&=&(\Gamma^{i}_{ml})_{0}\,x^{l}+(\Gamma^{i}_{lm})_{0}\,x^{l} \nonumber\\
&=&(\Gamma^{i}_{ml})_{0}\,x^{l}+(\Gamma^{i}_{ml})_{0}\,x^{l} \nonumber\\
&=& 2 (\Gamma^{i}_{ml})_{0}\,x^{l}.\nonumber
\end{eqnarray}
Dans ce cas, nous avons
\begin{equation}\label{trois 1}
  \frac{\partial x^{'i}}{\partial x^{m}}=\delta^{i}_{m}+\frac{1}{2}\,\frac{\partial (\Gamma^{i}_{kl})_{0}}{\partial x^{m}}\,x^{k}\,x^{l}+(\Gamma^{i}_{ml})_{0}\,x^{l}.
\end{equation}

Dans un deuxième temps, effectuons la dérivation de (\ref{trois 1})
\begin{eqnarray}
  \frac{\partial^{2} x^{'i}}{\partial x^{n} \partial x^{m}} &=& \frac{1}{2}\,\frac{\partial^{2} (\Gamma^{i}_{kl})_{0}}{\partial x^{n} \partial x^{m}}\,x^{k}\,x^{l}+\frac{\partial (\Gamma^{i}_{kl})_{0}}{\partial x^{m}}\,\left[\delta^{k}_{n}\,x^{l}+x^{k}\,\delta^{l}_{n}\right]+\frac{\partial (\Gamma^{i}_{ml})_{0}}{\partial x^{n}}\,x^{l}+(\Gamma^{i}_{ml})_{0}\,\delta^{l}_{n}\nonumber\\
  &=&\frac{1}{2}\,\frac{\partial^{2} (\Gamma^{i}_{kl})_{0}}{\partial x^{n} \partial x^{m}}\,x^{k}\,x^{l}+\left[\frac{\partial (\Gamma^{i}_{nl})_{0}}{\partial x^{m}}\,x^{l}+\frac{\partial (\Gamma^{i}_{kn})_{0}}{\partial x^{m}}\,x^{k}\right]+\frac{\partial (\Gamma^{i}_{ml})_{0}}{\partial x^{n}}\,x^{l}+(\Gamma^{i}_{mn})_{0} ,\nonumber\\
  &=&\frac{1}{2}\,\frac{\partial^{2} (\Gamma^{i}_{kl})_{0}}{\partial x^{n} \partial x^{m}}\,x^{k}\,x^{l}+2\,\frac{\partial (\Gamma^{i}_{nl})_{0}}{\partial x^{m}}\,x^{l}+\frac{\partial (\Gamma^{i}_{ml})_{0}}{\partial x^{n}}\,x^{l}+(\Gamma^{i}_{mn})_{0} ,\nonumber
\end{eqnarray}
qui s'écrit finalement sous la forme
\begin{equation}\label{quatre 1}
  \frac{\partial^{2} x^{'i}}{\partial x^{n} \partial x^{m}} = \frac{1}{2}\,\frac{\partial^{2} (\Gamma^{i}_{kl})_{0}}{\partial x^{n} \partial x^{m}}\,x^{k}\,x^{l}+\left[2\,\frac{\partial (\Gamma^{i}_{nl})_{0}}{\partial x^{m}}+\frac{\partial (\Gamma^{i}_{ml})_{0}}{\partial x^{n}}\right]\,x^{l}+(\Gamma^{i}_{mn})_{0}.
\end{equation}

La dernière étape consiste à contracter (\ref{quatre 1}) par $\partial x^{s}/\partial x^{'i}$
\begin{eqnarray}\label{cinq 1}
\frac{\partial^{2} x^{'i}}{\partial x^{n} \partial x^{m}}\,\frac{\partial x^{s}}{\partial x^{'i}} &=& \frac{1}{2}\,\frac{\partial^{2} (\Gamma^{i}_{kl})_{0}}{\partial x^{n} \partial x^{m}}\,\frac{\partial x^{s}}{\partial x^{'i}}\,x^{k}\,x^{l}+\left[2\,\frac{\partial (\Gamma^{i}_{nl})_{0}}{\partial x^{m}}+\frac{\partial (\Gamma^{i}_{ml})_{0}}{\partial x^{n}}\right]\,\frac{\partial x^{s}}{\partial x^{'i}}\,x^{l}+(\Gamma^{i}_{mn})_{0}\,\frac{\partial x^{s}}{\partial x^{'i}}.\nonumber\\
\end{eqnarray}
Au point arbitraire $O$, qu'on choisira comme origine des coordonnées ($\forall\; k,\hspace{0.2cm}(x^{k})_{O}=0$), la dérivée (\ref{trois 1}) se réduit au symbole de kronecker
\begin{eqnarray}\label{otrois 1}
  \left(\frac{\partial x^{'i}}{\partial x^{m}}\right)_{0}&=&\delta^{i}_{m}+\frac{1}{2}\,\frac{\partial (\Gamma^{i}_{kl})_{0}}{\partial x^{m}}\,(x^{k})_{0}\,(x^{l})_{0}+(\Gamma^{i}_{ml})_{0}\,(x^{l})_{0} = \delta^{i}_{m}.
\end{eqnarray}
Nous déduisons que la transformation de coordonnées (\ref{deux 1}) est inversible au point $O$ de telle sorte que le jacobien de la transformation inverse soit donné par
\begin{equation}\label{zsz}
    \left(\frac{\partial x^{m}}{\partial x^{'i}}\right)_{0}=\delta^{m}_{i}.
\end{equation}
Ainsi, à l'origine, l'expression (\ref{cinq 1})
\begin{eqnarray}
\left(\frac{\partial^{2} x^{'i}}{\partial x^{n} \partial x^{m}}\,\frac{\partial x^{s}}{\partial x^{'i}}\right)_{0} &=& (\Gamma^{i}_{mn})_{0}\,\left(\frac{\partial x^{s}}{\partial x^{'i}}\right)_{0} = (\Gamma^{i}_{mn})_{0}\,\delta^{s}_{i},
\end{eqnarray}
se réduit finalement à
\begin{equation}\label{sept 1}
    \left(\frac{\partial^{2} x^{'i}}{\partial x^{n} \partial x^{m}}\,\frac{\partial x^{s}}{\partial x^{'i}}\right)_{0} = (\Gamma^{s}_{mn})_{0}.
\end{equation}

En vertu de (\ref{sept 1}), (\ref{zsz}) et (\ref{otrois 1}), la transformation (\ref{un 1}), exprimée à l'origine des coordonnées, se met sous la forme
\begin{eqnarray}\label{huit 1}
\left(\Gamma^{i}_{kl}\right)_{0}&=&\delta^{i}_{m}\;\delta^{n}_{k}\;\delta^{p}_{l}\;\left(\Gamma^{'\,m}_{np}\right)_{0}
+\left(\frac{\partial^{2} x^{'m}}{\partial x^{k} \partial x^{l}}\;\frac{\partial x^{i}}{\partial x^{'m}}\right)_{0} = \left(\Gamma^{'i}_{kl}\right)_{0}+\left(\Gamma^{i}_{kl}\right)_{0},
\end{eqnarray}
et conduit au résultat final suivant
\begin{equation}\label{neuf 1}
\left(\Gamma^{'i}_{kl}\right)_{0}=0.
\end{equation}

Dans le nouveau système de coordonnées $\{x^{'\,i}\}$, obtenu par la transformation (\ref{deux 1}), tous les symboles de Christoffel sont nuls au point arbitraire $O$, pris comme origine des coordonnées; un tel système de coordonnées est dit localement géodésique.

\subsection{Dérivation covariante}
\subsubsection{Motivation} La transformation inverse des composantes contravariantes d'un vecteur $\overrightarrow{A}$ entre deux bases $\{\overrightarrow{e_{i}}\}\rightarrow\{\overrightarrow{e_{i}}^{'}\}$ est donnée par l'expression
\begin{equation}\label{aiaprimj cov}
       A^{i}=\displaystyle\frac{\partial x^{i}}{\partial
    x^{'k}}\;A^{'k}=\alpha^{i}_{k}\;A^{'k}.
\end{equation}
Dans le cas d'une base fixe $\{\overrightarrow{e_{i}}\}$,  les $dA^{i}$ sont les composantes contravariantes de $d\overrightarrow{A}$
\begin{eqnarray*}
  d\overrightarrow{A} = d\left(A^{i}\,\overrightarrow{e_{i}}\right) = dA^{i}\,\overrightarrow{e_{i}}+A^{i}\,\underbrace{d\overrightarrow{e_{i}}}_{\overrightarrow{0}} = dA^{i}\,\overrightarrow{e_{i}}.
\end{eqnarray*}
Dans ce cas les $\alpha^{i}_{k}$ sont des constantes\footnote{transformation entre deux bases fixes} et la différentielle ordinaire de $A^{i}$ se comporte ainsi comme un tenseur
\begin{equation}\label{aiaprimj deriv fixe}
       dA^{i}=\alpha^{i}_{k}\;dA^{'k}.
\end{equation}

Par contre dans le cas où les $\alpha^{i}_{k}$ ne sont plus des constantes\footnote{transformation entre deux bases locales}, la différentielle ordinaire de $A^{i}$ ne se comporte comme plus comme un tenseur
\begin{equation}\label{aiaprimj deriv local}
       dA^{i}=\alpha^{i}_{k}\;dA^{'k}+d\alpha^{i}_{k}\;A^{'k}=\displaystyle\frac{\partial x^{i}}{\partial
    x^{'k}}\;dA^{'k}+\displaystyle\frac{\partial^{\,2}x^{i}}{\partial x^{p}\,\partial x^{k}}\,dx^{'\,p}\;A^{'k},
\end{equation}
à cause de la présence du deuxième terme.

Dans le cas d'une base locale $\{\overrightarrow{e_{i}}\}$,  les $dA^{i}$ ne sont plus les composantes contravariantes de $d\overrightarrow{A}$
\begin{eqnarray*}
  d\overrightarrow{A} = d\left(A^{i}\,\overrightarrow{e_{i}}\right) = dA^{i}\,\overrightarrow{e_{i}}+A^{i}\,d\overrightarrow{e_{i}}
\end{eqnarray*}

Le but de ce qui suit est de déterminer, dans le cas d'une base locale, les "nouvelles" composantes contravairaintes $DA^{i}$ de $\overrightarrow{A}$
\begin{eqnarray}
   d\overrightarrow{A} &=& DA^{i}\;\overrightarrow{e_{i}},
\end{eqnarray}
qui se comportent comme les composantes d'un tenseur
\begin{equation}
    DA^{i} = \alpha^{i}_{k}\;DA^{'\,k}.
\end{equation}

\subsubsection{Tenseur dérivée covariante d'un vecteur $A^{i}$} Soit un vecteur
$\overrightarrow{A}$ qui se développe dans la base $\{\overrightarrow{e_{i}}\}_{i=1,\cdots,n}$,
$$\overrightarrow{A}=A^{i}\;\overrightarrow{e_{i}}.$$
Calculons la différentielle absolue d'un tel vecteur
\begin{eqnarray}
    d\overrightarrow{A}&=&dA^{i}\;\overrightarrow{e_{i}}+
       A^{i}\;d\overrightarrow{e_{i}}\nonumber\\
       &=&dA^{i}\;\overrightarrow{e_{i}}+
       A^{j}\;d\overrightarrow{e_{j}}\nonumber\\
       &=&dA^{i}\;\overrightarrow{e_{i}}+A^{j}\left(\Gamma_{j\,k}^{i}\;dx^{k}
       \;\overrightarrow{e_{i}}\right)\nonumber\\
       &=&\left(dA^{i}+A^{j}\;\Gamma_{j\,k}^{i}\;dx^{k}\right)
       \overrightarrow{e_{i}},
\end{eqnarray}
et définissons les composantes contravariantes de $d\overrightarrow{A}$, par: $$DA^{i}= dA^{i}+A^{j}\;\Gamma_{j\,k}^{i}\;dx^{k},$$
de telle sorte que
\begin{equation}\label{da}
    d\overrightarrow{A}=DA^{i}\;\overrightarrow{e_{i}}.
\end{equation}
En réécrivant $DA^{i}$
\begin{equation}\label{daicontrav}
   DA^{i}=\partial_{k}A^{i}\;dx^{k}+A^{j}\;\Gamma_{j\,k}^{i}\;dx^{k},
\end{equation}
tout en définissant
\begin{equation}\label{dkaicov}
    D_{k}A^{i}=\partial_{k}A^{i}+A^{j}\;\Gamma_{j\,k}^{i},
\end{equation}
il vient que
$$DA^{i}=D_{k}A^{i}\;dx^{k}.$$
Il est possible de montrer que les $D_{k}A^{i}$ sont les composantes d'un tenseur
d'ordre 2, appelées: Dérivée covariante de $\overrightarrow{A}$.

\subsubsection{Tenseur dérivée covariante d'un vecteur $A_{i}$} Dans le but de déterminer le tenseur dérivée covariante d'un vecteur $A_{i}$, on aura recours à un
artifice de calcul qui consiste à considérer un champ $\overrightarrow{B}$ uniforme,
tel que $d\overrightarrow{B}=\overrightarrow{0}$. D'après (\ref{da}), la différentielle d'un tel vecteur est donnée par
\begin{eqnarray}
     d\overrightarrow{B}&=&DB^{i}\;\overrightarrow{e_{i}}=
     \overrightarrow{0}\nonumber\\
     &\Rightarrow&\hspace{1cm}DB^{i}=0\hspace{0.5cm}\textrm{pour:}\;\;i=1,\cdots,
     n\nonumber\\
     &\Rightarrow&\hspace{1cm}dB^{i}+B^{j}\;\Gamma_{j\,k}^{i}\;dx^{k}=0\hspace{0.5cm}
       \textrm{pour:}\;\;i=1,\cdots,n\nonumber\\
       &\Rightarrow&\hspace{1cm}dB^{i}=-B^{j}\;\Gamma_{j\,k}^{i}\;dx^{k}\hspace{0.5cm}
       \textrm{pour:}\;\;i=1,\cdots,n
\end{eqnarray}

Soit un vecteur quelconque $\overrightarrow{A}$. Le produit scalaire de $\overrightarrow{B}$ par la différentielle de $\overrightarrow{A}$  se calcule d'une part comme suit
\begin{eqnarray}\label{daicov1}
    \overrightarrow{B}\,.\,d\overrightarrow{A}&=&d(\overrightarrow{B}.\overrightarrow{A})
    -\underbrace{d\overrightarrow{B}}_{\overrightarrow{0}}.\overrightarrow{A}\nonumber\\
    &=&d\left[\left(B^{i}\;\overrightarrow{e_{i}}\right).
       \left(A^{j}\;\overrightarrow{e_{j}}\right)\right]\nonumber\\
       &=&d\left(g_{ij}B^{i}\;A^{j}\right)
       =d\left(B^{i}\;A_{i}\right)
       \nonumber\\
       &=&dA_{i}\;B^{i}+
       A_{i}\;dB^{i}
       =dA_{i}\;B^{i}+
       A_{j}\;dB^{j}\nonumber\\
       &=&dA_{i}\;B^{i}+
       A_{j}\;\Big(-B^{i}\;\Gamma_{i\,k}^{j}\;dx^{k}\Big)\nonumber\\
       &=&B^{i}\big(dA_{i}-\Gamma_{i\,k}^{j}\;dx^{k}\;
       A_{j}\big),
\end{eqnarray}
et d'autre part
\begin{eqnarray}\label{daicov2}
   \overrightarrow{B}\,.\,d\overrightarrow{A}&=&\left(B^{i}\;
   \overrightarrow{e_{i}}\right).\left(DA^{j}\;\overrightarrow{e_{j}}
   \right)\nonumber\\
   &=&B^{i}\;DA^{j}\;g_{ij}
   =B^{i}\left(D_{k}A^{j}\;dx^{k}\right)g_{ij}\nonumber\\
   &=&B^{i}\left(D_{k}A^{j}\;g_{ij}\right)dx^{k}
   =B^{i}\left(
   D_{k}A_{i}\right)dx^{k}\nonumber\\
   &=&B^{i}\left(D_{k}A_{i}\;dx^{k}\right)\nonumber\\
   &=&B^{i}\;DA_{i},
\end{eqnarray}
où $DA_{i}=D_{k}A_{i}\;dx^{k}.$
En comparant les équations (\ref{daicov1}) et (\ref{daicov2}), on déduit l'expression
\begin{equation}\label{davor}
    DA_{i}=dA_{i}-\Gamma_{i\,k}^{j}\;dx^{k}\;
       A_{j},
\end{equation}
qui se met encore sous la forme
\begin{equation}\label{daicovx y}
    DA_{i}=\left(\partial_{k}A_{i}-
       \Gamma_{i\,k}^{j}\;
       A_{j}\right)dx^{k}.
\end{equation}
Nous déduisons finalement l'expression des composantes
\begin{equation}\label{daicovx}
  D_{k}A_{i}=\partial_{k}A_{i}-\Gamma_{i\,k}^{j}\;A_{j}.
\end{equation}
Il est possible de montrer que les $D_{k}A_{i}$ sont les composantes d'un tenseur.

En coordonnées galiléennes\footnote{Dans le cas où $g_{\mu\nu}=\eta_{\mu\nu}$, il est clair que d'après (\ref{importantegij}) les symboles de Christoffel sont nuls $\Gamma_{i\,k}^{j}=0$} les dérivées covariantes (\ref{dkaicov}) et (\ref{daicovx}) co\"{\i}ncident avec les dérivées ordinaires. De plus, la dérivée covariante d'un scalaire $\phi$ est, par définition
\begin{equation}\label{deriv cov scalaire}
    D_{k}\phi=\partial_{k}\phi,
\end{equation}
égale à la dérivée ordinaire.

\subsubsection{Tenseur dérivée covariante d'un Tenseur}
Considérons un tenseur d'ordre deux $\widetilde{T}$ de composantes contravariantes $T^{ij}$. Un tel tenseur se développe de façon unique dans la base $E_{ij}=\overrightarrow{e_{i}}\otimes\overrightarrow{e_{j}}$ (produit tensoriel) $$\widetilde{T}=T^{ij}\,\overrightarrow{e_{i}}\otimes\overrightarrow{e_{j}}.$$
La différentielle de ce tenseur est donnée par l'expression
\begin{eqnarray*}
  d\widetilde{T} &=& dT^{ij}\,\overrightarrow{e_{i}}\otimes\overrightarrow{e_{j}}
  +T^{ij}\left(d\overrightarrow{e_{i}}\right)\otimes\overrightarrow{e_{j}}+T^{ij}\,\overrightarrow{e_{i}}\otimes\left(d\overrightarrow{e_{j}}\right).
\end{eqnarray*}
Pour déterminer la dérivée covariante $DT^{ij}=D_{k}T^{ij}\,dx^{k}$ du tenseur, de telle sorte à avoir
\begin{equation}
    d\widetilde{T}=DT^{ij}\,\overrightarrow{e_{i}}\otimes\overrightarrow{e_{j}},
\end{equation}
utilisons (\ref{etoile filante}), avec une redéfinition adéquate des indices muets,
\begin{eqnarray*}
  d\widetilde{T} &=& \partial_{k}T^{ij}\,dx^{k}\,\overrightarrow{e_{i}}\otimes\overrightarrow{e_{j}}
  +T^{ij}\left(\Gamma_{i\,k}^{\ell}\;dx^{k}\;\overrightarrow{e_{\ell}}\right)\otimes\overrightarrow{e_{j}}
  +T^{ij}\,\overrightarrow{e_{i}}\otimes\left(\Gamma_{j\,k}^{\ell}\;dx^{k}\;\overrightarrow{e_{\ell}}\right) \\
  &=&\left(\partial_{k}T^{ij}+T^{\ell j}\,\Gamma^{i}_{\ell k}+T^{i \ell}\,\Gamma^{j}_{\ell k}\right)dx^{k}\,\overrightarrow{e_{i}}\otimes\overrightarrow{e_{j}}.
\end{eqnarray*}
Nous aboutissons finalement à l'expression
\begin{equation}\label{deriv cov tij}
    D_{k}T^{ij}=\partial_{k}T^{ij}+T^{\ell j}\,\Gamma^{i}_{\ell k}+T^{i \ell}\,\Gamma^{j}_{\ell k}
\end{equation}
du tenseur dérivée covariante d'un tenseur contravariant d'ordre 2.

Il est ainsi possible de généraliser l'expression de la dérivée covaiante pour un tenseur quelconque et ce en appliquant respectivement (\ref{dkaicov}) et (\ref{daicovx}) à chaque indice contravariant et covariant
\begin{eqnarray}\label{deriv cov tenseur quelconque}
    D_{k}T^{i_{1}i_{2}\cdots i_{s}}_{j_{1}j_{2}\cdots j_{p}}&=&\partial_{k}T^{i_{1}i_{2}\cdots i_{s}}_{j_{1}j_{2}\cdots j_{p}}+\left(T^{\ell\, i_{2}\cdots i_{s}}_{j_{1}j_{2}\cdots j_{p}}\,\Gamma^{i_{1}}_{\ell k}+T^{i_{1}\ell\cdots i_{s}}_{j_{1}j_{2}\cdots j_{p}}\,\Gamma^{i_{2}}_{\ell k}+\cdots+T^{i_{1}i_{2}\cdots \ell}_{j_{1}j_{2}\cdots j_{p}}\,\Gamma^{i_{s}}_{\ell k}\right)\nonumber\\
    &&\hspace{1.4cm}-\left(T_{\ell\, j_{2}\cdots j_{p}}^{i_{1}i_{2}\cdots i_{s}}\,\Gamma^{\ell}_{j_{1}\, k}+T_{j_{1}\ell\cdots j_{p}}^{i_{1}i_{2}\cdots i_{s}}\,\Gamma^{\ell}_{j_{2}\, k}+\cdots+T_{j_{1}j_{2}\cdots\ell}^{i_{1}i_{2}\cdots i_{s}}\,\Gamma^{\ell}_{j_{p}\, k}\right).
\end{eqnarray}

Il est possible de montrer, à partir de la définition (\ref{deriv cov tenseur quelconque}), que la dérivée covariante du produit de deux tenseurs\footnote{Le produit tensoriel de $A\in \textbf{T}^{p}_{q}$ et $B\in \textbf{T}^{r}_{s}$ est un tenseur $C=A\otimes B\in \textbf{T}^{p+r}_{q+s}$ de composantes qui vérifient $C^{i_{1}i_{2}\cdots i_{p+r}}_{j_{1}j_{2}\cdots j_{q+s}}=A^{i_{1}i_{2}\cdots i_{p}}_{j_{1}j_{2}\cdots j_{q}}\;B^{i_{p+1}i_{2}\cdots i_{p+r}}_{j_{q+1}j_{2}\cdots j_{q+s}}$} $A$ et $B$ obéit à la même règle de dérivation ordinaire
\begin{equation}\label{deriv cov produit tensor}
    D_{k}\left(A\otimes B\right)=\left(D_{k}A\right)\otimes B+A\otimes\left(D_{k}B\right),
\end{equation}
notée aussi
\begin{equation}\label{deriv cov produit tensor bis}
    D_{k}\left(AB\right)=\left(D_{k}A\right)B+A\left(D_{k}B\right),
\end{equation}
pour alleger les écritures.

\subsubsection{Théorème de Ricci}
La dérivée covariante du tenseur métrique est nulle. En effet, en utilisant (\ref{deriv cov tenseur quelconque}), (\ref{christoffel1}) et (\ref{dkgij bis}), nous avons
\begin{eqnarray}\label{ricci th cov}
  D_{k}g_{ij} &=& \partial_{k}g_{ij}-\Gamma_{jk}^{\ell}\,g_{i\ell}-\Gamma_{ik}^{\ell}\,g_{j\ell} = \partial_{k}g_{ij}-\Gamma_{jk,i}-\Gamma_{ik,j} = 0,
\end{eqnarray}
autrement dit
\begin{equation}\label{ricci th cov bis}
    Dg_{ij}\equiv D_{k}g_{ij}\,dx^{k}=0.
\end{equation}
De la même manière, il est aussi possible de montrer que
\begin{eqnarray}\label{ricci th contrav}
  D_{k}g^{ij} &=&  0,
\end{eqnarray}
et que
\begin{equation}\label{ricci th contrav bis}
    Dg^{ij}\equiv D_{k}g^{ij}\,dx^{k}=0.
\end{equation}

La dérivée covariante de la composantes covariante $A_{i}$ d'un vecteur, compte tenu de (\ref{passage cov vers contrav}) et (\ref{deriv cov produit tensor bis}), est donnée par l'expression
\begin{eqnarray}
  DA_{i} &=& D\left(g_{ij}\,A^{j}\right)=\left(Dg_{ij}\right)A^{j}+g_{ij}\left(DA^{j}\right).
\end{eqnarray}
En vertu du théorème de Ricci, le premier terme est nul de sorte à avoir finalement
\begin{eqnarray}
  DA_{i} &=& g_{ij}\left(DA^{j}\right).
\end{eqnarray}
Ce résultat est bien compatible avec la règle de passage (\ref{passage cov vers contrav gle}), des composantes contravariantes vers les composantes covariantes d'un tenseur. Bien sûr, il est possible de généraliser ce résultat à un tenseur quelconque $B\in \textbf{T}^{p}_{q}$
\begin{eqnarray}\label{label}
  D\left(g_{ki_{1}}\,B_{j_{1}j_{2}\cdots j_{p}}^{i_{1}i_{2}\cdots i_{p}}\right) = g_{ki_{1}}\left(DB_{j_{1}j_{2}\cdots j_{p}}^{i_{1}i_{2}\cdots i_{p}}\right).
\end{eqnarray}

\subsubsection{Transport parallèle}
Considérons deux points infiniment voisins $P(x^{i})$ et $Q(x^{i}+dx^{i})$ et un champ de vecteur qui prend les valeurs $A^{k}$ au point $P$ et $A^{k}+dA^{k}$ au point $Q$. Dans un système de coordonnées curvilignes $dA^{k}=A^{k}(x^{i}+dx^{i})-A^{k}(x^{i})$ n'est pas un vecteur, car la soustraction des deux vecteurs est donnée en deux points différents avec des transformations de coordonnées différentes. Pour pouvoir calculer la dérivée covariante entre deux vecteurs, définis à des points différents, il faut les ramener au même point. Pour ce faire, transportons $A^{k}$ de façon parallèle à lui-même du point $P$ au point $Q$ de telle sorte que le vecteur au point final soit $A^{k}+\delta A^{k}$, avec $\delta A^{k}$ représente son accroissement après transport \cite{Nikodem}. Dans ce cas, la différence entre les deux vecteurs, se trouvant désormais au même point, est
\begin{equation}\label{accrroissement deriv cov}
    DA^{k}=dA^{k}-\delta A^{k}.
\end{equation}
Une identification entre (\ref{accrroissement deriv cov}) et (\ref{daicontrav}) permet d'avoir
\begin{equation}\label{accroissement vect contrav}
    \delta A^{k}=-A^{j}\;\Gamma_{j\,i}^{k}\;dx^{i}.
\end{equation}
Remarquons que l'accroissement  $\delta A^{k}$ des composantes du vecteur, dans un transport parallèle infiniment petit possède les caractéristiques suivantes:
\begin{itemize}
  \item Il dépend des composantes même du vecteur et cette dépendance est linéaire pour garantir à la somme de deux vecteurs de se transformer selon la même loi pour chacun des vecteurs. De plus, dans le cas d'un vecteur nul $A^{k}=0$, l'accroissement  $\delta A^{k}=0$.
  \item La proportionnalité en $dx^{i}$ fait que si on est en même point, $dx^{i}=0$, alors les composantes du vecteur ne subissent aucun accroissement.
  \item Les "coefficients" de proportionnalité $\Gamma_{j\,i}^{k}$ sont des fonctions des coordonnées qui doivent s'annuler dans un système de coordonnées galiléen, de telle sorte que la dérivée covariante (\ref{accrroissement deriv cov}) se réduise, dans ce cas, à la différentielle.
\end{itemize}

Le produit scalaire de deux vecteurs\footnote{Le produit scalaire est invariant lors d'une transformation de coordonnées inversible $A^{'}_{k}B^{'\,k}=\Big(A_{i}\;\frac{\partial x^{i}}{\partial
    x^{'k}}\Big)\Big(B^{j}\;\frac{\partial x^{'k}}{\partial
    x^{j}}\Big)=A_{i}\,B^{j}\delta^{i}_{j}=A_{i}\,B^{i}$}, comme n'importe quel scalaire, est invariant lors d'un transport parallèle. Ainsi, à partir de $\delta\left(A_{k}B^{k}\right)=0$, il vient que $$B^{k}\delta A_{k}=-A_{k}\delta B^{k}=A_{k}\,B^{j}\;\Gamma_{j\,i}^{k}\;dx^{i},$$
or, du fait que $B^{k}$ sont arbitraires, nous retrouvons l'expression des accroissements
\begin{equation}\label{accroissement vect cov}
    \delta A_{k}=\Gamma_{k\,i}^{j}\;A_{j}\;dx^{i},
\end{equation}
compatible avec (\ref{daicovx y}).

\subsection{Espace de Riemann}
Dans un espace euclidien, il est toujours possible de définir un système d'axes rectilignes et orthonormés $\{x^{(0)i}\}$ de façon que la distance entre deux points quelconques $(x^{(0)1},x^{(0)2},\cdots,x^{(0)n})$ et $(y^{(0)1},y^{(0)2},\cdots,y^{(0)n})$, est telle que
\begin{equation}\label{carre distance quelconque euclid}
\bigtriangleup s^{2}=\delta_{k\ell}\,\left(x^{(0)k}-y^{(0)k}\right)\left(x^{(0)\ell}-y^{(0)\ell}\right)=\left(x^{(0)1}-y^{(0)1}\right)^{2}
+\cdots+\left(x^{(0)n}-y^{(0)n}\right)^{2}.
\end{equation}
Pour des points infiniment voisins $(x^{(0)1},x^{(0)2},\cdots,x^{(0)n})$ et $(x^{(0)1}+dx^{(0)1},x^{(0)2}+dx^{(0)2},\cdots,x^{(0)n}+dx^{(0)n})$,  le carré de la distance infinitésimale est donné par
\begin{eqnarray}\label{carre distance infinit euclid}
  ds^{2} =\delta_{k\ell}\,dx^{(0)k}\,dx^{(0)\ell} = \left(dx^{(0)1}\right)^{2}+\left(dx^{(0)2}\right)^{2}+\cdots+\left(dx^{(0)n}\right)^{2}.
\end{eqnarray}

Lors d'un passage au système de coordonnées curvilignes $\{x^{i}\}$, l'expression précédente (\ref{carre distance infinit euclid}) devient ainsi
\begin{eqnarray*}
  ds^{2} &=& \delta_{k\ell}\left(\displaystyle\frac{\partial x^{(0)k}}{\partial x^{i}}\,dx^{i}\right)\left(\displaystyle\frac{\partial x^{(0)\ell}}{\partial x^{j}}\,dx^{j}\right)= \left(\delta_{k\ell}\,\displaystyle\frac{\partial x^{(0)k}}{\partial x^{i}}\,\displaystyle\frac{\partial x^{(0)\ell}}{\partial x^{j}}\right)dx^{i}\,dx^{j},
\end{eqnarray*}
et se met finalement sous la forme quadratique fondamentale
\begin{equation}\label{forme quadratique fondamentale}
    ds^{2}=g_{ij}\,dx^{i}\,dx^{j}.
\end{equation}

Contrairement à un espace euclidien, dans un espace de Riemann, il n'est jamais possible de définir un système d'axes rectilignes de telle sorte à diagonaliser le tenseur métrique $g_{ij}$ pour garantir à la forme quadratique fondamentale (\ref{forme quadratique fondamentale}) de se réduire à la forme (\ref{carre distance infinit euclid}) pour tous les points de l'espace\footnote{En fait, il est toujours possible de définir \underline{localement} un système d'axes rectilignes de telle sorte à se ramener à (\ref{carre distance infinit euclid}) mais pas globalement.}.

Pour munir un espace de Riemann de propriétés géométriques, le plus simple , est de l'identifier localement avec un espace euclidien de telle manière que l'algèbre tensorielle euclidienne toute entière s'étende, sans modification, aux vecteurs et tenseurs attachés à un même point de l'espace riemannien \cite{Boudenot}.

\subsection{Tenseur de Courbure}
Dans le cas général d'un espace de Riemann, le transport parallèle infinitésimal d'un vecteur est définit comme un déplacement tel que les composantes du vecteur ne varient pas dans le cas d'un système de coordonnées galiléen. Dans le cas d'un système de coordonnées curviligne, le transport parallèle d'un vecteur d'un point initial à un autre point final donne des résultats différents si le chemin suivi est différent.  Ainsi, les composantes de la dérivée covariante seconde d'un vecteur, calculées selon deux chemins différents, ne sont pas égales.

\subsubsection{Définition}
Soit un vecteur $\overrightarrow{A}$ de composantes covariantes $A_{i}$. Calculons la dérivée covariante seconde de $A_{i}$ en utilisant deux chemins différents. Pour le premier chemin, appliquons d'abord une dérivée covariante au tenseur $T_{ki}=D_{k}A_{i}$, en utilisant (\ref{deriv cov tenseur quelconque}),
 \begin{eqnarray*}
   D_{\ell}\left(D_{k}A_{i}\right) &=& \partial_{\ell}\left(D_{k}A_{i}\right)-\Gamma_{k\ell}^{p}\;\left(D_{p}A_{i}\right)
   -\Gamma_{i\ell}^{p}\;\left(D_{k}A_{p}\right).
   \end{eqnarray*}
   D'après (\ref{daicovx}), on aboutit au résultat suivant
   \begin{eqnarray}\label{double deriv cov chemin1}
   D_{\ell}\left(D_{k}A_{i}\right)&=& \partial_{\ell}\left(\partial_{k}A_{i}-\Gamma_{ik}^{p}\;A_{p}\right)-\Gamma_{k\ell}^{p}\left(\partial_{p}A_{i}-\Gamma_{ip}^{j}\;A_{j}\right)
   -\Gamma_{i\ell}^{p}\left(\partial_{k}A_{p}-\Gamma_{pk}^{j}\;A_{j}\right).
 \end{eqnarray}
 Pour le deuxième chemin, intervertissons l'ordre des dérivées covariantes
 \begin{eqnarray}\label{double deriv cov chemin2}
   D_{k}\left(D_{\ell}A_{i}\right)&=& \partial_{k}\left(\partial_{\ell}A_{i}-\Gamma_{i\ell}^{p}\;A_{p}\right)-\Gamma_{\ell k}^{p}\;\left(\partial_{p}A_{i}-\Gamma_{ip}^{j}\;A_{j}\right)
   -\Gamma_{ik}^{p}\;\left(\partial_{\ell}A_{p}-\Gamma_{p\ell}^{j}\;A_{j}\right).
 \end{eqnarray}
La soustraction membre à membre de (\ref{double deriv cov chemin1}) et (\ref{double deriv cov chemin2})
\begin{equation}
    D_{\ell}D_{k}A_{i}- D_{k}D_{\ell}A_{i}=\left(\partial_{k}\Gamma_{i\ell}^{j}-\partial_{\ell}\Gamma_{ik}^{j}+\Gamma_{i\ell}^{p}\,\Gamma_{pk}^{j}
    -\Gamma_{ik}^{p}\,\Gamma_{p\ell}^{j}\right)A_{j},
\end{equation}
montre clairement que le résultat dépend de l'ordre des dérivées covariantes, contrairement au cas des dérivées ordinaires. Comme la différence de deux tenseurs $D_{\ell}D_{k}A_{i}- D_{k}D_{\ell}A_{i}$ est un tenseur et que $A_{i}$ est un vecteur, on déduit que l'expression entre parenthèses
\begin{equation}\label{tenseur riemann christoffel}
    R^{\,j}_{\;\;ik\ell}=\partial_{k}\Gamma_{i\ell}^{j}-\partial_{\ell}\Gamma_{ik}^{j}+\Gamma_{i\ell}^{p}\,\Gamma_{pk}^{j}
    -\Gamma_{ik}^{p}\,\Gamma_{p\ell}^{j},
\end{equation}
est un tenseur; celui-ci est dit Tenseur de Courbure de Riemann-Christoffel.

Il en résulte qu'un vecteur transporté parallèlement le long d'un contour fermé ne co\"{\i}ncide plus avec le vecteur initial; c'est la manifestation de la courbure de l'espace de Riemann. Dans le cas où les composantes du tenseur métrique $g_{ij}$ sont partout stationnaires\footnote{Dans le cas par exemple de l'espace-temps plat de Minkowski} (i.e $\partial_{k}g_{ij}=0$), conformément à (\ref{importantegij}), les symboles de Christoffel sont tous identiquement nuls. Dans ce cas, toutes les composantes de tenseur de courbure (\ref{tenseur riemann christoffel}) sont nulles (on parle d'espace plat) et il en résulte que le transport parallèle d'un vecteur le long d'un contour fermé co\"{\i}ncide avec le vecteur initial.

\subsubsection{Propriétés}
Le tenseur de Courbure est antisymétrique sur les indices $k$ et $\ell$
\begin{equation}\label{symetrie tens courb}
    R^{\,j}_{\;\;ik\ell}=-R^{\,j}_{\;\;i\ell k},
\end{equation}
et vérifie la relation cyclique suivante
\begin{equation}\label{cyclicite tens courb}
    R^{\,j}_{\;\;ik\ell}+R^{\,j}_{\;\;\ell ik}+R^{\,j}_{\;\;k\ell i}=0.
\end{equation}

De plus, le tenseur covariant
\begin{equation}\label{tenseur covariant courbure}
     R_{ik\ell m}=g_{ip}R^{\,p}_{\;\;k\ell m},
\end{equation}
est antisymétrique sur les indices $ik$ et $\ell m$,
\begin{equation}
     R_{ik\ell m}=- R_{ki\ell m}=-R_{ikm\ell}=R_{kim\ell},
\end{equation}
et est symétrique par rapport à la permutation de ces couples,
\begin{equation}
    R_{ik\ell m}=R_{\ell mik}.
\end{equation}
Il vérifie aussi la condition cyclique suivante
\begin{equation}
    R_{ik\ell m}+R_{imk\ell}+R_{i\ell mk}=0.
\end{equation}

\subsection{Tenseur de Ricci}
Le tenseur de Ricci est définit par la contraction suivante
\begin{equation}\label{tenseur de Ricci chap rappel}
    R_{ij}=g^{kp}R_{kipj}=R^{\,k}_{\;\;ikj}
\end{equation}
du tenseur de courbure. Il est possible de montrer que le tenseur de Ricci est symétrique
\begin{equation}
    R_{ij}=R_{ji}.
\end{equation}

\subsection{Courbure scalaire}
La contraction du tenseur de Ricci permet d'obtenir un invariant $R$,
\begin{equation}\label{courbure scalaire chap rappel}
    R=g^{ij}\,R_{ij}=R^{i}_{\;\;i},
\end{equation}
appelé Courbure scalaire riemannienne.

\subsection{Identités de Bianchi}
Du fait du caractère intrinsèque des tenseurs, les relations tensorielles sont exprimées \\indépendamment du système de coordonnées. Ainsi, si une relation tensorielle est démontrée dans un système de coordonnées particulier, elle sera valable dans tous les référentiels.

En se plaçant dans un système de coordonnées localement géodésique, dans lequel tous les symboles de Christoffel sont annulés en un point arbitraire $M$, $$(\Gamma_{ij}^{k})_{M}=0,$$
les dérivées covariantes se réduisent dans ce cas aux dérivées ordinaires $D_{p}=\partial_{p}$ et le tenseur de Courbure se met sous la forme suivante
\begin{equation}\label{tenseur riemann christoffel sys localement geodes}
    R^{\,i}_{\;\;k\ell m}=\partial_{\ell}\Gamma_{km}^{j}-\partial_{m}\Gamma_{k\ell}^{i}.
\end{equation}
Nous avons dans ce cas, d'une part
\begin{equation}\label{bianchi one}
    D_{p}R^{\,i}_{\;\;k\ell m}=\partial_{p}R^{i}_{\;\;k\ell m}=\partial_{p}\partial_{\ell}\Gamma_{km}^{j}-\partial_{p}\partial_{m}\Gamma_{k\ell}^{i}.
\end{equation}
D'autre part, une permutation circulaire sur les indices $p$, $\ell$ et $m$ permet d'aboutir aux relations
\begin{eqnarray}
   D_{m}R^{\,i}_{\;\;kp\ell}&=&\partial_{m}\partial_{p}\Gamma_{k\ell}^{j}-\partial_{m}\partial_{\ell}\Gamma_{kp}^{i},\label{bianchi two} \\
   D_{\ell}R^{\,i}_{\;\;kmp}&=&\partial_{\ell}\partial_{m}\Gamma_{kp}^{j}-\partial_{\ell}\partial_{p}\Gamma_{km}^{i}.\label{bianchi three}
\end{eqnarray}
L'addition membre à membre des trois relations (\ref{bianchi one}), (\ref{bianchi two}) et (\ref{bianchi three}) permet d'aboutir à l'identité de Bianchi
\begin{equation}\label{identite de bianchi}
    D_{p}R^{\,i}_{\;\;k\ell m}+D_{m}R^{\,i}_{\;\;kp\ell}+D_{\ell}R^{\,i}_{\;\;kmp}=0,
\end{equation}
qui reste valable dans n'importe quel référentiel.

\subsection{Tenseur d'Einstein}
Une contraction particulière de l'identité de Bianchi (\ref{identite de bianchi}), en posant $i=p$, permet d'avoir la relation
\begin{equation}\label{identite de bianchi contracte}
    D_{i}R^{\,i}_{\;\;k\ell m}+D_{m}R^{\,i}_{\;\;ki\ell}+D_{\ell}R^{\,i}_{\;\;kmi}=0.
\end{equation}
En contractant les membres de (\ref{identite de bianchi contracte}) par $g^{kp}$ tout en exploitant "l'insensibilité" du tenseur métrique à la dérivée covariante (\ref{label}), conséquence d'application du Théorème de Ricci, il vient que
\begin{eqnarray}
  0 &=& D_{i}\left(g^{kp}\,R^{\,i}_{\;\;k\ell m}\right)+D_{m}\left(g^{kp}R^{\,i}_{\;\;ki\ell}\right)+D_{\ell}\left(g^{kp}R^{\,i}_{\;\;kmi}\right) \nonumber\\
  &=& D_{i}\left(g^{kp}\,R^{\,i}_{\;\;k\ell m}\right)+D_{m}\left(g^{kp}R_{k\ell}\right)-D_{\ell}\left(g^{kp}R^{\,i}_{\;\;kim}\right) \nonumber\\
  &=& D_{i}\left(g^{kp}\,R^{\,i}_{\;\;k\ell m}\right)+D_{m}\left(g^{kp}R_{k\ell}\right)-D_{\ell}\left(g^{kp}R_{km}\right),
\end{eqnarray}
compte tenu des propriétés du tenseur de Courbure et de la définition du tenseur de Ricci.

Dans le cas particulier où $p=\ell$, la dernière relation devient
\begin{eqnarray*}
 0 &=& D_{i}\left(g^{k\ell}\,R^{\,i}_{\;\;k\ell m}\right)+D_{m}\left(g^{k\ell}R^{\,i}_{\;\;ki\ell}\right)+D_{\ell}\left(g^{k\ell}R^{\,i}_{\;\;kmi}\right) \\
   &=& D_{i}\left[g^{k\ell}\left(g^{is}R_{\;\;sk\ell m}\right)\right]+D_{m}\left(g^{k\ell}R^{\,i}_{\;\;ki\ell}\right)+D_{\ell}\left(g^{k\ell}R^{\,i}_{\;\;kmi}\right)\\
   &=& D_{i}\left[g^{is}\left(-g^{k\ell}R_{\;\;ks\ell m}\right)\right]+D_{m}\left(g^{k\ell}R^{\,i}_{\;\;ki\ell}\right)+D_{\ell}\left(g^{k\ell}R^{\,i}_{\;\;kmi}\right)\\
   &=& D_{i}\left[g^{is}\left(-R_{sm}\right)\right]+D_{m}\left(g^{k\ell}R^{\,i}_{\;\;ki\ell}\right)+D_{\ell}\left(g^{k\ell}R^{\,i}_{\;\;kmi}\right) \\
  &=& -D_{i}R^{\,i}_{\;\;m}+D_{m}R-D_{\ell}R^{\,\ell}_{\;\;m}\\
  &=& -2\,D_{i}R^{\,i}_{\;\;m}+D_{m}R.
\end{eqnarray*}
Du fait que $D_{m}R\equiv \delta^{i}_{m}D_{i}R$, on arrive finalement à la relation\footnote{Le symbole de Kronecker est un tenseur à composantes constantes $\delta^{'\,i}_{j}=\frac{\partial x^{'\,i}}{\partial x^{k}}\,\frac{\partial x^{p}}{\partial x^{'\,j}}\,\delta^{k}_{p}=\frac{\partial x^{'\,i}}{\partial x^{p}}\,\frac{\partial x^{p}}{\partial x^{'\,j}}=\delta^{i}_{j}$ et de dérivée covariante nulle $D_{k}\delta^{i}_{j}=\partial_{k}\delta^{i}_{j}+\Gamma^{i}_{pk}\delta^{p}_{j}-\Gamma^{p}_{kj}\delta^{i}_{p}=0+\Gamma^{i}_{jk}-\Gamma^{i}_{kj}=0$.}
\begin{equation}
    D_{i}\left(R^{\,i}_{\;\;m}-\frac{1}{2}\,\delta^{i}_{m}\,R\right)=0,
\end{equation}
qui nous permet de définir le tenseur d'Einstein
\begin{equation}\label{tenseur einstein chap rappel}
    G^{\,i}_{\;\;m}=R^{\,i}_{\;\;m}-\frac{1}{2}\,\delta^{i}_{m}\,R,
\end{equation}
de divergence nulle $$D_{i}G^{\,i}_{\;\;m}=0.$$

\section{Rappels sur la Gravitation}
La gravité est l'une des quatre interactions fondamentales au moyen duquel toute la panoplie des phénomènes naturels diversifiée est expliquée; elle est responsable de l'attraction universelle entre corps matériels et constitue l'une des interactions les plus connues car se manifestant à notre échelle macroscopique. En effet, la gravité est la force qui empêche les projectiles massiques, lancés au voisinage de la terre, d'avoir un mouvement rectiligne et uniforme par rapport aux référentiels inertiels\footnote{Un référentiel inertiel ou bien galiléen est un système de coordonnées dans lequel un corps qui ne
subit aucune force extérieure, se meut de façon rectiligne et uniforme \cite{einsteinenfeld}.} de telle sorte à retomber suivant des trajectoires courbes  \cite{Choquet}.

La gravitation est l'interaction la plus faible\footnote{Les quatres interactions fondamentales sont par ordre décroissant de grandeur: l'interaction forte, électromagnétique, faible et gravitationnelle.} de la nature; elle intervient significativement seulement pour des masses importantes et à grande échelle. Ceci est du essentiellement à sa portée infinie et au fait qu'elle est universellement attractive. En effet, contrairement au champ électromagnétique qui n'interagit pas avec les corps neutres, le champ gravitationnel "ressent" toujours la présence des masses non nulles. En plus de ça, contrairement à tous les autres champs, le champ gravitationnel se distingue par une propriété le caractérisant exclusivement; le mouvement d'un corps test sous l'effet d'un champ de gravitation est indépendant de sa masse et de sa composition. Cette propriété trouve son origine dans l'égalité entre la masse grave et inerte d'un corps matériel.

\subsection{Gravitation newtonienne}
Pour élaborer sa théorie de la gravitation, Newton s'est inspiré des travaux de ses prédécesseurs. Il fallait, d'une part, élaborer une équation qui décrit comment la gravitation est générée par la matière (force de gravitation), et d'autre part, décrire la réponse de la matière à cette gravité (le principe fondamental de dynamique).

En postulant que la force d'attraction gravitationnelle s'exerçant entre le soleil et une planète quelconque est dirigée suivant la droite joignant leurs centres de masses et de module, à la fois, proportionnel à leurs masses graves respectives mais inversement proportionnel au carré de l'interdistance, il arrive à retrouver les trois lois de Kepler selon lesquelles
\begin{itemize}
  \item Toute planète décrit une trajectoire elliptique dont le soleil occupe l'un des deux foyers.
  \item Le rayon vecteur dont l'origine est le soleil et l'extrémité est la planète en question balaye des aires égales pendant des durées égales. Cette loi se traduit par un ralentissement de la planète quand elle s'éloigne du soleil et par une augmentation de sa vitesse en se rapprochant du foyer.
  \item Le rapport du carré de la période de révolution par le cube du demi grand axe de la trajectoire elliptique est une constante pour toutes les planètes du système solaire $T^{2}/a^{3}=cte$.
\end{itemize}

\subsubsection{Loi de Gravitation universelle}
La force gravitationnelle de Newton est dite universelle car elle est responsable de l'attraction de tous les corps ayant une masse non nulle; c'est une force qui s'exerce à travers une interaction mutuelle entre deux corps massiques. La force de gravitation avec laquelle la masse $M$ attire une autre masse $m$ est donnée par
\begin{equation}\label{force gravitation universelle}
    \overrightarrow{F}_{M/m}=-\frac{GMm}{r^{2}}\,\overrightarrow{e_{r}},
\end{equation}
alors que la force avec laquelle $m$ agit sur $M$, en vertu de la loi d'action et de réaction, est $\overrightarrow{F}_{m/M}=-\overrightarrow{F}_{M/m}$, tel que la constante de proportionnalité $G\cong6.67\times 10^{-11}\text{m}^{3}\text{kg}^{-1}\text{s}^{-2}$ est une donnée expérimentale dite constante de gravitation universelle et $\overrightarrow{e_{r}}=\overrightarrow{r}/r$ est le vecteur unitaire radial dirigé suivant la droite reliant les deux masses et orienté de $M$ vers $m$.

La force gravitationnelle qui s'exerce sur une masse ponctuelle $m$, repérée par $\overrightarrow{r}$, due aux masses ponctuelles $M_{1}$, $M_{2}$,$\dots$, $M_{n}$ de positions respectives $\overrightarrow{r_{1}}^{'}$, $\overrightarrow{r_{2}}^{'}$,$\dots$, $\overrightarrow{r_{n}}^{'}$ est donnée par la superposition
\begin{equation}\label{force gravitation dicret plusieurs}
    \overrightarrow{F}(\overrightarrow{r})=-Gm\displaystyle\sum^{n}_{i=1}M_{i}\,
    \displaystyle\frac{\left(\overrightarrow{r}-\overrightarrow{r_{i}}^{'}\right)}{\|\overrightarrow{r}-\overrightarrow{r_{i}}^{'}\|^{3}},
\end{equation}
alors que pour une distribution continue, de densité de masse locale $\rho_{\text{\tiny{m}}}(\overrightarrow{r}^{'})$ au point $P^{'}(x^{'},y^{'},z^{'})$ de la distribution de sorte que  $dM(\overrightarrow{r}^{'})=\rho_{\text{\tiny{m}}}(\overrightarrow{r}^{'})dx^{'}dy^{'}dz^{'}$, la force de gravitation qui d'exerce sur la masse ponctuelle $m$, repérée par $\overrightarrow{r}$, est donnée plutôt par l'intégrale
\begin{equation}\label{force gravitation dicret plusieurs}
    \overrightarrow{F}(\overrightarrow{r})=-Gm\displaystyle\int \rho_{\text{\tiny{m}}}(\overrightarrow{r}^{'})\,
    \displaystyle\frac{\left(\overrightarrow{r}-\overrightarrow{r}^{'}\right)}{\|\overrightarrow{r}-\overrightarrow{r}^{'}\|^{3}}\,dx^{'}dy^{'}dz^{'},
\end{equation}
étendue à tous les points $P^{'}(\overrightarrow{r}^{'})$ de la distribution continue.

\subsubsection{Forme locale de la Loi de Gravitation}
Le champ gravitationnel $\overrightarrow{g_{M}}(A)$ et le potentiel gravitationnel $\phi_{M}(A)$ crées par une masse ponctuelle $M$ au point $A$, repéré par le vecteur position $\overrightarrow{r}$, sont définis de façon que si on place au point $A$ une autre masse ponctuelle $m$, elle subira une force
\begin{equation}\label{force champ relation}
    \overrightarrow{F}=m\,\overrightarrow{g_{M}}(A),
\end{equation}
et aura une énergie potentielle
\begin{equation}\label{energie potent et potentiel}
    E_{p}=m\,\phi_{M}(A).
\end{equation}
Dans le cas statique, compte tenu de (\ref{force gravitation universelle}) et du fait que $\overrightarrow{F}=-\overrightarrow{\nabla}E_{p}$, nous avons les expressions respectives
\begin{equation}\label{champ gravit ponctuel}
    \overrightarrow{g_{M}}(A)=-\frac{GM}{r^{2}}\,\overrightarrow{e_{r}},
\end{equation}
\begin{equation}\label{pot gravit ponctuel}
    \phi_{M}(A)=-\frac{GM}{r},
\end{equation}
du champ et du potentiel gravitationnels, crées par une masse ponctuelle $M$.  Il est clair que les deux expressions précédentes (\ref{champ gravit ponctuel}) et (\ref{pot gravit ponctuel}) sont reliées $\overrightarrow{g_{M}}(A)=-\overrightarrow{\nabla}\phi_{M}(A)$, d'une façon analogue que la force et l'énergie potentielle.

En exploitant l'analogie entre la force coulombienne, $\overrightarrow{F}_{Q/q}=\left(KQq/r^{2}\right)\overrightarrow{e_{r}}$ qui s'exerce entre les deux charges ponctuelles $Q$ et $q$, et la force de Newton par le recours à la prescription\footnote{Dans notre cas $K=1/4\pi\epsilon_{0}$} $1/4\pi\epsilon_{0}\rightarrow -G$, il est possible d'énoncer un théorème de Gauss pour un champ gravitationnel $\overrightarrow{g}$. Dans ce cas, le flux de $\overrightarrow{g}$ à travers une surface fermée $S$ est proportionnel à la masse interne à $S$
\begin{equation}\label{th de gauss gravite}
    \oint_{S} \overrightarrow{g}.\overrightarrow{ds}=-4\pi G m_{int}.
\end{equation}
En utilisant le théorème d'Ostrogradsky, pour transformer l'intégrale de surface en une intégrale portant sur le volume qu'elle renferme, sachant que $m_{int}=\int\rho_{\text{\tiny{m}}} dV$, il est possible d'aboutir à la forme locale du théorème de Gauss
\begin{equation}\label{th de gauss gravite local}
     \overrightarrow{\nabla}.\overrightarrow{g}=-4\pi G \rho_{\text{\tiny{m}}}.
\end{equation}
En introduisant le potentiel gravitationnel, la relation précédente (\ref{th de gauss gravite local}) se réduit à l'équation de Poisson
\begin{equation}\label{eq de poisson}
    \overrightarrow{\nabla}^{2}\phi=4\pi G \rho_{\text{\tiny{m}}}.
\end{equation}
Pour calculer le potentiel en un point donné, il faut connaître d'une part la densité de masse en ce point et d'autre part les conditions aux limites du champ gravitationnel.

La théorie de gravitation de Newton peut être résumée comme suit: Les masses génèrent le champ gravitationnel en vertu de l'équation de Poisson, et le champ de gravitation génère, à son tour, l'accélération en vertu du principe fondamental de la dynamique.

\subsection{Gravitation newtonienne et Relativité restreinte}
La force de gravitation newtonienne (\ref{force gravitation universelle}) est en désaccord avec la relativité restreinte, car son utilisation s'appuie implicitement sur une transmission instantanée de l'influence gravitationnelle entre les deux masses mises en jeux. Face à une situation similaire pour la force de Coulomb, la description de phénomènes non statiques ne s'est fait que par l'ajout du champ magnétique, i.e par une unification électricité-magnétisme.  Dans ce cas, Le champ électromagnétique se propage à vitesse finie conformément aux équations de Maxwell. Il est a noté qu'un premier pas a été franchis dans ce sens pour la gravitation en formulant des équations locales de type poisson pour le potentiel gravitationnel.

Il est à noter que dans l'équation (\ref{eq de poisson}), l'absence de temps fait que le potentiel répond instantanément à toute variation de la densité de masse $\rho_{\text{\tiny{m}}}$. Même en essayant de remplacer le laplacien par l'opérateur d'alembertien\footnote{L'opérateur d'Alembertien $\Box=\eta^{\mu\nu}\partial{\mu}\partial{\nu}=\frac{1}{c^{2}}\frac{\partial^{\,2}}{\partial t^{2}}-\overrightarrow{\nabla}^{2}$ est un invariant relativiste qui se réduit à $(-\overrightarrow{\nabla}^{2})$ quand $c\rightarrow +\infty$}, néanmoins la densité de masse $\rho_{\text{\tiny{m}}}$ n'est pas invariante par transformation de Lorentz, suite à la contraction de longueur du volume infinitésimale $dV$ selon la direction du mouvement. Plusieurs tentatives ont été avancées pour essayer de rendre conforme la gravitation de Newton à la relativité restreinte sans réel succès.

Au lieu de cela, Einstein a compris qu'il ne fallait restreindre le Principe de Relativité aux référentiels d'inertie, animés les uns par à port aux autres de mouvement rectiligne et uniforme, mais qu'il fallait plutôt étendre ce principe à tous les référentiels, sans la moindre restriction quant à leurs états de mouvement relatifs.

\subsection{Gravitation et Inertie}
En relativité restreinte, les référentiels d'inertie animés d'un mouvement rectiligne et uniforme les uns par rapport aux autres sont tous équivalents pour décrire les phénomènes naturels. Par contre, les référentiels aimés de mouvement accélérés sont le siège de force à caractère inertiel laissant croire qu'ils sont dotés d'un caractère absolu. En effet, en supposant deux masses en mouvement relatif rectiligne et uniforme, il est tout à fait équivalent de dire que la première masse et au repos alors que la deuxième se déplace ou vis-versa. Par contre, si l'une des deux est accélérée par rapport à l'autre, la situation n'est plus symétrique car l'apparition d'une force d'inertie agissant sur la particule accélérée confère au mouvement accéléré un caractère absolu.

Pour étendre le principe de Relativité aux référentiels accélérés, Einstein a remarqué qu'il pouvait attribuer les forces d'inertie à l'existence d'un champ de gravitation. Cette équivalence est rendue possible grâce à la propriété du champ de gravitation d'accélérer tous les corps à la même accélération. Cette propriété trouve ses origines profondes dans l'égalité des masses pesante et inertielle d'un même corps.

La masse inertielle $m_{i}$, qui intervient dans le principe fondamental de la dynamique $\overrightarrow{F}=m_{i}\,\overrightarrow{a}$, est la grandeur qui exprime la capacité d'un corps à résister à toute modification de son état de mouvement. Par contre, la masse gravitationnelle $m_{g}$, qui intervient dans la loi de gravitation de Newton $\overrightarrow{F}=\left(-GM_{g}m_{g}/r^{2}\right)\overrightarrow{e_{r}}$, exprime la capacité du même corps à interagir avec un autre corps $M_{g}$ situé à une certaine distance de lui. Ainsi, l'accélération du corps plongé dans un champ gravitationnel $\overrightarrow{g}=-\overrightarrow{\nabla}\phi$ est donnée par
\begin{equation}\label{accel masse grav inertie}
    \overrightarrow{a}=-\left(\frac{m_{g}}{m_{i}}\right)\overrightarrow{\nabla}\phi.
\end{equation}

\subsection{Principe d'Equivalence}
Il est bien connu expérimentalement, déjà depuis Galilée jusqu'aux versions récentes de l'experience d'Eötvös , que le rapport $m_{g}/m_{i}$ est le même pour toutes les particlues. Par un choix approprié d'unités, il est toujours possible de se ramener à la situation d'égalité entre les masses inertielle et gravitationnelle de sorte que le rapport soit égal à l'unité. Dans ce cas, il est bien évident que le mouvement d'une particule sous l'effet d'un champ gravitationnel est indépendant de la nature de cette particule
\begin{equation}\label{accel grav equiv}
    \overrightarrow{a}=\overrightarrow{g}.
\end{equation}

Alors que l'égalité des masses pesante et inerte était purement "accidentelle" (une co\"{\i}ncidence) dans la Mécanique Classique de Newton, elle revet un caractère fondamental dans la nouvelle théorie de gravitation; elle constitue l'argument clé qui a incité Einstein d'énoncer le Principe d'Equivalence en vue d'étendre le Principe de Relativité à tous les référentiels.

Dans la formulation faible du Principe d'Equivalence, l'égalité entre la masse inerte et pesante d'un même corps, vérifiée expérimentalement, est désormais postulée comme exacte \cite{Groen}.

Une autre conséquence du Principe d'Equivalence découle de l'utilisation de la propriété du champ gravitationnel à communiquer la même accélération à tous les corps (\ref{accel grav equiv}). En effet, pour un corps en chute libre dans un champ gravitationnel homogène, la force inertielle due à l'accélération est contrebalancée par la force gravitationnelle, de sorte que pour un référentiel galiléen en chute libre, tous les corps libres, se déplaçant conformément à l'équation \cite{Groen}
\begin{equation}
    m_{i}\frac{d^{\,2}\overrightarrow{r}^{'}}{dt^{2}}=\left(m_{g}-m_{i}\right)\overrightarrow{g}=\overrightarrow{0},
\end{equation}
possèdent des accélérations relatives nulles. Dans ce référentiel, tous ces corps ne sont soumis à aucune force et se déplacent suivant des lignes droites à vitesses constantes. Einstein postule ainsi l'équivalence de tels référentiels d'inertie avec les référentiels galiléens, situés dans des régions suffisamment éloignés de toute influence d'une distribution matérielle.

Inversement, un observateur au repos dans un champ gravitationnel homogène et un observateur accéléré dans un référentiel galiléen éloigné de toute source masse obtiennent les mêmes résultats en effectuant des expériences similaires.

Le Principe d'Equivalence peut s'énoncer comme suit \cite{boratav} \textbf{"Un référentiel uniformément accéléré est localement équivalent à un référentiel inertiel plongé dans un champ de gravitation"}.

Il est à noter que cette équivalence n'est valable que localement, pour garantir au champ gravitationnel newtonien d'être constant; ceci n'a lieu que dans la mesure que les inhomogé\"{\i}tés, en module et en direction, du champ de gravitation peuvent être négligées.

\subsection{Postulat de Relativité Générale}

En relativité restreinte, la classe des référentiels galiléens, en mouvement rectiligne et uniforme les uns par rapport autres, est privilégiée pour décrire les phénomènes physiques. Guidé par l'intuition selon laquelle les phénomènes physiques sont indépendants du référentiel dans lequel sont exprimées leurs lois, et par le Principe d'Equivalence, Einstein étend le principe de Relativité à tous les référentiels, sans aucune restriction sur leurs états de mouvement relatifs. Pour lui \cite{relativrestreinte} \textbf{"Tous les corps de référence, quels que soient leurs états de mouvement, sont équivalents pour la description des lois de la nature"}.

Pour que les lois physiques gardent la même forme lors du passage d'un référentiel de coordonnées à un autre, il suffit de les exprimer sous forme tensorielle. Dans ce cas, les lois sont covariantes par transformations arbitraires de coordonnées, autrement dit, elles possèdent une forme indépendante du référentiel dans lequel elles sont exprimées.

Conformément au Principe d'Equivalence, les termes inertiels qui apparaissent lors du passage d'un référentiel galiléen à un référentiel accéléré peuvent être expliqués par l'introduction d'un champ gravitationnel. Ainsi, la relativité restreinte serait un cas particulier de la nouvelle théorie de la relativité générale quand les champs de gravitation sont très faibles, voir inexistants.
\newpage
\subsection{La Gravitation selon Einstein}
L'adoption du Principe d'Equivalence, vu le caractère universel de l'interaction gravitationnelle, conduit Einstein a ne pas voir la gravitation comme une force, au sens conventionnel, mais plutôt comme la manifestation de la courbure de l'espace-temps, due à la présence d'une distribution matérielle ou bien d'énergie\footnote{Suite à l'équivalence masse-énergie.}. Une particule test répond à cette courbure de l'espace-temps en suivant une trajectoire dite géodésique de façon à minimiser la distance de son parcours.

\subsubsection{Interprétation géométrique de la Gravité}
\begin{figure}[\here]
   \centering
  \includegraphics[width=8cm]{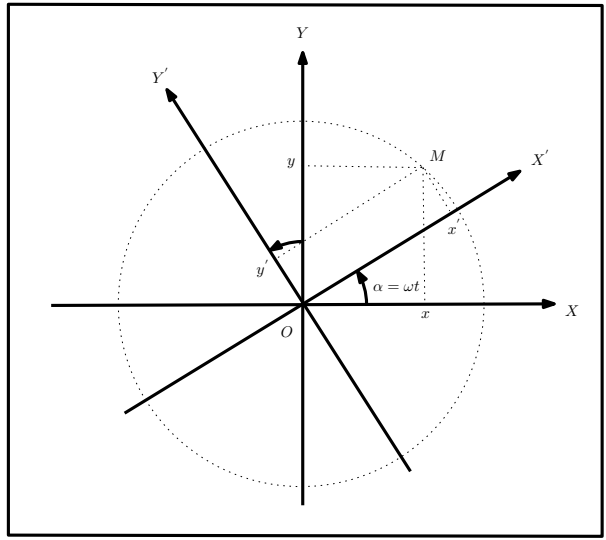}\\
  \caption{Deux repères en rotation}\label{figrotation}
\end{figure}

Soit un repère un référentiel galiléen muni d'un repère $R(OXYZ)$, et considérons un deuxième référentiel $R^{'}(OX^{'}Y^{'}Z^{'})$ en rotation autour de l'axe commune $(OZ)$ avec la vitesse angulaire constante $\omega$ (voir figure \ref{figrotation}).

On se place dans le cas où le temps s'écoule de la même manière dans les deux référentiels. Dans ce cas, un point quelconque est repéré par les coordonnées $M(x,y,z)$ dans le référentiel galiléen $R(OXYZ)$, alors que ses coordonnées dans le repère non galiléen $R^{'}(OX^{'}Y^{'}Z)$ sont $M(x^{'},y^{'},z^{'})$ de telle sorte que
\begin{eqnarray}\label{transform rotation}
  \left\{
  \begin{array}{ll}
    x=x^{'}\,\cos(\omega t)-y^{'}\,\sin(\omega t) \\
    y=x^{'}\,\sin(\omega t)+y^{'}\,\cos(\omega t)  \\
    z=z^{'}\\
    t=t^{'}
  \end{array}
\right.
\end{eqnarray}

Conformément à (\ref{carre interval inertiel}), qui donne
\begin{equation}\label{carre interval inertiel explicit}
    ds^{2}=(c\,dt)^{2}-dx^{2}-dy^{2}-dz^{2},
\end{equation}
et de la transformation (\ref{transform rotation}), le carrée de l'intervalle dans le référentiel non galiléen est donné par \cite{Boudenot}
\begin{equation}
    ds^{2}=\left[c^{2}-\omega^{2}\left(x^{'\,2}+y^{'\,2}\right)\right]dt^{'\,2}-dx^{'\,2}-dy^{'\,2}-dz^{'\,2}-\left(2\,\omega\,x^{'}\right)dy^{'}dt^{'}
+\left(2\,\omega\,y^{'}\right)dx^{'}dt^{'}.
\end{equation}
Il est claire que, quelque soit la transformation du temps, cette expression ne peut jamais se réduire à la forme (\ref{carre interval inertiel explicit}) de l'expression du carré d'interval infinitésimal d'un référentiel galiléen; dans un référentiel non galiléen, l'expression du carré de l'intervalle est plutôt de la forme générale
\begin{equation}\label{ds carre non galileen}
    ds^{2}=g_{\mu\nu}^{'}\,dx^{'\mu}dx^{'\nu},
\end{equation}
tel que les sommes sont effectuées pour $\mu,\nu=0,1,2,3$ et la métrique $g_{\mu\nu}^{'}=g_{\mu\nu}^{'}(x^{'\sigma})$ dépend des coordonnées de l'espace-temps.

En vertu du Principe d'Equivalence, les effet inertiels qui se manifestent dans le référentiel non galiléen, à travers les termes non diagonaux $(\sim dy^{'}dt^{'})$ et $(\sim dx^{'}dt^{'})$, peuvent être attribués à un champ de gravitation, à condition de considérer $R^{'}(OX^{'}Y^{'}Z)$ au repos. Les composantes du tenseur métrique $g_{\mu\nu}^{'}(x^{'\sigma})$ qui figurent dans (\ref{ds carre non galileen}) décrivent les effets de ce champ gravitationnel; ces effet se manifestent à travers la géométrie de l'espace-temps.

\subsubsection{Pourquoi une géométrie riemannienne}
L'espace-temps de la Relativité Générale est pseudo-Rimannien de telle sorte qu'en l'absence de gravité, il est toujours possible de se ramener à un référentiel d'inertie pour lequel l'expression du carré d'intervalle est donnée par (\ref{carre interval inertiel}). En imposant l'invariance de la signature de l'espace-temps \cite{Plebanski}, pour ne pas avoir des discontinuités au niveau des composantes de la métrique de l'espace-temps $g_{\mu\nu}$ conduisant à l'apparition de singularités, dans ce cas, la signature d'un tel espace-temps de la relativité générale est la même que celle de l'espace-temps de Minkowski $(+,-,-,-)$. Par contre, il faut savoir qu'un espace riemannien possède une signature $(+,+,+,+)$ car il est localement équivalent à un espace euclidien pour lequel l'expression du carré de l'interval est donnée par $ds^{2}\simeq\left(dx^{0}\right)^{2}+\left(dx^{1}\right)^{2}+\left(dx^{2}\right)^{2}+\left(dx^{3}\right)^{2}$.

La définition d'un espace pseudo-riemannien est analogue à celle de l'espace riemannien. En effet, dans un espace pseudo-euclidien à 4 dimensions, il est toujours possible de définir un système d'axes rectilignes et orthonormés $\{X^{\mu}\}$ de façon que la distance entre deux points infiniment voisins $(X^{0},X^{1},X^{2},X^{3})$ et $(X^{0}+dX^{0},X^{1}+dX^{1},X^{2}+dX^{2},X^{3}+dX^{3})$ est telle que
\begin{eqnarray}\label{carre distance infinit pseudo euclid}
  ds^{2} = \left(dX^{0}\right)^{2}-\left(dX^{1}\right)^{2}-\left(dX^{2}\right)^{2}-\left(dX^{3}\right)^{2}.
\end{eqnarray}
Lors d'un passage au système de coordonnées curvilignes $\{x^{\mu}\}$, l'expression précédente (\ref{carre distance infinit pseudo euclid}) se met sous la forme quadratique fondamentale
\begin{equation}\label{forme quadratique fondamentale pseudo}
    ds^{2}=g_{\mu\nu}\,dx^{\mu}\,dx^{\nu}.
\end{equation}

Contrairement à un espace pseudo-euclidien, dans un espace pseudo-riemannien, il n'est jamais possible de définir un système d'axes rectilignes de telle sorte à garantir à la forme quadratique fondamentale (\ref{forme quadratique fondamentale pseudo}) de se réduire à la forme (\ref{carre distance infinit pseudo euclid}) pour tous les points de l'espace. La forme (\ref{carre distance infinit pseudo euclid}) peut être vérifiée localement, mais pas globalement.

Les définitions, exprimées dans la section précédente pour un espace riemannien, par exemple du tenseur de courbure, tenseur de Ricci, tenseur d'Einstein, $\cdots$, ect, restent valables pour l'espace-temps pseudo-riemannien à dimension 4 de la relativité générale.

\subsection{Equations de mouvement d'une particule}
\subsubsection{Particule libre}
Considérons une particule libre de position $\overrightarrow{r}$ et de vitesse $\overrightarrow{v}$ dans un référentiel galiléen. L'action de cette particule est par définition l'intégrale du lagrangien
\begin{equation}\label{action libre restreinte}
    S=\int_{t_{1}}^{t_{2}}\textit{L}(\overrightarrow{r},\overrightarrow{v})dt.
\end{equation}
En introduisant le temps propre\footnote{C'est un invariant relativiste} $d\tau=dt\displaystyle\sqrt{1-v^{2}/c^{2}}$, l'action précédente se met alors sous la forme
\begin{equation}
    S=\int_{\tau_{1}}^{\tau_{2}}\frac{\textit{L}}{\sqrt{1-\frac{v^{2}}{c^{2}}}}\,d\tau.
\end{equation}
La façon la plus simple d'imposer à l'action d'être un invariant relativiste est de chercher l'intégrant $\textit{L}/\sqrt{1-\frac{v^{2}}{c^{2}}}$ sous forme d'une constante. Le passage à la limite classique permet de fixer cette dernière à $-m\,c^{2}$, de sorte à avoir
\begin{equation}\label{action libre rest}
    S=-m\,c^{2}\int_{t_{1}}^{t_{2}}\sqrt{1-\frac{v^{2}}{c^{2}}}\,dt
\end{equation}

Dans le but d'exprimer (\ref{action libre rest}) en fonction de l'interval infinitésimal $ds$ entre deux événements infiniment voisins $(x,y,z,t)$ et $(x+dx,y+dy,z+dz,t+dt)$, nous avons par définition
\begin{eqnarray}
ds^{2}&=&\eta_{\mu\nu}\,dx^{\mu}dx^{\nu}=(c\,dt)^{2}-(dx)^{2}-(dy)^{2}-(dz)^{2}\nonumber\\
      &=&c^{2}\,dt^{2}\left\{1-\frac{1}{c^{2}}\left[\left(\frac{dx}{dt}\right)^{2}+\left(\frac{dx}{dt}\right)^{2}+\left(\frac{dx}{dt}\right)^{2}\right]\right\}
      =c^{2}\,dt^{2}\left(1-\frac{v^{2}}{c^{2}}\right).\nonumber
\end{eqnarray}
Ainsi
\begin{equation}\label{ds infint vitesse}
    ds=c\,\sqrt{1-\frac{v^{2}}{c^{2}}}\,dt=c\,d\tau.
\end{equation}
En tenant compte de (\ref{ds infint vitesse}), l'expression de l'action (\ref{action libre rest}) d'une particule libre, évoluant entre deux événements $E_{1}(\overrightarrow{r_{1}},t_{1})$ et $E_{2}(\overrightarrow{r_{2}},t_{2})$, se met finalement sous la forme covariante
\begin{equation}\label{action libre rest ds}
    S=-m\,c\int_{E_{1}}^{E_{2}}ds.
\end{equation}

Pour déterminer l'équation de mouvement de la particule libre, appliquons le principe de moindre action à (\ref{action libre rest ds})
\begin{equation}\label{moindre action restreint}
    \delta S=-m\,c\int_{E_{1}}^{E_{2}}\delta ds=0.
\end{equation}
Pour ce faire, calculons la variation du carré de l'interval infinitésimal dans un espace-temps de Minkowski
\begin{eqnarray}
  \delta ds^{2}&=&\delta \left(\eta_{\mu\nu}\,dx^{\mu}dx^{\nu}\right)\nonumber \\
  2\,ds\,\delta ds &=& \eta_{\mu\nu}\left(\delta dx^{\mu}\right)dx^{\nu}+\eta_{\mu\nu}\,dx^{\mu}\left(\delta dx^{\nu}\right).
\end{eqnarray}
La symétrie de la métrique de Minkowski ($\eta_{\mu\nu}=\eta_{\nu\mu}$) ainsi que l'interversion des indices muets $\mu$ et $\nu$ nous permet de déduire que
\begin{equation}
    \delta ds=\eta_{\mu\nu}\left(\delta dx^{\mu}\right)\frac{dx^{\nu}}{ds}=\frac{dx_{\mu}}{ds}\,\delta dx^{\mu}.
\end{equation}
Compte tenu de la définition du quadrivecteur vitesse de la particule $U_{\mu}=dx_{\mu}/d\tau$ et de la relation (\ref{ds infint vitesse}), nous aboutissons finalement à la relation
\begin{equation}\label{delta ds restreinte}
    \delta ds=\frac{U_{\mu}}{c}\,\delta dx^{\mu}.
\end{equation}
En remplaçant (\ref{delta ds restreinte}) dans (\ref{moindre action restreint})
\begin{eqnarray*}
  \delta S &=& -m\,c\int_{E_{1}}^{E_{2}}\frac{U_{\mu}}{c}\,\delta dx^{\mu}=0,
\end{eqnarray*}
il est ainsi possible, par une simple intégration par parties et sachant que $\delta dx^{\mu}= d (\delta x^{\mu})$, d'avoir
\begin{eqnarray*}
  -m\int_{E_{1}}^{E_{2}}d\left(U_{\mu}\,\delta x^{\mu}\right)+m\int_{E_{1}}^{E_{2}}dU_{\mu}\,\delta x^{\mu} &=& 0 \\
  -m\underbrace{\Big[U_{\mu}\,\delta x^{\mu}\Big]_{E_{1}}^{E_{2}}}_{0}+m\int_{E_{1}}^{E_{2}}\frac{dU_{\mu}}{ds}\,ds\,\delta x^{\mu} &=& 0 \hspace{1cm}{\forall \;\;\delta x^{\mu}}\\
\end{eqnarray*}
Pour des $\delta x^{\mu}$ arbitraires mais nuls aux frantières\footnote{La condition $\delta x^{\mu}(E_{1})=\delta x^{\mu}(E_{2})=0$ permet d'annuler le premier terme de l'intégrale précédente}, le principe de moindre action permet finalement d'aboutir à l'équation de mouvement
\begin{equation}\label{eq geodes restreinte}
    \frac{dU_{\mu}}{ds}=0,
\end{equation}
pour une particule libre.

\subsubsection{Equation des géodésiques}

Le mouvement d'une particule qui évolue dans un champ de gravitation entre deux événements $E_{1}(\overrightarrow{r_{1}},t_{1})$ et $E_{2}(\overrightarrow{r_{2}},t_{2})$ peut être décrit en utilisant l'action
\begin{equation}\label{delta ds rappels}
    \delta\int_{E_{1}}^{E_{2}}ds=0.
\end{equation}
Le champ de gravitation agit sur la particule à travers la métrique $g_{\mu\nu}$ en modifiant la structure de l'espace-temps. Pour tenir compte de la gravitation, nul besoin d'ajouter un terme d'interaction de la particule avec le champ de gravitation, il suffit pour cela de prendre $g_{\mu\nu}$ au lieu de $\eta_{\mu\nu}$ dans la définition de l'interval infinitésimal
\begin{equation}\label{ds carre rappels}
    ds^{2}=g_{\mu\nu}\,dx^{\mu}dx^{\nu}.
\end{equation}

La variation de $ds^{2}$
\begin{eqnarray*}
  \delta ds^{2} &=& \delta\left(g_{\mu\nu}\,dx^{\mu}dx^{\nu}\right)\\
  2\,ds\,\delta ds &=& \delta g_{\mu\nu}\,dx^{\mu}dx^{\nu}+2\,g_{\mu\nu}\left(\delta dx^{\mu}\right)dx^{\nu}
\end{eqnarray*}
nous permet de déduire que
\begin{eqnarray*}
  \delta ds &=& \left[\frac{1}{2}\,\delta g_{\mu\nu}\,\frac{dx^{\mu}}{ds}\,\frac{dx^{\nu}}{ds}+g_{\mu\nu}\frac{\delta dx^{\mu}}{ds}\,\frac{dx^{\nu}}{ds}\right]ds. \\
\end{eqnarray*}
En intégrant par parties entre les événements $E_{1}$ et $E_{2}$, sachant que $\delta x^{\mu}(E_{1})=\delta x^{\mu}(E_{2})=0$ aux frontières, et en intervertissant entre la différentielle et la variation\footnote{$\frac{d}{ds}(\delta x^{mu})=\delta\frac{dx^{mu}}{ds}$}, il est possible d'écrire
\begin{eqnarray}
  0 &=&\int_{E_{1}}^{E_{2}}\delta ds\nonumber\\
   &=& \int_{E_{1}}^{E_{2}}\left\{\frac{1}{2}\,\delta g_{\mu\nu}\,\frac{dx^{\mu}}{ds}\,\frac{dx^{\nu}}{ds}+\frac{d}{ds}\left(g_{\mu\nu}\,\frac{dx^{\nu}}{ds}\,\delta x^{\mu}\right)-\frac{d}{ds}\left(g_{\mu\nu}\,\frac{dx^{\nu}}{ds}\right)\delta x^{\mu}\right\}ds \nonumber\\
   &=& \int_{E_{1}}^{E_{2}}\left\{\frac{1}{2}\,\delta g_{\mu\nu}\,\frac{dx^{\mu}}{ds}\,\frac{dx^{\nu}}{ds}-\frac{d}{ds}\left(g_{\mu\nu}\,\frac{dx^{\nu}}{ds}\right)\delta x^{\mu}\right\}ds+\left[g_{\mu\nu}\,\frac{dx^{\nu}}{ds}\,\delta x^{\mu}\right]_{E_{1}}^{E_{2}}\nonumber \\
   &=& \int_{E_{1}}^{E_{2}}\left\{\frac{1}{2}\,\delta g_{\mu\nu}\,\frac{dx^{\mu}}{ds}\,\frac{dx^{\nu}}{ds}-\frac{d}{ds}\left(g_{\mu\nu}\,\frac{dx^{\nu}}{ds}\right)\delta x^{\mu}\right\}ds \nonumber\\
   &=& \int_{E_{1}}^{E_{2}}\left\{\frac{1}{2}\left(\frac{\partial g_{\mu\nu}}{\partial x^{\ell}}\,\delta x^{\ell}\right)\frac{dx^{\mu}}{ds}\,\frac{dx^{\nu}}{ds}-\frac{d}{ds}\left(g_{\ell\nu}\,\frac{dx^{\nu}}{ds}\right)\delta x^{\ell}\right\}ds \nonumber\\
   &=& \int_{E_{1}}^{E_{2}}\left\{\frac{1}{2}\left(\frac{\partial g_{\mu\nu}}{\partial x^{\ell}}\,\delta x^{\ell}\right)\frac{dx^{\mu}}{ds}\,\frac{dx^{\nu}}{ds}-\left(\frac{dg_{\ell\nu}}{ds}\,\frac{dx^{\nu}}{ds}+g_{\ell\nu}\,\frac{d^{\,2}x^{\nu}}{ds^{2}}\right)\delta x^{\ell}\right\}ds \nonumber\\
   &=& \int_{E_{1}}^{E_{2}}\left\{\frac{1}{2}\left(\frac{\partial g_{\mu\nu}}{\partial x^{\ell}}\,\delta x^{\ell}\right)\frac{dx^{\mu}}{ds}\,\frac{dx^{\nu}}{ds}-\left(\frac{\partial g_{\ell\nu}}{\partial x^{\mu}}\,\frac{dx^{\mu}}{ds}\,\frac{dx^{\nu}}{ds}+g_{\ell\nu}\,\frac{d^{\,2}x^{\nu}}{ds^{2}}\right)\delta x^{\ell}\right\}ds. \label{moindre action 1}
\end{eqnarray}
Sachant que les indices de sommation $\mu$ et $\nu$ sont muets\footnote{ils peuvent être intervertis}, nous avons
\begin{eqnarray}\label{muets}
  \frac{\partial g_{\ell\nu}}{\partial x^{\mu}}\,\frac{dx^{\mu}}{ds}\,\frac{dx^{\nu}}{ds} &=& \frac{1}{2}\left(\frac{\partial g_{\ell\nu}}{\partial x^{\mu}}\,\frac{dx^{\mu}}{ds}\,\frac{dx^{\nu}}{ds}+\frac{\partial g_{\ell\mu}}{\partial x^{\nu}}\,\frac{dx^{\nu}}{ds}\,\frac{dx^{\mu}}{ds} \right),
\end{eqnarray}
de sorte que (\ref{moindre action 1}) se réduit à
\begin{eqnarray*}
  0 &=& \int_{E_{1}}^{E_{2}}\left\{\frac{1}{2}\left(\frac{\partial g_{\mu\nu}}{\partial x^{\ell}}-\frac{\partial g_{\ell\nu}}{\partial x^{\mu}}-\frac{\partial g_{\ell\mu}}{\partial x^{\nu}}\right)\frac{dx^{\mu}}{ds}\,\frac{dx^{\nu}}{ds}-g_{\ell\nu}\,\frac{d^{\,2}x^{\nu}}{ds^{2}}\right\}\,\delta x^{\ell}ds.
\end{eqnarray*}
Pour des $\delta x^{\ell}$ arbitraires et compte tenu de (\ref{importantegij}), le principe de moindre action conduit finalement à l'équation des géodésiques suivante
\begin{equation}\label{eq geodes}
    \frac{d^{\,2}x^{\nu}}{ds^{2}}+\Gamma^{\nu}_{\mu\rho}\,\frac{dx^{\mu}}{ds}\,\frac{dx^{\rho}}{ds}=0.
\end{equation}

Ce résultat est en parfait accord avec le Principe d'Equivalence, en vertu du duquel, il est toujours possible de passer d'un système de coordonnées galiléen $\{x^{(0)\lambda}\}$, dans lequel le mouvement d'une particule libre est une ligne droite tel que
\begin{equation}\label{accel nulle}
    \frac{d^{\,2}x^{(0)\lambda}}{d\tau^{2}}=0,
\end{equation}
vers un nouveau système de coordonné $\{x^{\lambda}\}$ quelconque de sorte que (\ref{accel nulle}) se met la forme
\begin{eqnarray*}
  0 &=& \frac{d}{d\tau}\left(\frac{\partial x^{(0)\lambda}}{\partial x^{\rho}}\,\frac{dx^{\rho}}{d\tau}\right) = \frac{\partial x^{(0)\lambda}}{\partial x^{\rho}}\,\frac{d^{2\,}x^{\rho}}{d\tau^{2}}
   +\frac{\partial}{\partial x^{\mu}}\left(\frac{\partial x^{(0)\lambda}}{\partial x^{\rho}}\right)\frac{dx^{\mu}}{d\tau}\,\frac{dx^{\rho}}{d\tau},
\end{eqnarray*}
et donne, après contraction par $(\partial x^{\nu}/\partial x^{(0)\lambda})$
\begin{eqnarray}
  0 &=& \underbrace{\frac{\partial x^{\nu}}{\partial x^{(0)\lambda}}\,\frac{\partial x^{(0)\lambda}}{\partial x^{\rho}}}_{\delta^{\nu}_{\rho}}\frac{d^{2\,}x^{\rho}}{d\tau^{2}}
  +\left(\frac{\partial x^{\nu}}{\partial x^{(0)\lambda}}\,\frac{\partial^{\,2} x^{(0)\lambda}}{\partial x^{\mu}\partial x^{\rho}}\right)\frac{dx^{\mu}}{d\tau}\,\frac{dx^{\rho}}{d\tau},
\end{eqnarray}
l'équation des géodésiques
\begin{equation}\label{eq geodes bis}
    \frac{d^{\,2}x^{\nu}}{d\tau^{2}}+\Gamma^{\nu}_{\rho\mu}\,\frac{dx^{\mu}}{d\tau}\,\frac{dx^{\rho}}{d\tau}=0.
\end{equation}

Ainsi, le passage direct de l'équation de mouvement d'une particule libre en relativité restreinte $dU^{\nu}/ds=0$ vers l'équation de la même particule en présence d'un champ de gravitation se fait par le remplacement de la dérivée ordinaire $dU^{\nu}$ du quadrivecteur vitesse par sa dérivée covariante 
\begin{eqnarray}\label{du cov ds}
   0 = \frac{DU^{\nu}}{ds}=
    \frac{dU^{\nu}}{ds}+\Gamma^{\nu}_{\mu\rho}\,U^{\mu}\,\frac{dx^{\rho}}{ds}=
    \frac{d}{ds}\left(c\,\frac{dx^{\nu}}{ds}\right)+\Gamma^{\nu}_{\mu\rho}\,\left(c\,\frac{dx^{\mu}}{ds}\right)\frac{dx^{\rho}}{ds},
\end{eqnarray}
de sorte à retrouver l'équation (\ref{eq geodes}).

Les particules libres se déplacent en lignes droites dans un espace-temps plat de Minkowski, par contre en présence de gravité, qui courbe la structure de l'espace-temps, ces particules test se déplacent suivant des géodésiques de telle manière à minimiser leurs distances parcourues.

\subsubsection{Approximation newtonienne pour les géodésiques}
Dans ce qui suit, on va s'intéresser à l'équation de mouvement d'une particule dans un champ de gravitation faible et stationnaire. Dans ce cas, l'écart de la métrique de l'espace-temps par rapport à la métrique plate de Minkowski $\eta_{\mu\nu}=(1,-1,-1,-1)$ est négligeable de telle sorte à avoir
\begin{equation}\label{minkowski plus perturb 1}
    g_{\mu\nu}=\eta_{\mu\nu}+h_{\mu\nu},
\end{equation}
avec $|h_{\mu\nu}|\ll 1$.

Nous assumons aussi que la vitesse de la particule est négligeable devant celle de la lumière, de sorte que les composantes de vitesse vérifient $dx^{i}/dt \ll c\;\;(i=1,2,3)$, où $t$ est défini par $x^{0}=c\,t$. Ceci revient à imposer que $$\frac{dx^{i}}{d\tau} \ll \frac{dx^{0}}{d\tau},$$
pour $i=1,2,3$, vu que dans ce contexte il n'y a pas de différence entre le temps dans le repère et le temps propre\footnote{En vertu de (\ref{ds infint vitesse}), nous avons pour des vitesses faibles $c^{2}(d\tau/dt)^{2}=c^{2}-v^{2}\simeq c^{2}$.}
\begin{equation}\label{temp egale temp propre}
    \frac{d\tau}{dt}\simeq 1.
\end{equation}
Avec toutes ces approximations, les deux derniers termes de l'équation de géodésiques (\ref{eq geodes}) $$\frac{d^{\,2}x^{\mu}}{d\tau^{2}}+\Gamma^{\mu}_{00}\left(\frac{dx^{0}}{d\tau}\right)^{2}+2\,\Gamma^{\mu}_{0i}\,\frac{dx^{i}}{d\tau}\,\frac{dx^{0}}{d\tau}
+\Gamma^{\mu}_{ij}\,\frac{dx^{i}}{d\tau}\,\frac{dx^{i}}{d\tau}=0,$$
sont négligés pour avoir finalement
\begin{equation}\label{geodesique newton 1}
    \frac{d^{\,2}x^{\mu}}{d\tau^{2}}\simeq -c^{2}\,\Gamma^{\mu}_{00}\left(\frac{dt}{d\tau}\right)^{2}\simeq -c^{2}\,\Gamma^{\mu}_{00}.
\end{equation}
Compte tenu de la définition des symboles de Christoffel (\ref{importantegij}), l'équation précédente donne dans le cas où $\mu=0$ un résultat $c\,d^{\,2}t/d\tau^{2}=0$,
compatible avec (\ref{temp egale temp propre}), alors que pour $\mu=i$, nous avons
\begin{equation}\label{geodesique spatiale newton 1}
    \frac{d^{\,2}x^{\mu}}{d\tau^{2}}\simeq -c^{2}\,\Gamma^{i}_{00}=\frac{c^{2}}{2}\,\eta^{ij}\partial_{j}h_{00},
\end{equation}
avec $i,j=1,2,3$. Dans un premier temps, remarquons que cette dernière équation peut être réécrite sous forme vectorielle
\begin{equation}\label{geodesique spatiale vecteur newton 1}
    \frac{d^{\,2}\overrightarrow{r}}{dt^{2}}=-\frac{c^{2}}{2}\,\overrightarrow{\nabla}h_{00}.
\end{equation}
En deuxième lieu, l'équation de Newton d'une particule\footnote{Le principe fondamental de dynamique s'écrit $m\,\frac{d^{\,2}\overrightarrow{r}}{dt^{2}}=-\overrightarrow{\nabla}V$, où l'énergie potentielle gravitationnelle est reliée au potentiel par $V=m\phi$.} soumise à un potentiel gravitationnel $\phi=-GM/r$ est donnée par
\begin{equation}\label{equation de newton 1}
    \frac{d^{\,2}\overrightarrow{r}}{dt^{2}}=-\,\overrightarrow{\nabla}\phi.
\end{equation}
La comparaison entre (\ref{geodesique spatiale vecteur newton 1}) et (\ref{equation de newton 1}), tout en exigeant que la métrique de l'espace-temps devienne minkowskienne quand la particule $m$ s'éloigne de la source du champs $M$ (i.e $h_{00}\rightarrow 0$ quand $r\rightarrow +\infty$), permet de retrouver la relation importante
\begin{equation}\label{metrique 00 et phi}
    g_{00}=1+\frac{2\phi}{c^{2}},
\end{equation}
qui exprime clairement une identification des effets gravitationnels avec la géométrie de l'espace-temps.

\subsection{Equations d'Einstein}
Après avoir déterminé les équations des géodésiques qui expriment comment la courbure de l'espace-temps agit sur la matière pour se manifester sous forme d'un champ gravitationnel, il faut à présent déterminer les équations qui permettent de décrire la génération du champ gravitationnel par une source matérielle; ces équations relativistes sont les Equations d'Einstein; elles se réduisent à la limite non relativiste à l'équation de Poisson.

De façon analogue aux équations de Maxwell qui déterminent comment les champs électrique et magnétique répondent aux courants et charges, les équations d'Einstein montrent comment la métrique de l'espace-temps répond à l'énergie et l'impulsion de la source matérielle. De plus, pour être manifestement covariantes, elles doivent être exprimées sous forme tensorielle.

\subsubsection{De la Loi de Newton universelle aux Equations d'Einstein}

Dans la théorie de Newton, le champ de gravitation $\overrightarrow{g}=-\overrightarrow{\nabla}\phi$ est définit par la distribution de masse conformément à l'équation de Poisson (\ref{eq de poisson}), le terme de gauche fait intervenir le potentiel par ses dérivées secondes, alors que le terme à droite est proportionnel à la densité de masse $\rho_{\text{\tiny{m}}}$. En relativité générale, le potentiel gravitationnel est remplacé par les composantes de la métrique\footnote{voir \ref{metrique 00 et phi}} $g_{\mu\nu}$, alors que la densité doit être remplacée par le tenseur énergie-impulsion $T_{\mu\nu}$, dont la divergence nulle exprime la conservation de l'énergie et l'impulsion de la distribution matérielle qui génère le champ gravitationnel.

De façon analogue à l'équation de Poisson, les Equations d'Einstein doivent comporter un membre sous forme d'un tenseur purement géométrique $K_{\mu\nu}$ et un autre membre proportionnel au tenseur énergie-impulsion $T_{\mu\nu}$ de telle sorte à avoir le profil suivant \cite{Carroll1}
\begin{equation}\label{profil einstein eq}
   \overrightarrow{\nabla}^{2}\left(g_{\mu\nu}\right)\propto T_{\mu\nu}.
\end{equation}
En particulier, le calcul du laplacien de la composante $g_{00}$ qui figure dans (\ref{metrique 00 et phi}), compte tenu de l'équation de Poisson, donne
\begin{eqnarray*}
    \overrightarrow{\nabla}^{2}g_{00}&=&\overrightarrow{\nabla}^{2}\left(1+\frac{2\phi}{c^{2}}\right)\\
                                     &=&\frac{2}{c^{2}}\left(\overrightarrow{\nabla}^{2}\phi\right)\\
                                     &=&\frac{2}{c^{2}}\left(4\pi G\,\rho_{\text{\tiny{m}}}\right)
\end{eqnarray*}
En supposant que $T_{00}$ du tenseur énergie-impulsion représente la densité d'énergie $\rho_{\text{\tiny{m}}}\,c^{2}$, dans ce cas nous aboutissons à
\begin{equation}\label{g00 proportion t00}
\overrightarrow{\nabla}^{2}g_{00}=\frac{8\pi G}{c^{4}}\,T_{00}.
\end{equation}

En plus d'être un tenseur qui s'exprime en fonction des secondes dérivées des composantes de la métrique, le tenseur géométrique $K_{\mu\nu}$ doit aussi avoir une divergence nulle $D_{\mu}K^{\mu\nu}=0$. Le candidat qui répond à ces deux exigences est le tenseur d'Einstein (\ref{tenseur einstein chap rappel})
\begin{equation}\label{tenseur einstein espace temps}
    G_{\mu\nu}=R_{\mu\nu}-\frac{1}{2}\,g_{\mu\nu}\,R
\end{equation}
où $\mu, \nu=0,1,2,3.$

Dans ce cas, les équations d'Einstein possèdent la forme suivante
\begin{equation}\label{eq einstein espace temps}
    R_{\mu\nu}-\frac{1}{2}\,g_{\mu\nu}\,R=\chi\,T_{\mu\nu},
\end{equation}
avec une constante d'Einstein qui doit être égale à $\chi=\frac{8\pi G}{c^{4}}$ pour pouvoir retrouver l'équation de Poisson, à la limite non relativiste.

Il faut savoir que le choix du tenseur géométrique $K_{\mu\nu}$ n'est pas unique. En effet, à cause du théorème de Ricci $D_{\mu}g^{\mu\nu}=0$ et de la symétrie $g_{\mu\nu}=g_{\nu\mu}$, Einstein et Eddington ont ajouté le terme $\Lambda\,g_{\mu\nu}$ au tenseur $S_{\mu\nu}$, de sorte à avoir l'équation
\begin{equation}\label{eq einstein espace temps}
    R_{\mu\nu}-\frac{1}{2}\,g_{\mu\nu}\,R+\Lambda\,g_{\mu\nu}=\frac{8\pi G}{c^{4}}\,T_{\mu\nu}.
\end{equation}
L'usage de ce nouveau terme ne fait pas l'unanimité des physiciens\footnote{même Einstein s'est montré reticent après l'avoir introduit, car il conduit à l'apparition d'une gravité repulsive \cite{Hobson}}; la valeur de la constante cosmologique $\Lambda$ est très faible et n'intervient que pour tenir compte de quelques problèmes cosmologiques. Touts les tests classiques des équations d'Einstein (Avance du périhélie des planètes, Déflexion des rayons lumineux par des objets massifs, Décalage des raies spectrales, $\cdots$) se font quand $\Lambda=0$.

\subsubsection{Equations du champ gravitationnel dans le vide}
Une région dans laquelle le tenseur énergie-impulsion est nul $T_{\mu\nu}=0$ est appelée vide. Pour obtenir les équations d'Einstein dans le vide, réecrivons d'abord les équations d'Einstein (\ref{eq einstein espace temps}) sous une forme équivalente par une contraction par $g^{\mu\nu}$
\begin{eqnarray}
  g^{\mu\nu}R_{\mu\nu}-\frac{1}{2}\,g^{\mu\nu}g_{\mu\nu}\,R &=& \chi\,g^{\mu\nu}T_{\mu\nu}\nonumber\\
  R-\frac{4}{2}\,R &=& \chi\,T. \nonumber
\end{eqnarray}
Dans ce cas, la courbure scalaire est donnée par
\begin{equation}\label{ricci fonction trace}
    R = -\chi\,T,
\end{equation}
où $T=T^{\mu}_{\mu}$ représente la trace du tenseur énergie-impulsion.

En remplaçant (\ref{ricci fonction trace}) dans (\ref{eq einstein espace temps}) nous obtenons les équations d'Einstein suivantes
\begin{equation}\label{einstein eq equivalente}
    R_{\mu\nu}=\chi\left(T_{\mu\nu}-\frac{1}{2}\,g_{\mu\nu}\,T\right).
\end{equation}

Dans le cas extérieur à la source $T_{\mu\nu}=0$, les équations d'Einstein (\ref{einstein eq equivalente}) se mettent finalement sous la forme
\begin{equation}\label{einstein eq vide}
    R_{\mu\nu}=0.
\end{equation}

\subsubsection{Tenseur énergie-impulsion}
Pour les équations d'Einstein, le tenseur énergie-impulsion joue un rôle analogue du quadrivecteur densité de courant $J^{\mu}(c\,\rho,\overrightarrow{j})$, pour les équations de Maxwell. En effet, soit l'équation de Maxwell
$$\partial_{\mu}F^{\mu\nu}=\mu_{0}J^{\nu},$$
où $F^{\mu\nu}=\partial^{\mu}A^{\nu}-\partial^{\nu}A^{\mu}=-F^{\nu\mu}$ est le champ électromagnétique. En appliquant à cette équation la dérivée $\partial_{\nu}$, de telle sorte à exploiter l'antisymétrie de $F^{\mu\nu}$, nous aboutissons à l'équation de continuité
$$\partial_{\nu}J^{\nu}=0.$$
De façon analogue, en appliquant la dérivée covariante $D_{\mu}$ sur les deux membres de l'équation d'Einstein nous avons aussi $D_{\mu}T^{\mu\nu}=0$.

Le tenseur énergie-impulsion $T_{\mu\nu}$ représente  la caractérisation éssentielle d'un milieu matériel dans un espace-temps, dont les composantes, dans un référetiel minkowskien, ont la siginfication physique suivante \cite{Linet}
\begin{eqnarray}
  T^{\mu\nu} = \left(
                 \begin{tabular}{c|c}
                   $T^{00}$ & $T^{oj}$\\
                   \hline
                   $T^{i0}$ & $T^{ij}$
                 \end{tabular}
               \right) =\left(
                 \begin{tabular}{c|c}
                   Densité d'énergie & (Flux d'énergie)/c\\
                   \hline
                   \begin{tabular}{c}
                                                                     c$\times$(Densité de quantité \\
                                                                      de mouvement) \\
                                                                   \end{tabular} & \begin{tabular}{c}
                                                                     Flux de quantité \\
                                                                      de mouvement \\
                                                                   \end{tabular}
                 \end{tabular}
               \right)
\end{eqnarray}
Dans un référentiel galiléen, la conservation du tenseur énergie-impulsion s'exprime en fonction d'une dérivée ordinaire
$\partial_{\mu}T^{\mu\nu}=0$. Cette condition exprime, à la fois, la conservation d'énergie $\partial_{t}T^{00}+c\,\partial_{j}T^{0j}=0$ et la conservation de la quantité de mouvement $\partial_{t}T^{i0}+c\,\partial_{j}T^{ij}=0$.

Pour une distribution spatialement bornée de fluide parfait, le tenseur énergie-impulsion est donnée par
\begin{equation}\label{t mu nu parfait}
    T^{\mu\nu}=(\rho_{\text{\tiny{m}}}+p)U^{\mu}\,U^{\nu}+p\,g^{\mu\nu},
\end{equation}
où $U^{\mu}=dx^{\mu}/d\tau=(\gamma_{u}c,\gamma_{u}\overrightarrow{u})$ représente le quadrivecteur vitesse du fluide et $p$ la pression interne isotrope.

Considérons une distribution de masses ponctuelles identiques sans interactions mutuelles, communément appelé nuage de poussière en littérature. Dans ca cas, les pressions sont négligés de sorte à avoir
\begin{equation}\label{t mu nu dust}
    T^{\mu\nu}=\rho_{\text{\tiny{m}}}\,U^{\mu}\,U^{\nu}.
\end{equation}
Dans un référentiel galiléen, nous avons les composantes suivantes \cite{Hobson}
\begin{eqnarray}
  T^{00} &=& \rho_{\text{\tiny{m}}}\,U^{0}\,U^{0} = (\gamma_{u})^{2}\rho_{\text{\tiny{m}}}c^{2} \label{t00 chap1} \\
  T^{0i} &=& \rho_{\text{\tiny{m}}}\,U^{0}\,U^{i} = (\gamma_{u})^{2}\rho_{\text{\tiny{m}}}c\,u^{i} \label{t0i chap1}\\
  T^{ij} &=& \rho_{\text{\tiny{m}}}\,U^{i}\,U^{j} = (\gamma_{u})^{2}\rho_{\text{\tiny{m}}}u^{i}\,u^{j}\label{tij chap1}
\end{eqnarray}
avec $\overrightarrow{u}(u^{1},u^{2},u^{3})$ et $\gamma_{u}=1$ dans le repère où le fluide est au repos.

Nous avons $T^{00}\sim c^{2}$, $T^{0i}\sim cu^{i}$, $T^{ij}\sim u^{i}u^{j}$, ainsi dans l'approximation des vitesses faibles, la densité d'énergie est la composante prépondérante $T^{00} \gg T^{0i} \gg T^{ij}$.

Soulignons finalement le fait que le tenseur énergie-impulsion ne peut pas rendre compte des effets quantiques qui interviennent au niveau atomique car il a été défini de façon à permettre la description de la matière à une échelle macroscopique.

\subsubsection{Limite newtonienne pour les Equations d'Einstein}
L'approximation newtonienne s'obtient à la limite non relativiste dans laquelle le champ gravitationnel \textbf{stationnaire} ($\partial_{0}g_{\mu\nu}=0$) est si \textbf{faible} qu'il est vu comme une perturbation, \textbf{au premier ordre}, de la métrique minkowskienne ($g_{\mu\nu}=\eta_{\mu\nu}+h_{\mu\nu}$ avec $|h_{\mu\nu}|\ll 1$). Dans ce cas, les particules répondent à ce champ avec des \textbf{vitesses faible} par rapport à la vitesse de la lumière ($v\ll c$).

Le calcul préliminaire qui a conduit à (\ref{g00 proportion t00}), nous suggère de ne prendre que la composante $00$ du tenseur de Ricci qui figure dans l'équation d'Einstein (\ref{einstein eq equivalente}) pour avoir
\begin{equation}\label{prepoderante r00 ricci enj impul}
    R_{00}=\chi\left(T_{00}-\frac{1}{2}\,g_{00}\,T\right).
\end{equation}

En premier lieu, calculons $R_{00}$ à la limite non-relativite. Pour ce faire, contractons le tenseur de riemann (\ref{tenseur riemann christoffel}) de sorte à avoir
\begin{equation}
    R_{00}=\partial_{\mu}\Gamma^{\mu}_{00}-\partial_{0}\Gamma^{\mu}_{0\mu}+\Gamma^{\mu}_{\nu\mu}\Gamma^{\nu}_{00}-\Gamma^{\nu}_{0\mu}\Gamma^{\mu}_{\nu0}.
\end{equation}
Dans l'approximation newtonienne du champ faible, pour laquelle\footnote{En effet, nous avons au premier ordre de perturbation
\begin{eqnarray*}
  g^{\mu\nu}g_{\nu\sigma} &\approx& \left(\eta^{\mu\nu}-h^{\mu\nu}\right)\left(\eta_{\nu\sigma}+h_{\nu\sigma}\right)
                          = \eta^{\mu\nu}\eta_{\nu\sigma}-\eta_{\nu\sigma}h^{\mu\nu}+\eta^{\mu\nu}h_{\nu\sigma}+\underbrace{h^{\mu\nu}h_{\nu\sigma}}_{\text{négligé}}
                          \approx \delta^{\mu}_{\sigma}-h^{\mu}_{\;\sigma}+h^{\mu}_{\;\sigma}
                          \approx \delta^{\mu}_{\sigma}
\end{eqnarray*}}
$g^{\mu\nu}\approx\eta^{\mu\nu}-h^{\mu\nu}$, les symboles de Christoffel stationnaires ($\partial_{0}\Gamma^{\mu}_{0\mu}=\partial_{0}\Gamma^{0}_{00}=0$) sont définis au premier ordre\footnote{D'après (\ref{importantegij}) nous avons au premier ordre de perturbation $\Gamma^{\rho}_{\mu\nu} \approx \frac{1}{2}\,\eta^{\rho\lambda} \left(\partial_{\mu}h_{\nu\lambda}+\partial_{\nu}h_{\lambda\mu}-\partial_{\lambda}h_{\mu\nu}\right)$} de telle sorte à pouvoir négliger leurs produits d'ordre deux.  Dans ce cas, nous avons au premier ordre de perturbation
\begin{equation}\label{terme gauche ricci}
    R_{00}\approx\partial_{i}\Gamma^{i}_{00}\approx\frac{1}{2}\,\overrightarrow{\nabla}^{2}g_{00},
\end{equation}
où $i=1,2,3$.

D'autre part, en supposant que la densité d'énergie $T_{00}=\rho c^{2}$ soit la composante prépondérante du tenseur énergie-impulsion de sorte que   $T=T_{00}-T_{11}-T_{22}-T_{33}\approx T_{00}$, le membre droit de (\ref{prepoderante r00 ricci enj impul}) se réduit à
\begin{equation}\label{terme droit ricci}
    \chi\left(T_{00}-g_{00}\,T/2\right)\approx \chi\left(T_{00}-T_{00}/2\right)=\chi\,T_{00}/2
\end{equation}

En tenant compte de (\ref{terme droit ricci}) et (\ref{terme gauche ricci}), l'équation (\ref{prepoderante r00 ricci enj impul}) se réduit à
\begin{equation}\label{laplacien g00}
    \overrightarrow{\nabla}^{2}g_{00}=\chi\,T_{00}.
\end{equation}
La comparaison de (\ref{laplacien g00}) avec (\ref{g00 proportion t00}) permet finalement  de fixer la valeur de la constante d'Einstein à
\begin{equation}\label{constante einstein}
    \chi=\frac{8\pi G}{c^{4}},
\end{equation}
tel que $G$ est la constante de gravitation de Newton et $c$ est la vitesse de la lumière.

\subsubsection{Trajectoires de particules ponctuelles libres}
Considérons un ensemble de particules libres dans un espace-temps courbe de telle sorte qu'ils n'entrent pas en collisions; un tel système peut être décrit par le tenseur énergie-impulsion (\ref{t mu nu dust}). Compte tenu de la condition $D_{\nu}T^{\mu\nu}=0$, nous avons d'une part\footnote{pour alléger l'écriture la densité de masse est écrite $\rho$}
\begin{eqnarray}\label{cond 1 deriv cov}
  D_{\nu}\left(\rho\,U^{\mu}\,U^{\nu}\right) &=& \big[D_{\nu}\left(\rho\,U^{\nu}\right)\big]U^{\mu}+\rho\,U^{\nu}\left(D_{\nu}U^{\mu}\right) = 0
\end{eqnarray}
D'autre part, en effectuant la dérivée covariante de la condition de normalisation\footnote{La condition $ds^{2}=g_{\mu\nu}dx^{\mu}dx^{\nu}$ implique que $1 = g_{\mu\nu}\,\frac{dx^{\mu}}{ds}\,\frac{dx^{\nu}}{ds} =\frac{1}{c^{2}}\,g_{\mu\nu}\,\frac{dx^{\mu}}{d\tau}\,\frac{dx^{\nu}}{d\tau} = \frac{1}{c^{2}}\,g_{\mu\nu}\,U^{\mu}\,U^{\nu}= \frac{1}{c^{2}}\,U^{\mu}\,U_{\mu}$} $U^{\mu}\,U_{\mu}=c^{2}$, nous avons
\begin{eqnarray*}
  \left(D_{\nu}U^{\mu}\right)U_{\mu}+U^{\mu}\left(D_{\nu}U_{\mu}\right) &=& 0 \\
  2\left(D_{\nu}U^{\mu}\right)U_{\mu} &=& 0,
\end{eqnarray*}
ou encore finalement
\begin{equation}\label{cond 2 deriv cov}
    \left(D_{\nu}U^{\mu}\right)U_{\mu} = 0.
\end{equation}
Pour pouvoir utiliser la condition de normalisation ainsi que (\ref{cond 2 deriv cov}), contractons la relation (\ref{cond 1 deriv cov}) par $U_{\mu}$ de sorte à avoir
\begin{eqnarray*}
0  &=&  \big[D_{\nu}\left(\rho\,U^{\nu}\right)\big]\left(U^{\mu}U_{\mu}\right)+\rho\,U^{\nu}\big[\left(D_{\nu}U^{\mu}\right)U_{\mu}\big]\\
0  &=&  c^{2}\;D_{\nu}\left(\rho\,U^{\nu}\right)
\end{eqnarray*}
ou encore
\begin{equation}\label{cond 3 deriv cov}
    D_{\nu}\left(\rho\,U^{\nu}\right)=0.
\end{equation}
En remplaçant (\ref{cond 3 deriv cov}) dans (\ref{cond 1 deriv cov}) nous aboutissons finalement à l'équation \cite{Hobson}
\begin{equation}\label{equ geodesique camoufle}
   U^{\nu}(D_{\nu}U^{\mu}) = 0,
\end{equation}
car $\rho$ est non nulle. Cette dernière équation conduit à l'équation des géodésiques\\ $DU^{\mu}=\left(D_{\nu}U^{\mu}\right)dx^{\nu}=0$.

Le calcul précédent révèle l'une des différences fondamentales entre la gravitation est l'électro-magnétisme. En effet, les équations d'Einstein contiennent en fait, à la fois, les équations du champ gravitationnel lui-même ainsi que l'équation de mouvement des masses qui créent ce champ, contrairement aux équations de Maxwell qui ne conduisent qu'à l'équation de conservation de la charge totale $\partial_{\mu}J^{\mu}=\frac{\partial \rho}{\partial t}+\overrightarrow{\nabla}.\overrightarrow{j}=0$, et ne donne ainsi aucune information sur le mouvement des charges qui créent le champ électromagnétique \cite{landau}. Autrement dit, en électromagnétisme, l'équation de mouvement de charges électriques sous l'action de la force de Lorentz n'est pas une conséquence des équations de Maxwell, contrairement à ce qui se passe pour la gravitation où les équations de mouvement de particules test selon des géodésiques sont issues des équations d'Einstein elles-même\footnote{Ou de façon plus précise, les équations des géodésiques découlent de la conservation du tenseur énergie-impulsion. Si on postule les équations d'Einstein $G_{\mu\nu}=\chi\,T_{\mu\nu}$ avec l'exigence $D^{\mu}G_{\mu\nu}=0$, nous pouvons voir la conservation du tenseur énergie-impulsion $D^{\mu}T_{\mu\nu}=0$ comme une conséquence des équations d'Einstein.}.

Une autre différence fondamentale entre les interactions électromagnétique et gravitationnelle est que les équations de Maxwell sont linéaires alors que les équations d'Einstein sont non linéaires. Pour cette raison, le principe de superposition des champs applicable pour le champ électro-magnétique ne s'applique plus pour le champ gravitationnel, sauf dans le domaine linéaire des champs faibles où la métrique est vue comme une perturbation, au premier ordre, de la métrique de Minkowski $g_{\mu\nu}=\eta_{\mu\nu}+h_{\mu\nu}$ avec $|h_{\mu\nu}|\ll 1$.

\subsubsection{Source et particule d'épreuve}

Pour l'interaction gravitationnelle, en se donnant une distribution particulière de matière "dense", la courbure de l'espace-temps est fixée par les équations du champ $g_{\mu\nu}$ à l'extérieur de cette source. Le mouvement des particules test est aussi fixé, car la réponse de ces particules d'épreuves à cette courbure se fait en suivant des géodésiques. Cette situation est bien résumée par Wheeler \cite{Ryder} "La matière dicte à l'espace comment se courber alors que l'espace dicte à la matière comment se déplacer". Bien sûr il faut savoir que cette situation est bien particulière car dans le cas où on n'a plus affaire à des particules test, la perturbation apportée par ces particules au champ gravitationnel de la source matérielle ne peut plus être négligée. De plus, il va falloir tenir compte des effets de l'interaction de ses particules avec leurs propres champ gravitationnel (self-interaction), car dans ce cas le mouvement ne se fait plus selon des géodésiques \cite{wald1}-\cite{wald4}. Cette situation peut avoir lieu quand la masse totale de la première distribution matérielle est du même ordre de grandeur que la masse de la seconde distribution matérielle, de sorte que la distinction "source" et "particule test" n'a plus lieu d'exister.

\subsection{La solution de Schwarzschild}
La résolution des équation d'Einstein se fait par la détermination des composantes de la métrique $g_{\mu\nu}$ en se donnant une certaine distribution matérielle. La recherche de solutions à l'intérieur de la source nécessite l'utilisation d'une forme particulière du tenseur énergie-impulsion, par contre la détermination de solutions à l'extérieur de cette distribution matérielle se fait en annulant toutes les composantes de $T_{\mu\nu}$. La non-linéarité des équations d'Einstein complique considérablement toute détermination de solutions analytiques, encourageant ainsi le recours à la recherche de solutions pour des situations bien particulières. La première résolution, proposée dans ce sens, a été effectuée par Karl Schwarzschild en 1916.

\subsubsection{Symétries du problème et forme générale de l'invariant relativiste}
En se donnant une distribution matérielle \textbf{statique} dotée d'une \textbf{symétrie sphérique}, il se propose de déterminer la métrique de l'espace-temps à l'extérieur de cette source. La démarche utilisée consiste donc à résoudre l'équation d'Einstein dans le cas extérieur $R_{\mu\nu}=0$, de telle sorte à exploiter les symétries du problème pour restreindre la forme de la solution $g_{\mu\nu}$.

Pour une distribution de masse statique, il est possible de montrer, qu'en première approximation, le potentiel newtonien en un point située très loin du centre de masse est proportionnel l'inverse de cette distance $\phi\approx-G(\int\rho_{m}\,dV)/\textsc{r}$. En effet, en adoptant un repérage par rapport au centre de masse $\int \rho_{m}\overrightarrow{r}\,dV=\overrightarrow{0}$, de telle sorte à annuler identiquement le moment dipolaire tout en négligeant les moments multipolaires d'ordres supérieurs, on montre que le champ gravitationnel extérieur d'une telle distribution matérielle statique est dotée d'une symétrie sphérique à l'ordre le plus bas \cite{landau}.

Pour un champ de gravitation extérieur isotrope et stationnaire, la forme des composantes de la métrique est soumise à quelques restrictions. L'isotropie s'exprime par le fait que le carré de l'intervalle $ds^{2}=g_{\mu\nu}dx^{\mu}dx^{\nu}$ soit invariant pour tous les points équidistants au centre, de sorte que sa dependence spatiale ne se manifeste qu'à travers les termes $\textsc{r}^{2}$, $\textsc{r}\,d\textsc{r}$ et $d\textsc{r}^{2}$ \cite{Hobson}. Le régime stationnaire implique des composantes de métrique $g_{\mu\nu}$ sans dépendance explicite en temps, de plus l'invariance de $ds^{2}$ par renversement du temps $x^{0}\rightarrow-x^{0}$ conduit à l'annulation des composantes $g_{0i}=g_{0i}=0$ \cite{Ryder}.

Pour déterminer la forme la plus générale des composantes de la métrique qui satisfait toutes ces exigences, plaçons-nous, d'abord dans le cas d'une métrique minkowskienne plate\\ $ds^{2}=(c\,dt)^{2}-dx^{2}-dy^{2}-dz^{2}$ qui s'exprime en coordonnées sphériques
\begin{eqnarray*}
  x = \textsc{r}\,\sin\theta\,\cos\phi \hspace{0.3cm},\hspace{0.3cm}
  y = \textsc{r}\,\sin\theta\,\sin\phi \hspace{0.3cm},\hspace{0.3cm}
  z = \textsc{r}\,\cos\theta
\end{eqnarray*}
sous la forme
\begin{equation}\label{minkowski spherique}
    ds^{2}=(c\,dt)^{2}-d\textsc{r}^{2}-\textsc{r}^{2}\left(d\theta^{2}+\sin^{2}\theta\,d\phi^{2}\right).
\end{equation}
La relation (\ref{minkowski spherique}) suggère d'adopter, en coordonnées sphériques $x^{\mu}(c\,t,\textsc{r},\theta,\phi)$, la forme la plus générale suivante \cite{Groen}
\begin{equation}\label{metrique isotrope et statique forme general}
    ds^{2}=c^{2}\,F(\textsc{r})\,dt^{2}-E(\textsc{r})\,d\textsc{r}^{2}-H(\textsc{r})\,\textsc{r}^{2}\left(d\theta^{2}+\sin^{2}\theta\,d\phi^{2}\right).
\end{equation}
d'une métrique isotrope et statique , tel que $F(\textsc{r})$, $E(\textsc{r})$ et $H(\textsc{r})$ sont des fonctions, pour l'instant arbitraires, mais qu'il va falloir déterminer plus tard.

Dans le but de réduire le nombre de fonctions arbitraires à deux, introduisons la nouvelle variable radiale $r=\textsc{r}\sqrt{H(\textsc{r})}$, avec laquelle la relation (\ref{metrique isotrope et statique forme general}) prend la nouvelle forme suivante
\begin{equation}\label{metrique isotrope et statique forme general nouvelle variable}
    ds^{2}=c^{2}\,A(r)\,dt^{2}-B(r)\,dr^{2}-r^{2}\left(d\theta^{2}+\sin^{2}\theta\,d\phi^{2}\right),
\end{equation}
où $A(r)$ et $B(r)$ sont les nouvelles fonctions arbitraires.

\subsubsection{Exploitation des équations d'Einstein dans le vide}
Après avoir déterminé, en coordonnées de Schwarzschild $x^{\mu}(c\,t,r,\theta,\phi)$, la plus générale de l'expression de l'invariant relativiste dans un espace isotrope et stationnaire, il reste à présent de déterminer les deux fonctions $A(r)$ et $B(r)$ figurant dans (\ref{metrique isotrope et statique forme general nouvelle variable}) par l'exploitation des équations d'Einstein en dehors de la source $R_{\mu\nu}=0$.

La démarche de calcul est résumée comme suit
\begin{enumerate}
  \item Déterminer les composantes du tenseur métrique.
  \item Calculer les symboles de Christoffel non nuls par (\ref{importantegij}).
  \item Calculer les composantes non nulles du tenseur de Ricci (\ref{tenseur de Ricci chap rappel}) en contractant le tenseur de courbure (\ref{tenseur riemann christoffel}).
  \item Résoudre les équations d'Einstein $R_{\mu\nu}=0$, de telle sorte à assurer la convergence asymptotique à l'infini.
\end{enumerate}

D'après (\ref{metrique isotrope et statique forme general nouvelle variable}), il est clair que les seules composantes non nulles de la métrique sont
\begin{eqnarray}
  g_{tt} &=& A(r) \\
  g_{rr} &=& -B(r) \\
  g_{\theta\theta} &=& -r^{2} \\
  g_{\phi\phi} &=& -r^{2}\,\sin^{2}\theta
\end{eqnarray}
alors que les composantes de la métrique inverse sont plutôt données par
\begin{eqnarray}
  g^{tt} &=& 1/A(r) \\
  g^{rr} &=& -1/B(r) \\
  g^{\theta\theta} &=& -1/r^{2} \\
  g^{\phi\phi} &=& -1/(r^{2}\,\sin^{2}\theta)
\end{eqnarray}
du fait que $g_{\mu\nu}g^{\nu\sigma}=\delta_{\mu}^{\sigma}$.

D'après (\ref{importantegij}), les neuf symboles de Christoffel non nuls sont donnés par \cite{BoudaRG}

\begin{eqnarray}
  \begin{tabular}{lll}
  $\Gamma_{tt}^{r} = A^{'}/(2B)$\hspace{2cm} & $\Gamma_{r\theta}^{\theta} = 1/r$\hspace{2cm} & $\Gamma_{\phi\phi}^{\theta} = -\sin\theta \cos\theta$ \\\\
  $\Gamma_{rr}^{r} = B^{'}/(2B)$ & $\Gamma_{r\phi}^{\phi} = 1/r$ & $\Gamma_{\phi\theta}^{\phi} = \cos\theta/\sin\theta$ \\\\
  $\Gamma_{tr}^{t} = A^{'}/(2A)$ & $\Gamma_{\theta\theta}^{r} = -r/B$ & $\Gamma_{\phi\phi}^{r} = -r\,\sin^{2}\theta/B$ \\
\end{tabular}
\end{eqnarray}
où $A^{'}$ et $B^{'}$ désignent les dérivées respectives de $A$ et $B$ par rapport à $r$.

La contraction du tenseur de courbure (\ref{tenseur riemann christoffel}) permet de calculer les composantes non nulles du tenseur de Ricci
 \begin{equation}
    R_{\mu\nu}=R^{\,\sigma}_{\;\;\mu\sigma\nu}=\partial_{\sigma}\Gamma_{\mu\nu}^{\sigma}-\partial_{\nu}\Gamma_{\mu\sigma}^{\sigma}+\Gamma_{\mu\nu}^{\ell}\,\Gamma_{\ell\sigma}^{\sigma}
    -\Gamma_{\mu\sigma}^{\ell}\,\Gamma_{\ell\nu}^{\sigma},
\end{equation}
de telle sorte à aboutir aux équations d'Einstein suivantes \cite{BoudaRG}
\begin{eqnarray}
  R_{tt} &=& \frac{A^{''}}{2B}-\frac{A^{'}}{4B}\left(\frac{A^{'}}{A}+\frac{B^{'}}{B}\right)+\frac{1}{r}\,\frac{A^{'}}{B}=0 \label{ricci 1}\\
  R_{rr} &=& -\frac{A^{''}}{2A}+\frac{A^{'}}{4A}\left(\frac{A^{'}}{A}+\frac{B^{'}}{B}\right)+\frac{1}{r}\,\frac{B^{'}}{B}=0\label{ricci 2} \\
  R_{\theta\theta} &=& 1+\frac{r}{2B}\left(\frac{B^{'}}{B}-\frac{A^{'}}{A}\right)-\frac{1}{B}=0 \label{ricci 3}\\\nonumber \\
  R_{\phi\phi} &=& \sin^{2}\theta\,R_{\theta\theta}=0 \label{ricci 4}
\end{eqnarray}

Commençons à présent l'exploitation des quatre équations d'Einstein précédentes (de \ref{ricci 1} à \ref{ricci 4}). Notons que l'équation (\ref{ricci 4}) ne contient aucune nouvelle information par rapport à (\ref{ricci 3}), réduisant ainsi le nombre d'équations d'Einstein utiles à trois. Au lieu d'utiliser les deux équations $R_{tt}=0$ et $R_{rr}=0$ de manière individuelle, calculons plutôt l'expression
\begin{equation*}
    0=\frac{R_{rr}}{B}+\frac{R_{tt}}{A}=\frac{1}{rB}\left(\frac{A^{'}}{A}+\frac{B^{'}}{B}\right),
\end{equation*}
dont la résolution par séparation de variables donne
\begin{equation}
    AB=K,
\end{equation}
où $K$ est une certaine constante à déterminer moyennant une condition aux limites. On fixe la valeur de cette constante, en exigeant que la métrique de l'espace-temps tend à devenir minkowskienne très loin de la source; autrement dit, pour que la métrique (\ref{metrique isotrope et statique forme general nouvelle variable}) devienne sous la forme (\ref{minkowski spherique}) à l'infini, nous imposons que
\begin{eqnarray}
  \left\{
    \begin{array}{ll}
      \displaystyle\lim_{r\rightarrow +\infty}A(r) = +1 \\\\
      \displaystyle\lim_{r\rightarrow +\infty}B(r) = +1
    \end{array}
  \right.
\end{eqnarray}
de sorte que $K=1$. Dans ce cas, nous avons donc la relation
\begin{equation}\label{relation entre A et B}
    B(r)=1/A(r).
\end{equation}
A présent, il ne reste plus qu'à exploiter la dernière équation d'Einstein $R_{\theta\theta}=0$, qui donne
\begin{equation}
    R_{\theta\theta}=1-A-rA^{'}=1-\frac{d}{dr}\left(rA\right)=0,
\end{equation}
compte de (\ref{relation entre A et B}). La résolution de l'équation précédente permet d'avoir la solution
\begin{equation}\label{solution A}
    A(r)=1+k/r,
\end{equation}
où $k$ est une constante d'intégration.

Il ne reste plus qu'à exiger une identification, quand $r\rightarrow +\infty$, entre la composante $g_{tt}$ donnée par (\ref{solution A}) et la composante (\ref{metrique 00 et phi}) écrite en fonction du potentiel newtonien $\phi(r)=-GM/r$ de telle sorte à pouvoir écrire la condition
\begin{equation}
   g_{tt}=1+\frac{k}{r}=1+\frac{2(-GM/r)}{c^{2}},
\end{equation}
qui fixe finalement la valeur de la constante d'intégration à $k=-2GM/c^{2}$. Ainsi, compte tenu des relations (\ref{solution A}) et (\ref{relation entre A et B}), nous obtenons
\begin{eqnarray}
  A(r) &=& \left(1-\frac{2GM}{c^{2}r}\right)\label{A} \\
  B(r) &=& \left(1-\frac{2GM}{c^{2}r}\right)^{-1}\label{B}
\end{eqnarray}
Rappelons que $M$ représente la masse de la distribution de masse sphérique qui crée le champ de gravitation.

La solution de Schwarzschild s'obtient en remplaçant les deux fonctions données par (\ref{A}) et (\ref{B}) dans (\ref{metrique isotrope et statique forme general nouvelle variable}) pour avoir finalement
\begin{equation}\label{metrique isotrope et statique forme general nouvelle variable shwrzscild}
    ds^{2}=c^{2}\left(1-\frac{2GM}{c^{2}r}\right)dt^{2}-\left(1-\frac{2GM}{c^{2}r}\right)^{-1}\,dr^{2}-r^{2}\left(d\theta^{2}+\sin^{2}\theta\,d\phi^{2}\right).
\end{equation}

Le même résultat (\ref{metrique isotrope et statique forme general nouvelle variable shwrzscild}) est obtenu même si la condition de métrique statique $\partial_{0}g_{\mu\nu}$ dans (\ref{metrique isotrope et statique forme general nouvelle variable}) est relaxée, i.e $A(r,t)$ et $B(r,t)$; autrement dit la solution est nécessairement statique (Théorème de Birkhoff). Une conséquence de ce résultat est que pour une étoile s'effondrant de façon radiale en trou noir, de sorte que sa vitesse est dotée d'une symétrie sphérique, elle aura toujours  une solution de Schwarzschild comme métrique extérieure \cite{Beuthe}.

\section{Approche de C.C.Barros}
\subsection{Introduction}

Dans un travail récent, C.C.Barros \cite{Barros1,Barros2,Barros3} suppose que, d'une manière similaire à la gravité, les interactions non gravitationnelles  peuvent affecter les propriétés de l'espace-temps. A partir de cette hypothèse, apparemment non justifiée, il arrive à décrire l'atome d'hydrogène de façon tout à fait inédite. En effet, en raison de la symétrie sphérique du problème, il adopte une métrique similaire à celle de Schwarzschild

\begin{equation}\label{metrique etoile}
    ds^{2}=c^{2}\xi(r)dt^{2}-r^{2}(d\theta^{2}+\sin^{2}\theta
    d\phi^{2})-[\xi(r)]^{-1}dr^{2},
\end{equation}
et de façon analogue à ce qui se fait en Relativité Générale (RG), il incorpore l'interaction électrostatique $V(r)$ "proton-électron" dans la métrique de l'espace à travers la relation \cite{Barros1}

\begin{equation}\label{xi}
    \xi(r)=\left(1+\frac{V(r)}{m_{0}c^{2}}\right)^{2},
\end{equation}
avec $m_{0}c^{2}$ et $V(r)$ sont respectivement l'énergie au repos et l'énergie potentielle de l'électron.

Pour ce faire, Barros utilise l'équation de Dirac d'une particule libre dans laquelle il substitue les dérivées ordinaires par des dérivées covariantes, construites sur la base de la métrique de Schwarzschild. C'est ainsi qu'il arrive à reproduire le spectre de l'atome d'hydrogène.


\subsection{Principe}
L'idée fondamentale consiste à vouloir décrire une particule soumise à un potentiel non gravitationnel qui affecterait la métrique de l'espace-temps. Comme première étape, un système à symétrie sphérique est considéré, mais l'idée de base pourrait être généralisée à des systèmes soumis à des potentiels arbitraires.

\subsubsection{Quadrivecteur énergie-impulsion et invariant relativiste}
Ainsi,  en utilisant une métrique similaire à celle de Schwarzschild (\ref{metrique etoile}), et à partir de l'expression de l'action d'une particule dont l'interaction, non gravitationnelle, est absorbée dans la métrique \cite{landau}
\begin{eqnarray}
   S\equiv\int L\,dt
    =-m_{0}c\int ds
    =-(m_{0}c^{2})\int \frac{dt}{\gamma_{s}},\nonumber
\end{eqnarray}
il est possible de construire le lagrangien,
\begin{eqnarray}\label{lagrangien}
    \hspace*{-1cm}L=\frac{-(m_{0}c^{2})}{\gamma_{s}}&=&-(m_{0}c^{2})\displaystyle\sqrt{\xi(r)-[\xi(r)]^{-1}\bigg(\frac{dr}{cdt}\bigg)^{2}
    -r^{2}\bigg[\bigg(\frac{d\theta}{cdt}\bigg)^{2}+\bigg(\frac{d\phi}{cdt}\bigg)^{2}\sin^{2}\theta\bigg]}.
\end{eqnarray}
Celui-ci permettra de déterminer, d'une part, les principales grandeurs dynamiques relatives à la particule, évoluant dans l'espace courbe, à savoir les impulsions
\begin{eqnarray}\label{composantesimpulsioncov}
    P_{r}&\equiv&\displaystyle\frac{\partial L}{\partial\bigg(\displaystyle \frac{dr}{dt}\bigg)}=m_{0}\gamma_{s}[\xi(r)]^{-1}\frac{dr}{dt}=-P_{1}\\\nonumber\\
    P_{\theta}&\equiv&\displaystyle\frac{\partial L}{\partial\bigg( \displaystyle\frac{d\theta}{dt}\bigg)}=m_{0}\gamma_{s}r^{2}\frac{d\theta}{dt}=-P_{2}\\\nonumber\\
    P_{\phi}&\equiv&\displaystyle\frac{\partial L}{\partial\bigg(\displaystyle \frac{d\phi}{dt}\bigg)}=m_{0}\gamma_{s}r^{2}\sin^{2}\theta\frac{d\phi}{dt}=-P_{3},
\end{eqnarray}
l'énergie
\begin{eqnarray}\label{energie}
    E&=&(m_{0}c^{2})\gamma_{s}\,\xi(r)=cP_{0},
\end{eqnarray}
de sorte à définir le quadrivecteur énergie-impulsion $P_{\mu}=(P_{0},P_{1},P_{2},P_{3})$. Nous avons, d'autre part, l'expression de l'invariant relativiste $P_{\mu}P^{\mu}= g^{\mu\nu}P_{\mu}P_{\nu}=m_{0}^{2}c^{2}$ qui se met sous la forme explicite\footnote{Compte tenu de de la relation (\ref{lienmetriqueavecplatte})}
\begin{equation}\label{invariantrelativiste}
    \displaystyle\frac{E^{2}}{\xi(r)}=\Bigg[\xi(r)P_{r}^{2}
    +\displaystyle\frac{P_{\theta}^{2}}{r^{2}}+\displaystyle\frac{P_{\phi}^{2}}{r^{2}\sin^{2}\theta}\Bigg]c^{2}+m_{0}^{2}c^{4}.
\end{equation}

\subsubsection{Principe de correspondance pour une métrique de Schwarzschild}

Pour pouvoir établir des équations d'ondes quantiques, il faut adapter les principes de correspondance d'énergie et d'impulsion, initialement formulés dans un espace-temps plat de Minkowski
\begin{eqnarray}
    E&\rightarrow&i\hbar\,\frac{\partial}{\partial t},\label{correspondancenrjclassique}\\
    \overrightarrow{p}&\rightarrow&-i\hbar\left[\overrightarrow{e_{x}}\frac{\partial}{\partial x}+\overrightarrow{e_{y}}\frac{\partial}{\partial y}+\overrightarrow{e_{z}}\frac{\partial}{\partial z}\right],\label{gradient}
\end{eqnarray}
à un espace temps courbe doté d'une métrique qu'on définit à partir de (\ref{metrique etoile}) et qu'on écrit sous la forme
\begin{equation}\label{lienmetriqueavecplatte}
     g^{\mu\nu}=h_{\mu}^{-2}\,\eta^{\mu\nu},
\end{equation}
(sans sommation sur l'indice $\mu$) où la métrique de Minkowski est par définition $\eta^{\mu\nu}=(1,-1,-1,-1)$ et les coefficients $h_{\mu}$ sont donnés par
\begin{eqnarray}\label{coefhmunu}
    &&h_{0}=\sqrt{\xi(r)},\\
    &&h_{1}=1/\sqrt{\xi(r)},\\
    &&h_{2}=r,\\
    &&h_{3}=r\sin\theta.
\end{eqnarray}
Pour ce faire, il suffit de remplacer les dérivées spatiales et temporelle figurant dans (\ref{correspondancenrjclassique}) et (\ref{gradient}) par des dérivées covariantes \cite{Weinberg, Belabbas}
\begin{eqnarray}
E&\rightarrow&i\hbar\,\nabla_{0}=i\hbar\,h_{0}^{-1}\,\frac{\partial}{\partial t},\label{correspondancenrjcourbe}\\
\overrightarrow{P}&\rightarrow&-i\hbar\Bigg[\overrightarrow{e_{1}}h_{1}^{-1}\frac{\partial}{\partial x^{1}}+\overrightarrow{e_{2}}h_{2}^{-1}\frac{\partial}{\partial x^{2}}+\overrightarrow{e_{3}}h_{3}^{-1}\frac{\partial}{\partial x^{3}}\Bigg]\nonumber\\
&&\hspace{3cm}=-i\hbar\left[\overrightarrow{e_{r}}\,\sqrt{\xi(r)}\,\frac{\partial}{\partial r}+
    \overrightarrow{e_{\theta}}\frac{1}{r}\frac{\partial}{\partial \theta}+
    \overrightarrow{e_{\phi}}\frac{1}{r\sin\theta}\frac{\partial}{\partial
    \phi}\right].\label{impulscorres}
\end{eqnarray}

\subsubsection{Relation entre la métrique et l'énergie potentielle de la particule}
Le référentiel du centre de masse est un référentiel où
$\overrightarrow{P}=\overrightarrow{0}$ ($P_{r}=P_{\theta}=P_{\phi}=0$). Dans un tel
référentiel, l'expression de l'invariant relativiste (\ref{invariantrelativiste}) devient
\begin{equation}\label{cous}
    E(\overrightarrow{P}=\overrightarrow{0})=(m_{0}c^{2})\sqrt{\xi(r)}.
\end{equation}

D'autre part l'énergie totale, dans le référentiel de centre de masse, s'écrit comme
la somme de l'énergie au repos et de l'énergie potentielle
\begin{equation}\label{couscous}
  E(\overrightarrow{P}=\overrightarrow{0})=m_{0}c^{2}+V(r),
\end{equation}
étant donné que l'énergie cinétique est nulle.

La combinaison de (\ref{cous}) et (\ref{couscous}) permet d'écrire
\begin{equation}\label{racinxi}
 \sqrt{\xi(r)}=\displaystyle\frac{m_{0}c^{2}+V(r)}{m_{0}c^{2}}=1+\frac{V(r)}{m_{0}c^{2}},
\end{equation}
ou encore finalement \cite{Barros1}
\begin{equation}\label{xi}
    \xi(r)=\Bigg(1+\frac{V(r)}{m_{0}c^{2}}\Bigg)^{2}.
\end{equation}
Pour les faibles potentiels, caractérisés par
$V(r)/(m_{0}c^{2})\ll1$, l'expression (\ref{xi}) se réduit à la relation
\begin{equation}
 \xi(r)\simeq1+\frac{2V(r)}{m_{0}c^{2}}.
\end{equation}
Avec (\ref{xi}), on retrouve au premier ordre l'expression usuelle en relativité générale
\begin{equation}\label{xigravit}
    \xi_{G}=\Bigg(1-\displaystyle\frac{GM}{rc^{2}}\Bigg)^{2}\simeq1-\displaystyle\frac{2GM}{rc^{2}}.
\end{equation}
pour l'interaction gravitationnelle (voir (\ref{metrique isotrope et statique forme general nouvelle variable shwrzscild}) pour comparaison).

\subsubsection{Equation d'onde quantique pour une particule de spin 1/2}
Pour un espace-temps plat de Minkowski, l'équation de Dirac libre est donnée par
\begin{eqnarray}\label{dirac libre eq}
\big(\overrightarrow{\alpha}.{\overrightarrow{p}}c+\beta\,m_{0}c^{2}\big)\Psi(\overrightarrow{r},t)&=&i\hbar\;{\frac{\partial}{\partial t}}\,\Psi(\overrightarrow{r},t)
\end{eqnarray}
où les quatre matrices de Dirac figurant dans l'hamiltonien
\begin{eqnarray}
    \overrightarrow{\alpha}=\Bigg(
\begin{array}{cc}
0 & \overrightarrow{\sigma}\\
\overrightarrow{\sigma} & 0\\
\end{array}
\Bigg) \hspace{2cm} \beta=\Bigg(
\begin{array}{cc}
\mathbf{1} & 0 \\
0 & -\mathbf{1}\\
\end{array}
\Bigg),
\end{eqnarray}
sont construites par les éléments de la base des matrices complexes d'ordre $(2\times2)$; la matrice identité $\mathbf{1}$ et les trois matrices
de Pauli
\begin{equation}\label{matpauli}
    \sigma_{1}=\Bigg(
\begin{array}{cc}
0 & 1 \\
1 & 0 \\
\end{array}
\Bigg) \hspace{1cm}\sigma_{2}=\Bigg(
\begin{array}{cc}
0 & -i \\
i & 0 \\
\end{array}
\Bigg) \hspace{1cm}\sigma_{3}=\Bigg(
\begin{array}{cc}
1 & 0 \\
0 & -1 \\
\end{array}
\Bigg),
\end{equation}
tel que $\overrightarrow{\sigma}=\overrightarrow{e_{1}}\;\sigma_{1}+
\overrightarrow{e_{2}}\;\sigma_{2}+\overrightarrow{e_{3}}\;\sigma_{3}$.

\noindent Pour déterminer l'équation de Dirac correspondant à l'espace-temps courbe, il faut substituer les dérivées ordinaires figurant dans (\ref{dirac libre eq}) par les dérivées covariantes (\ref{correspondancenrjcourbe}) et (\ref{impulscorres}), construites sur la base de la métrique de Schwarzschild,
\begin{eqnarray}
 \left(\overrightarrow{\alpha}.{\overrightarrow{P}}c+\beta\,m_{0}c^{2}\right)\Psi(\overrightarrow{r},t)&=&i\hbar\,{\nabla_{0}}\;\Psi(\overrightarrow{r},t),
\end{eqnarray}
pour obtenir l'équation d'onde suivante
\begin{eqnarray}\label{eqdiracourbe}
&&\Bigg\{\displaystyle\frac{i\hbar}{\sqrt{\xi(r)}}\frac{\partial}{\partial
t}\Bigg\}\;\Psi(\overrightarrow{r},t)=\nonumber\\
&&\hspace{0.5cm}\Bigg\{-i\hbar
c\overrightarrow{\alpha}.\bigg[\overrightarrow{e_{r}}\sqrt{\xi(r)}\frac{\partial}{\partial
r}+\overrightarrow{e_{\theta}}\frac{1}{r}\frac{\partial}{\partial
\theta}+\overrightarrow{e_{\phi}}\frac{1}{r\sin\theta}\frac{\partial}{\partial
\phi}\bigg]+\beta
m_{0}c^{2}\Bigg\}\;\Psi(\overrightarrow{r},t).
\end{eqnarray}

\subsubsection{Solution stationnaire}
Dans le cas stationnaire, la solution
\begin{equation}\label{foncsepdirac}
    \Psi(\overrightarrow{r},t)=\psi(r,\theta,\phi)\,\displaystyle
    e^{-\frac{iEt}{\hbar}},
\end{equation}
est remplacée dans (\ref{eqdiracourbe}) pour donner l'équation d'onde
\begin{eqnarray}\label{bigequation}
\left(
\begin{array}{cc}
m_{0}c^{2}-\displaystyle\frac{E}{\sqrt{\xi(r)}} & \overrightarrow{\sigma}.\overrightarrow{P}\,c\\
\overrightarrow{\sigma}.\overrightarrow{P}\,c & -m_{0}c^{2}-\displaystyle\frac{E}{\sqrt{\xi(r)}}\\
\end{array}
\right)\left(
\begin{array}{c}
\varphi\\
\\
\\
\chi\\
\end{array}
\right)=\left(
\begin{array}{c}
0\\
\\
\\
0\\
\end{array}
\right),
\end{eqnarray}
où
\begin{eqnarray}
\chi=\left(\begin{array}{c}
             \psi_{1} \\
             \psi_{2}
           \end{array}
\right) \hspace{0.4cm}\varphi=\left(\begin{array}{c}
             \psi_{3} \\
             \psi_{4}
           \end{array}
\right).
\end{eqnarray}

En utilisant la conservation du moment cinétique orbital (par rapport au centre du champ $V(r)$ pris comme origine des coordonnées) et de la parité, il est possible de déterminer complètement la partie angulaire des solutions stationnaires qui s'écrivent dans la représentation standard sous la forme \cite{landau}
\begin{equation}\label{spineurlandau}
    \psi=\Bigg(
\begin{array}{c}
\varphi\\
\chi\\
\end{array}
\Bigg)=\Bigg(
\begin{array}{c}
f(r)\,\Omega_{j\,\ell\,m}\\
(-1)^{\frac{1+\ell-\ell\,^{'}}{2}}g(r)\,\Omega_{j\,\ell\,^{'}\,m}\\
\end{array}
\Bigg),
\end{equation}
Les spineurs sphériques (partie angulaire de la solution) figurant dans (\ref{spineurlandau}) sont reliés par l'expression suivante
\begin{equation}\label{yahya}
    \Omega_{j\,\ell\,^{'}\,m}=i^{\ell-\ell\,^{'}}\bigg(\overrightarrow{\sigma}.
    \frac{\overrightarrow{r}}{r}\bigg)\Omega_{j\,\ell\,m},
\end{equation}
de sorte que les moments cinétiques orbitaux soient reliés au moment cinétique total $j$ par la condition
\begin{eqnarray}
    \Bigg\{
\begin{array}{c}
\ell=j\pm1/2,\\
\ell\,^{'}=2j-\ell.\\
\end{array}
\end{eqnarray}
Dans le but de déterminer complètement la partie radiale de la solution, injectons (\ref{spineurlandau}) dans (\ref{bigequation}) pour arriver au système d'équations différentielles couplées \cite{Belabbas}
\begin{eqnarray}\label{sysdiff}
\left\{%
\begin{array}{ll}
    \displaystyle\frac{df}{dr}+\displaystyle\frac{(1+x)}{r}\;f=\;\;\,\displaystyle\frac{1}{\hbar\,c}
    \left(\displaystyle\frac{E}{\sqrt{\xi(r)}}+m_{0}\,c^{2}\right)\;g(r)
    \\
    \\
    \displaystyle\frac{dg}{dr}+\displaystyle\frac{(1-x)}{r}\;g=-\displaystyle\frac{1}{\hbar\,c}
    \left(\displaystyle\frac{E}{\sqrt{\xi(r)}}-m_{0}\,c^{2}\right)\;f(r)\\
\end{array}%
\right.
\end{eqnarray}
tel que
\begin{eqnarray}\label{xcondense}
    x=\Bigg\{
\begin{array}{l}
-\left(j+1/2\right)=-(\ell+1)\hspace{1cm}\textrm {pour}:\,j=\ell+1/2,\\
+\left(j+1/2\right)=\ell\hspace{2.3cm}\;\textrm {pour}:\,j=\ell-1/2.\\
\end{array}
\end{eqnarray}

Attirons l'attention sur le fait que C.C.Barros ne retrouve pas exactement le système (\ref{sysdiff}). Il retrouve plutôt le système d'équations suivant \cite{Barros1}
\begin{eqnarray}\label{sysdiffbarros}
\left\{%
\begin{array}{ll}
    \sqrt{\xi(r)}\;\,\displaystyle\frac{df}{dr}+\displaystyle\frac{(1+x)}{r}\;f=
    \;\;\,\displaystyle\frac{1}{\hbar\,c}
    \left(\displaystyle\frac{E}{\sqrt{\xi(r)}}+m_{0}\,c^{2}\right)\;g(r)
    \\
    \\
    \sqrt{\xi(r)}\;\,\displaystyle\frac{dg}{dr}+\displaystyle\frac{(1-x)}{r}\;g=
    -\displaystyle\frac{1}{\hbar\,c}
    \left(\displaystyle\frac{E}{\sqrt{\xi(r)}}-m_{0}\,c^{2}\right)\;f(r)\\
\end{array}%
\right.
\end{eqnarray}

\subsection{Application à l'atome d'hydrogène}
L'atome d'hydrogène est décrit de façon originale. En effet, au lieu d'introduire l'interaction à laquelle est soumise la particule sous forme d'un champ extérieur, par des considérations de symétrie (couplage minimal), l'interaction sera contenue dans la métrique de l'espace-temps. L'électron de l'atome est soumis à un potentiel\footnote{En fait c'est plutôt une énergie potentielle coulombienne entre l'électron (-e) et le proton (+e), qui est le produit de la charge de l'électron par le potentiel du proton $V(r)=(-e)\phi(r)=\left(\frac{-e^{2}}{4\pi\varepsilon_{0}}\right)\frac{1}{r}$} $V(r)=-\alpha/r$ . Ainsi en utilisant (\ref{xi}), on détermine la fonction
\begin{equation}\label{xiep}
    \xi_{ep}(r)=\bigg(1-\displaystyle\frac{\alpha}{m_{0}c^{2}\,r}\bigg)^{2}.
\end{equation}
En tenant compte de (\ref{xiep}), le système d'équations différentielles (\ref{sysdiff}) s'écrit sous la forme
\begin{eqnarray}\label{vignt}
\left\{
  \begin{array}{ll}
    \displaystyle\frac{df}{dr}+\displaystyle\frac{(1+x)}{r}\,f=\;\;\,\frac{1}{\hbar\,c}\;
\left[\displaystyle\frac{E}
{1+\frac{V(r)}{m_{0}c^{2}}}+m_{0}\,c^{2}\right]g(r), \\\\
 \displaystyle\frac{dg}{dr}+\displaystyle\frac{(1-x)}{r}\,g=-\frac{1}{\hbar\,c}\;
\left[\displaystyle\frac{E}{1+\frac{V(r)}{m_{0}c^{2}}}-m_{0}\,c^{2}\right]f(r),
  \end{array}
\right.
\end{eqnarray}
Etant donné que $\varepsilon=\frac{V(r)}{m_{0}c^{2}}\ll 1,$ à l'ordre 1 en puissances de $\varepsilon,$ le système (\ref{vignt}) s'écrit comme suit
\begin{eqnarray}\label{vingt et un}
  \hspace*{-0.7cm}\left\{
    \begin{array}{ll}
      \displaystyle\frac{df}{dr}+\displaystyle\frac{(1+x)}{r}\,f \simeq \;\;\,\frac{1}{\hbar\,c}\;
\Bigg[E\bigg(1-\frac{V(r)}{m_{0}c^{2}}\bigg)+m_{0}\,c^{2}\Bigg]g(r)\\
\\
      \displaystyle\frac{dg}{dr}+\displaystyle\frac{(1-x)}{r}\,g \simeq -\frac{1}{\hbar\,c}\;
\Bigg[E\bigg(1-\frac{V(r)}{m_{0}c^{2}}\bigg)-m_{0}\,c^{2}\Bigg]f(r)
    \end{array}
  \right.
\end{eqnarray}
Du fait qu'on s'intéresse à des énergies $E$ d'ordre de grandeur comparable à l'énergie au repos $m_{0}c^{2}$, de sorte que $E/m_{0}c^{2}\sim 1,$ le système d'équations (\ref{vingt et un}) se réduit identiquement au système d'équations de la théorie de Dirac \cite{landau}
\begin{eqnarray}\label{sysdiffdirac}
\left\{%
\begin{array}{ll}
    \displaystyle\frac{df}{dr}+\displaystyle\frac{(1+x)}{r}\,f=\;\;\,\frac{1}{\hbar\,c}\;
    \bigg(E-V(r)+m_{0}\,c^{2}\bigg)g(r),\\\\
    \displaystyle\frac{dg}{dr}+\displaystyle\frac{(1-x)}{r}\,g=-\frac{1}{\hbar\,c}\;
    \bigg(E-V(r)-m_{0}\,c^{2}\bigg)f(r),
\end{array}%
\right.
\end{eqnarray}
dans lequel l'énergie potentielle non gravitationnelle  $V(r)$ retrouve sa place habituelle.

La méthode standard de Dirac pour déterminer les niveaux d'énergie de l'atome
d'hydrogène consiste à résoudre le système d'équations
différentielles couplées (\ref{sysdiffdirac}), par rapport à la nouvelle variable $\rho =\beta\;r$, en choisissant des solutions sous forme de séries de type Frobinius \cite{Barros1}
\begin{eqnarray}
f(\rho)=\sum_{n=0}^{N}a_{n}\;\rho^{n+s}\;e^{-\rho},\label{frobinius1}\\
g(\rho)=\sum_{n=0}^{N}b_{n}\;\rho^{n+s}\;e^{-\rho}.\label{frobinius2}
\end{eqnarray}
Les valeurs des paramètres ajustables $s$ et $\beta$, introduits dans le but d'avoir une plus grande
flexibilité dans le choix de la forme des solutions, sont données, après calculs (voir \cite{Belabbas}), par
\begin{equation}
s=-1+\sqrt{x^{2}-\left(\frac{\alpha}{\hbar\,c}\right)^{2}},\label{sracine}
\end{equation}
et
\begin{equation}\label{betaracine}
\beta=\frac{1}{\hbar\,c}\;\sqrt{\left(\,m_{e}\,c^{\,2}\,\right)^{2}-E^{2}}.
\end{equation}
Le spectre discret des niveaux d'énergie $E=E_{N}$ de l'atome d'hydrogène est donné par
\begin{eqnarray}
    E_{N}=(m_{0}c^{2})\;\left[1+
\displaystyle\frac{\left(\displaystyle\frac{\alpha}{\hbar\,c}\right)^{2}}
{\left(\;N+\displaystyle\sqrt{x^{2}-\left(\frac{\alpha}{\hbar\,c}\right)^{2}}\;
\right)^{2}}\right]^{-1/2},\label{nivnrjatomhydrog}
\end{eqnarray}
dans le cas de la théorie de Dirac.

Précisons que C.C.Barros prétend apporter une correction insignifiante aux niveaux d'énergie (\ref{nivnrjatomhydrog}), en se basant sur le système d'équations (\ref{sysdiffbarros}); il retrouve le
spectre d'énergie suivant \cite{Barros1}
\begin{equation}\label{spectrebarros}
    E_{N}=m_{0}\,c^{2}\;\displaystyle\sqrt{\frac{1}{2}-\frac{N^{2}}{8\alpha^{2}}
    +\frac{N}{4\alpha}\;\displaystyle\sqrt{\frac{N^{2}}{4\alpha^{2}}+2}},
\end{equation}
et affirme que ses résultats sont plus réalistes car l'écart des
niveaux d'énergie (\ref{spectrebarros}) avec l'expérience est de
l'ordre de $0,005\%$, alors que l'écart des niveaux
d'énergie de Dirac (\ref{nivnrjatomhydrog}) est de l'ordre de
$0.027\%$ \cite{Barros1}.

 Dans notre travail \cite{Belabbas}, on a montré que son approche permettait juste de reproduire la théorie de Dirac dans le cadre de l'approximation du champ faible $V(r)\ll m_{0}c^{2}$.

Ces résultats spectaculaires nous interpellent et nous incitent à nous poser des questions sur le rôle du Principe d'Equivalence en RG : Est-il indispensable pour formuler la Théorie de la RG ? Est-ce que l'exigence de covariance des lois physiques, à elle seule, est suffisante pour servir de base à la RG ? Faut-il penser à reformuler le Principe d'Equivalence pour l'étendre aux interactions non gravitationnelles \cite{ozer}, \cite{Iliev1} ?

\section{Vers un nouveau Principe d'Equivalence "généralisé"}
Le Principe d'Equivalence a été énoncé, par Einstein, exclusivement pour le champ de gravitation. Il stipule qu'un référentiel uniformément accéléré est localement équivalent à un référentiel inertiel plongé dans un champ de gravitation. Cette localité n'est valable que dans la mesure où le champ de gravitation peut être considéré homogène.

Dans ce qui suit, quelques arguments en faveur de l'extension du Principe d'Equivalence sont présentés, suivis de quelques remarques sur la manière avec laquelle nous envisageons d'étendre ce principe aux interactions non gravitationnelles en général. Une attention particulière est portée sur l'interaction électromagnétique pour laquelle les contours du nouveau Principe d'Equivalence sont révélés.

\subsection{Arguments en faveur d'une extension du Principe d'Equivalence}

\subsubsection{Cadre de validité du Principe d'Equivalence}
Dans un premier temps, nous allons voir dans quelle mesure un champ gravitationnel, généré par une masse ponctuelle ou bien une distribution de masse sphérique, peut être considéré homogène. Dans ce cas, les champs de gravitation qui règnent en deux points différents de l'espace peuvent être assimilés au même vecteur constant, si les deux \textbf{directions} et \textbf{modules} sont presque les mêmes.

\begin{itemize}
  \item La première exigence a lieu dans la mesure où deux trajectoires de deux masses, convergentes vers le centre du champ de gravitation, peuvent être considérées comme parallèles (voir figure \ref{figlocalite}).
  \item Pour définir les limites de la seconde exigence, considérons une masse $m$ qui se déplace sous l'action du champ de gravitation de la terre $M$. Compte tenu de la loi de gravitation universelle et du principe fondamental de dynamique, appliqué à $m$, il est clair que le champ de gravitation de la terre
\begin{equation}
    \overrightarrow{g_{M}}(r)=-\left(\frac{GM}{r^{2}}\right)\overrightarrow{e_{r}},
\end{equation}
dépend de la distance $r$ de la particule considérée au centre de la terre. Dans ce cas, le module de la différence des champs gravitationnels qui règnent à des distances respectives $r_{1}$ et $r_{2}$ du centre de la terre est donné par
\begin{equation}
    \|\overrightarrow{g_{M}}\|= \|\overrightarrow{g_{M}}(r_{1})-\overrightarrow{g_{M}}(r_{2})\|=GM\left[\frac{|r_{1}-r_{2}|(r_{1}+r_{2})}{\left(r_{1}\,r_{2}\right)^{2}}\right].
\end{equation}
Il est clair que $ \|\overrightarrow{g_{M}}\|\rightarrow 0$ quand la distance entre le point initial et final tend à être nulle $r_{2}\rightarrow r_{1}$, ou bien  dans la mesure où les distances respectives $R_{1}=r_{1}-R$ et $R_{2}=r_{2}-R$ des points initial et final de la surface de la terre soient négligeables devant le rayon de la terre\footnote{Dans ce cas $\|\overrightarrow{g_{M}}\|\simeq 2|R_{1}-R_{2}|/R^{3}\rightarrow 0$.}, i.e $R_{1}\ll R$ et $R_{2}\ll R$ (voir figure \ref{figlocalite}).

\begin{figure}[\here]
   \centering
  \includegraphics[width=8cm]{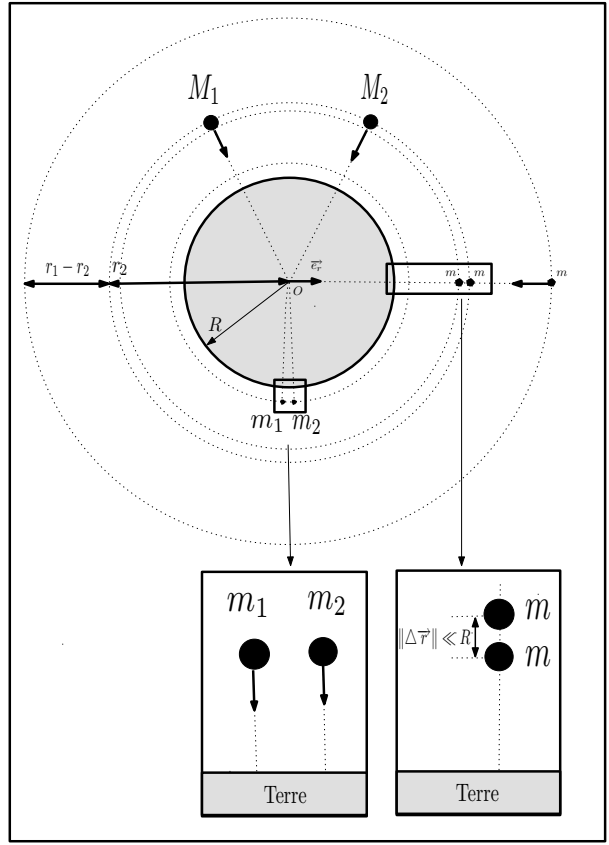}\\
  \caption{Représentation schématique de la validité locale du principe
   d'équivalence}\label{figlocalite}
\end{figure}
  \item Il est aussi possible d'ajouter la condition de faible interaction entre la source du champ de gravitation et la particule qui évolue dans ce champ. En effet, le champ gravitationnel, créé par la source $M$ au niveau de la particule $m$, $\overrightarrow{g_{M}}(r)=-(GM/r^{2})\overrightarrow{e_{r}}$ est fortement influencé par le champ gravitationnel $\overrightarrow{g_{m}}(r)=(Gm/r^{2})\overrightarrow{e_{r}}$ de $m$ qui agit sur $M$ dans le cas où la distance mutuelle est très petite $r\rightarrow 0$ et aussi dans le cas où les deux masses sont du même ordre de grandeur $M\simeq m$. Dans ce cas, la distinction "source" et "particule test" ne peut plus avoir lieu; on a plutôt deux masses en interaction mutuelle.
\end{itemize}

\begin{figure}[\here]
   \centering
  \includegraphics[width=4cm]{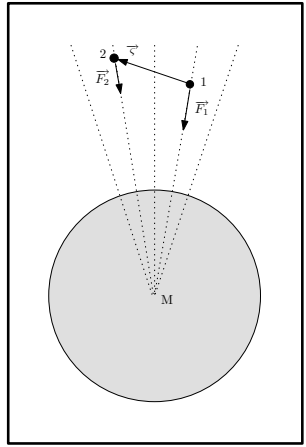}\\
  \caption{Forces de marées}\label{tidal}
\end{figure}

Les inhomogé\"{\i}tés du champ de gravitation d'une masse sphérique $M$ entraînent l'apparition d'un nouveau type de forces dit de marées. Ces forces sont dues à la différence entre les champs de gravitations qui règnent en deux points voisins dans l'espace; autrement dit, à l'existence de gradients non nuls du champ de gravitation dans certaines régions de l'espace. En effet, dans le but de déterminer l'expression de la force de marées qui s'exercerait entre deux masses identiques $m$, situées aux points voisins 1 et 2 repérés avec les vecteurs positions respectifs $\overrightarrow{r}$ et $\overrightarrow{r}+\overrightarrow{\varsigma}$ (voir figure \ref{tidal}), plaçons-nous dans le cas où le vecteur de séparation $\overrightarrow{\varsigma}$ est tel que $\|\overrightarrow{\varsigma}\|\ll\|\overrightarrow{r}\|$. Dans ce cas, la force de marées qui s'exerce entre les deux masses situées aux points 2 et 1 est par définition égale à la différence entre les forces de gravitations $\overrightarrow{F_{2}}$ et $\overrightarrow{F_{1}}$
\begin{eqnarray}
 \overrightarrow{f} &=& \overrightarrow{F}(\overrightarrow{r}+\overrightarrow{\varsigma})-\overrightarrow{F}(\overrightarrow{r}).
 \end{eqnarray}
Un développement de Taylor à l'ordre 1, par rapport à $\|\overrightarrow{\varsigma}\|$ permet d'avoir l'expression explicite\footnote{Un calcul similaire est développé pour exprimer le champ électrique en fonction du moment dipolaire aux grandes distances \cite{landau}.} de  $\overrightarrow{f}$
\begin{eqnarray}
   \overrightarrow{f} &=& GmM\left[\frac{\overrightarrow{r}}{\|\overrightarrow{r}\|^{3}}-\frac{\overrightarrow{r}+\overrightarrow{\varsigma}}{\|\overrightarrow{r}
   +\overrightarrow{\varsigma}\|^{3}}\right] \nonumber\\
   &\simeq& GmM\left[\frac{\overrightarrow{r}}{r^{3}}
   -(\overrightarrow{r}+\overrightarrow{\varsigma})\left(\frac{1}{r^{3}}-3\,\frac{\overrightarrow{r}.\overrightarrow{\varsigma}}{r^{5}}+\cdots\right)\right]\nonumber\\
   &=& GmM\left[-\frac{\overrightarrow{\varsigma}}{r^{3}}+\frac{3\left(\overrightarrow{r}.\overrightarrow{\varsigma}\right)\overrightarrow{r}}{r^{5}}\right]\nonumber\\
   &=& GmM\left[\frac{3\left(\overrightarrow{\varsigma}.\overrightarrow{e_{r}}\right)\overrightarrow{e_{r}}
    -\overrightarrow{\varsigma}}{r^{3}}\right].
\end{eqnarray}
En introduisant le potentiel gravitationnel $\phi(\overrightarrow{r})=-GM\|\overrightarrow{r}\|^{-1}$, tel que$\overrightarrow{F}=-m\,\overrightarrow{\nabla}\phi$, nous avons d'autre part
\begin{eqnarray*}
  \overrightarrow{f} &=& -m\left[\overrightarrow{\nabla}\phi(\overrightarrow{r}+\overrightarrow{\varsigma})-\overrightarrow{\nabla}\phi(\overrightarrow{r})\right] =-m\overrightarrow{\nabla}\left[\phi(\overrightarrow{r}+\overrightarrow{\varsigma})-\phi(\overrightarrow{r})\right] \\
   &\simeq& -m\overrightarrow{\nabla}\left[\left(\overrightarrow{\nabla}\phi\right).\overrightarrow{\varsigma}\right] = GMm\overrightarrow{\nabla}\left[\overrightarrow{\varsigma}.\overrightarrow{\nabla}\left(\frac{1}{r}\right)\right]
\end{eqnarray*}
Sachant que $\overrightarrow{\varsigma}$ est indépendant de $\overrightarrow{r}$, la relation\footnote{Compte tenu des deux propriétés $$\overrightarrow{\nabla}\left[a(r)\right]=\left(\frac{da}{dr}\right)\frac{\overrightarrow{r}}{r},$$ et
$$\overrightarrow{\nabla}\left(\overrightarrow{A}.\overrightarrow{B}\right) = \overrightarrow{A}\times\left(\overrightarrow{\nabla}\times\overrightarrow{B}\right)
  +\left(\overrightarrow{A}.\overrightarrow{\nabla}\right)\overrightarrow{B}+\overrightarrow{B}\times\left(\overrightarrow{\nabla}\times\overrightarrow{A}\right)
  +\left(\overrightarrow{B}.\overrightarrow{\nabla}\right)\overrightarrow{A},$$
nous avons
\begin{eqnarray*}
  \left(\overrightarrow{\varsigma}.\overrightarrow{\nabla}\right)\left[\overrightarrow{\nabla}\left(\frac{1}{r}\right)\right] &=& \overrightarrow{\nabla}\left[\overrightarrow{\varsigma}.\overrightarrow{\nabla}\left(\frac{1}{r}\right)\right]
  -\overrightarrow{\varsigma}\times\underbrace{\left[\overrightarrow{\nabla}\times\overrightarrow{\nabla}\left(\frac{1}{r}\right)\right]}_{\overrightarrow{0}}
  -\overrightarrow{\nabla}\left(\frac{1}{r}\right)\times\underbrace{\left(\overrightarrow{\nabla}\times\overrightarrow{\varsigma}\right)}_{\overrightarrow{0}}
  -\left[\overrightarrow{\nabla}\left(\frac{1}{r}\right).\overrightarrow{\nabla}\right]\overrightarrow{\varsigma}\\
   &=&\overrightarrow{\nabla}\left[\overrightarrow{\varsigma}.\overrightarrow{\nabla}\left(\frac{1}{r}\right)\right]
   -\left[\left(\frac{-1}{r^{2}}\,\overrightarrow{e_{r}}\right).\left(\overrightarrow{e_{r}}\,\nabla_{r}
   +\overrightarrow{e_{\theta}}\,\nabla_{\theta}+\overrightarrow{e_{\phi}}\,\nabla_{\phi}\right)\right]\overrightarrow{\varsigma}\\
   &=&\overrightarrow{\nabla}\left[\overrightarrow{\varsigma}.\overrightarrow{\nabla}\left(\frac{1}{r}\right)\right]
   -\underbrace{\left(\frac{-1}{r^{2}}\,\frac{d\overrightarrow{\varsigma}}{dr}\right)}_{\overrightarrow{0}}\\
\end{eqnarray*}}
$$\left(\overrightarrow{\varsigma}.\overrightarrow{\nabla}\right)\left[\overrightarrow{\nabla}\left(\frac{1}{r}\right)\right]
=\overrightarrow{\nabla}\left[\overrightarrow{\varsigma}.\overrightarrow{\nabla}\left(\frac{1}{r}\right)\right],$$
permet de mettre la force de marées sous la forme
\begin{equation}
    \overrightarrow{f}=-m\left(\overrightarrow{\varsigma}.\overrightarrow{\nabla}\right)\left(\overrightarrow{\nabla}\phi\right).
\end{equation}
Il s'ensuit finalement que la $i^{\text{ème}}$ composante de l'accélération relative entre les deux masses situées aux points voisins 1 et 2 est donnée, dans un système de coordonnées cartésien, par l'expression \cite{Groen}
\begin{equation}\label{acceleration marree}
    \frac{d^{\,2}\varsigma_{i}}{dt^{2}}=-\varsigma^{j}\left(\frac{\partial^{\,2}\phi}{\partial x^{i}\partial x^{j}}\right)_{\overrightarrow{r}}.
\end{equation}

Ces forces de marées jouent un rôle important dans les régions pour lesquelles il y a une très grande variation du champ gravitationnel. En effet, pour un corps macroscopique ayant une extension spatiale importante, de telle sorte à "ressentir" la non-uniformité du champ gravitationnel dans lequel il est plongé, les forces gravitationnelles, non égales, qui s'exercent sur les différentes parties constitutives ont tendance à modifier la forme générale du corps.

Ajoutons finalement que ces forces sont du même ordre de grandeur que les composantes du tenseur de courbure. En effet, d'après (\ref{acceleration marree}), elles s'expriment en fonction des secondes dérivées du potentiel gravitationnel, d'autre part, en revenant aux définitions du tenseur de Courbure (\ref{tenseur riemann christoffel}), des symboles de Christoffel (\ref{importantegij}) et de la relation (\ref{metrique 00 et phi}), il est clair que les composantes du tenseur de courbure s'expriment sous forme d'une certaine combinaison des secondes dérivées du potentiel gravitationnel.  Dans ce cas, de la même manière que pour les forces de marées, l'existence de composantes non nulles du tenseur de courbure est due aussi aux inhomogé\"{\i}tés du champ gravitationnel.

Après avoir discuté les limites pour lesquelles le champ gravitationnel peut être considéré homogène, à présent, nous allons présenter deux arguments qui montrent clairement que la notion de localité, comme cadre de validité du Principe d'Equivalence, n'est qu'une approximation qu'il va falloir préciser si on veut espérer absorber l'effet des interactions non gravitationnelles, agissant sur une particule, dans la métrique de l'Espace-temps; autrement dit, se ramener à un système de coordonnées particulier dans lequel la particule est considérée comme libre.

\begin{enumerate}
  \item Il est bien connu en RG qu'il est possible d'annuler tous les symboles de Christoffel $\Gamma^{\lambda}_{\mu\nu}$, en un point arbitraire $O$, choisit comme origine des coordonnées. Pour ce faire, il suffit d'effectuer, au voisinage de ce point, la transformation de coordonnées suivante \cite{landau}
\begin{equation}\label{deux bis}
x^{'\lambda}=x^{\lambda}+\frac{1}{2}\,(\Gamma^{\lambda}_{\mu\nu})_{O}\,x^{\mu}\,x^{\nu}.
\end{equation}
En appliquant la transformation de coordonnées (\ref{deux bis}), l'équation de géodésique (\ref{eq geodes bis}) se réduit au point $O$ à l'équation d'une particule libre
\begin{equation}
    \left(\frac{d^{\,2}x^{'\nu}}{d\tau^{2}}\right)_{O}=0.
\end{equation}
D'autre part, les composantes non nulles du tenseur de courbure (\ref{tenseur riemann christoffel}) sont dues aux dérivées des symboles de Christoffel
\begin{equation}\label{tenseur riemann christoffel gamma nul}
    \left(R^{'\,j}_{\;\;ik\ell}\right)_{O}=\left(\partial^{'}_{k}\Gamma_{i\ell}^{'\,j}\right)_{O}-\left(\partial^{'}_{\ell}\Gamma_{ik}^{\,'j}\right)_{O}.
\end{equation}
Par contre, si on remplace le point $O$ par une petite région qui l'entoure, dans ce cas, les symboles de Christoffel nuls vont imposer des dérivées nulles des $\Gamma_{i\ell}^{'\,j}$, de sorte que toutes les composantes du tenseur de courbure $R^{'\,j}_{\;\;ik\ell}=0$ soient nulles dans la région qui entoure le point $O$. Or si le tenseur de courbure est nul dans un système de coordonnées particulier, il en sera de même pour n'importe quel autre référentiel. Cette conclusion est absurde car nous avons assumé, dès le début, la présence d'un champ de gravitation qui se manifeste par une courbure non nulle d'espace-temps.
  \item Même si nous considérons que l'égalité des masses inerte et gravitationnelle est rigoureusement exact, deux corps $A$ et $B$ ne tombent pas vers la terre avec la même accélération par rapport au même référentiel. En effet, les centres de masses respectifs par rapport auxquels, les accélérations sont mesurées sont différents; en supposant que $A$ et $B$ sont à la même hauteur $r$ du centre de la terre\footnote{On se place dans le cas $\overrightarrow{r}=\overrightarrow{OA}=\overrightarrow{OB}$ où les deux masses $A$ et $B$ se déplacent suivant la même direction.}, alors la distance entre leurs centres de masses respectifs
\begin{equation}
    \|\triangle \overrightarrow{R}\|=\frac{M|m_{A}-m_{B}|\,r}{\left(M+m_{A}\right)\left(M+m_{B}\right)},
\end{equation}
ne s'annule que dans les cas où les deux corps ont la même masse ($m_{A}=m_{B}$), ou bien dans le cas où les deux masses sont négligeables devant la masse de la terre ($M\gg m_{A}$ et $M\gg m_{B}$). Cette circonstance implique que la propriété selon laquelle tous les corps tombent avec la même accélération sous l'action du champ gravitationnel terrestre, n'est en fait qu'une approximation, appliquée dans certaines circonstances bien particulières.
\end{enumerate}

On s'aperçoit ainsi que la notion de localité, comme cadre de validité du Principe d'Equivalence, n'est rigoureusement exacte qu'en la réduisant à un point. Ainsi se présente la possibilité d'étendre le Principe d'Equivalence à toutes les interactions fondamentales, car en un point donné, il est possible d'annuler l'action ou bien l'effet de n'importe quel champ.

Attirons l'attention sur une différence fondamentale entre la gravitation et l'électromagnétisme. Compte tenu du principe fondamental de la dynamique, l'accélération d'une particule, de masse inertielle $m_{i}$ et de masse grave $m_{g}$, soumise au champ gravitationnel $g$, supposé constant, d'une seconde masse $M$ est donnée $$\ddot{x}=\left(\frac{m_{g}}{m_{i}}\right)g.$$ L'égalité entre la masse inertielle et gravitationnelle $m_{i}=m_{g}$, implique que la transformation de coordonnées
\begin{equation}\label{transf annulation grav const}
    x^{'}=x-\frac{1}{2}\,gt^{2},
\end{equation}
conduit à voir la particule avec une accélération nulle $\ddot{x}^{'}=0$ dans le nouveau référentiel. De même, pour annuler l'accélération d'une particule, de charge $q$ et de masse $m_{i}$, soumise à un champ électrostatique constant $E$  $$\ddot{x}=\left(\frac{q}{m_{i}}\right)E,$$ il faut plutôt appliquer la transformation de coordonnées suivante
\begin{equation}\label{transf annulation champ electrique const}
    x^{'}=x-\frac{1}{2}\left(\frac{q}{m_{i}}\right)Et^{2}.
\end{equation}
La différence fondamentale entre les deux transformations de coordonnées précédentes est que (\ref{transf annulation grav const}) est universelle, i.e indépendante des propriétés de la particule test, alors que (\ref{transf annulation champ electrique const}) dépend des propriétés de la charge électrique. L'égalité des masses inerte et grave implique que l'espace-temps de la relativité générale est une scène unique pour toutes les masses, alors que l'absence d'une propriété similaire pour les charges électriques fait que l'espace-temps, de la nouvelle approche, dépend des propriétés des charges électriques. Dans ce cas, l'espace-temps qui permetterait de décrire par exemple deux charges $q_{1}$ et $q_{2}$ est différent de celui qui permet de décrire les charges $q_{1}$ et $q_{3}$.

\subsubsection{Approche de Barros}
Essayons maintenant de montrer que les résultats surprenants de C.C.Barros sont dus à une adoption implicite d'un Principe d'Equivalence, étendu à l'interaction électrostatique. Pour ce faire, considérons une particule de charge $q$ et de masse inertielle $m_{i}$. De façon similaire au cas gravitationnel, Il est toujours possible de trouver un système de coordonnées  $\left\{x^{(0)i}\right\}$  dans lequel la particule est libre. A la limite newtonienne, il est possible de montrer que l'équation de géodésique se réduit à
\begin{equation}\label{newtonhij}
    \frac{d^{2}\overrightarrow{r}}{dt^{2}}=-\frac{c^{2}}{2}\;\overrightarrow{\nabla}h_{00}.
\end{equation}
où $g_{\mu\nu}=\eta_{\mu\nu}+h_{\mu\nu}$. D'autre part, l'équation de Newton pour une particule chargée, soumise à un champ électrostatique $\overrightarrow{E}$ est donnée par
\begin{equation}\label{newton}
    \frac{d^{2}\overrightarrow{r}}{dt^{2}}=\frac{q}{m_{i}}\;\overrightarrow{E}=-\frac{q}{m_{i}}\;
    \overrightarrow{\nabla}\phi.
\end{equation}
La comparaison de (\ref{newton}) et (\ref{newtonhij}), tout en exigeant au potentiel scalaire $\phi$ de tendre vers zéro à l'infini, nous permet de déduire l'élément de la métrique suivant
\begin{equation}\label{metrique ele 00 phi}
    g_{00}=1+\frac{2}{c^{2}}\,\frac{q}{m_{i}}\;\phi.
\end{equation}
Ce résultat est exactement celui utilisé par C.C.Barros, lorsqu'il adopte une métrique similaire à celle de Schwarzschild (\ref{metrique etoile}),
\begin{equation}
    g_{00}=1+\frac{2\,V(r)}{m_{i}c^{2}},
\end{equation}
où l'énergie potentielle $V(r)=q \phi(r)$. Ainsi, ce raisonnement confirme notre hypothèse que C.C.Barros adopte implicitement le Principe d'Equivalence étendu au champ électrostatique. Dans ce cas, il faut admettre alors l'existence d'équations de géodésiques pour la charge électrique $q$ soumise au champ électrostatique.

A présent, il serait intéressant de vérifier les possibilités d'étendre le Principe d'Equivalence au cas plus général de l'interaction électromagnétique.

\subsubsection{Approche de Iliev}
Dans sa contribution \cite{Iliev1}, Iliev parvient à annuler le champ électromagnétique $A_{\mu}$ en un point arbitraire $O$, par le biais d'une transformation de jauge
\begin{equation}\label{onze}
 A_{\mu}(M)\rightarrow A_{\mu}^{'}(M)= A_{\mu}(M)-\frac{\partial \phi(M)}{\partial x^{\mu}},
\end{equation}
qui correspond au choix particulier de la fonction scalaire
\begin{equation}\label{choix de phi iliev}
    \phi(M)=\Upsilon_{\rho}\,x^{\rho}(M)+x^{\rho}(M)\,x^{\sigma}(M)\left[\partial_{\sigma}A_{\rho}(O)\right],
\end{equation}
tel que $\Upsilon_{\rho}=A_{\rho}(O)$ et $\partial_{\sigma}A_{\rho}(O)=(\partial A_{\rho}/\partial x^{\sigma})_{O}$ représentent respectivement la valeur du champ électromagnétique et de sa dérivée au point $O$.

En effet, compte tenu de (\ref{choix de phi iliev}) qui conduit à la dérivée
\begin{eqnarray}\label{deriv phi iliev}
    \frac{\partial\phi(M)}{\partial x^{\mu}}&=&\left(\frac{\partial\Upsilon_{\rho}}{\partial x^{\mu}}\right)x^{\rho}(M)+x^{\rho}(M)\,x^{\sigma}(M)\frac{\partial}{\partial x^{\mu}}\Big[\partial_{\sigma}A_{\rho}(O)\Big]+\Upsilon_{\mu}\nonumber\\
    &&\hspace{4cm}+x^{\sigma}(M)\Big[\partial_{\sigma}A_{\mu}(O)\Big]+x^{\rho}(M)\Big[\partial_{\mu}A_{\rho}(O)\Big],
\end{eqnarray}
et compte tenu du développement au premier ordre du potentiel
\begin{equation}\label{premier ordre amu}
    A_{\mu}(M)=A_{\mu}(O)+x^{\lambda}(M)\Big[\partial_{\lambda}A_{\rho}(O)\Big],
\end{equation}
au voisinage du point arbitraire $O$ pris comme origine des coordonnées (i.e $x^{\ell}(O)=0\;\;\forall \ell$),
il est ainsi possible d'aboutir à l'expression du potentiel dans la nouvelle jauge
\begin{equation}\label{nouvelle jauge ptentiel general}
    A_{\mu}^{'}(M)=-\left(\frac{\partial\Upsilon_{\rho}}{\partial x^{\mu}}\right)x^{\rho}(M)-x^{\rho}(M)\,x^{\sigma}(M)\frac{\partial}{\partial x^{\mu}}\Big[\partial_{\sigma}A_{\rho}(O)\Big]-x^{\rho}(M)\Big[\partial_{\mu}A_{\rho}(O)\Big].
\end{equation}
Il est claire que ce potentiel est nul
\begin{equation}\label{nouvelle jauge ptentiel nul}
    A_{\mu}^{'}(O)=0,
\end{equation}
à l'origine des coordonnées.

Nous avons montré qu'une transformation de jauge ne peut pas être réduite ou assimilée à une transformation de coordonnées \cite{Belabbas3}. En effet, il est bien connu que sous l'action d'une transformation de coordonnées infinitésimale et arbitraire
\begin{equation}\label{un}
x^{i} \rightarrow x^{'i}=x^{i}+\varepsilon^{i},
\end{equation}
le champ électromagnétique se transforme, à l'ordre 1, comme suit
\begin{eqnarray}\label{dix}
     \hspace*{-1cm} A_{\mu}^{'}(x^{i})&\simeq& A_{\mu}(x^{i})-\frac{\partial \varepsilon^{\rho}}{\partial x^{\mu}}\,A_{\rho}(x^{i})-\varepsilon^{\rho}\,\frac{\partial A_{\mu}}{\partial x^{\rho}}.
\end{eqnarray}
La comparaison de (\ref{dix}) avec la transformation de jauge
\begin{eqnarray}\label{onze bis}
      A_{\mu}^{'}(x^{i})= A_{\mu}(x^{i})-\frac{\partial \phi(x^{i})}{\partial x^{\mu}}.
\end{eqnarray}
nous a permis d'établir l'identification suivante
\begin{eqnarray}
      \hspace*{-0.5cm}\frac{\partial \phi(x^{i})}{\partial x^{\mu}}&=&\frac{\partial \varepsilon^{\rho}}{\partial x^{\mu}}\,A_{\rho}(x^{i})+\varepsilon^{\rho}\,\frac{\partial A_{\mu}}{\partial x^{\rho}}\nonumber\\
      &=&\frac{\partial}{\partial x^{\mu}}\Big(A_{\rho}(x^{i})\,\varepsilon^{\rho}\Big)+\varepsilon^{\rho}
      F_{\rho \mu}(x^{i}),\label{douzebis}
\end{eqnarray}
où $F_{\rho \mu}(x^{i})=\frac{\partial A_{\mu}(x^{i})}{\partial x^{\rho}}-\frac{\partial A_{\rho}(x^{i})}{\partial x^{\mu}}.$ L'adoption de la solution $\phi(x^{i})=A_{\rho}(x^{i})\,\varepsilon^{\rho}$ impose la condition suivante
\begin{equation}
    \varepsilon^{\rho}
      F_{\rho \mu}(x^{i})=0,
\end{equation}
qui est un système d'équations linéaires sans second membre, dont la solution n'existe que si
\begin{equation}\label{determinant}
    det\left[F_{\rho \mu}(x^{i})\right]=0.
\end{equation}

N'étant pas conforme aux équations de Maxwell, la condition (\ref{determinant}) nous amène à conclure qu'une transformation de jauge ne peut pas être vue comme une transformation de coordonnées.

\subsection{Extension à l'interaction électromagnétique}
La critique précédente de l'approche de Iliev, nous amène à penser que pour étendre le Principe d'Equivalence à l'interaction électromagnétique, il ne faut pas donc chercher à annuler le 4-potentiel $A_{\mu}$, moyennant une transformation de coordonnées adéquate, au niveau du point $O$ où se trouve la charge électrique $q$, mais plutôt annuler l'effet du champ électromagnétique en $O$. Autrement dit, nous espérons définir des symboles de Christoffels qui génèrent la force de Lorentz qui agit sur la charge électrique, dans certaines conditions à préciser ultérieurement. De cette façon, le recours à la transformation de coordonnées (\ref{deux bis}), permettrait d'annuler ces nouveaux symboles, de telle sorte à annuler l'effet du champ électromagnétique sur la particule test chargée pour avoir une particule libre.

\section{Conclusion et Résultats}
Contrairement à C.C.Barros qui prétend apporter une correction insignifiante aux niveaux d'énergie de l'électron, en se plaçant dans le même contexte, nous avons reproduit le même résultat pour le spectre de l'atome d'hydrogène que celui obtenu dans le cadre de l'équation de Dirac.

Vouloir réduire l'essence du principe d'Equivalence à l'égalité entre la masse inertielle et gravitationnelle n'était pas de nature à faciliter l'extension de ce principe aux interactions non-gravitationnelles en général à cause d'une absence d'arguments similaires permettant de soumettre à la même accélération toutes les particules qui se déplacent uniquement sous l'action du champ en question. Au lieu de cela, il fallait plutôt chercher le moyen d'annuler l'effet du champ sur la particule de telle sorte à aboutir à un mouvement libre.

Pour commencer, on a discuté la validité du Principe d'Equivalence, tel qu'il a était formulé pour le champ de gravitation. On s'est aperçu que la localité, utilisée pour servir de cadre de validité du Principe, n'était rigoureusement utilisée que dans le cas limite d'un point. Cette remarque ouvre de grandes perspectives pour étendre le Principe d'Equivalence à toutes les interactions fondamentales, car en un point donné, il y a l'espoir d'annuler l'action de n'importe quel champ.

De plus, pour confirmer notre raisonnement, nous avons montré que les résultats surprenants de C.C.Barros sont dus à une utilisation implicite du Principe d'Equivalence, étendu au champ électrostatique.

Dans notre effort de recherche des possibilités d'étendre le Principe d'Equivalence au cas plus général de l'interaction électromagnétique on s'est penché sur l'approche de Iliev, selon laquelle le champ électromagnétique est annulé en un point quelconque grâce à une transformation de jauge particulière. En guise de critique, nous avons établi qu'une transformation de jauge ne peut pas être assimilée à une transformation de coordonnées. Ceci nous amène à la conclusion qu'il ne faut pas chercher à annuler le $A_{\mu}$ en un point quelconque par le biais d'une transformation de coordonnée particulière, mais plutôt à postuler l'existence d'équations de géodésiques pour une charge électrique soumise à un champ électromagnétique. Les symboles de Christoffel qui figurent dans ces équations permettraient, dans certaines conditions, de générer la force de Lorentz.

Il faut admettre que l'utilisation, par Barros, d'une solution similaire à celle de Schwarzschild (\ref{metrique etoile}) est tout à fait curieuse; elle nous amènent à soupçonner l'existence d'équations de type Einstein pour le champ électromagnétique qui doivent se réduire, dans certaines conditions à déterminer ultérieurement, aux équations de Maxwell.

\newpage

\pagestyle{fancy} \lhead{chapitre\;3}\rhead{La Gravité Linéaire}
\chapter{La Gravité Linéaire}

\section{Introduction}

L'analyse de l'approche de Barros \cite{Barros1}, pour décrire l'atome d'hydrogène, nous a permis de souligner deux points intéressants. Le premier point concerne l'utilisation d'une solution similaire à celle de Schwarzschild et va être à l'origine de notre hypothèse de l'existence d'équations de type Einstein pour l'interaction électromagnétique. Le second point concerne la possibilité de l'extension du Principe d'Equivalence à l'interaction électromagnétique et va nous permettre d'émettre l'hypothèse de l'existence d'équations de géodésiques pour des charges électrique soumises à un champ électromagnétique.

A la lumière des résultats surprenants de Barros, témoins d'une profonde unité entre la gravité et l'électromagnétisme, nous avons commencé nos investigations par la recherche de toutes les analogies qui pourraient conduire à une description géométrique unifiée de ces interactions fondamentales.

Dans ce chapitre, dans le but d'explorer les analogies qui puissent exister entre la gravité et l'électromagnétisme, nous allons nous pencher sur le domaine de la Gravité Linéaire où des phénomènes très intéressants ont lieu. Dans les régions où le champ gravitationnel est de faible intensité, il est possible d'adopter une approche perturbative de sorte à pouvoir étudier l'écart de la métrique de l'espace-temps par rapport à la métrique plate de Minkowski. L'appellation de Gravité Linéaire trouve son origine dans la volonté de limiter l'approche au premier ordre de la perturbation, de sorte à ne considérer que les termes linéaires et de pouvoir ainsi exclure les termes d'ordres supérieurs; bien évidemment, dans le cas où le champ gravitationnel est plus intense il est possible de pousser l'étude aux ordres supérieurs de la perturbation.

Dans un premier temps, un rappel des notions fondamentales de la Gravité Linéaire "standard" est présenté; une attention toute particulière est portée sur l'approche de P. Huei \cite{Huei} et celle de S. Carroll \cite{Carroll1}.

Dans l'approche de Huei, les équations d'Einstein se réduisent à des équations de type Maxwell et une particule test, astreinte à se mouvoir suivant des géodésiques, est soumise à une force gravitationnelle de type Lorentz. Néanmoins la partie "magnétique" de cette force contient un facteur 4 indésirable.

Dans l'approche de Carroll, les champs sont définis de telle manière à surmonter le problème lié à l'existence du facteur 4 indésirable dans la force gravitationnelle de type Lorentz, néanmoins les équations linéaires d'Einstein ne prennent plus la forme d'équations de type Maxwell.

Une critique des deux approches précédentes va nous permettre de souligner quelques imperfections dont la résolution va nous permettre de revisiter la Gravité linéaire de sorte à aboutir à une meilleure analogie entre la gravitation et l'électromagnétisme.

\section{Version standard}

\subsection{Définition de la limite du champs faible}

L'équation d'Einstein se réduit à l'équation de Newton dans le cas des champs faibles, statiques et pour des faibles vitesses des particules test. Dans ce qui suit, nous allons considérer une solution moins restrictive que la limite non relativiste, où le champ reste faible, mais n'est plus statique et où il n'y a pas de restriction sur la vitesse des particules test.

En effet, la présence d'une distribution matérielle crée une perturbation de la métrique dans son voisinage, et cette perturbation devient non stationnaire si les constituants de la source sont en mouvement. De plus, très loin de la source\footnote{A une distance largement supérieure au rayon de Schwarzschild $r\gg 2GM/c^{2}$} le champ gravitationnel est si faible que la métrique de l'espace-temps peut être vue comme une légère perturbation de la métrique de Minkowski
\begin{eqnarray}\label{plate plus perturbation}
    g_{\mu\nu}=\eta_{\mu\nu}+h_{\mu\nu},\hspace{1cm}\mid h_{\mu\nu}\mid \ll 1.
\end{eqnarray}
tel que $\eta_{\mu\nu}=(1,-1,-1,-1)$.

La perturbation est suffisamment faible pour pouvoir considérer qu'elle se propage sur une métrique "de fond" stationnaire, autrement dit, on se contente de perturber la métrique de Minkowski juste à l'ordre 1 de sorte à négliger tous les termes d'ordres supérieurs de cette perturbation $h_{\mu\nu}$. De plus, il est possible d'élever ou d'abaisser les indices en utilisant simplement $\eta_{\mu\nu}$ et $\eta^{\mu\nu}$, car l'erreur commise est d'un ordre supérieur de cette perturbation, donc négligeable. Dans ce cas, pour pouvoir satisfaire la condition $g_{\mu\nu}g^{\nu\sigma}=\delta^{\sigma}_{\mu}$, nous avons au premier ordre de la perturbation
\begin{eqnarray}\label{plate plus perturbation contrav}
    g^{\mu\nu}=\eta^{\mu\nu}-h^{\mu\nu},
\end{eqnarray}
avec $h^{\mu\nu}=\eta^{\mu\alpha}\eta^{\nu\beta}h_{\alpha\beta}$.

La propagation de la perturbation linéaire de la métrique est décrite par des équations d'Einstein qui ressemblent aux équations de Maxwell. De la même façon que les charges électriques en mouvement génèrent des ondes électromagnétiques, le mouvement des masses permet de générer des ondes gravitationnelles.

\subsection{Linéariation des équations d'Einstein}
Nous allons considérer une version des équations d'Einstein où les effets des ordres de $h_{\mu\nu}$ supérieurs au premier ordre sont négligés, une théorie où un champ de tenseurs symétrique $h_{\mu\nu}$ se propage dans un espace-temps plat de Minkowski.

\subsubsection{Symboles de Christoffel}
Conformément à (\ref{plate plus perturbation contrav}), les symboles de Christoffel (\ref{importantegij}) au premier ordre
\begin{eqnarray}
  \Gamma^{\rho}_{\mu\nu} &=& \frac{1}{2}\,g^{\rho\lambda} \left(\partial_{\mu}g_{\nu\lambda}+\partial_{\nu}g_{\lambda\mu}-\partial_{\lambda}g_{\mu\nu}\right)\nonumber\\
   &=& \frac{1}{2}\,(\eta^{\rho\lambda}-h^{\rho\lambda}) \left[\partial_{\mu}(\eta_{\nu\lambda}+h_{\nu\lambda})+\partial_{\nu}(\eta_{\lambda\mu}+h_{\lambda\mu})-\partial_{\lambda}(\eta_{\mu\nu}+h_{\mu\nu})\right]\nonumber\\
    &=& \frac{1}{2}\,\eta^{\rho\lambda} \left(\partial_{\mu}h_{\nu\lambda}+\partial_{\nu}h_{\lambda\mu}-\partial_{\lambda}h_{\mu\nu}\right)+\mathcal{O}(h^{2})\nonumber
\end{eqnarray}
sont donnés par l'expression
\begin{equation}\label{christoffel}
     \Gamma^{\rho}_{\mu\nu} \approx \frac{1}{2}\,\eta^{\rho\lambda} \Big(\partial_{\mu}h_{\nu\lambda}+\partial_{\nu}h_{\lambda\mu}-\partial_{\lambda}h_{\mu\nu}\Big).
\end{equation}

\subsubsection{Tenseur de Riemann}
A l'ordre un de la perturbation, le produit des symboles de Christoffel qui figurent dans l'expression du tenseur de courbure
\begin{equation}\label{tenseur de courbure}
    R_{\;\;\mu\rho\sigma}^{\,\lambda}=\partial_{\rho}\Gamma_{\mu\sigma}^{\lambda}-\partial_{\sigma}\Gamma_{\mu\rho}^{\lambda}
    +\underbrace{\Gamma_{\alpha\rho}^{\lambda}\Gamma_{\mu\sigma}^{\alpha}}_{\text{ordre 2}}-\underbrace{\Gamma_{\alpha\sigma}^{\lambda}\Gamma_{\mu\rho}^{\alpha}}_{\text{ordre 2}}
\end{equation}
sont négligés. Dans ce cas, d'après les relations (\ref{tenseur covariant courbure}) et (\ref{plate plus perturbation contrav}), le tenseur de Riemann est donné par
\begin{eqnarray}
  R_{\mu\nu\rho\sigma} &\approx& \eta_{\mu\lambda}\left[\partial_{\rho}\Gamma_{\nu\sigma}^{\lambda}-
  \partial_{\sigma}\Gamma_{\nu\rho}^{\lambda}\right] \nonumber\\
  &\approx& \frac{1}{2}\,\eta_{\mu\lambda}\;\partial_{\rho}\left[\eta^{\lambda\alpha}\left(\partial_{\nu}h_{\sigma\alpha}+\partial_{\sigma}h_{\alpha\nu}
  -\partial_{\alpha}h_{\nu\sigma}\right)\right]-\frac{1}{2}\,\eta_{\mu\lambda}\;\partial_{\sigma}\left[\eta^{\lambda\alpha}\left(\partial_{\nu}h_{\rho\alpha}+
  \partial_{\rho}h_{\alpha\nu}
  -\partial_{\alpha}h_{\nu\rho}\right)\right]\nonumber\\
  &\approx& \frac{1}{2}\,\delta_{\mu}^{\alpha}\left(\partial_{\rho}\partial_{\nu}h_{\sigma\alpha}+\partial_{\rho}\partial_{\sigma}h_{\alpha\nu}
  -\partial_{\rho}\partial_{\alpha}h_{\nu\sigma}\right)-\frac{1}{2}\,\delta_{\mu}^{\alpha}\left(\partial_{\sigma}\partial_{\nu}h_{\rho\alpha}+
  \partial_{\sigma}\partial_{\rho}h_{\alpha\nu}
  -\partial_{\sigma}\partial_{\alpha}h_{\nu\rho}\right)\nonumber\\
   &\approx& \frac{1}{2}\left(\partial_{\rho}\partial_{\nu}h_{\sigma\mu}+\partial_{\rho}\partial_{\sigma}h_{\mu\nu}-\partial_{\rho}\partial_{\mu}h_{\nu\sigma}-
   \partial_{\sigma}\partial_{\nu}h_{\rho\mu}-
  \partial_{\sigma}\partial_{\rho}h_{\mu\nu}
  +\partial_{\sigma}\partial_{\mu}h_{\nu\rho}\right)\nonumber
\end{eqnarray}
ou encore finalement
\begin{equation}\label{tenseur de riemann}
    R_{\mu\nu\rho\sigma}\approx \frac{1}{2}\Big(\partial_{\rho}\partial_{\nu}h_{\sigma\mu}+\partial_{\sigma}\partial_{\mu}h_{\nu\rho}-
   \partial_{\sigma}\partial_{\nu}h_{\rho\mu}-\partial_{\rho}\partial_{\mu}h_{\nu\sigma}
 \Big).
\end{equation}

\subsubsection{Tenseur de Ricci}
D'après les expressions du tenseur de Riemann (\ref{tenseur de riemann}) et de la métrique inverse (\ref{plate plus perturbation contrav}), les composantes du tenseur de Ricci (\ref{tenseur de Ricci chap rappel}) sont donnés au premier ordre par
\begin{eqnarray}
  R_{\mu\nu} &\approx& \eta^{\alpha\beta}\,R_{\alpha\mu\beta\nu}\nonumber\\
  &\approx& \frac{1}{2}\eta^{\alpha\beta}\Big(\partial_{\beta}\partial_{\mu}h_{\nu\alpha}+\partial_{\nu}\partial_{\alpha}h_{\mu\beta}-
   \partial_{\nu}\partial_{\mu}h_{\beta\alpha}-\partial_{\beta}\partial_{\alpha}h_{\mu\nu}\label{ricci tensor 1 electrom}
 \Big)
\end{eqnarray}
En introduisant la trace de la perturbation  $$h=\eta^{\alpha\beta}\,h_{\alpha\beta}=h^{\beta}_{\beta}=h_{00}-h_{11}-h_{22}-h_{33}$$
et le d'Alembertien $$\Box=\eta^{\alpha\beta}\,\partial_{\alpha}\partial_{\beta}=-\overrightarrow{\nabla}^{2}+\frac{1}{c^{\,2}}\frac{\partial^{\,2}}{\partial t^{\,2}},$$
nous avons finalement
\begin{equation}\label{tenseur ricci}
    R_{\mu\nu}\approx\frac{1}{2}\Big(\partial_{\sigma}\partial_{\mu}h_{\nu}^{\sigma}+\partial_{\nu}\partial_{\sigma}h_{\mu}^{\sigma}-
   \partial_{\nu}\partial_{\mu}h-\Box h_{\mu\nu}
 \Big)
\end{equation}

\subsubsection{Courbure scalaire}
La courbure scalaire (\ref{courbure scalaire chap rappel}), à l'ordre un de la perturbation, est obtenue en contractant le tenseur de Ricci (\ref{tenseur ricci}) par la métrique inverse (\ref{plate plus perturbation contrav})
\begin{eqnarray}
R &\approx& \eta^{\mu\nu}\,R_{\mu\nu}\nonumber\\
  &\approx& \frac{1}{2}\eta^{\mu\nu}\Big(\partial_{\sigma}\partial_{\mu}h_{\nu}^{\sigma}+\partial_{\nu}\partial_{\sigma}h_{\mu}^{\sigma}-
   \partial_{\nu}\partial_{\mu}h-\Box h_{\mu\nu}\Big)\nonumber\\
  &\approx&  \frac{1}{2}\Big(\partial_{\sigma}\partial_{\mu}h^{\sigma\mu}+\partial_{\nu}\partial_{\sigma}h^{\sigma\nu}-\eta^{\mu\nu}
   \partial_{\nu}\partial_{\mu}h-\eta^{\mu\nu}\Box h_{\mu\nu}\Big)\nonumber\\
   &\approx&  \frac{1}{2}\Big(\partial_{\nu}\partial_{\sigma}h^{\sigma\nu}+\partial_{\nu}\partial_{\sigma}h^{\sigma\nu}-\Box h-\Box h\Big)\nonumber\\
   &\approx&  \frac{1}{2}\Big(2\,\partial_{\nu}\partial_{\sigma}h^{\sigma\nu}-2\,\Box h\Big)\nonumber
\end{eqnarray}
pour avoir finalement
\begin{equation}\label{courbure scalaire}
    R \approx \partial_{\mu}\partial_{\nu}h^{\mu\nu}-\Box h.
\end{equation}

\subsubsection{Tenseur d'Einstein linéarisé}
Au premier ordre de la perturbation, le tenseur d'Einstein (\ref{tenseur einstein espace temps}) est obtenu grâce aux relations (\ref{courbure scalaire}) et (\ref{tenseur ricci}) d'ordres 1 et la métrique (\ref{plate plus perturbation}) à l'ordre 0,
\begin{eqnarray}
G_{\mu\nu} &\approx& R_{\mu\nu}-\frac{1}{2}\,\eta_{\mu\nu}\,R\nonumber\\
&\approx& \frac{1}{2}\Big(\partial_{\sigma}\partial_{\mu}h_{\nu}^{\sigma}+\partial_{\nu}\partial_{\sigma}h_{\mu}^{\sigma}-
   \partial_{\nu}\partial_{\mu}h-\Box h_{\mu\nu}
 \Big)-\frac{1}{2}\,\eta_{\mu\nu}\Big(\partial_{\alpha}\partial_{\beta}h^{\alpha\beta}-\Box h\Big),\nonumber
\end{eqnarray}
de sorte à avoir finalement
\begin{equation}\label{tenseur d'einstein linearise}
    G_{\mu\nu} \approx \frac{1}{2}\Big(\partial_{\sigma}\partial_{\mu}h_{\nu}^{\sigma}+\partial_{\nu}\partial_{\sigma}h_{\mu}^{\sigma}-
   \partial_{\nu}\partial_{\mu}h-\Box h_{\mu\nu}
 -\eta_{\mu\nu}\,\partial_{\alpha}\partial_{\beta}h^{\alpha\beta}+\eta_{\mu\nu}\,\Box h\Big).
\end{equation}

\subsubsection{Equations d'Einstein linéarisées}
Les équations d'Einstein, au premier ordre de la perturbation, sont données par
\begin{eqnarray}\label{equation einstein linearise}
  \frac{8\pi G}{c^{4}}\,T_{\mu\nu}&=& G_{\mu\nu}  \nonumber \\
  \frac{8\pi G}{c^{4}}\,T_{\mu\nu}&\approx&\frac{1}{2}\Big(\partial_{\sigma}\partial_{\mu}h_{\nu}^{\sigma}+\partial_{\nu}\partial_{\sigma}h_{\mu}^{\sigma}-
   \partial_{\nu}\partial_{\mu}h-\Box h_{\mu\nu}
 -\eta_{\mu\nu}\,\partial_{\alpha}\partial_{\beta}h^{\alpha\beta}+\eta_{\mu\nu}\,\Box h\Big).
\end{eqnarray}
où $T_{\mu\nu}$ représente le tenseur d'énergie-impulsion de la source du champ.

\subsection{Transformation de jauge}
La condition $g_{\mu\nu}=\eta_{\mu\nu}+h_{\mu\nu}$ ne spécifie pas complètement le système de coordonnées dans l'espace-temps. Il peut y avoir un autre système de coordonnées où la métrique peut toujours être écrite sous forme de la métrique de Minkowski plus une perturbation qui peut être différente de $h_{\mu\nu}$.

Soit la transformation de coordonnées infinitésimale suivante
\begin{equation}\label{transf coord infinit}
    x^{'\,\mu}=x^{\mu}+\xi^{\mu},
\end{equation}
où $\xi^{\mu}=\varepsilon\,\zeta^{\mu}$, avec $\varepsilon\ll 1$. La transformation inverse s'écrit alors
\begin{equation}\label{transf coord infinit inv}
    x^{\mu}=x^{'\,\mu}-\xi^{'\,\mu}
\end{equation}
de telle sorte que $\xi^{\mu}(x^{'\,\rho})=\xi^{'\,\,\mu}(x^{\rho})$. Les composantes du tenseur métrique, dans le nouveau système de coordonnées, sont données par
 $$g^{'}_{\mu\nu}(x^{'\,\sigma}) = \frac{\partial x^{\alpha}}{\partial x^{'\,\mu}}\frac{\partial x^{\beta}}{\partial x^{'\,\nu}}\;g_{\alpha\beta}(x^{\sigma}).$$
En se limitant aux termes du premier ordre, nous avons
\begin{eqnarray}
   \eta^{'}_{\mu\nu}+h^{'}_{\mu\nu} &\approx& \frac{\partial x^{\alpha}}{\partial x^{'\,\mu}}\frac{\partial x^{\beta}}{\partial x^{'\,\nu}}\left(\eta_{\alpha\beta}+h_{\alpha\beta}\right) \nonumber\\
   &\approx& \left(\delta^{\alpha}_{\mu}-\partial^{'}_{\mu}\xi^{'\,\alpha}\right)\left(\delta^{\beta}_{\nu}-\partial^{'}_{\nu}\xi^{'\,\beta}\right)
   \left(\eta_{\alpha\beta}+h_{\alpha\beta}\right) \nonumber\\
   &\approx& \Big[\delta^{\alpha}_{\mu}\delta^{\beta}_{\nu}-\delta^{\beta}_{\nu}\,\partial^{'}_{\mu}\xi^{'\,\alpha}
   -\delta^{\alpha}_{\mu}\,\partial^{'}_{\nu}\xi^{'\,\beta}+\underbrace{(\partial^{'}_{\mu}\xi^{'\,\alpha})(\partial^{'}_{\nu}\xi^{'\,\beta})}_{\text{ordre 2}}\Big]\left(\eta_{\alpha\beta}+h_{\alpha\beta}\right)
\end{eqnarray}
Les termes exprimés dans le nouveau système de coordonnées sont évalués au point $x^{'\,\sigma}$ alors que ceux exprimés dans l'ancien système de coordonnées sont évalués au point $x^{\sigma}$, de sorte que
   $$\eta^{'}_{\mu\nu}(x^{'\,\sigma})+h^{'}_{\mu\nu}(x^{'\,\sigma}) \approx \eta_{\mu\nu}(x^{\sigma})+h_{\mu\nu}(x^{\sigma})-\left(\eta_{\mu\beta}+h_{\mu\beta}\right)[\partial^{'}_{\nu}\xi^{'\,\beta}(x^{\sigma})]
   -\left(\eta_{\alpha\nu}+h_{\alpha\nu}\right)[\partial^{'}_{\mu}\xi^{'\,\alpha}(x^{\sigma})]$$
En admettant que $\eta^{'}_{\mu\nu}(x^{'\,\sigma})=\eta_{\mu\nu}(x^{\sigma})$, l'équation précédente se simplifie
$$h^{'}_{\mu\nu}(x^{'\,\sigma})\approx h_{\mu\nu}(x^{\sigma})-\eta_{\mu\beta}[\partial^{'}_{\nu}\xi^{'\,\beta}(x^{\sigma})]-\eta_{\alpha\nu}[\partial^{'}_{\mu}\xi^{'\,\alpha}(x^{\sigma})],
$$
pour donner lieu finalement à
\begin{eqnarray}\label{transf perturbation inifint}
  h^{'}_{\mu\nu}(x^{'\,\sigma})&\approx& h_{\mu\nu}(x^{\sigma})-\partial^{'}_{\nu}\xi^{'}_{\mu}(x^{\sigma})-\partial^{'}_{\mu}\xi^{'}_{\nu}(x^{\sigma}).
\end{eqnarray}
Mais à l'approximation linéaire en $h_{\mu\nu}$, nous pouvons d'une part prendre toutes les quantités figurant dans (\ref{transf perturbation inifint}) au point $x^{\sigma}$. En effet, un développement de Taylor au premier ordre de $\xi^{\sigma}$
\begin{eqnarray}
  h^{'}_{\mu\nu}(x^{'\,\sigma}) = h^{'}_{\mu\nu}(x^{\sigma}+\xi^{\sigma}) =  h^{'}_{\mu\nu}(x^{\sigma})+\underbrace{\xi^{\rho}[\partial_{\rho}h^{'}_{\mu\nu}(x^{\sigma})]}_{\text{ordre 2}}+\mathcal{O}(\xi^{2}),
\end{eqnarray}
permet de montrer que\footnote{En posant $h_{\mu\nu}=\epsilon\,\mathbf{h_{\mu\nu}}$ et $\xi^{\rho}=\epsilon\,\zeta^{\rho}$, avec $\epsilon\ll 1$, il vient que le terme $\xi^{\rho}[\partial_{\rho}h^{'}_{\mu\nu}(x^{\sigma})]=\epsilon^{2}\zeta^{\rho}[\partial_{\rho}\mathbf{h^{'}_{\mu\nu}}(x^{\sigma})]$ est du second ordre en $\epsilon$.}
\begin{equation}\label{meme point}
     h^{'}_{\mu\nu}(x^{'\,\sigma})\approx h^{'}_{\mu\nu}(x^{\sigma}),
\end{equation}
de sorte qu'en remplaçant (\ref{meme point}) dans (\ref{transf perturbation inifint}) nous obtenons
\begin{eqnarray}\label{jauge temporaire}
 h^{'}_{\mu\nu}(x^{\sigma}) \approx  h_{\mu\nu}(x^{\sigma})-\partial^{'}_{\nu}\xi^{'}_{\mu}(x^{\sigma})-\partial^{'}_{\mu}\xi^{'}_{\nu}(x^{\sigma}).
\end{eqnarray}
D'autre part, il est aussi possible de montrer qu'au premier ordre $\partial^{\,'}_{\alpha}\xi^{'\,\mu}\approx \partial_{\alpha}\xi^{\mu}$. En effet,
\begin{eqnarray}\label{deriv epsil epsilprim}
  \frac{\partial \xi^{'\,\mu}}{\partial x^{'\,\alpha}}= \frac{\partial\xi^{\mu}}{\partial x^{\beta}}\,\frac{\partial x^{\beta}}{\partial x^{'\,\alpha}} =\frac{\partial\xi^{\mu}}{\partial x^{\beta}}\left(\delta^{\beta}_{\alpha}-\frac{\partial\xi^{'\,\beta}}{\partial x^{'\,\alpha}}\right)\approx \frac{\partial\xi^{\mu}}{\partial x^{\alpha}}.
\end{eqnarray}
Donc compte tenu de (\ref{deriv epsil epsilprim}), l'équation (\ref{jauge temporaire}) se réduit finalement à la transformation de jauge
\begin{eqnarray}\label{jauge}
 h^{'}_{\mu\nu}(x^{\sigma}) \approx  h_{\mu\nu}(x^{\sigma})-\partial_{\nu}\xi_{\mu}(x^{\sigma})-\partial_{\mu}\xi_{\nu}(x^{\sigma}).
\end{eqnarray}
soit \begin{eqnarray}
       \delta h_{\mu\nu}(x^{\sigma}) &\equiv& h^{'}_{\mu\nu}(x^{\sigma})-h_{\mu\nu}(x^{\sigma}) = -\partial_{\nu}\xi_{\mu}(x^{\sigma})-\partial_{\mu}\xi_{\nu}(x^{\sigma}).
     \end{eqnarray}
Sous l'action des transformations linéaires et infinitésimales arbitraires (\ref{transf coord infinit}), le tenseur de courbure et le tenseur d'énergie-impulsion (évalués au premier odre) sont invariants. L'invariance de la théorie sous de telles transformations est analogue à l'invariance de jauge traditionnelle en électromagnétisme par la transformation $A_{\mu}\rightarrow A_{\mu}^{'}=A_{\mu}-\partial_{\mu}\chi$.

\subsection{Invariance du tenseur de Riemann par transformation de jauge}
Vérifions l'invariance du tenseur de Riemann linéarisé (\ref{tenseur de riemann}) sous la transformation de coordonnées infinitésimale (\ref{transf coord infinit}), ou de manière équivalente sous la transformation de jauge (\ref{jauge}). Pour ce faire, calculons la variation
\begin{eqnarray}
  \delta R_{\mu\nu\rho\sigma} &\equiv& R_{\mu\nu\rho\sigma}^{'}(x^{\beta})-R_{\mu\nu\rho\sigma}(x^{\beta}) \nonumber\\
  &=& \frac{1}{2}\left[\partial_{\rho}\partial_{\nu}h_{\sigma\mu}^{'}+\partial_{\sigma}\partial_{\mu}h_{\nu\rho}^{'}
   -\partial_{\sigma}\partial_{\nu}h_{\rho\mu}^{'}-\partial_{\rho}\partial_{\mu}h_{\nu\sigma}^{'}\right]\nonumber\\
   &&\hspace{5cm}-\frac{1}{2}\Big[\partial_{\rho}\partial_{\nu}h_{\sigma\mu}+\partial_{\sigma}\partial_{\mu}h_{\nu\rho}-
   \partial_{\sigma}\partial_{\nu}h_{\rho\mu}-\partial_{\rho}\partial_{\mu}h_{\nu\sigma}
 \Big]\nonumber\\
 &=& \Bigg(R_{\mu\nu\rho\sigma}+\frac{1}{2}\Big[\partial_{\rho}\partial_{\nu}\left(-\partial_{\sigma}\xi_{\mu}-\partial_{\mu}\xi_{\sigma}\right)
 +\partial_{\sigma}\partial_{\mu}\left(-\partial_{\nu}\xi_{\rho}-\partial_{\rho}\xi_{\nu}\right)\nonumber\\
  &&\hspace{3cm}-\partial_{\sigma}\partial_{\nu}\left(-\partial_{\rho}\xi_{\mu}-\partial_{\mu}\xi_{\rho}\right)-
  \partial_{\rho}\partial_{\mu}\left(-\partial_{\nu}\xi_{\sigma}-\partial_{\sigma}\xi_{\nu}\right)\Big]\Bigg)-R_{\mu\nu\rho\sigma}\nonumber
\end{eqnarray}
qui se réduit finalement à
\begin{equation}\label{invariance de jauge riemann}
    \delta R_{\mu\nu\rho\sigma}= 0.
\end{equation}

Il est claire qu'une telle variation du tenseur de Riemann conduit, bien évidemment, à l'invariance de jauge du tenseur d'Einstein par la transformation (\ref{jauge}). L'invariance du tenseur énergie-impulsion est due à la limitation au premier ordre de la perturbation de sorte que les forces gravitationnelles sont beaucoup plus faibles que les autres forces dues aux tensions dans le milieu, par exemple un solide en rotation. Dans ce cas, les équations de mouvement du milieu matériel ne sont pas affectées par le champ de gravitation engendré par celui-ci. Elles se réduisent donc aux équations minkowskiennes du mouvement, caractérisé par $\partial_{\mu}T^{\mu\nu}=0$ (au lieu de $\nabla_{\mu}T^{\mu\nu}=0$).

\subsection{La jauge harmonique}
Face à un système présentant une invariance par rapport à une transformation de jauge, le premier réflexe consiste à fixer une jauge. Ce choix de jauge va être conditionné par la simplification des équations dans une situation bien particulière.

\subsubsection{Définition de la jauge harmonique}
La jauge harmonique\footnote{Dite aussi jauge de Hilbert ou bien jauge de Donder.} est spécifiée par les quatre conditions suivantes
\begin{equation}
g^{\mu\nu}\,\Gamma^{\rho}_{\mu\nu}=0,
\end{equation}
une relation "temporelle" pour $\rho=0$ et trois relations "spatiales" pour $\rho=i=1,2,3$.

Au premier ordre de la perturbation $h_{\mu\nu}$, la condition précédente est donnée par
\begin{eqnarray}
  \frac{1}{2}\,\eta^{\mu\nu}\eta^{\lambda\rho} (\partial_{\mu}h_{\nu\lambda}+\partial_{\nu}h_{\lambda\mu}-\partial_{\lambda}h_{\mu\nu})&=& 0 \nonumber\\
  \frac{1}{2} (\partial_{\mu}h^{\mu\rho}+\partial_{\nu}h^{\rho\nu}-\eta^{\lambda\rho}\,\partial_{\lambda}h)&=&0\nonumber\\
  \frac{1}{2}(2\,\partial_{\mu}h^{\mu\rho}-\eta^{\lambda\rho}\,\partial_{\lambda}h)&=&0\nonumber\\
  \partial_{\mu}h^{\mu\rho}- \frac{1}{2}\,\eta^{\lambda\rho}\,\partial_{\lambda}h&=&0,\nonumber
\end{eqnarray}
soit encore en appliquant une contraction par $\eta_{\rho\beta}$
\begin{eqnarray*}
\eta_{\rho\beta}\,\partial_{\mu}h^{\mu\rho}- \frac{1}{2}\,\eta^{\lambda\rho}\eta_{\rho\beta}\,\partial_{\lambda}h&=&0\\
\partial_{\mu}h^{\mu}_{\beta}- \frac{1}{2}\,\delta^{\lambda}_{\beta}\,\partial_{\lambda}h&=&0,
\end{eqnarray*}
nous aboutissons finalement à la condition de jauge harmonique dans la limite du champ faible
\begin{equation}\label{jauge harmonique}
    \partial_{\mu}h^{\mu}_{\beta}- \frac{1}{2}\,\partial_{\beta}h=0.
\end{equation}
Cette dernière condition est équivalente aux quatre conditions
\begin{eqnarray}
  \left\{
  \begin{array}{ll}
    \partial_{\mu}h^{\mu}_{0}-\displaystyle\frac{1}{2}\,\partial_{0}h=0, \\\\
    \partial_{\mu}h^{\mu}_{i}-\displaystyle\frac{1}{2}\,\partial_{i}h=0.
  \end{array}
\right.
\end{eqnarray}

En utilisant la jauge harmonique (\ref{jauge harmonique}), les équations d'Einstein linéarisées (\ref{equation einstein linearise}) se mettent sous la forme
\begin{eqnarray}
  \frac{8\pi G}{c^{4}}\,T_{\mu\nu}&\approx&\frac{1}{2}\bigg(\partial_{\mu}(\partial_{\sigma}h_{\nu}^{\sigma})+\partial_{\nu}(\partial_{\sigma}h_{\mu}^{\sigma})-
   \partial_{\nu}\partial_{\mu}h-\Box h_{\mu\nu}
 -\eta_{\mu\nu}\,\partial_{\alpha}\partial_{\beta}(\eta^{\alpha\sigma}\,h^{\beta}_{\sigma})+\eta_{\mu\nu}\,\Box h\bigg)\nonumber\\
 &\approx&\frac{1}{2}\left(\frac{1}{2}\,\partial_{\mu}\partial_{\nu}h+\frac{1}{2}\,\partial_{\nu}\partial_{\mu}h-
   \partial_{\nu}\partial_{\mu}h-\Box h_{\mu\nu}
 -\eta_{\mu\nu}\,\eta^{\alpha\sigma}\,\partial_{\alpha}(\partial_{\beta}\,h^{\beta}_{\sigma})+\eta_{\mu\nu}\,\Box h\right)\nonumber\\
 &\approx&\frac{1}{2}\left(-\Box h_{\mu\nu}
 -\frac{1}{2}\,\eta_{\mu\nu}\,\eta^{\alpha\sigma}\,\partial_{\alpha}\,\partial_{\sigma}\,h+\eta_{\mu\nu}\,\Box h\right)\nonumber\\
 &\approx&\frac{1}{2}\left(-\Box h_{\mu\nu}
 -\frac{1}{2}\,\eta_{\mu\nu}\,\Box h+\eta_{\mu\nu}\,\Box h\right)\nonumber\\
 &\approx&\frac{1}{2}\left(-\Box h_{\mu\nu}
 +\frac{1}{2}\,\eta_{\mu\nu}\,\Box h\right)\nonumber\\
 &\approx&-\frac{1}{2}\left(\Box h_{\mu\nu}
 -\frac{1}{2}\,\eta_{\mu\nu}\,\Box h\right)\nonumber\\
 &\approx&-\frac{1}{2}\,\Box\left( h_{\mu\nu}\label{equation einstein lineaire harmonique bis}
 -\frac{1}{2}\,\eta_{\mu\nu}\, h\right),
\end{eqnarray}
soit
\begin{equation}\label{equation einstein lineaire harmonique}
    \Box\left( h_{\mu\nu}
 -\frac{1}{2}\,\eta_{\mu\nu}\, h\right)\approx -\frac{16\pi G}{c^{4}}\,T_{\mu\nu}.
\end{equation}

\subsubsection{Changement de métrique}
Posons par commodité
\begin{equation}\label{changement de metrique}
    \overline{h}_{\mu\nu}=h_{\mu\nu}-\frac{1}{2}\,\eta_{\mu\nu}\,h.
\end{equation}
Dans le but d'inverser cette relation, i.e exprimer $h_{\mu\nu}$, il faut déterminer la trace de $\overline{h}_{\mu\nu}$. Pour ce faire, appliquons $\eta^{\mu\nu}$ à (\ref{changement de metrique})
\begin{eqnarray}
  \overline{h} &\equiv& \eta^{\mu\nu}\,\overline{h}_{\mu\nu}
  =  \eta^{\mu\nu}\,h_{\mu\nu}-\frac{1}{2}\,\eta^{\mu\nu}\eta_{\mu\nu}\,h =  h-2h  \nonumber
\end{eqnarray}
pour avoir la relation entre les traces
\begin{equation}\label{trace metrique}
    \overline{h} =-h.
\end{equation}
A cause de cette propriété, la nouvelle métrique $\overline{h}_{\mu\nu}$, définie par (\ref{changement de metrique}), est dite la perturbation à trace inversée. En remplaçant (\ref{trace metrique}) dans (\ref{changement de metrique}), on aboutit à la relation inverse
\begin{equation}\label{changement de metrique inverse}
    h_{\mu\nu}=\overline{h}_{\mu\nu}-\frac{1}{2}\,\eta_{\mu\nu}\,\overline{h}.
\end{equation}
De même en appliquant $\eta^{\mu\rho}$ à (\ref{changement de metrique inverse}) nous obtenons aussi
\begin{eqnarray}\label{metrique nouvelle mixte}
  \eta^{\mu\rho}\,h_{\mu\nu} &=&  \eta^{\mu\rho}\,\overline{h}_{\mu\nu}-\frac{1}{2}\,\eta^{\mu\rho}\eta_{\mu\nu}\,\overline{h} \nonumber\\
  h_{\nu}^{\rho} &=&  \overline{h}_{\nu}^{\,\rho}-\frac{1}{2}\,\delta_{\nu}^{\rho}\,\overline{h}.
\end{eqnarray}
En utilisant la "nouvelle" métrique $\overline{h}_{\mu\nu}$, les équations d'Einstein linéarisées (\ref{equation einstein linearise}) deviennent
\begin{eqnarray}
  \frac{16\pi G}{c^{4}}\,T_{\mu\nu}&\approx& \partial_{\sigma}\partial_{\mu}h_{\nu}^{\sigma}+\partial_{\nu}\partial_{\sigma}h_{\mu}^{\sigma}-
   \partial_{\nu}\partial_{\mu}h-\Box h_{\mu\nu}
 -\eta_{\mu\nu}\,\partial_{\alpha}\partial_{\beta}h^{\alpha\beta}+\eta_{\mu\nu}\,\Box h\nonumber\\
  &\approx& \partial_{\sigma}\partial_{\mu}\left(\overline{h}_{\nu}^{\sigma}-\frac{1}{2}\,\delta_{\nu}^{\sigma}\,\overline{h}\right)
 +\partial_{\nu}\partial_{\sigma}\left(\overline{h}_{\mu}^{\sigma}-\frac{1}{2}\,\delta_{\mu}^{\sigma}\,\overline{h}\right)+
   \partial_{\nu}\partial_{\mu}\overline{h}\nonumber\\
 &&\hspace{2.5cm}-\Box \left(\overline{h}_{\mu\nu}-\frac{1}{2}\,\eta_{\mu\nu}\,\overline{h}\right)
 -\eta_{\mu\nu}\,\partial_{\alpha}\partial_{\beta}\left(\overline{h}^{\alpha\beta}-\frac{1}{2}\,\eta^{\alpha\beta}\,\overline{h}\right)-\eta_{\mu\nu}\,\Box \overline{h} \nonumber\\
 \nonumber\\
 &\approx& \partial_{\sigma}\partial_{\mu}\,\overline{h}_{\nu}^{\sigma}-\frac{1}{2}\,\partial_{\nu}\partial_{\mu}\,\overline{h}
 +\partial_{\nu}\partial_{\sigma}\,\overline{h}_{\mu}^{\sigma}-\frac{1}{2}\,\partial_{\nu}\partial_{\mu}\,\overline{h}+
   \partial_{\nu}\partial_{\mu}\overline{h}\nonumber\\
 &&\hspace{2cm}-\Box \overline{h}_{\mu\nu}+\frac{1}{2}\,\eta_{\mu\nu}\,\Box\overline{h}
 -\eta_{\mu\nu}\,\partial_{\alpha}\partial_{\beta}\,\overline{h}^{\alpha\beta}+
 \frac{1}{2}\,\eta_{\mu\nu}\,\eta^{\alpha\beta}\partial_{\alpha}\partial_{\beta}\,\,\overline{h}-\eta_{\mu\nu}\,\Box \overline{h} \nonumber\\
 \nonumber\\
 &\approx& \partial_{\sigma}\partial_{\mu}\,\overline{h}_{\nu}^{\sigma}-\frac{1}{2}\,\partial_{\nu}\partial_{\mu}\,\overline{h}
 +\partial_{\nu}\partial_{\sigma}\,\overline{h}_{\mu}^{\sigma}-\frac{1}{2}\,\partial_{\nu}\partial_{\mu}\,\overline{h}+
   \partial_{\nu}\partial_{\mu}\overline{h}\nonumber\\
 &&\hspace{3.3cm}-\Box \overline{h}_{\mu\nu}+\frac{1}{2}\,\eta_{\mu\nu}\,\Box\overline{h}
 -\eta_{\mu\nu}\,\partial_{\alpha}\partial_{\beta}\,\overline{h}^{\alpha\beta}+
 \frac{1}{2}\,\eta_{\mu\nu}\,\Box\overline{h}-\eta_{\mu\nu}\,\Box \overline{h} \nonumber
\end{eqnarray}
Finalement

\begin{eqnarray}\label{der prim}
   -\Box \overline{h}_{\mu\nu}-\eta_{\mu\nu}\,\partial_{\alpha}\partial_{\beta}\,\overline{h}^{\,\alpha\beta}
  +\partial_{\sigma}\partial_{\mu}\,\overline{h}_{\nu}^{\,\sigma}+\partial_{\nu}\partial_{\sigma}\,\overline{h}_{\mu}^{\,\sigma}\approx\frac{16\pi G}{c^{4}}\,T_{\mu\nu}
\end{eqnarray}
L'utilisation de la nouvelle métrique (\ref{changement de metrique}) n'a d'intérêt pratique que combinée au choix de jauge harmonique .

\subsubsection{La jauge harmonique combinée à un changement de métrique}
L'utilisation de la jauge harmonique (\ref{jauge harmonique}), combinée à la nouvelle métrique (\ref{changement de metrique}) va permettre de simplifier considérablement les calculs. En effet, l'équation (\ref{equation einstein lineaire harmonique}) se réduit dans ce cas à l'équation de propagation
\begin{equation}\label{equation einstein lineaire harmonique hbar}
    \Box \overline{h}_{\mu\nu}\approx -\frac{16\pi G}{c^{4}}\,T_{\mu\nu}.
\end{equation}
A présent, il est possible de déterminer les équations du champs de gravitation sous forme d'équations de Maxwell, mais avant de le faire, essayons d'exprimer la condition de jauge harmonique en utilisant la nouvelle métrique (\ref{changement de metrique}).

A partir de (\ref{metrique nouvelle mixte}) et (\ref{trace metrique}) nous pouvons montrer que
\begin{eqnarray}\label{metrique nouvelle mixte inverse}
  \overline{h}_{\nu}^{\rho} =  h_{\nu}^{\,\rho}-\frac{1}{2}\,\delta_{\nu}^{\rho}\,h,
\end{eqnarray}
ce qui permet de mettre la condition (\ref{jauge harmonique}) sous la forme
\begin{eqnarray}
    \partial_{\mu}h^{\mu}_{\beta}- \frac{1}{2}\,\delta^{\mu}_{\beta}\partial_{\mu}h&=&0\nonumber\\
    \partial_{\mu}\left(h^{\mu}_{\beta}- \frac{1}{2}\,\delta^{\mu}_{\beta}h\right)&=&0,\nonumber
\end{eqnarray}
ou encore en fonction de la nouvelle métrique
\begin{equation}\label{jauge harmonique nouvelle}
    \partial_{\mu}\overline{h}^{\,\mu}_{\beta}=0.
\end{equation}
Montrons qu'il est toujours possible de choisir une jauge (\ref{jauge harmonique nouvelle}), i.e. déterminons la transformation de coordonnées particulière (\ref{transf coord infinit}) permettant de vérifier la condition de jauge harmonique dans la nouvelle métrique. Pour ce faire, combinons (\ref{jauge}) et (\ref{changement de metrique}) pour exprimer le changement de jauge $\overline{h}_{\mu\nu}\longrightarrow\overline{h}^{\,'}_{\mu\nu}$ tel que
\begin{eqnarray}
   \overline{h}^{\,'}_{\mu\nu}&=&h_{\mu\nu}^{\,'}-\frac{1}{2}\,\eta_{\mu\nu}\,h^{\,'}\nonumber\\
   &=&\left(h_{\mu\nu}-\partial_{\nu}\xi_{\mu}-\partial_{\mu}\xi_{\nu}\right)-\frac{1}{2}\,\eta_{\mu\nu}\,\left(\eta^{\alpha\beta}\,h_{\alpha\beta}^{'}\right)\nonumber\\
  &=&\left(h_{\mu\nu}-\partial_{\nu}\xi_{\mu}-\partial_{\mu}\xi_{\nu}\right)-
  \frac{1}{2}\,\eta_{\mu\nu}\,\left[\eta^{\alpha\beta}\,\left(h_{\alpha\beta}-\partial_{\alpha}\xi_{\beta}-\partial_{\beta}\xi_{\alpha}\right)\right]\nonumber\\
    &=&\left(h_{\mu\nu}-\partial_{\nu}\xi_{\mu}-\partial_{\mu}\xi_{\nu}\right)-
  \frac{1}{2}\,\eta_{\mu\nu}\,\left(h-2\,\eta^{\alpha\beta}\,\partial_{\alpha}\xi_{\beta}\right)\nonumber\\
  &=&\left(h_{\mu\nu}-\frac{1}{2}\,\eta_{\mu\nu}\,h\right)-\partial_{\nu}\xi_{\mu}-\partial_{\mu}\xi_{\nu}
  +\eta_{\mu\nu}\,\eta^{\alpha\beta}\,\partial_{\alpha}\xi_{\beta},\nonumber
\end{eqnarray}
soit encore finalement
\begin{equation}
  \overline{h}_{\mu\nu}\hspace{0.3cm}\longrightarrow\hspace{0.3cm}\overline{h}^{\,'}_{\mu\nu}=\overline{h}_{\mu\nu}-\partial_{\nu}\xi_{\mu}-\partial_{\mu}\xi_{\nu}
  +\eta_{\mu\nu}\,\partial_{\alpha}\xi^{\alpha}.
\end{equation}
Une double contraction par la métrique de Minkowski permet d'obtenir les composantes contravariantes au premier ordre $$\overline{h}^{\,'\,\rho\sigma}=\overline{h}^{\,\rho\sigma}
  -\eta^{\mu\rho}\eta^{\nu\sigma}\left(\partial_{\nu}\xi_{\mu}+\partial_{\mu}\xi_{\nu}\right)
  +\eta^{\mu\rho}\eta^{\nu\sigma}\,\eta_{\mu\nu}\,\partial_{\alpha}\xi^{\alpha}$$
  de sorte à avoir
\begin{eqnarray}
\overline{h}^{\,\rho\sigma}\hspace{0.3cm}\longrightarrow\hspace{0.3cm}
  \overline{h}^{\,'\,\rho\sigma}&=&\overline{h}^{\,\rho\sigma}
  -\eta^{\nu\sigma}\,\partial_{\nu}\xi^{\rho}-\eta^{\mu\rho}\,\partial_{\mu}\xi^{\sigma}
   +\eta^{\rho\sigma}\,\partial_{\alpha}\xi^{\alpha}.
   \end{eqnarray}
   Appliquons à présent une dérivation (au premier ordre, nous avons $\partial^{'}_{\sigma}\approx\partial_{\sigma}$)
   \begin{eqnarray*}
     \partial^{'}_{\sigma}\,\overline{h}^{\,'\,\rho\sigma}
  &=&\partial_{\sigma}\,\overline{h}^{\,\rho\sigma}
  -\eta^{\nu\sigma}\,\partial_{\sigma}\,\partial_{\nu}\xi^{\rho}-\underbrace{\eta^{\mu\rho}\,\partial_{\sigma}\,\partial_{\mu}\xi^{\sigma}}_{\mu\rightarrow \sigma \;\text {et}\; \sigma\rightarrow \alpha}
   +\eta^{\rho\sigma}\,\partial_{\sigma}\,\partial_{\alpha}\xi^{\alpha}\\
   &=&\partial_{\sigma}\,\overline{h}^{\,\rho\sigma}
  -\eta^{\nu\sigma}\,\partial_{\sigma}\,\partial_{\nu}\xi^{\rho}-\eta^{\sigma\rho}\,\partial_{\alpha}\,\partial_{\sigma}\,\xi^{\alpha}
   +\eta^{\rho\sigma}\,\partial_{\sigma}\,\partial_{\alpha}\xi^{\alpha}\\\\
   &=&\partial_{\sigma}\,\overline{h}^{\,\rho\sigma}
  -\eta^{\nu\sigma}\,\partial_{\sigma}\,\partial_{\nu}\xi^{\rho},
   \end{eqnarray*}
pour avoir finalement
\begin{eqnarray}\label{derivee nouvelle metrique nulle}
\partial_{\sigma}\,\overline{h}^{\,\rho\sigma}\hspace{0.3cm}\longrightarrow\hspace{0.3cm}
  \partial^{'}_{\sigma}\,\overline{h}^{\,'\,\rho\sigma} &=& \partial_{\sigma}\,\overline{h}^{\,\rho\sigma}
  -\Box\xi^{\rho}.
\end{eqnarray}
La transformation de coordonnées (\ref{transf coord infinit}) pour laquelle $\xi^{\mu}$ vérifie la condition
\begin{equation}
    \Box\xi^{\mu}=\partial_{\sigma}\,\overline{h}^{\,\mu\sigma},
\end{equation}
permet de passer à un système de coordonnées où la condition de jauge (\ref{jauge harmonique nouvelle}) est vérifiée
 \begin{eqnarray}\label{der}
\partial_{\sigma}\,\overline{h}^{\,\rho\sigma}\hspace{0.3cm}\longrightarrow\hspace{0.3cm}
  \partial^{'}_{\sigma}\,\overline{h}^{\,'\,\rho\sigma} &=& 0.
\end{eqnarray}

Remarquons finalement que l'application de la condition de jauge (\ref{jauge harmonique nouvelle}) dans (\ref{der prim}) permet aussi d'aboutir à l'équation de propagation (\ref{equation einstein lineaire harmonique hbar}).

\subsection{Equations d'Einstein linéarisées sous forme d'équations de type Maxwell: Approche de P. Huei}
Dans ce qui suit nous allons montrer que, sous certaines conditions, les équations d'Einstein linéarisées peuvent se réduirent à un ensemble d'équations similaire au groupe d'équations de Maxwell.

Dans un premier temps, nous supposerons que la composante dominante du tenseur énergie-impulsion $T^{oo}\propto c^{2}$ (\ref{t00 chap1}) est la seule composante non nulle. Ainsi, dans le cadre de l'approximation de Newton, l'accélération de la particule est proportionnelle au gradient d'un certain potentiel scalaire $\phi_{g}$ produisant un effet radial sur une particule test (il permettra de définir le champs gravitoélectrique).

Dans un deuxième temps, nous allons considérer que le mouvement des particules n'est pas suffisamment lent pour pouvoir négliger les composantes $T^{0i}\propto c\,v^{i}$ (\ref{t0i chap1}). Par contre, les contraintes à l'intérieur de la source du champs de gravitation sont supposées négligeables, ainsi les composantes $T^{ij}\propto v^{i}\,v^{j}$ sont nulles (\ref{tij chap1}). Dans ce cas de figure, l'équation de mouvement d'une particule est similaire à l'équation d'une charge électrique soumise à une force de Lorentz. Il y a apparition d'un nouveau type d'effet de gravitation similaire à l'effet du champs magnétique sur une charge électrique. Celui-ci est du à un potentiel vecteur permettant de définir un champ gravitomagnétique qui agit de manière orthoradiale sur la particule test.

\subsubsection{Champ Gravitoélectrique}
Dans le cas de l'approximation de Newton\footnote{champ faible, particules lentes, régime stationnaire et contraintes à l'intérieur de la source négligeables}, la seule composante non nulle du tenseur énergie-impulsion est $T^{oo}=\rho_{\text{\tiny{m}}}\,c^{2}$.

Nous cherchons des solutions stationnaires de (\ref{equation einstein lineaire harmonique hbar}) pour lesquelles $\partial_{t}\overline{h}_{\mu\nu}$ sont négligés
\begin{eqnarray}
  \underbrace{\frac{1}{c^{2}}\,\frac{\partial^{2}}{\partial t^{2}}\,\overline{h}_{\mu\nu}}_{\text{négligé}}-\overrightarrow{\nabla}^{2}\,\overline{h}_{\mu\nu} \approx -\frac{16\pi G}{c^{4}}\,T_{\mu\nu}\nonumber
\end{eqnarray}
\begin{eqnarray}\label{deux equ}
\Rightarrow\hspace{0.5cm}\left\{
                             \begin{array}{ll}
                              \overrightarrow{\nabla}^{2}\,\overline{h}_{\mu\nu}\approx 0 \hspace{0.5cm}\forall\; \mu,\nu \;\;(\text{sauf quand}\; \mu=\nu=0) \\\\
                               \overrightarrow{\nabla}^{2}\,\overline{h}_{00}\approx\displaystyle\frac{16\pi\,G\,\rho_{\text{\tiny{m}}}}{c^{2}}.
                             \end{array}
                           \right.
\end{eqnarray}
La première équation de (\ref{deux equ}) n'admet de solutions régulières compatibles avec l'exigence de la convergence asymptotique, à l'infini spacial, vers une métrique de Minkowski que si $\overline{h}_{ij}=0$ et $\overline{h}_{i0}=0$.

Dans le but de réduire la deuxième équation de (\ref{deux equ}) à une équation de type Poisson, introduisons le potentiel gravitationnel
\begin{equation}\label{potentiel gravitation}
    \phi_{g}=\frac{c^{2}}{4}\,\overline{h}_{00}
\end{equation}
pour avoir
\begin{equation}\label{poisson}
    \overrightarrow{\nabla}^{2}\,\phi_{g}\approx 4\pi\,G\,\rho_{\text{\tiny{m}}}.
\end{equation}

A l'approximation newtonienne, où $\tau\simeq t$, l'équation des géodésiques (\ref{eq geodes bis}) d'une particule test
\begin{eqnarray}
\frac{d^{2}x^{\mu}}{d\tau^{2}} = -\Gamma^{\mu}_{00}\,\frac{dx^{0}}{d\tau}\,\frac{dx^{0}}{d\tau}-2\,\Gamma^{\mu}_{0i}\,\frac{dx^{0}}{d\tau}\,\frac{dx^{i}}{d\tau}
 -\Gamma^{\mu}_{ij}\,\frac{dx^{i}}{d\tau}\,\frac{dx^{j}}{d\tau}\nonumber
\end{eqnarray}
se réduit à
\begin{eqnarray}\label{eq geodes approxim newton}
\frac{d^{2}x^{\mu}}{dt^{2}} \simeq -\Gamma^{\mu}_{00}\,\frac{c\,dt}{dt}\,\frac{c\,dt}{dt}-2\,\Gamma^{\mu}_{0i}\,\frac{c\,dt}{dt}\,\frac{dx^{i}}{dt}
-\Gamma^{\mu}_{ij}\,\frac{dx^{i}}{dt}\,\frac{dx^{j}}{dt}\simeq-c^{2}\,\Gamma^{\mu}_{00}-2c\,v^{i}\,\Gamma^{\mu}_{oi}-v^{i}\,v^{j}\,\Gamma^{\mu}_{ij}
\end{eqnarray}
En négligeant les deux derniers termes par rapport au premier nous avons
\begin{equation}\label{accel gammaoo}
    \frac{d^{2}x^{\mu}}{dt^{2}}\approx -c^{2}\,\Gamma^{\mu}_{00}.
\end{equation}
Or d'après (\ref{christoffel}), le symbole de Christoffel apparaissant dans l'expression précédente
\begin{eqnarray}
  \Gamma^{\mu}_{00} \approx \frac{1}{2}\,\eta^{\mu\lambda}(\partial_{0}h_{0\lambda}+\partial_{0}h_{\lambda 0}-\partial_{\lambda}h_{00}) \nonumber
\end{eqnarray}
se réduit dans le cas stationnaire à
\begin{equation}
\Gamma^{\mu}_{00} \approx -\frac{1}{2}\,\eta^{\mu\lambda}\,\partial_{\lambda}h_{00}.
\end{equation}
Pour $\mu=i=0,1,3$, nous avons
\begin{equation}\label{gamma oo}
    \Gamma^{i}_{00}\approx \frac{1}{2}\,\partial_{i}h_{00},
\end{equation}
de sorte que les composantes spatiales de l'accélération de la particule soient données par
\begin{equation}\label{accel gammaoo bis}
    \frac{d^{2}x^{i}}{dt^{2}}\approx -\frac{c^{2}}{2}\,\partial_{i}h_{00}.
\end{equation}
Il ne reste plus à présent que d'exprimer cette accélération en fonction du potentiel gravitationnel $\phi_{g}$. Pour ce faire, rappelons que dans le cadre de l'approximation newtonienne
\begin{eqnarray}
  \overline{h}_{ij} &=& 0 \\
  \overline{h}_{i0} &=& 0 \\
  \overline{h}_{00} &=& 4\phi_{g}/c^{2},
\end{eqnarray}
de sorte que (\ref{changement de metrique inverse}) permet de déduire les composantes de la perturbation de la métrique
\begin{eqnarray}
  h_{ij} &=& \overline{h}_{ij}-\frac{1}{2}\,\eta_{ij}\,\overline{h}=0 \\
  h_{i0} &=& 0 \\
  h_{00} &=& \overline{h}_{00}-\frac{1}{2}\,\eta_{00}\,\overline{h}=\overline{h}_{00}-\frac{1}{2}\,(\overline{h}_{00}-\overline{h}_{11}-\overline{h}_{22}-\overline{h}_{33})
  =\frac{1}{2}\,\overline{h}_{00}=\frac{2}{c^{2}}\,\phi_{g}.\label{hoo kz}
\end{eqnarray}
En remplaçant (\ref{hoo kz}) dans (\ref{accel gammaoo bis})
\begin{equation}\label{pfd i}
    \frac{d^{2}x^{i}}{dt^{2}}\approx -\partial_{i}\phi_{g}, \hspace{0.2cm}i=1,2,3
\end{equation}
nous aboutissons à l'expression finale
\begin{center}
\begin{equation}\label{pfd}
    \overrightarrow{a}\approx -\overrightarrow{\nabla}\phi_{g}.
\end{equation}
\textbf{C.Q.F.D}\vspace{0.5cm}\end{center}

A l'approximation newtonnienne, le champ de gravitation radial $\overrightarrow{g}$, créé par une source matérielle, dérive du potentiel scalaire $\phi_{g}$
de telle sorte que la force qui agit sur une masse ponctuelle $m_{g}$ est donnée par
\begin{equation}
    m_{g}\,\overrightarrow{g}\approx m_{i}(-\overrightarrow{\nabla}\phi_{g}).
\end{equation}
Ce champ de gravitation est analogue au champ électrostatique $\overrightarrow{E}$ qui agit sur une charge ponctuelle $q$
\begin{equation}
    q\,\overrightarrow{E}=m_{i}(-\overrightarrow{\nabla}\phi).
\end{equation}
Pour cette raison, le champ $\overrightarrow{g}$ est appelé champ Gravitoélectrique.

\subsubsection{Champ Gravitomagnétique et équations pour la gravitation de type Maxwell}
De façon analogue au champ électromagnétique $F^{\mu\nu}=\partial^{\mu}A^{\nu}-\partial^{\nu}A^{\mu}$, P. Huei \cite{Huei} a introduit la combinaison linéaire suivante
\begin{equation}
\mathcal{G}^{\mu\nu\lambda}\equiv\frac{1}{4}\left(\partial^{\lambda}\overline{h}^{\,\mu\nu}-\partial^{\nu}\overline{h}^{\,\mu\lambda}+\eta^{\mu\nu}\,\partial_{\alpha}\overline{h}^{\,\lambda\alpha}
    -\eta^{\mu\lambda}\,\partial_{\alpha}\overline{h}^{\,\nu\alpha}\right)
\end{equation}
des dérivées de la perturbation de la métrique pour déterminer les effets des corps en mouvement. Les dérivations peuvent être écrites de façon plus simple en adoptant la convention $\partial_{\mu}W=W_{,\mu}$ et $\partial^{\mu}W=W^{,\mu}$ de sorte que
\begin{equation}\label{bmunulambda}
    \mathcal{G}^{\mu\nu\lambda}\equiv \frac{1}{4}\left(\overline{h}^{\,\mu\nu,\lambda}-\overline{h}^{\,\mu\lambda,\nu}+\eta^{\mu\nu}\,\overline{h}^{\,\lambda\alpha}_{\;\;\;\;\;,\alpha}
    -\eta^{\mu\lambda}\,\overline{h}^{\,\nu\alpha}_{\;\;\;\;\;,\alpha}\right).
\end{equation}
\begin{enumerate}
  \item \textbf{Quelques propriétés de $\mathcal{G}^{\mu\nu\lambda}$}

  Dans la jauge harmonique
\begin{equation}\label{jauge harmonique en notation nouvelle}
    \overline{h}^{\mu\nu}_{\;\;\;\;,\nu}=0,
\end{equation}
montrons les trois propriétés suivantes
\begin{eqnarray}
  &&\mathcal{G}^{\mu\nu\lambda} = -\;\mathcal{G}^{\mu\lambda\nu} \label{antisym}\\
  &&\mathcal{G}^{\mu\nu\lambda}+\mathcal{G}^{\nu\lambda\mu}+\mathcal{G}^{\lambda\mu\nu} = 0 \label{cyclicte}\\
  &&\mathcal{G}^{\alpha\mu\nu,\lambda}+\mathcal{G}^{\alpha\nu\lambda,\mu}+\mathcal{G}^{\alpha\lambda\mu,\nu} = 0\label{cyclicte complique}
\end{eqnarray}

La propriété d'antisymétrie (\ref{antisym}) découle de l'utilisation de la condition de jauge harmonique dans la définition (\ref{bmunulambda})
\begin{center}
\begin{eqnarray}
  \mathcal{G}^{\mu\nu\lambda} &=& \frac{1}{4}\Big(\overline{h}^{\,\mu\nu,\lambda}-\overline{h}^{\,\mu\lambda,\nu}+\eta^{\mu\nu}\,\underbrace{\overline{h}^{\,\lambda\alpha}_{\;\;\;\;\;,\alpha}}_{0}
    -\eta^{\mu\lambda}\,\underbrace{\overline{h}^{\,\nu\alpha}_{\;\;\;\;\;,\alpha}}_{0}\Big) = -\frac{1}{4}\left(\overline{h}^{\,\mu\lambda,\nu}-\overline{h}^{\,\mu\nu,\lambda}\right) = -\;\mathcal{G}^{\mu\lambda\nu}\nonumber
\end{eqnarray}
\textbf{C.Q.F.D}\vspace{0.5cm}\end{center}

Pour montrer la propriété de cyclicité (\ref{cyclicte}), il suffit de sommer membre à membre les expressions
\begin{eqnarray}
   &&\mathcal{G}^{\mu\nu\lambda} = \frac{1}{4}\left(\overline{h}^{\,\mu\nu,\lambda}-\overline{h}^{\,\mu\lambda,\nu}\right) \nonumber\\
   &&\mathcal{G}^{\nu\lambda\mu} = \frac{1}{4}\left(\overline{h}^{\,\nu\lambda,\mu}-\overline{h}^{\,\nu\mu,\lambda}\right) \nonumber\\
   &&\mathcal{G}^{\lambda\mu\nu} = \frac{1}{4}\left(\overline{h}^{\,\lambda\mu,\nu}-\overline{h}^{\,\lambda\nu,\mu}\right) \nonumber
\end{eqnarray}
pour avoir finalement
$$\mathcal{G}^{\mu\nu\lambda}+\mathcal{G}^{\nu\lambda\mu}+\mathcal{G}^{\lambda\mu\nu} = 0\nonumber$$
\begin{center}\textbf{C.Q.F.D}\vspace{0.5cm}\end{center}

De même pour montrer la cyclicité par rapport aux trois derniers indices de $G^{\alpha\mu\nu,\lambda}$ figurant dans (\ref{cyclicte complique}), il suffit de sommer membre à membre
\begin{eqnarray}
  &&\mathcal{G}^{\alpha\mu\nu,\lambda} = \eta^{\lambda\sigma}\,\mathcal{G}^{\alpha\mu\nu}_{\;\;\;\;\;\;,\sigma}=\frac{1}{4}\,\eta^{\lambda\sigma}\,\partial_{\sigma}
  \left(\overline{h}^{\,\alpha\mu,\nu}-\overline{h}^{\,\alpha\nu,\mu}\right)=\frac{1}{4}
  \left(\overline{h}^{\,\alpha\mu,\nu\lambda}-\overline{h}^{\,\alpha\nu,\mu\lambda}\right) \nonumber\\
 &&\mathcal{G}^{\alpha\nu\lambda,\mu} = \frac{1}{4}
  \left(\overline{h}^{\,\alpha\nu,\lambda\mu}-\overline{h}^{\,\alpha\lambda,\nu\mu}\right) \nonumber\\
  &&\mathcal{G}^{\alpha\lambda\mu,\nu} = \frac{1}{4}
  \left(\overline{h}^{\,\alpha\lambda,\mu\nu}-\overline{h}^{\,\alpha\mu,\lambda\nu}\right) \nonumber
\end{eqnarray}
pour avoir finalement
$$\mathcal{G}^{\alpha\mu\nu,\lambda}+\mathcal{G}^{\alpha\nu\lambda,\mu}+\mathcal{G}^{\alpha\lambda\mu,\nu} = 0$$
\begin{center}\textbf{C.Q.F.D}\vspace{0.5cm}\end{center}\newpage
  \item \textbf{Une expression de l'équation d'Einstein linéarisée dans la jauge harmonique}

  A partir de (\ref{der prim}) l'équation d'Einstein linéarisée, dans la jauge harmonique, devient
\begin{eqnarray}
   \frac{4\pi G}{c^{4}}\;\eta^{\mu\rho}\eta^{\nu\gamma}\,T_{\mu\nu}&\approx& \frac{1}{4}\left[
  \eta^{\mu\rho}\eta^{\nu\gamma}\left(
  \partial_{\nu}\partial_{\sigma}\,\overline{h}_{\mu}^{\,\sigma}+\partial_{\sigma}\partial_{\mu}\,\overline{h}_{\nu}^{\,\sigma}\right)-\eta^{\mu\rho}\eta^{\nu\gamma}\,\Box \overline{h}_{\mu\nu}-\eta^{\mu\rho}\eta^{\nu\gamma}\,\eta_{\mu\nu}\,\partial_{\alpha}\partial_{\beta}\,\overline{h}^{\,\alpha\beta}\right]\nonumber\\
  \frac{4\pi G}{c^{4}}\;T^{\rho\gamma}&\approx&\frac{1}{4}\left[\eta^{\nu\gamma}\,\partial_{\nu}\partial_{\sigma}\,\overline{h}^{\,\sigma\rho}
  +\eta^{\mu\rho}\,\partial_{\sigma}\partial_{\mu}\,\overline{h}^{\,\sigma\gamma}-\Box \overline{h}^{\,\rho\gamma}-\eta^{\,\rho\gamma}\,\partial_{\alpha}\partial_{\beta}\,\overline{h}^{\,\alpha\beta}\right]\nonumber\\
  \frac{4\pi G}{c^{4}}\;T^{\rho\gamma}&\approx&\frac{1}{4}\left[\eta^{\beta\gamma}\,\partial_{\beta}\partial_{\alpha}\,\overline{h}^{\,\alpha\rho}
  +\eta^{\alpha\rho}\,\partial_{\beta}\partial_{\alpha}\,\overline{h}^{\,\beta\gamma}-\eta^{\alpha\beta}\,\partial_{\alpha}\partial_{\beta} \overline{h}^{\,\rho\gamma}-\eta^{\,\rho\gamma}\,\partial_{\alpha}\partial_{\beta}\,\overline{h}^{\,\alpha\beta}\right]\nonumber\\
   -\frac{4\pi G}{c^{4}}\;T^{\rho\gamma}&\approx&\frac{1}{4}\left[\left(\eta^{\alpha\beta}\,\partial_{\alpha}\partial_{\beta}\,\overline{h}^{\,\rho\gamma}-
  \eta^{\beta\gamma}\,\partial_{\beta}\partial_{\alpha}\,\overline{h}^{\,\alpha\rho}\right)+
  \left(\eta^{\,\rho\gamma}\,\partial_{\alpha}\partial_{\beta}\,\overline{h}^{\,\alpha\beta}
  -\eta^{\alpha\rho}\,\partial_{\beta}\partial_{\alpha}\,\overline{h}^{\,\beta\gamma}\right)\right]\nonumber\\
  -\frac{4\pi G}{c^{4}}\;T^{\rho\gamma}&\approx&\partial_{\alpha}\;\left\{\frac{1}{4}\left[\left(\eta^{\alpha\beta}\,\partial_{\beta}\,\overline{h}^{\,\rho\gamma}-
  \eta^{\beta\gamma}\,\partial_{\beta}\,\overline{h}^{\,\rho\alpha}\right)+
  \left(\eta^{\,\rho\gamma}\,\partial_{\beta}\,\overline{h}^{\,\alpha\beta}
  -\eta^{\rho\alpha}\,\partial_{\beta}\,\overline{h}^{\,\gamma\beta}\right)\right]\right\}\nonumber\\
  -\frac{4\pi G}{c^{4}}\;T^{\rho\gamma}&\approx&\partial_{\alpha}\;\left\{\frac{1}{4}\left[\left(\overline{h}^{\,\rho\gamma,\alpha}-
  \overline{h}^{\,\rho\alpha,\gamma}\right)+
  \left(\eta^{\,\rho\gamma}\,\overline{h}^{\,\alpha\beta}_{\;\;\;\;,\beta}
  -\eta^{\rho\alpha}\,\overline{h}^{\,\gamma\beta}_{\;\;\;\;,\beta}\right)\right]\right\}\nonumber
\end{eqnarray}
pour pouvoir enfin l'écrire sous la forme
\begin{equation}\label{eq base}
  \partial_{\alpha}\mathcal{G}^{\rho\gamma\alpha} \approx -\frac{4\pi G}{c^{4}}\;T^{\rho\gamma}.
\end{equation}
L'intérêt d'introduire le tenseur (\ref{bmunulambda}), c'est de pouvoir mettre les équations d'Einstein linéaires dans la jauge harmonique (\ref{eq base}) sous une forme analogue aux équations de Maxwell
\begin{equation}
    \partial_{\mu}F^{\mu\nu}=\mu_{0}J^{\nu},
\end{equation}
avec source.
  \item \textbf{Groupe d'équations de la gravité sous forme d'équations de Maxwell}

Introduisons le champ Gravitoélectrique $\overrightarrow{g}=(g^{1},g^{2},g^{3})$ tel que \cite{Huei}
\begin{eqnarray}
    &&\left\{
      \begin{array}{ll}
        g^{i}=c^{2}\;\mathcal{G}^{00i} \\
        i=1,2,3
      \end{array}
    \right. 
    \label{gooi}
\end{eqnarray}
le potentiel vecteur $\overrightarrow{A_{g}}=(A_{g}^{\;1},A_{g}^{\;2},A_{g}^{\;3})$ dont les composantes sont définies par \cite{Huei}
\begin{eqnarray}
    &&\left\{
      \begin{array}{ll}
        A_{g}^{\;i}=c\,\overline{h}^{\,0i}/4 \\
        i=1,2,3
      \end{array}
    \right. 
    \label{hoi}
    \end{eqnarray}
et le champ Gravitomagnétique $\overrightarrow{B_{g}}=(B_{g}^{\;1},B_{g}^{\;2},B_{g}^{\;3})$, avec \cite{Huei}
\begin{eqnarray}
    &&\left\{
      \begin{array}{ll}
        B_{g}^{\;1}=c\;\mathcal{G}^{023} \\\\
        B_{g}^{\;2}=c\;\mathcal{G}^{031}\\\\
        B_{g}^{\;3}=c\;\mathcal{G}^{012}
      \end{array}
    \right. 
    \label{bg}
\end{eqnarray}
Utilisons (\ref{bg}) et (\ref{bmunulambda}) pour montrer que $\overrightarrow{B_{g}}$ est défini à partir du potentiel vecteur $\overrightarrow{A_{g}}$ de façon analogue au champ magnétique. En effet, nous avons pour chaque composante
\begin{eqnarray}
  B_{g}^{\;1}&=&c\;\mathcal{G}^{023}=\frac{c}{4}\left(\overline{h}^{\,02,3}-\overline{h}^{\,03,2}\right)
=\partial^{3}\,A_{g}^{\;2}-\partial^{2}\,A_{g}^{\;3} \nonumber\\
B_{g}^{\;2}&=&c\;\mathcal{G}^{031}=\frac{c}{4}\left(\overline{h}^{\,03,1}-\overline{h}^{\,01,3}\right)
=\partial^{1}\,A_{g}^{\;3}-\partial^{3}\,A_{g}^{\;1} \nonumber\\
B_{g}^{\;3}&=&c\;\mathcal{G}^{012}=\frac{c}{4}\left(\overline{h}^{\,01,2}-\overline{h}^{\,02,1}\right)
=\partial^{2}\,A_{g}^{\;1}-\partial^{1}\,A_{g}^{\;2} \nonumber
\end{eqnarray}
ou encore de façon équivalente
\begin{eqnarray}
    \left\{
      \begin{array}{ll}
        c\;\mathcal{G}^{0ij}=\partial^{j}\,A_{g}^{\;i}-\partial^{i}\,A_{g}^{\;j}=A_{g}^{\;i,j}-A_{g}^{\;j,\,i} \\
        i,j=1,2,3
      \end{array}
    \right.
\end{eqnarray}
ce qui permet finalement de montrer que le champ Gravitomagnétique dérive du potentiel vecteur selon la relation
\begin{center}
\begin{equation}\label{champ gravit mag deriv pot vect}
\overrightarrow{B_{g}}=\overrightarrow{\nabla}\times\overrightarrow{A_{g}}.
\end{equation}
\textbf{C.Q.F.D}\vspace{0.5cm}\end{center}

Déterminons à présent le groupe d'équations de la gravité analogue au groupe d'équations de Maxwell:
\begin{enumerate}
\item Compte tenu de la définition des composantes du champ Gravitomagnétique (\ref{bg}), l'équation (\ref{cyclicte complique}) permet d'avoir
\begin{eqnarray}
  \mathcal{G}^{012,3}+\mathcal{G}^{023,1}+\mathcal{G}^{031,2} &=& 0 \nonumber\\
 \frac{1}{c}\left(\partial^{3}\,B_{g}^{\;3}+\partial^{1}\,B_{g}^{\;1}+\partial^{2}\,B_{g}^{\;2}\right) &=& 0 \nonumber
\end{eqnarray}
ou encore de façon équivalente
\begin{center}
\begin{equation}\label{divbg}
    \overrightarrow{\nabla}.\overrightarrow{B_{g}} = 0.
\end{equation}
\textbf{C.Q.F.D}\vspace{0.5cm}\end{center}
Cette relation est compatible avec la définition (\ref{champ gravit mag deriv pot vect}); elle exprime le fait que le flux du champ Gravitomagnétique à travers une surface quelconque est nul. L'équation de Maxwell analogue à (\ref{divbg}) est $$\text{div} \overrightarrow{B}=0.$$
\item En s'intéressant à la composante $T^{00}$, l'équation d'Einstein (\ref{eq base})
\begin{eqnarray}
 -\frac{4\pi G}{c^{4}}\;T^{00} &\approx&\displaystyle\frac{\partial \mathcal{G}^{00\lambda}}{\partial x^{\lambda}} \nonumber\\
  -\frac{4\pi G}{c^{4}}(\rho_{\text{\tiny{m}}}\,c^{2})&\approx&  \partial_{0}\,\underbrace{\mathcal{G}^{000}}_{0}+\partial_{1}\,\mathcal{G}^{001}+\partial_{2}\,\mathcal{G}^{002}+\partial_{3}\,\mathcal{G}^{003} \nonumber
\end{eqnarray}
permet de retrouver l'équation locale qui décrit la génération du champ Gravitoélectrique à partir d'une densité de masse
\begin{center}
\begin{equation}\label{divg}
    \overrightarrow{\nabla}.\overrightarrow{g} \approx -4\pi G\rho_{\text{\tiny{m}}}
\end{equation}
\textbf{C.Q.F.D}\vspace{0.5cm}\end{center}

Cette relation est analogue à l'équation de Maxwell $$\text{div} \overrightarrow{E}=\rho/\varepsilon_{0}.$$
\item En s'intéressant à la composante $T^{i0}$, l'équation d'Einstein (\ref{eq base}) permet aussi d'avoir
\begin{eqnarray}
\hspace*{-1.5cm}-\frac{4\pi G}{c^{4}}\;T^{i0} &\approx&   \displaystyle\frac{\partial \mathcal{G}^{i0\lambda}}{\partial x^{\lambda}} \nonumber\\
\hspace*{-1.5cm}  -\frac{4\pi G}{c^{4}}\left[(\rho_{\text{\tiny{m}}}\,c^{2})\frac{v^{i}}{c}\right]&\approx&    \underbrace{\partial_{0}\,\mathcal{G}^{0i0}}_{-\partial_{0}\,\mathcal{G}^{00i}}+\partial_{1}\,\mathcal{G}^{0i1}+\partial_{2}\,\mathcal{G}^{0i2}
+\partial_{3}\,\mathcal{G}^{0i3}\nonumber\\
 \hspace*{-1.5cm}-\frac{4\pi G\rho_{\text{\tiny{m}}}}{c^{3}}\,v^{i} &\approx& -\frac{1}{c}\,\frac{\partial}{\partial t}\left(\frac{g^{i}}{c^{2}}\right)+\frac{1}{c}\big[\partial_{1}\left(A_{g}^{\;i,1}-A_{g}^{\;1,\,i}\right)+\partial_{2}\left(A_{g}^{\;i,2}-A_{g}^{\;2,\,i}\right)\nonumber\\
&&+\partial_{3}\left(A_{g}^{\;i,3}-A_{g}^{\;3,\,i}\right)\big].\nonumber
\end{eqnarray}
Pour chaque valeur $i=1,2,3$ de l'indice spatial libre, nous obtenons une équation
\begin{eqnarray}
\left\{
                             \begin{array}{ll}
                               \partial_{2}\left(A_{g}^{\;1,2}-A_{g}^{\;2,\,1}\right)
+\partial_{3}\left(A_{g}^{\;1,3}-A_{g}^{\;3,\,1}\right)\approx -\displaystyle\frac{4\pi G\rho_{\text{\tiny{m}}}}{c^{2}}\,v^{1}+\displaystyle\frac{1}{c^{2}}\,\frac{\partial g^{1}}{\partial t} \nonumber\\\\
                               \partial_{1}\left(A_{g}^{\;2,1}-A_{g}^{\;1,\,2}\right)
+\partial_{3}\left(A_{g}^{\;2,3}-A_{g}^{\;3,\,2}\right)\approx -\displaystyle\frac{4\pi G\rho_{\text{\tiny{m}}}}{c^{2}}\,v^{2}+\displaystyle\frac{1}{c^{2}}\,\frac{\partial g^{2}}{\partial t} \nonumber\\\\
                               \partial_{1}\left(A_{g}^{\;3,1}-A_{g}^{\;1,\,3}\right)
+\partial_{2}\left(A_{g}^{\;3,2}-A_{g}^{\;2,\,3}\right)\approx -\displaystyle\frac{4\pi G\rho_{\text{\tiny{m}}}}{c^{2}}\,v^{3}+\displaystyle\frac{1}{c^{2}}\,\frac{\partial g^{3}}{\partial t}.\nonumber\\
                             \end{array}
                           \right.\nonumber
\end{eqnarray}
Compte tenu de la définition des composantes du champ Gravitomagnétique (\ref{bg}), les trois équations précédentes se mettent sous la forme
\begin{eqnarray}
\left\{
                             \begin{array}{ll}
                               +\partial_{2}\,B_{g}^{\;3}
-\partial_{3}\,B_{g}^{\;2}\approx -\displaystyle\frac{4\pi G\rho_{\text{\tiny{m}}}}{c^{2}}\,v^{1}+\displaystyle\frac{1}{c^{2}}\,\frac{\partial g^{1}}{\partial t} \\\\
                               -\partial_{1}\,B_{g}^{\;3}
+\partial_{3}\,B_{g}^{\;1}\approx -\displaystyle\frac{4\pi G\rho_{\text{\tiny{m}}}}{c^{2}}\,v^{2}+\displaystyle\frac{1}{c^{2}}\,\frac{\partial g^{2}}{\partial t} \\\\
                               +\partial_{1}\,B_{g}^{\;2}
-\partial_{2}\,B_{g}^{\;1}\approx -\displaystyle\frac{4\pi G\rho_{\text{\tiny{m}}}}{c^{2}}\,v^{3}+\displaystyle\frac{1}{c^{2}}\,\frac{\partial g^{3}}{\partial t},\\
                             \end{array}
                           \right.\nonumber
\end{eqnarray}
ou de façon équivalente
\begin{center}
\begin{equation}\label{rotbg}
    \overrightarrow{\nabla}\times\overrightarrow{B_{g}}\approx -\displaystyle\frac{4\pi G}{c^{2}}\,\left(\rho_{\text{\tiny{m}}}\overrightarrow{v}\right)+\displaystyle\frac{1}{c^{2}}\,\frac{\partial \overrightarrow{g}}{\partial t}.
\end{equation}
\textbf{C.Q.F.D}\vspace{0.5cm}\end{center}

Cette relation est analogue à l'équation de Maxwell $$\text{rot}\overrightarrow{B}=\mu_{0}\overrightarrow{j}+\mu_{0}\varepsilon_{0}\,\frac{\partial \overrightarrow{E}}{\partial t},$$
où $\mu_{0}\varepsilon_{0}=c^{-2}$ et la densité de courant est donnée par $\overrightarrow{j}=\rho\overrightarrow{v}$, en fonction de la densité de charge $\rho$.
\item En utilisant la définition des composantes du champ Gravitoélectrique (\ref{gooi}) et du potentiel vecteur ( \ref{hoi}), la relation (\ref{cyclicte complique}) permet d'avoir
\begin{eqnarray}
        0 &=& \mathcal{G}^{00i,j}+\mathcal{G}^{0ij,0}+\mathcal{G}^{0j0,i}  \nonumber\\
        0 &=& \partial^{j}\,\mathcal{G}^{00i}+\partial^{0}\,\mathcal{G}^{0ij}+\partial^{i}\,\underbrace{\mathcal{G}^{0j0}}_{-\mathcal{G}^{00j}} \nonumber\\
        0 &=& \partial^{j}\left(g^{i}/c^{2}\right)+\displaystyle\frac{1}{c}\,\partial^{0}\left(A_{g}^{\;i,j}-A_{g}^{\;j,\,i}\right)-
\partial^{i}\left(g^{j}/c^{2}\right) \nonumber\\
0 &=& \displaystyle\frac{1}{c^{2}}\left(\partial^{j}\,g^{i}-\partial^{i}\,g^{j}\right)
+\displaystyle\frac{1}{c^{2}}\partial_{t}\left(A_{g}^{\;i,j}-A_{g}^{\;j,\,i}\right).
\end{eqnarray}
En explicitant les valeurs des indices libres $i,j=1,2,3$ nous aboutissons aux trois relations
\begin{eqnarray}
\left\{
  \begin{array}{ll}
    \left(\partial^{2}\,g^{1}-\partial^{1}\,g^{2}\right)=-\partial_{t}\left(A_{g}^{\;1,2}-A_{g}^{\;2,\,1}\right)\\\nonumber\\
    \left(\partial^{3}\,g^{1}-\partial^{1}\,g^{3}\right)=-\partial_{t}\left(A_{g}^{\;1,3}-A_{g}^{\;3,\,1}\right) \\\nonumber\\
    \left(\partial^{2}\,g^{3}-\partial^{3}\,g^{2}\right)=-\partial_{t}\left(A_{g}^{\;3,2}-A_{g}^{\;2,\,3}\right)
  \end{array}
\right.
\end{eqnarray}
qui expriment les composantes spatiales du produit vectoriel du champ Gravitoélectrique
\begin{eqnarray}
\left\{
  \begin{array}{ll}
    \left(\overrightarrow{\nabla}\times \overrightarrow{g}\right)^{3}=-\partial_{t}\left(A_{g}^{\;1,2}-A_{g}^{\;2,\,1}\right)\\
    \left(\overrightarrow{\nabla}\times \overrightarrow{g}\right)^{2}=-\partial_{t}\left(A_{g}^{\;1,3}-A_{g}^{\;3,\,1}\right) \\
    \left(\overrightarrow{\nabla}\times \overrightarrow{g}\right)^{1}=-\partial_{t}\left(A_{g}^{\;3,2}-A_{g}^{\;2,\,3}\right)
  \end{array}
\right.\nonumber
      \end{eqnarray}
      de sorte à avoir finalement
\begin{center}
\begin{equation}\label{rotg}
    \overrightarrow{\nabla}\times \overrightarrow{g}=-\frac{\partial \overrightarrow{B_{g}}}{\partial t}.
\end{equation}
\textbf{C.Q.F.D}\vspace{0.5cm}\end{center}

L'équation (\ref{rotg}) est analogue à l'équation de Maxwell $$\text{rot} \overrightarrow{E}=-\frac{\partial \overrightarrow{B}}{\partial t}.$$
\item Montrons aussi la relations suivante
\begin{equation}\label{propagag}
    \Box \overrightarrow{A_{g}}\approx -\frac{4\pi G}{c^{2}}\;\left(\rho_{\text{\tiny{m}}}\overrightarrow{v}\right)
\end{equation}

L'utilisation de la composante définie par $\mu=0$ et $\nu=i$ de l'équation d'Einstein linéaire (\ref{equation einstein lineaire harmonique hbar}) permet d'avoir
\begin{eqnarray}
  \frac{c}{4}\;\Box \overline{h}^{\,0i}&\approx& -\frac{4\pi G}{c^{3}}\;T^{0i} \nonumber\\
 \Box A_{g}^{\,i}&\approx& -\frac{4\pi G}{c^{3}}\left[(\rho_{\text{\tiny{m}}}\,c^{2})\frac{v^{i}}{c}\right] \nonumber
\end{eqnarray}
ce qui conduit à la relation
\begin{center}
\begin{eqnarray}
\left\{
    \begin{array}{ll}
      \Box A_{g}^{\,i}\approx -\displaystyle\frac{4\pi G}{c^{2}}\left(\rho_{\text{\tiny{m}}}\,v^{i}\right) \nonumber\\\nonumber\\
      i=1,2,3
    \end{array}
  \right.
\end{eqnarray}
\textbf{C.Q.F.D}\vspace{0.5cm}\end{center}
\end{enumerate}

  \item \textbf{Equation de mouvement d'une particule soumise à une force de gravité de type Lorentz}

Reprenons l'équation des géodésiques (\ref{eq geodes approxim newton}), écrite à faibles vitesses où $\tau\simeq t$,
\begin{eqnarray}
\frac{d^{2}x^{\mu}}{dt^{2}}
&\simeq&-c^{2}\,\Gamma^{\mu}_{00}-2\,c\,v^{i}\,\Gamma^{\mu}_{oi}-v^{i}\,v^{j}\,\Gamma^{\mu}_{ij}
\end{eqnarray}
En se plaçant dans le cas où les contraintes à l'intérieur de la source matérielle sont négligeables de sorte à pouvoir négliger le dernier terme par rapport aux deux premiers ($T^{ij}\propto v^{i}v^{j}$ est négligé devant la densité d'énergie $T^{00}\propto c^{2}$ et la densité de quantité de mouvement $T^{0i}\propto cv^{i}$), nous pouvons écrire
\begin{equation}\label{eq geod gravitomag}
    \frac{d^{2}x^{\mu}}{dt^{2}}\approx -c^{2}\,\Gamma^{\mu}_{00}-2\,c\,v^{i}\,\Gamma^{\mu}_{oi}
\end{equation}
Dans le cas stationnaire et conformément à (\ref{christoffel}) nous avons, d'une part,
\begin{equation}\label{gamma oo gravitom}
    \Gamma^{l}_{00}\approx \frac{1}{2}\,\partial_{l}h_{00},
\end{equation}
et d'autre part,
\begin{eqnarray}\label{gamma i0j}
  \Gamma^{l}_{0i} &\approx& \frac{1}{2}\,\eta^{l\lambda}(\partial_{0}h_{i\lambda}+\partial_{i}h_{\lambda 0}-\partial_{\lambda}h_{0i})\approx \frac{1}{2}\,\underbrace{\eta^{ll}}_{-1}(\underbrace{\partial_{0}h_{il}}_{0}+\partial_{i}h_{l0}-\partial_{l}h_{0i})\nonumber\\\nonumber\\
  \Gamma^{l}_{0i} &\approx& -\frac{1}{2}(\partial_{i}h_{l0}-\partial_{l}h_{0i})
\end{eqnarray}
En tenant compte de (\ref{gamma i0j}), (\ref{gamma oo gravitom}) et (\ref{eq geod gravitomag}), nous obtenons pour chaque valeur de l'indice libre $l$ de l'équation (\ref{eq geod gravitomag}), les trois composantes de l'accélération
\begin{enumerate}
  \item Pour $l=1$
\begin{eqnarray}
  \frac{d^{2}x^{1}}{dt^{2}}&\approx& -\frac{c^{2}}{2}\,\partial_{1}h_{00}-2\,c\,\left(-\frac{1}{2}\right)(\partial_{i}h_{01}-\partial_{1}h_{0i})v^{i} \nonumber\\
   &\approx& -\frac{c^{2}}{2}\,\partial_{1}h_{00}+c\left(\partial_{1}h_{01}-\partial_{1}h_{01}\right)v^{1}+c\left(\partial_{2}h_{01}-\partial_{1}h_{02}\right)v^{2} \nonumber\\\nonumber\\
&&+c\left(\partial_{3}h_{01}-\partial_{1}h_{03}\right)v^{3}\nonumber\\\nonumber\\
 \frac{d^{2}x^{1}}{dt^{2}}  &\approx& -\frac{c^{2}}{2}\,\partial_{1}h_{00}+4\left[c\left(\partial_{2}h_{01}-\partial_{1}h_{02}\right)\frac{v^{2}}{4}
+c\left(\partial_{3}h_{01}-\partial_{1}h_{03}\right)\frac{v^{3}}{4}\right]
\end{eqnarray}
  \item  Pour $l=2$
\begin{eqnarray}
  \frac{d^{2}x^{2}}{dt^{2}}&\approx& -\frac{c^{2}}{2}\,\partial_{2}h_{00}-2\,c\,\left(-\frac{1}{2}\right)(\partial_{i}h_{02}-\partial_{2}h_{0i})v^{i} \nonumber\\\nonumber\\
 \frac{d^{2}x^{2}}{dt^{2}}  &\approx& -\frac{c^{2}}{2}\,\partial_{2}h_{00}+4\left[c\left(\partial_{1}h_{02}-\partial_{2}h_{01}\right)\frac{v^{1}}{4}
+c\left(\partial_{3}h_{02}-\partial_{2}h_{03}\right)\frac{v^{3}}{4}\right]
\end{eqnarray}
  \item Pour $l=3$
\begin{eqnarray}
  \frac{d^{2}x^{3}}{dt^{2}}&\approx& -\frac{c^{2}}{2}\,\partial_{3}h_{00}-2\,c\,\left(-\frac{1}{2}\right)(\partial_{i}h_{03}-\partial_{3}h_{0i})v^{i} \nonumber\\\nonumber\\
 \frac{d^{2}x^{2}}{dt^{2}}  &\approx& -\frac{c^{2}}{2}\,\partial_{3}h_{00}+4\left[c\left(\partial_{1}h_{03}-\partial_{3}h_{01}\right)\frac{v^{1}}{4}
+c\left(\partial_{2}h_{03}-\partial_{3}h_{02}\right)\frac{v^{2}}{4}\right]
\end{eqnarray}
\end{enumerate}
Dans la jauge harmonique nous avons
\begin{equation}\label{facteur 4 origine}
A_{g}^{\;i}=\eta^{i\sigma}\,(A_{g})_{\sigma}=\eta^{ii}\,(A_{g})_{\sigma}=-(A_{g})_{i}\hspace{0.5cm}\Rightarrow\hspace{0.5cm}(A_{g})_{i}=-\frac{c}{4}\,\overline{h}^{0i}
=\frac{c}{4}\,\overline{h}_{0i}=\frac{c}{4}\,h_{0i}
\end{equation}

de sorte que les composantes de l'accélération précédentes deviennent
\begin{eqnarray}
  \frac{d^{2}x^{1}}{dt^{2}}  &\approx& -\left(\overrightarrow{\nabla}\phi_{g}\right)_{1}+\mathbf{4}\,\Bigg\{\bigg[\partial_{2}(A_{g})_{1}-\partial_{1}(A_{g})_{2}\bigg]v^{2}
+\bigg[\partial_{3}(A_{g})_{1}-\partial_{1}(A_{g})_{3}\bigg]v^{3}\Bigg\} \nonumber\\
  &\approx& -\left(\overrightarrow{\nabla}\phi_{g}\right)_{1}+\mathbf{4}\,\bigg[-(B_{g})_{3}\,v^{2}\label{a1}
+(B_{g})_{2}\,v^{3}\bigg] \nonumber\\
&\approx& -\left(\overrightarrow{\nabla}\phi_{g}\right)_{1}+\mathbf{4}\,\left(\overrightarrow{v}\times\overrightarrow{B_{g}}\right)_{1} \\\nonumber\\
\frac{d^{2}x^{2}}{dt^{2}}  &\approx& -\left(\overrightarrow{\nabla}\phi_{g}\right)_{2}+\mathbf{4}\,\left(\overrightarrow{v}\times\overrightarrow{B_{g}}\right)_{2} \label{a2} \\\nonumber\\
\frac{d^{2}x^{3}}{dt^{2}}  &\approx&-\left(\overrightarrow{\nabla}\phi_{g}\right)_{3}+\mathbf{4}\,\left(\overrightarrow{v}\times\overrightarrow{B_{g}}\right)_{3} \label{a3}
\end{eqnarray}
Les trois expressions précédentes (\ref{a1}), (\ref{a2}) et (\ref{a3}) se mettent sous la forme vectorielle équivalente
\begin{center}
\begin{equation}\label{eq lorentz grav}
    \frac{d^{2}\overrightarrow{x}}{dt^{2}}\approx -\overrightarrow{\nabla}\phi_{g}+\mathbf{4}\,\left(\overrightarrow{v}\times\overrightarrow{B_{g}}\right)
\end{equation}
\textbf{C.Q.F.D}\vspace{0.5cm}\end{center}
\end{enumerate}

Cette dernière équation représente l'accélération d'une particule ponctuelle soumise à une force gravitationnelle de type Lorentz
\begin{equation}\label{eq lorentz grav force}
    m_{i}\frac{d^{2}\overrightarrow{x}}{dt^{2}}\approx m_{g}\bigg[\overrightarrow{g}+\mathbf{4}\,\Big(\overrightarrow{v}\times\overrightarrow{B_{g}}\Big)\bigg],
\end{equation}
à faible vitesse.

En plus de l'effet radial dû au champ Gravitoélectrique $\overrightarrow{g}$, il y a apparition d'un effet orthoradial sur la particule test, dû au champ Gravitomagnétique $\overrightarrow{B_{g}}$.

La présence d'un facteur 4 indésirable dans la partie magnétique de la force de type Lorentz (\ref{eq lorentz grav force}) trouve son origine dans la définition des composantes du potentiel vecteur en fonction des composantes de la perturbation de la métrique (\ref{facteur 4 origine}); soulignons que ce résultat a aussi été retrouvé par Wald \cite{Wald}.


\subsection{Génération des Ondes Gravitationnelles}
Dans la jauge harmonique, l'équation d'Einstein linéaire (\ref{equation einstein lineaire harmonique hbar}),
\begin{equation}\label{equation einstein lineaire harmonique hbar gen onde grav}
    \Box \overline{h}_{\mu\nu}\approx -\frac{16\pi G}{c^{4}}\,T_{\mu\nu},
\end{equation}
est analogue à l'équation de propagation du 4-potentiel électromagnétique $A^{\mu}(\phi/c,\overrightarrow{A})$
\begin{equation}\label{eq propag electro}
    \Box A^{\mu}=\mu_{0}\,J^{\mu}
\end{equation}
dans la jauge de Lorentz $\partial_{\mu}A^{\mu}=0$, où $J^{\mu}(\rho\,c,\rho\overrightarrow{v})$ représente le 4-courant. Ces équations de Maxwell impliquent que le mouvement des charges électriques de la source génère des ondes électromagnétique. De façon analogue à l'électromagnétisme, les équations d'Einstein linéaires, dans la jauge harmonique impliquent que le mouvement des masses constituant la distribution matérielle entraîne l'apparition d'ondes gravitationnelles \cite{Woodhouse}.

\subsubsection{Solution des potentiels retardés}
La présence du d'Alembertien dans l'équation d'Einstein linéaire (\ref{equation einstein lineaire harmonique hbar gen onde grav}) montre clairement que la perturbation de la métrique, ou de façon équivalente l'onde gravitationnelle, se propage à la vitesse de la lumière.

Une conséquence de la linéarisation de l'équation d'Einstein, dans la jauge harmonique, est le fait de pouvoir découpler le mouvement des composantes de la métrique de la perturbation de sorte que la propagation de chaque composante $\overline{h}_{\mu\nu}$ est générée par la composante correspondante $T_{\mu\nu}$ du tenseur énergie-impulsion. Dans le régime non linéaire, caractérisé par des champs gravitationnels intenses, aucun choix de jauge ne permet d'aboutir à des équations découplées; désormais les différents modes de la perturbation sont fortement interagissant de manière que la propagation des composantes $\overline{h}_{\mu\nu}$ ne se fait plus de façon indépendante des autres composantes.

D'une manière générale, une équation de propagation linéaire avec second membre
\begin{equation}\label{eq propag second membre}
    \Box \psi(\overrightarrow{x},t)=4\pi\;f(\overrightarrow{x},t)
\end{equation}
admet une solution sous forme d'une somme étendue à toutes les contributions individuelles et retardées\footnote{Seule la solution retardée est retenue et la solution avancée est ignorée car non physique} de tous les points de la source \cite{Jackson}
$$\psi(\overrightarrow{x},t)=\int d^{3}x^{'}\;\frac{\left[f(\overrightarrow{x}^{'},t^{'})\right]_{\text{ret}}}{\|\overrightarrow{x}-\overrightarrow{x}^{'}\|}$$
où $\left[f(\overrightarrow{x^{'}},t^{'})\right]_{\text{ret}}=f\left(\overrightarrow{x}^{'},t-\frac{\|\overrightarrow{x}-\overrightarrow{x}^{'}\|}{c}\right)$.

Ainsi, de manière analogue à la solution électromagnétique des potentiels retardés
\begin{eqnarray}
    \phi(\overrightarrow{x},t)&=&\frac{1}{4\pi\varepsilon_{0}}\int d^{3}x^{'}\;\frac{\Big[\rho(\overrightarrow{x}^{'},t^{'})\Big]_{\text{ret}}}{\|\overrightarrow{x}-\overrightarrow{x}^{'}\|}\\
    \overrightarrow{A}(\overrightarrow{x},t)&=&\frac{\mu_{0}}{4\pi}\int d^{3}x^{'}\;\frac{\left[\overrightarrow{j}(\overrightarrow{x}^{'},t^{'})\right]_{\text{ret}}}{{\|\overrightarrow{x}-\overrightarrow{x}^{'}\|}},
\end{eqnarray}
l'équation d'Einstein linéaire (\ref{equation einstein lineaire harmonique hbar gen onde grav}) admet comme solution \cite{Plebanski}
\begin{eqnarray}\label{poichiche}
\overline{h}_{\mu\nu}(\overrightarrow{x},t) &=& -\frac{\chi}{2\pi}\;\int d^{3}x^{'}
\frac{1}{\|\overrightarrow{x}-\overrightarrow{x}^{'}\|}\;T_{\mu\nu}\left(\overrightarrow{x}^{'},t-\frac{\|\overrightarrow{x}-\overrightarrow{x}^{'}\|}{c}\right)
\end{eqnarray}
où $\frac{\chi}{2\pi}=\frac{4G}{c^{4}}$.

\subsubsection{Comparaison des rayonnements Gravitationnel et Electromagnétique}
Dans le cas des radiations gravitationnelles émises par une source isolée, suffisamment éloignée\footnote{Pour une source isolée et suffisamment éloignée, il est possible d'effectuer le développement limité suivant \cite{Stephani} $$\frac{1}{\|\overrightarrow{x}-\overrightarrow{x}^{'}\|}=\frac{1}{\|\overrightarrow{x}\|}+\frac{\overrightarrow{x}.\overrightarrow{x}^{\,'}}{\|\overrightarrow{x}\|^{3}}
+\sum_{i=1}^{3}\sum_{j=1}^{3}\frac{x^{i}x^{j}}{2\|\overrightarrow{x}\|^{5}}\left(3\,x^{'i}x^{'j}-\|\overrightarrow{x}^{\,'}\|^{2}\,\delta^{ij}\right)+\cdots$$ } et constituée d'une matière doté d'un mouvement non relativiste\footnote{La matière voyage dans la source beaucoup moins vite que la lumière.}, il est possible de montrer que la solution (\ref{poichiche}) mène dans le cas où $\mu=i$ et $\nu=j$ à \cite{Carroll1, Plebanski}
\begin{equation}
    \overline{h}_{ij}(\overrightarrow{x},t)\approx-\frac{\chi}{2\pi}\left(\frac{1}{2c^{2}\|\overrightarrow{x}\|}\right)\frac{d^{\,2}I_{ij}}{dt^{2}}
\end{equation}
où
\begin{equation}
    I_{ij}=\int d^{3}x^{\,'}\; x_{i}^{\,'}x_{j}^{\,'}\,T_{00}(\overrightarrow{x}^{\,'},t)
\end{equation}
est le tenseur moment quadrupolaire.

En électromagnétisme, la contribution principale au rayonnement électromagnétique provient du changement du moment dipolaire de la densité de charge; la puissance rayonnée à travers une sphère de grand rayon $\|\overrightarrow{x}\|$ est donnée \cite{Griffiths}
\begin{equation}
    P=\frac{\Big[\ddot{p}(t-\|\overrightarrow{x}\|/c)\Big]^{2}}{6\pi c}
\end{equation}
en fonction du moment dipolaire $p(t)=q\,d(t)$. Il est clair que le rayonnement électromagnétique est étroitement lié à l'accélération $\ddot{d}=\ddot{p}/q$ des charges électriques.

En électromagnétisme, le déplacement du centre de densité de charge, relié à une variation du moment dipolaire, est permis, alors que le moment dipolaire gravitationnel est identiquement nul. 
Le moment quadripolaire, qui mesure l'écart du système de la forme sphérique, est généralement plus petit que le moment dipolaire, de plus la gravité se couple très faiblement avec la matière\footnote{A cause de la faible valeur de la constante de Gravitation universelle $G$}; pour toutes ces raisons le rayonnement gravitationnel est beaucoup plus faible que le rayonnement électromagnétique \cite{Carroll1}.


\subsection{Ondes Gravitationnelles: Approche de S. Carroll}

\subsubsection{Introduction}
Dans le cadre de l'approximation du champ faible, nous allons nous intérresser à la description du rayonnement gravitationnel. Dans ce qui suit, il est question d'étudier les degrés de liberté effectifs du champ de gravitation $h_{\mu\nu}$, libres de se propager et dotés d'une existence indépendante de la source qui les produit; autrement dit, une fois crées les ondes gravitationnelles se propagent dans l'espace même si la source qui les produit est supprimée.

Dans le contexte de la Gravité Linéaire, l'invariance du tenseur d'Einstein (\ref{tenseur d'einstein linearise}) par transformation de jauge
(\ref{jauge}) devrait permettre d'effectuer un choix de jauge particulier pour pouvoir résoudre les équations d'Einstein. Néanmoins, avant d'exploiter cette liberté de jauge il serait instructif de choisir un système de coordonnées inertiel dans l'espace de Minkowski et de décomposer les composantes de la perturbation de la métrique $h_{\mu\nu}$ comme suit \cite{Bertschinger} 

\begin{eqnarray}\label{hmunu decompos degre liberte}
  h^{\mu\nu}\left(
                                           \begin{array}{c|ccc}
                                             h^{00} & h^{01} & h^{02} & h^{03} \\
                                             \hline
                                             h^{01} & h^{11} & h^{12} & h^{13} \\
                                             h^{02} & h^{12} & h^{22} & h^{23} \\
                                             h^{03} & h^{13} & h^{23} & h^{33} \\
                                           \end{array}
                                         \right) &\longrightarrow& \left\{
                                 \begin{array}{ll}
                                   h^{00},& \hbox{Scalaire.}  \\
                                   h^{0i}(i=1,2,3), & \hbox{Vecteur.} \\
                                   h^{ij}(i,j=1,2,3)
                                   , & \hbox{Tenseur symétrique.}
                                 \end{array}
                               \right.
\end{eqnarray}
A leurs tours les $h^{ij}$ sont décomposés, d'une part, en une partie à trace nulle et, d'autre part, en une partie à trace non nulle, tel que
\begin{eqnarray}
  h_{00} &=& 2\,\phi/c^{2}  \label{h00 degre liberte}\\
   h_{0i} &=& -\omega_{i}/c  \label{h0i degre liberte}\\
   h_{ij} &=& 2\delta_{ij}\,\psi-2S_{ij}  \label{hij degre liberte}
\end{eqnarray}
où\footnote{La partie à trace nulle est définie de telle sorte que $\delta^{ij}\,S_{ij}=\frac{1}{2}\big(\delta^{ij}\,h_{ij}-\frac{1}{3}\,\delta^{km}\,h_{km}\,\underbrace{\delta^{ij}\delta_{ij}}_{3}\big)=0$}
\begin{eqnarray}
  \psi &=& \frac{1}{6}\,\delta^{ij}\,h_{ij} \label{psi degre liberte}\\
  S_{ij} &=& \frac{1}{2}\left(\frac{1}{3}\,\delta^{kl}\,h_{kl}\,\delta_{ij}-h_{ij}\right). \label{sij degre liberte}
\end{eqnarray}

L'esprit de cette décomposition est analogue à la décomposition du champ électromagnétique $F^{\mu\nu}$ en fonction du champ électrique $\overrightarrow{E}$ et du champ magnétique $\overrightarrow{B}$.



Les définitions précédentes permettent de calculer le carré de l'interval infinitésimal
\begin{eqnarray*}
  ds^{2} &=& g_{\mu\nu}\,dx^{\mu}dx^{\nu} = g_{00}(dx^{00})^{2} + 2\,g_{oi}\,dx^{0}dx^{i} + g_{ij}\,dx^{i}dx^{j}\\
         &=& \left(1+2\,\phi/c^{2}\right)(c\,dt)^{2} - 2\,\frac{\omega_{i}}{c}(c\,dt)dx^{i} + \left(-\delta_{ij}+h_{ij}\right)\,dx^{i}dx^{j}\\
         &=& \left(1+2\,\phi/c^{2}\right)(c\,dt)^{2} - 2\,\omega_{i}dtdx^{i} + \left[-\delta_{ij}+2\left(\delta_{ij}\,\psi-S_{ij}\right)\right]\,dx^{i}dx^{j}
\end{eqnarray*}
dont l'expression finale est donnée par\footnote{Le résultat dans \cite{Carroll1} est donné par un facteur (-1) global du fait que S. Carroll utilise la métrique  $\eta_{\mu\nu}=(-1,+1,+1,+1)$}
\begin{equation}
  ds^{2} = c^{2}\left(1+\frac{2\,\phi}{c^{2}}\right)dt^{2} - 2\,\omega_{i}dtdx^{i} - \bigg[\delta_{ij}\big(1-2\,\psi\big)+2\,S_{ij}\bigg]\,dx^{i}dx^{j}.
\end{equation}



\subsubsection{Equation de mouvement d'une particule soumise à la gravité}
Dans le but de déterminer le sens physique de chaque composante de $h_{\mu\nu}$, considérons le mouvement d'une particule test décrivant une géodésique. Avec les définition des champs (\ref{h00 degre liberte}), (\ref{h0i degre liberte}) et (\ref{hij degre liberte}), Carroll arrive à reproduire une force gravitationnelle de type Lorentz, sans le facteur 4 indésirable \cite{Carroll1}. Néanmoins les seuls degrés de liberté effectifs, capables de décrire une réalité physique dans la jauge transverse, sont représentés par la partie à trace nulle $S_{ij}$, de sorte que le potentiel scalaire $\phi$, le potentiel vecteur $\omega_{i}$ et la trace spatiale $\psi$ sont vus comme des degrés de liberté qui ne se propagent pas et qui vérifient des équations de contraintes.

\begin{enumerate}
  \item \textbf{Equation de Géodésiques}

Le 4-vecteur énergie-impulsion est défini

    $$P^{\mu}=m\,\frac{dx^{\mu}}{d\tau}$$
de sorte que
\begin{eqnarray}
  P^{0} &=& m\,\frac{dx^{0}}{d\tau}= mc\,\frac{dt}{d\tau}=m \gamma c=\frac{E}{c}\\
  P^{i} &=& m\,\frac{dx^{i}}{d\tau}= m\,\frac{dx^{i}}{dt}\,\frac{dt}{d\tau}=m \gamma \,\frac{dx^{i}}{dt}=\frac{E}{c^{2}}\,v^{i}.
\end{eqnarray}

L'équation des géodésiques
 \begin{equation}\label{geodesique eq degre liberte}
    \frac{d^{\,2}x^{\mu}}{d\tau^{2}}+\Gamma^{\mu}_{\rho\sigma}\,\frac{dx^{\rho}}{d\tau}\,\frac{dx^{\sigma}}{d\tau}=0
\end{equation}
peut se mettre sous la forme
\begin{eqnarray}
  m\,\frac{d}{d\tau}\left(m\,\frac{dx^{\mu}}{d\tau}\right)&=&-\Gamma^{\mu}_{\rho\sigma}\left(m\,\frac{dx^{\rho}}{d\tau}\right)\left(m\,\frac{dx^{\sigma}}{d\tau}\right) \nonumber\\
  m\,\frac{dP^{\mu}}{d\tau} &=& -\Gamma^{\mu}_{\rho\sigma}\,P^{\rho}\,P^{\sigma}. \label{eq geodes degre liberte} 
\end{eqnarray}

Nous pouvons calculer la variation, par rapport au temps, du 4-vecteur énergie-impulsion, à partir de l'équation de géodésique (\ref{eq geodes degre liberte})
\begin{eqnarray}
  m\,\frac{dP^{\mu}}{d\tau} &=& -\Gamma^{\mu}_{\rho\sigma}\,P^{\rho}\,P^{\sigma} \nonumber\\
  m\,\frac{dP^{\mu}}{dt}\,\frac{dt}{d\tau} &=& -\Gamma^{\mu}_{\rho\sigma}\,P^{\rho}\,P^{\sigma}\nonumber\\
  \frac{m\,\gamma\,c^{2}}{c^{2}}\,\frac{dP^{\mu}}{dt} &=& -\Gamma^{\mu}_{\rho\sigma}\,P^{\rho}\,P^{\sigma}\nonumber\\
  \frac{E}{c^{2}}\,\frac{dP^{\mu}}{dt} &=& -\Gamma^{\mu}_{\rho\sigma}\,P^{\rho}\,P^{\sigma}\nonumber
\end{eqnarray}

Finalement
\begin{equation}\label{variation pmu}
    \frac{dP^{\mu}}{dt} = -c^{2}\,\Gamma^{\mu}_{\rho\sigma}\,\frac{P^{\rho}\,P^{\sigma}}{E}.
\end{equation}
  \item \textbf{Symboles de Christoffel}

Au premier ordre nous avons
\begin{eqnarray}
  \Gamma^{\rho}_{\mu\nu}
    &=& \frac{1}{2}\,\eta^{\rho\lambda} \left(\partial_{\mu}h_{\nu\lambda}+\partial_{\nu}h_{\lambda\mu}-\partial_{\lambda}h_{\mu\nu}\right).
\end{eqnarray}

\begin{enumerate}
  \item Calcul de $\mathbf{\Gamma^{0}_{00}}$
\begin{eqnarray}
  \Gamma^{0}_{00}
    &=& \frac{1}{2}\,\underbrace{\eta^{00}}_{1} \left(\partial_{0}h_{00}+\partial_{0}h_{00}-\partial_{0}h_{00}\right)=\frac{1}{2}\,\partial_{0}\left(2\,\phi/c^{2}\right)\nonumber
\end{eqnarray}
\begin{equation}\label{christoffel 000 degre liberte}
     \Gamma^{0}_{00}=\frac{1}{c^{2}}\,\partial_{0}\phi
\end{equation}

  \item Calcul de $\mathbf{\Gamma^{i}_{00}}$
\begin{eqnarray}
  \Gamma^{i}_{00}
    &=& \frac{1}{2}\,\eta^{i\lambda} \left(\partial_{0}h_{0\lambda}+\partial_{0}h_{\lambda0}-\partial_{\lambda}h_{00}\right)=\frac{1}{2}\,\underbrace{\eta^{ii}}_{-1} \left(\partial_{0}h_{0i}+\partial_{0}h_{i0}-\partial_{i}h_{00}\right)\nonumber\\
    &=&-\frac{1}{2}\left(\partial_{0}h_{0i}+\partial_{0}h_{i0}-\partial_{i}h_{00}\right)
    =-\frac{1}{2}\,2\,\partial_{0}h_{i0}+\frac{1}{2}\partial_{i}h_{00}\nonumber\\
    &=&-\partial_{0}\left(-\frac{\omega_{i}}{c}\right)-\frac{1}{2}\partial_{i}\left(\frac{2\,\phi}{c^{2}}\right)\nonumber
\end{eqnarray}
\begin{equation}\label{christoffel i00 degre liberte}
     \Gamma^{i}_{00}=\partial_{0}\left(\omega_{i}/c\right)+\partial_{i}\left(\phi/c^{2}\right)
\end{equation}

  \item Calcul de $\mathbf{\Gamma^{0}_{j0}}$
\begin{eqnarray}
  \Gamma^{0}_{j0}
    &=& \frac{1}{2}\,\eta^{00} \left(\partial_{j}h_{00}+\partial_{0}h_{0j}-\partial_{0}h_{j0}\right)=\frac{1}{2}\,\partial_{j}h_{00} \nonumber\\
    &=&\frac{1}{2}\,\partial_{j}\left(\frac{2\,\phi}{c^{2}}\right)\nonumber
\end{eqnarray}
\begin{equation}\label{christoffel 0j0 degre liberte}
     \Gamma^{0}_{j0}=\frac{1}{c^{2}}\partial_{j}\phi
\end{equation}

  \item Calcul de $\mathbf{\Gamma^{i}_{j0}}$
\begin{eqnarray}
  \Gamma^{i}_{j0}
    &=& \frac{1}{2}\,\eta^{ii} \left(\partial_{j}h_{0i}+\partial_{0}h_{ij}-\partial_{i}h_{j0}\right)\nonumber\\
    &=& -\frac{1}{c}\underbrace{\left[-\frac{1}{2}\left(\partial_{j}\omega_{i}-\partial_{i}\omega_{j}\right)\right]}_{
    -\partial_{[j}\omega_{i]}}-\frac{1}{2}\,\partial_{0}h_{ij}\nonumber
\end{eqnarray}
\begin{equation}\label{christoffel ij0 degre liberte}
     \Gamma^{0}_{j0}=\frac{1}{2c}\bigg(\partial_{j}\omega_{i}-\partial_{i}\omega_{j}\bigg)-\frac{1}{2}\,\partial_{0}h_{ij}
\end{equation}

  \item Calcul de $\mathbf{\Gamma^{0}_{jk}}$
\begin{eqnarray}
  \Gamma^{0}_{jk}
    &=& \frac{1}{2}\,\eta^{00} \left(\partial_{j}h_{k0}+\partial_{k}h_{0j}-\partial_{0}h_{jk}\right)\nonumber\\
    &=& -\frac{1}{c}\underbrace{\left[\frac{1}{2}\left(\partial_{j}\omega_{k}+\partial_{k}\omega_{j}\right)\right]}_{
    \partial_{(j}\omega_{k)}}-\frac{1}{2}\,\partial_{0}h_{jk}\nonumber
\end{eqnarray}
\begin{equation}\label{christoffel ij0 degre liberte}
     \Gamma^{0}_{jk}=-\frac{1}{2c}\bigg(\partial_{j}\omega_{k}+\partial_{k}\omega_{j}\bigg)-\frac{1}{2}\,\partial_{0}h_{jk}
\end{equation}

  \item Calcul de $\mathbf{\Gamma^{i}_{jk}}$
\begin{eqnarray}
  \Gamma^{i}_{jk}
    &=& \frac{1}{2}\,\eta^{ii} \left(\partial_{j}h_{ki}+\partial_{k}h_{ij}-\partial_{i}h_{jk}\right)\nonumber\\
    &=& -\frac{1}{2}\bigg(\underbrace{\partial_{j}h_{ki}+\partial_{k}h_{ji}}_{
    2\,\partial_{(j}h_{k)i}}\bigg)+\frac{1}{2}\,\partial_{i}h_{jk}\nonumber
\end{eqnarray}
\begin{equation}\label{christoffel ijk degre liberte}
     \Gamma^{i}_{jk}=-\frac{1}{2}\bigg(\partial_{j}h_{ki}+\partial_{k}h_{ji}\bigg)+\frac{1}{2}\,\partial_{i}h_{jk}
\end{equation}
\end{enumerate}

\newpage
\item \textbf{Evolution de l'Impulsion}

La composante spatiale $\mu=i$ de l'équation (\ref{variation pmu}) est

\begin{eqnarray*}
  \frac{dP^{i}}{dt} &=& -c^{2}\bigg(\Gamma^{i}_{00}\,\frac{P^{0}\,P^{0}}{E}
  +2\,\Gamma^{i}_{j0}\,\frac{P^{o}\,P^{j}}{E}+\Gamma^{i}_{jk}\,\frac{P^{j}\,P^{k}}{E}\bigg) \\
   &=&c^{2}\left\{-\Gamma^{i}_{00}\,\displaystyle\frac{\left(\displaystyle\frac{E}{c}\right)\left(\displaystyle\frac{E}{c}\right)}{E}
  -2\,\Gamma^{i}_{j0}\,\displaystyle\frac{\left(\displaystyle\frac{E}{c}\right)\left(\displaystyle\frac{E\,v^{j}}{c^{2}}\right)}{E}
  -\Gamma^{i}_{jk}\,\displaystyle\frac{\left(\displaystyle\frac{E\,v^{j}}{c^{2}}\right)\left(\displaystyle\frac{E\,v^{k}}{c^{2}}\right)}{E}\right\} \\
   &=& -\frac{E}{c}\bigg[\partial_{0}\omega_{i}+\partial_{i}(\phi/c)\bigg]-2\,\frac{E\,v^{j}}{c}\left[\frac{1}{c}\,\partial_{[j}\omega_{i]}
   -\frac{1}{2}\,\partial_{0}h_{ij}\right]\\
  &&\hspace{7.5cm}-\frac{E\,v^{j}\,v^{k}}{c^{2}}\left[-\frac{1}{c}\partial_{(j}h_{k)i}+\frac{1}{2}\,\partial_{i}h_{jk}\right] \\
   &=& \frac{E}{c^{2}}\Bigg[\bigg(-\partial_{i}\phi-c\,\partial_{0}\omega_{i}\bigg)-\bigg(\partial_{j}\omega_{i}-\partial_{i}\omega_{j}\bigg)v^{j}
   +c\,\partial_{0}h_{ij}\,v^{j}\\
  &&\hspace{8cm}-\left(\partial_{(j}h_{k)i}-\frac{1}{2}\,\partial_{i}h_{jk}\right)v^{j}\,v^{k}\Bigg]
\end{eqnarray*}

Définissons les champs "Gravitoélectrique" $\overrightarrow{G}(G^{1},G^{2},G^{3})$
\begin{eqnarray}
  \left\{
    \begin{array}{ll}
      G^{i} \equiv -\partial_{i}\phi-c\,\partial_{0}\omega_{i},  \\
      i=1,2,3
    \end{array}
  \right.
\end{eqnarray}
et "Gravitomagnétique" $\overrightarrow{H}(H^{1},H^{2},H^{3})$
\begin{eqnarray}
 \left\{
   \begin{array}{ll}
     H^{i} \equiv \left(\overrightarrow{\nabla}\times\overrightarrow{\omega}\right)^{i}=\epsilon^{ijk}\,\partial_{j}\omega_{k},  \\
     i=1,2,3
   \end{array}
 \right.
\end{eqnarray}

Or nous avons d'une part \begin{eqnarray*}
     \big(\partial_{j}\omega_{i}-\partial_{i}\omega_{j}\big)v^{j} &=& \big(\partial_{1}\omega_{i}-\partial_{i}\omega_{1}\big)v^{1}+\big(\partial_{2}\omega_{i}-\partial_{i}\omega_{2}\big)v^{2}
     +\big(\partial_{3}\omega_{i}-\partial_{i}\omega_{3}\big)v^{3} \\
      &=& \left\{
            \begin{array}{ll}
              \big(\partial_{2}\omega_{1}-\partial_{1}\omega_{2}\big)v^{2}+\big(\partial_{3}\omega_{1}-\partial_{1}\omega_{3}\big)v^{3}, & \hbox{pour i=1;} \\
              \big(\partial_{1}\omega_{2}-\partial_{2}\omega_{1}\big)v^{1}+\big(\partial_{3}\omega_{2}-\partial_{2}\omega_{3}\big)v^{3}, & \hbox{pour i=2;} \\
              \big(\partial_{1}\omega_{3}-\partial_{3}\omega_{1}\big)v^{1}+\big(\partial_{2}\omega_{3}-\partial_{3}\omega_{2}\big)v^{2}, & \hbox{pour i=3.}
            \end{array}
          \right.
   \end{eqnarray*}
et d'autre part
\begin{eqnarray*}
  \overrightarrow{v}\times\overrightarrow{H} &=& \overrightarrow{v}\times\left(\overrightarrow{\nabla}\times\overrightarrow{\omega}\right) \\
   &=& \overrightarrow{i}\bigg[v^{2}\left(\partial_{1}\omega_{2}-\partial_{2}\omega_{1}\right)-v^{3}\left(\partial_{3}\omega_{1}-\partial_{1}\omega_{3}\right)\bigg]\\
   &&\hspace{2cm}+ \overrightarrow{j}\bigg[v^{3}\left(\partial_{2}\omega_{3}-\partial_{3}\omega_{2}\right)-v^{1}\left(\partial_{1}\omega_{2}-\partial_{2}\omega_{1}\right)\bigg] \\
   &&\hspace{4cm}+ \overrightarrow{k}\bigg[v^{1}\left(\partial_{3}\omega_{1}-\partial_{1}\omega_{3}\right)-v^{2}\left(\partial_{2}\omega_{3}-\partial_{3}\omega_{2}\right)\bigg],
\end{eqnarray*}

ce qui nous permet de déduire que
\begin{equation}\label{gravito magnetique degre liberte}
    -\big(\partial_{j}\omega_{i}-\partial_{i}\omega_{j}\big)v^{j}=\left(\overrightarrow{v}\times\overrightarrow{H}\right)^{i}.
\end{equation}

Finalement
\begin{equation}\label{variation pi degre liberte}
    \frac{dP^{i}}{dt}=\frac{E}{c^{2}}\Bigg[G^{i}+\left(\overrightarrow{v}\times\overrightarrow{H}\right)^{i}
   -c\,\partial_{0}h_{ij}\,v^{j}
  -\left(\partial_{(j}h_{k)i}-\frac{1}{2}\,\partial_{i}h_{jk}\right)v^{j}\,v^{k}\Bigg]
\end{equation}

Les deux premiers termes $G^{i}$ et $(\overrightarrow{v}\times\overrightarrow{H})^{i}$ décrivent comment la particule test, se déplaçant sur une géodésique, répond aux perturbations du scalaire $\phi$ et du vecteur $\overrightarrow{\omega}$ de façon similaire à la force de Lorentz pour l'électromagnétisme.

\hspace{0.5cm}Le champ Gravitoélectrique $\overrightarrow{G}$ est responsable de l'apparition d'un effet radial de la particule test similaire à l'effet du champ électrique sur une charge $q$, alors que Le champ Gravitomagnétique $\overrightarrow{H}$ est responsable de l'apparition d'un effet ortho-radial similaire à l'effet du champ magnétique sur une charge électrique.

\hspace{0.5cm}Les perturbations spatiales $h_{ij}$ sont couplées linéairement et quadratiquement aux composantes de vitesses de la particule test
\begin{itemize}
  \item Linéairement ($\propto v^{j}$) $\rightarrow$ $\partial_{0}h_{ij}$; ce terme peut s'annuler, à priori, dans le cas où les $h_{ij}$ sont stationnaires.
  \item Quadratiquement ($\propto v^{j}\,v^{k}$) $\rightarrow$ termes négligés si on suppose que les contraintes raignant à l'intérieur de la source du champ (fluide parfait) sont négligeables ($T_{ij}\propto v_{i}\,v_{j}\ll T_{0i}\propto c\,v_{i}\ll T_{00}\propto c^{2}$) (cas des vitesses faibles)
\end{itemize}

Nous allons voir plus tard que seule la partie décrite par $S_{ij}$ est responsable d'un effet physique (rayonnement gravitationnel) et que la partie décrite par $\psi$ ne constitue pas un degré de liberté qui se propage, donc sans le moindre effet physique décelable.

\end{enumerate}

\subsubsection{Tenseur d'Einstein (au premier ordre)}
Au premier ordre de la perturbation, les composantes du tenseur d'Einstein sont données par
\begin{equation}\label{tenseur d'einstein linearise degre liberte}
    G_{\mu\nu} \approx \frac{1}{2}\left[\partial_{\sigma}\partial_{\mu}h_{\nu}^{\sigma}+\partial_{\nu}\partial_{\sigma}h_{\mu}^{\sigma}-
   \partial_{\nu}\partial_{\mu}h-\Box h_{\mu\nu}
 -\eta_{\mu\nu}\,\partial_{\alpha}\partial_{\beta}h^{\alpha\beta}+\eta_{\mu\nu}\,\Box h\right].
\end{equation}
La trace de la perturbation est donnée par
$$h=\eta^{\alpha\beta}\,h_{\alpha\beta}=h_{00}-h_{11}-h_{22}-h_{33}=h_{00}-\delta^{ij}h_{ij}=h_{00}-6\psi.$$

\begin{enumerate}
  \item \textbf{Calcul de $\mathbf{G_{00}}$}

\begin{eqnarray*}
 \hspace*{-1.5cm} G_{00} &=& \frac{1}{2}\bigg[\partial_{\sigma}\partial_{0}h_{0}^{\sigma}+\partial_{0}\partial_{\sigma}h_{0}^{\sigma}-
   \partial_{0}^{2}h-\Box h_{00}
 -\eta_{00}\,\partial_{\alpha}\partial_{\beta}h^{\alpha\beta}+\eta_{00}\,\Box h\bigg]\\
        &=& \frac{1}{2}\bigg[2\,\partial_{\sigma}\partial_{0}h_{0}^{\sigma}+\big(-\partial_{0}^{2}+\partial_{0}^{2}-\overrightarrow{\nabla}^{2}\big) h
        -\big(\partial_{0}^{2}h_{00}-\overrightarrow{\nabla}^{2}h_{00}\big)-\big(\partial^{\,2}_{0}h^{00}+2\partial_{0}\partial_{i}h^{0i}
        +\partial_{i}\partial_{j}h^{ij}\big)\bigg]\\
        &=& \bigg(\partial_{0}\partial_{0}h_{0}^{0}+\partial_{i}\partial_{0}h_{0}^{i}\bigg)-\frac{1}{2}\,\overrightarrow{\nabla}^{2}h
        -\frac{1}{2}\,\bigg(\partial_{0}^{2}h_{00}-\overrightarrow{\nabla}^{2}h_{00}\bigg)
        -\frac{1}{2}\bigg(\partial^{\,2}_{0}h^{00}+2\partial_{0}\partial_{i}h^{0i}
        +\partial_{i}\partial_{j}h^{ij}\bigg)\\
       &=& -\frac{1}{2}\,\overrightarrow{\nabla}^{2}\left(h-h_{00}\right)-\frac{1}{2}\,\partial_{i}\partial_{j}h^{ij}\\
       &=& -\frac{1}{2}\overrightarrow{\nabla}^{2}\bigg(\underbrace{h_{11}+h_{22}+h_{33}}_{-6\,\psi}\bigg)-\frac{1}{2}\,\partial_{i}\partial_{j}
       \bigg[2\big(\delta^{ij}\,\psi-S^{ij}\big)\bigg]\\
       &=& 3\overrightarrow{\nabla}^{2}\psi-\overrightarrow{\nabla}^{2}\psi
       +\partial_{i}\partial_{j}S^{ij}
\end{eqnarray*}

\begin{eqnarray}
    G_{00} &=& 2\,\overrightarrow{\nabla}^{2}\psi+\partial_{i}\partial_{j}S^{ij}.  \label{G00 degre liberte}
\end{eqnarray}

  \item \textbf{Calcul de $\mathbf{G_{0i}}$}

\begin{eqnarray*}
 \hspace*{-1.5cm} G_{0i} &=& \frac{1}{2}\bigg[\partial_{\sigma}\partial_{i}h_{0}^{\sigma}+\partial_{0}\partial_{\sigma}h_{i}^{\sigma}-
   \partial_{0}\partial_{i}h-\Box h_{0i}
 -\underbrace{\eta_{0i}}_{0}\,\partial_{\alpha}\partial_{\beta}h^{\alpha\beta}+\underbrace{\eta_{0i}}_{0}\,\Box h\bigg]\\
        &=& \frac{1}{2}\bigg[\big(\partial_{0}\partial_{i}h_{0}^{0}+\partial_{j}\partial_{i}h_{0}^{j}\big)+\big(\partial_{0}^{2}h_{i}^{0}+\partial_{0}\partial_{j}h_{i}^{j}\big)
        -\partial_{0}\partial_{i}\big(h_{00}-6\,\psi\big)-\big(\partial_{0}^{2}h_{0i}-\overrightarrow{\nabla}^{2}\underbrace{h_{0i}}_{-\omega_{i}/c}\big)\bigg]\\
        &=& -\frac{1}{2c}\,\overrightarrow{\nabla}^{2}\omega_{i}+\frac{1}{2}\,\partial_{i}(\partial^{j}h_{j0})
        +\frac{1}{2}\,\partial_{0}\partial^{j}\bigg[2\big(\delta_{ij}\,\psi-S_{ij}\big)\bigg]+3\,\partial_{0}\partial_{i}\psi\\
        &=& -\frac{1}{2c}\,\overrightarrow{\nabla}^{2}\omega_{i}-\frac{1}{2c}\,\partial_{i}\partial^{j}\omega_{j}
        -\underbrace{(-\delta_{ij})\,\partial^{j}\partial_{0}\psi}_{\eta_{ij}\,\partial^{j}\partial_{0}\psi}-\partial_{0}\partial^{j}S_{ij}+3\,\partial_{0}\partial_{i}\psi\\
  &=& -\frac{1}{2c}\,\overrightarrow{\nabla}^{2}\omega_{i}-\frac{1}{2c}\,\partial_{i}\partial^{j}\omega_{j}
        -\partial_{i}\partial_{0}\psi-\partial_{0}\partial^{j}S_{ij}+3\,\partial_{0}\partial_{i}\psi\\
\end{eqnarray*}

\begin{eqnarray}
    G_{0i} &=& -\frac{1}{2c}\,\overrightarrow{\nabla}^{2}\omega_{i}-\frac{1}{2c}\,\partial_{i}\partial_{j}\omega^{j}
        -\partial_{0}\partial^{j}S_{ij}+2\,\partial_{0}\partial_{i}\psi.  \label{G0i degre liberte}
\end{eqnarray}

  \item \textbf{Calcul de $\mathbf{G_{ij}}$}

\begin{eqnarray*}
  \hspace*{-1.5cm} G_{ij} &=& \frac{1}{2}\bigg[\partial_{\sigma}\partial_{j}h_{i}^{\sigma}+\partial_{i}\partial_{\sigma}h_{j}^{\sigma}-
   \partial_{i}\partial_{j}h-\Box h_{ij}
 +\delta_{ij}\left(\partial_{\alpha}\partial_{\beta}h^{\alpha\beta}\right)-\delta_{ij}\,\Box h\bigg]\\
         &=& \frac{1}{2}\bigg[\big(\partial_{0}\partial_{j}h_{i}^{0}+\partial_{k}\partial_{j}h_{i}^{k}\big)
         +\big(\partial_{i}\partial_{0}h_{j}^{0}+\partial_{i}\partial_{k}h_{j}^{k}\big)
         -\partial_{i}\partial_{j}\big(h_{00}-6\,\psi\big)\\
         &&\hspace{0.5cm}+2\,\Box \big(S_{ij}-\delta_{ij}\,\psi\big)
 +\delta_{ij}\big(\partial_{0}^{2}h^{00}+2\,\partial_{0}\partial_{k}h^{0k}+\partial_{k}\partial_{\ell}h^{k\ell}\big)
 -\delta_{ij}\,\big(\partial_{0}^{2}-\overrightarrow{\nabla}^{2}\big)\big(h_{00}-6\,\psi\big)\bigg]\\
        &=& \partial_{0}\bigg[\frac{1}{2}\bigg(\underbrace{\partial_{j}h_{i0}+\partial_{i}h_{j0}}_{2\,\partial_{(i}h_{j)0}}\bigg)\bigg]
            +\partial^{k}\left[\frac{1}{2}\bigg(\partial_{j}h_{ki}+\partial_{i}h_{kj}\bigg)\right]
            -\frac{1}{2}\,\partial_{i}\partial_{j}\underbrace{h_{00}}_{2\,\phi/c^{2}}+3\,\partial_{i}\partial_{j}\psi\\
            &&+\Box S_{ij}-\delta_{ij}\Box\psi+\frac{1}{2}\,\delta_{ij}\,\bigg(\partial_{0}^{2}h^{00}
            +2\,\partial_{0}\partial^{k}h_{0k}+\partial_{k}\partial_{\ell}h^{k\ell}\bigg)\\
            &&-\frac{1}{2}\,\delta_{ij}\bigg(\partial_{0}^{2}h_{00}-6\,\partial_{0}^{2}\psi-\overrightarrow{\nabla}^{2}\underbrace{h_{00}}_{2\,\phi/c^{2}}
            +6\,\overrightarrow{\nabla}^{2}\psi\bigg)\\
        &=& -\frac{1}{c}\,\partial_{0}\partial_{(i}\omega_{j)}+\partial^{k}\bigg[\partial_{j}\big(\delta_{ki}\,\psi-S_{ki}\big)
        +\partial_{i}\big(\delta_{kj}\,\psi-S_{kj}\big)\bigg]+\bigg(\delta_{ij}\,\overrightarrow{\nabla}^{2}
        -\partial_{i}\partial_{j}\bigg)\frac{\phi}{c^{2}}+3\,\partial_{i}\partial_{j}\psi\\
        &&+\Box S_{ij}-\delta_{ij}\Box\psi+\delta_{ij}\,\partial_{0}\partial^{k}\left(-\frac{\omega_{k}}{c}\right)
        +\delta_{ij}\,\partial_{k}\partial_{\ell}\bigg(\delta^{k\ell}\,\psi-S^{k\ell}\bigg)
        +3\,\delta_{ij}\,\partial_{0}^{2}\psi-3\,\delta_{ij}\,\overrightarrow{\nabla}^{2}\psi
\end{eqnarray*}

Nous avons d'une part
\begin{eqnarray*}
        \partial^{k}\left[\partial_{j}\big(\delta_{ki}\,\psi-S_{ki}\big)
        +\partial_{i}\big(\delta_{kj}\,\psi-S_{kj}\big)\right]
    &=& -\partial^{k}\partial_{j}S_{ki}-\partial^{k}\partial_{i}S_{kj} \\
    &&-[\partial^{k}(-\delta_{ki})]\partial_{j}\,\psi-[\partial^{k}(-\delta_{kj})]\,\partial_{i}\psi\\
        &=& -\partial^{k}\left(\partial_{j}S_{ki}+\partial_{i}S_{kj}\right)\\
        &&-\left(\partial^{k}\eta_{ki}\right)\partial_{j}\psi-\left(\partial^{k}\eta_{kj}\right)\partial_{i}\psi\\
     &=& -2\,\partial_{k}\partial_{(j}S_{i)}^{k}-\partial_{i}\partial_{j}\,\psi-\partial_{j}\partial_{i}\psi\\
      &=& -2\,\partial_{k}\partial_{(j}S_{i)}^{k}-2\,\partial_{i}\partial_{j}\psi
  \end{eqnarray*}

  et d'autre part
   \begin{eqnarray*}
   -\delta_{ij}\,\Box\psi+\delta_{ij}\big(\partial_{k}\partial_{\ell}\delta^{k\ell}\big)\,\psi+3\,\delta_{ij}\,\partial_{0}^{2}\psi
    -3\,\delta_{ij}\,\overrightarrow{\nabla}^{2}\psi  &=&
   -\delta_{ij}\big(\partial_{0}^{2}-\overrightarrow{\nabla}^{2}\big)\psi+\delta_{ij}\,\overrightarrow{\nabla}^{2}\psi\\
   &&+3\,\delta_{ij}\,\partial_{0}^{2}\psi
    -3\,\delta_{ij}\,\overrightarrow{\nabla}^{2}\psi\\
      &=&2\,\delta_{ij}\,\partial_{0}^{2}\psi-\delta_{ij}\,\overrightarrow{\nabla}^{2}\psi.
  \end{eqnarray*}

  Ainsi

  \begin{eqnarray*}
  G_{ij} &=& -\frac{1}{c}\,\partial_{0}\partial_{(i}\omega_{j)}-2\,\partial_{k}\partial_{(j}S_{i)}^{k}-2\,\partial_{i}\partial_{j}\psi+\bigg(\delta_{ij}\,\overrightarrow{\nabla}^{2}
        -\partial_{i}\partial_{j}\bigg)\frac{\phi}{c^{2}}+3\,\partial_{i}\partial_{j}\psi\\
        &&+\Box S_{ij}-\frac{1}{c}\,\delta_{ij}\,\partial_{0}\partial^{k}\omega_{k}-\delta_{ij}\,\partial_{k}\partial_{\ell}S^{k\ell}
        +\big(-2\,\delta_{ij}\,\partial_{0}^{2}\psi+\delta_{ij}\,\overrightarrow{\nabla}^{2}\psi\big).
\end{eqnarray*}

Finalement
\begin{eqnarray}
    G_{ij} &=& -\frac{1}{c}\,\partial_{0}\partial_{(i}\omega_{j)}-2\,\partial_{k}\partial_{(j}S_{i)}^{k}+\bigg(\delta_{ij}\,\overrightarrow{\nabla}^{2}
        -\partial_{i}\partial_{j}\bigg)\bigg(\frac{\phi}{c^{2}}-\psi\bigg)\nonumber\\
        &&+\Box S_{ij}-\frac{1}{c}\,\delta_{ij}\,\partial_{0}\partial^{k}\omega_{k}-\delta_{ij}\,\partial_{k}\partial_{\ell}S^{k\ell}
        +2\,\delta_{ij}\,\partial_{0}^{2}\psi.  \label{Gij degre liberte}
\end{eqnarray}

\end{enumerate}

\subsubsection{Degrés de liberté}

L'utilisation des expressions $G_{00}$, $G_{0i}$ et $G_{ij}$ dans les équations d'Einstein $G_{\mu\nu}=(8\pi\,G/c^{4})T_{\mu\nu}$ va révéler que seulement une petite partie des composantes de $h_{\mu\nu}$ sont des degrés de liberté effectifs du champ gravitationnel; le reste des composantes obéit à des équations permettant de les relier aux degrés de liberté effectifs (équations de contraintes).

\begin{enumerate}
  \item \textbf{La trace spatiale $\mathbf{\psi}$}

D'après (\ref{G00 degre liberte}), nous avons
\begin{eqnarray*}
  \frac{8\pi G}{c^{4}}\,T_{00} &=& G_{00} \\
  \frac{8\pi G}{c^{4}}\,T_{00} &=& 2\,\overrightarrow{\nabla}^{2}\psi+\partial_{i}\partial_{j}S^{ij},
\end{eqnarray*}
ce qui permet d'isoler le laplacien de la trace spatiale
\begin{equation}\label{eq contrainte psi}
    \overrightarrow{\nabla}^{2}\psi=\frac{4\pi G}{c^{4}}\,T_{00}-\frac{1}{2}\,\partial_{i}\partial_{j}S^{ij}.
\end{equation}


Du fait qu'il n'y a pas de termes contenant des dérivées "temporelles" de $\psi$ ($\partial_{0}\psi$ ou bien $\partial_{0}^{2}\psi$) nous déduisons que $\psi$ est un degré de liberté qui ne se propage pas. La connaissance de la composante $T_{00}$ et des composantes $S^{ij}$ (avec les conditions aux limites à l'infini) permet de déterminer complètement la trace spatiale $\psi$ \cite{Carroll1}.

  \item \textbf{Les composantes $\mathbf{\omega_{k}}$ du potentiel vecteur}

  D'après (\ref{G0i degre liberte}), nous avons aussi
\begin{eqnarray*}
  \frac{8\pi G}{c^{4}}\,T_{0i} &=& G_{0i} \\
  \frac{8\pi G}{c^{4}}\,T_{0i} &=& -\frac{1}{2c}\,\overrightarrow{\nabla}^{2}\omega_{i}-\frac{1}{2c}\,\partial_{i}\partial_{j}\omega^{j}
        -\partial_{0}\partial^{j}S_{ij}+2\,\partial_{0}\partial_{i}\psi\\
  &=&-\frac{1}{2c}\,\delta^{k}_{i}\,\overrightarrow{\nabla}^{2}\omega_{k}
  -\frac{1}{2c}\,\partial_{i}\partial^{k}\omega_{k}
        -\partial_{0}\partial^{j}S_{ij}+2\,\partial_{0}\partial_{i}\psi,
\end{eqnarray*}
ce qui permet d'avoir
\begin{equation}\label{eq contrainte omega k}
    \bigg(\delta^{k}_{i}\,\overrightarrow{\nabla}^{2}+\partial_{i}\partial^{k}\bigg)\omega_{k}=-\frac{16\pi G}{c^{3}}\,T_{0i}+4c\,\partial_{0}\partial_{i}\psi
    -2c\,\partial_{0}\partial^{k}S_{ik}.
\end{equation}

Du fait qu'il n'y a pas de termes contenant des dérivées "temporelles" de $\omega^{k}$,  nous déduisons aussi que les $\omega^{k}$ sont des degrés de liberté qui ne se propagent pas. En connaissant $T_{0i}$ et les $S^{ij}$ ainsi que $\psi$ déterminé précédemment (avec les conditions aux limites à l'infini), il est ainsi possible de  déterminer complètement les $\omega_{k}$ \cite{Carroll1}.

  \item \textbf{Le potentiel scalaire $\mathbf{\phi}$}

D'après (\ref{Gij degre liberte}), nous avons
\begin{eqnarray*}
  \frac{8\pi G}{c^{4}}\,T_{ij} &=& G_{ij} \\
  \frac{8\pi G}{c^{4}}\,T_{ij} &=& -\frac{1}{c}\,\partial_{0}\partial_{(i}\omega_{j)}+\bigg(\delta_{ij}\,\overrightarrow{\nabla}^{2}
        -\partial_{i}\partial_{j}\bigg)\bigg(\frac{\phi}{c^{2}}-\psi\bigg)-2\,\partial_{k}\partial_{(j}S_{i)}^{k}\nonumber\\
        &&\hspace{3cm}+\Box S_{ij}-\frac{1}{c}\,\delta_{ij}\,\partial_{0}\partial^{k}\omega_{k}-\delta_{ij}\,\partial_{k}\partial_{\ell}S^{k\ell}
        +2\,\delta_{ij}\,\partial_{0}^{2}\psi,
\end{eqnarray*}
ce qui donne
\begin{eqnarray}\label{eq contrainte phi}
    \bigg(\delta_{ij}\,\overrightarrow{\nabla}^{2}-\partial_{i}\partial_{j}\bigg)\frac{\phi}{c^{2}}&=&\frac{8\pi G}{c^{4}}\,T_{ij}
    +\bigg(\delta_{ij}\,\overrightarrow{\nabla}^{2}-\partial_{i}\partial_{j}-2\,\delta_{ij}\,\partial_{0}^{2}\bigg)\psi\nonumber\\
    &&+\frac{1}{c}\,\delta_{ij}\,\partial_{0}\partial^{k}\omega_{k}+\frac{1}{c}\,\partial_{0}\partial_{(i}\omega_{j)}\nonumber\\ \nonumber\\&&+2\,\partial_{k}\partial_{(j}S_{i)}^{k}
    +\delta_{ij}\,\partial_{k}\partial_{\ell}S^{k\ell}.
\end{eqnarray}

Du fait qu'il n'y a pas de termes contenant des dérivées "temporelles" de $\phi$, nous déduisons que le potentiel scalaire $\phi$ est un degré de liberté qui ne se propage pas. En connaissant $T_{\mu\nu}$ et les $S^{ij}$, $\psi$ et $\omega_{k}$ étant déjà déterminés par des conditions aux limites à l'infini, il est ainsi possible de déterminer complètement $\phi$ \cite{Carroll1}.

\end{enumerate}

Les seuls degrés de liberté effectifs dans les équations d'Einstein sont les $S_{ij}$; ils seront utilisés pour décrire les ondes gravitationnelles.


\subsubsection{La jauge transverse}
La liberté de jauge va être exploitée; les équations d'Einstein (\ref{eq contrainte psi}), (\ref{eq contrainte omega k}) et (\ref{eq contrainte phi}) seront réécrites alors dans une jauge particulière dite: jauge transverse.

La transformation linéaire
\begin{equation}\label{transf coord infinit ondes gravit}
    x^{'\,\mu}=x^{\mu}+\xi^{\mu},
\end{equation}
va générer une transformation de jauge 
\begin{eqnarray}\label{jauge ondes gravit}
 h_{\mu\nu}\hspace{0.2cm}\longrightarrow\hspace{0.2cm} h^{'}_{\mu\nu} =  h_{\mu\nu}-\partial_{\nu}\xi_{\mu}-\partial_{\mu}\xi_{\nu},
\end{eqnarray}
sous laquelle le tenseur de courbure est invariant. 

\begin{enumerate}
  \item \textbf{Transformation de jauge des composantes du tenseur de perturbation}

  Selon (\ref{transf phi ondes grav index titre}), (\ref{transf omegai ondes grav index titre}) (\ref{transf psi ondes grav index titre}) et (\ref{transf sij ondes grav index titre}), les composantes du tenseur de perturbation $\phi$, $\omega_{i}$, $\psi$ et les $S_{ij}$ se transforment via (\ref{jauge ondes gravit}) comme suit \cite{Carroll1}
\begin{eqnarray}
    \phi &\longrightarrow& \phi-c^{2}\,\partial_{0}\xi^{0}\label{transf phi ondes grav}\\
    \omega_{i} &\longrightarrow& \omega_{i}+c\left(\partial_{0}\xi_{i}+\partial_{i}\xi_{0}\right)\label{transf omegai ondes grav}\\
    \psi &\longrightarrow& \psi+(\partial_{i}\xi^{i})/3\label{transf psi ondes grav}\\
        S_{ij} &\longrightarrow&  S_{ij}+\left(\partial_{i}\xi_{j}+\partial_{j}\xi_{i}\right)/2+\left(\partial_{k}\xi^{k}\,\delta_{ij}\right)/3.\label{transf sij ondes grav}
\end{eqnarray}

 \item \textbf{Conditions de jauge}

En pratique, fixer une jauge pour écrire les équations d'Einstein revient à déterminer les quatre composantes de $\xi^{\mu}$ figurant dans (\ref{transf coord infinit ondes gravit}); autrement dit, imposer quatre relations pour déterminer $\xi^{0}$ et $\xi^{i}=(\xi^{1},\xi^{2},\xi^{3})$.

\hspace{0.5cm}La jauge transverse est similaire à la jauge de Coulomb $\partial_{i}A^{i}=0$ en électromagnétisme. Nous allons commencer par imposer aux $S_{ij}$ d'être spacialement transverses \cite{Carroll1}
\begin{equation}\label{transverse spatial sij}
    \partial^{i}S_{ij}=0,
\end{equation}
en choisissant les $\xi^{j}$ de sorte que (voir \ref{fixation xi i})
\begin{equation}
    \overrightarrow{\nabla}^{2}\xi_{j}-\frac{1}{3}\,\partial_{j}\left(\partial_{i}\xi^{i}\right)=2\,\partial^{i}S_{ij}.
\end{equation}

Ces trois relations permettent de déterminer les trois $\xi^{j}$, à condition d'imposer des conditions aux limites. Néanmoins, $\xi^{0}$ demeure encore indéterminé, donc nous pouvons exploiter cette liberté pour rendre le vecteur de perturbation transverse \cite{Carroll1}
\begin{equation}\label{transv omegai ondes grav}
    \partial^{i}\omega_{i}=0,
\end{equation}
et ce en imposant à $\xi^{0}$ de satisfaire (voir \ref{fixation xi 0})
\begin{equation}\label{condition xi zero}
    \overrightarrow{\nabla}^{2}\xi^{0}=\frac{1}{c}\,\partial^{i}\omega_{i}+\partial_{0}\left(\partial^{i}\xi_{i}\right).
\end{equation}

Notons aussi que (\ref{condition xi zero}) permet de déterminer $\xi^{0}$ moyennant des conditions aux limites.
La condition (\ref{transverse spatial sij}) fournit trois équations, alors que la condition (\ref{transv omegai ondes grav}) fournit une équation (en tout quatre équations) de telle sorte que la transformation de coordonnées (\ref{transf coord infinit ondes gravit}) soit complètement déterminée pour satisfaire la jauge transverse.
\end{enumerate}

\subsubsection{Equations d'Einstein linéarisées dans la jauge transverse}

Les équations d'Einstein, exprimées en fonction de $\phi$, $\omega^{i}$,  $S^{ij}$ et $\psi$ sont données par

\begin{eqnarray}
    G_{00} &=& 2\,\overrightarrow{\nabla}^{2}\psi+\partial_{i}\left(\mathbf{\partial_{j}S^{ij}}\right)=\frac{8\pi G}{c^{4}}\,T_{00}  \label{G00 onde gravit}\\
    G_{0i} &=& -\frac{1}{2c}\,\overrightarrow{\nabla}^{2}\omega_{i}-\frac{1}{2c}\,\partial_{i}\left(\mathbf{\partial_{j}\omega^{j}}\right)
        -\partial_{0}\left(\mathbf{\partial^{j}S_{ij}}\right)+2\,\partial_{0}\partial_{i}\psi=\frac{8\pi G}{c^{4}}\,T_{0i}  \label{G0i onde gravit}\\
    G_{ij} &=& \bigg(\delta_{ij}\,\overrightarrow{\nabla}^{2}-\partial_{i}\partial_{j}\bigg)\bigg(\frac{\phi}{c^{2}}+\psi\bigg)-\frac{1}{c}\,\partial_{0}\partial_{(i}\omega_{j)}
    -2\,\mathbf{\partial_{k}\partial_{(j}S_{i)}^{k}}\nonumber\\
    &&+\Box S_{ij}-\frac{1}{c}\,\delta_{ij}\,\partial_{0}\left(\mathbf{\partial^{k}\omega_{k}}\right)-\delta_{ij}\,\partial_{k}\left(\mathbf{\partial_{\ell}S^{k\ell}}\right)
        +2\,\delta_{ij}\,\partial_{0}^{2}\psi=\frac{8\pi G}{c^{4}}\,T_{ij}.  \label{Gij onde gravit}
\end{eqnarray}

Le choix des conditions (\ref{transverse spatial sij}) et (\ref{transv omegai ondes grav}) est motivé par le besoin de simplifier les équations d'Einstein dans le vide
\begin{eqnarray}
    0 &=& 2\,\overrightarrow{\nabla}^{2}\psi+\partial_{i}\left(\mathbf{\partial_{j}S^{ij}}\right)  \label{G00 vide onde gravit}\\
    0 &=& -\frac{1}{2c}\,\overrightarrow{\nabla}^{2}\omega_{i}-\frac{1}{2c}\,\partial_{i}\left(\mathbf{\partial_{j}\omega^{j}}\right)
        -\partial_{0}\left(\mathbf{\partial^{j}S_{ij}}\right)+2\,\partial_{0}\partial_{i}\psi  \label{G0i vide onde gravit}\\
    0 &=& \bigg(\delta_{ij}\,\overrightarrow{\nabla}^{2}-\partial_{i}\partial_{j}\bigg)\bigg(\frac{\phi}{c^{2}}+\psi\bigg)-\frac{1}{c}\,\partial_{0}\partial_{(i}\omega_{j)}
    -2\,\mathbf{\partial_{k}\partial_{(j}S_{i)}^{k}}\nonumber\\
    &&+\Box S_{ij}-\frac{1}{c}\,\delta_{ij}\,\partial_{0}\left(\mathbf{\partial^{k}\omega_{k}}\right)-\delta_{ij}\,\partial_{k}\left(\mathbf{\partial_{\ell}S^{k\ell}}\right)
        +2\,\delta_{ij}\,\partial_{0}^{2}\psi
         \label{Gij vide onde gravit}
\end{eqnarray}

En effet, dans la jauge transverse, nous avons:
\begin{enumerate}
  \item L'équation (\ref{G00 vide onde gravit}) se réduit à
\begin{eqnarray}
    \overrightarrow{\nabla}^{2}\psi=0.
\end{eqnarray}
Avec des conditions aux limites appropriées\footnote{$\psi=0$ à l'infini}, l'équation précédente possède une solution identiquement nulle \cite{Carroll1}
\begin{equation}\label{psi egale zero}
    \psi=0.
\end{equation}
  \item L'équation (\ref{G0i vide onde gravit}), sachant (\ref{psi egale zero}), se réduit à
\begin{eqnarray}
    \overrightarrow{\nabla}^{2}\omega_{i}=0.
\end{eqnarray}
De même, en imposant à $\omega_{i}$ d'être nul à l'infini, nous déduisons que \cite{Carroll1}
\begin{equation}\label{omegai egale zero}
    \omega_{i}=0.
\end{equation}
  \item En utilisant (\ref{Gij vide onde gravit}), (\ref{psi egale zero}), (\ref{omegai egale zero}) et sachant que $\delta^{ij}\,S_{ij}=0$, le calcul de la trace
  $$\delta^{ij}\,G_{ij}=0,$$
  nous permet d'aboutir à l'équation
  \begin{equation}\label{phi egale zero}
    \overrightarrow{\nabla}^{2}\phi=0.
\end{equation}
Pour garantir à $\phi$ d'être nul à l'infini, il faut que
\begin{equation}\label{phi egale zero}
    \phi=0,
\end{equation}
partout \cite{Carroll1}.
  \item Finalement, en tenant compte de (\ref{psi egale zero}), (\ref{omegai egale zero}) et (\ref{phi egale zero}), l'équation (\ref{Gij vide onde gravit}) se réduit à l'équation de propagation suivante
\begin{equation}\label{eq propag sij}
    \Box S_{ij}=0.
\end{equation}

Les $S_{ij}$ sont des degrés de liberté effectifs qui permettent de décrire le rayonnement gravitationnel dans la jauge transverse.
\end{enumerate}

\subsubsection{Conclusion}
Les définitions du champ (\ref{h00 degre liberte}), (\ref{h0i degre liberte}) et (\ref{hij degre liberte}) ont permis à Carroll \cite{Carroll1} de reproduire une force gravitationnelle de type Lorentz, sans le facteur 4 indésirable.

La jauge transverse est particulièrement appropriée pour étudier le rayonnement gravitationnel dans le cas extérieur à la source; les équations d'Einstein dans le vide s'y expriment de façon très simple. De plus, dans cette approche seule une petite partie des composantes de la perturbation $h_{\mu\nu}$ sont des degrés de liberté effectifs; les $S_{ij}$ sont utilisés pour décrire le rayonnement gravitationnel, alors que $h_{00}\propto\phi$, $h_{0i}\propto\omega_{i}$ et $h_{i}^{i}\propto\psi$ sont identiquement nuls dans la jauge transverse.

Il est à noter que cette situation est totalement différente pour l'électromagnétisme où le rayonnement est décrit par les potentiels scalaire et vectoriel. De plus, nous avons vu que le rayonnement gravitationnel est généralement plus faible, donc plus difficile à détecter, que le rayonnement électromagnétique et ce pour plusieurs raisons
\begin{itemize}
  \item Le rayonnement gravitationnel est dû à la variation du moment quadrupolaire alors que le rayonnement électromagnétique est essentiellement de nature dipolaire. La contribution dipolaire gravitationnelle est nulle du fait de l'adoption d'un repérage relatif au référentiel de centre de masse de la source, alors que la contribution dipolaire électromagnétique ne s'annule que dans le cas où la charge totale de la source est nulle (corps neutre).
  \item L'interaction gravitationnelle est faiblement couplée à la matière du fait de la petite valeur de la constante de gravitation universelle $G$.
\end{itemize}

Dans l'approche proposée par Carroll \cite{Carroll1}, les champs sont définis de manière à reproduire la force gravitationnelle de type Lorentz sans le facteur 4 indésirable, néanmoins les équations d'Einstein ne se réduisent plus à des équations de type Maxwell.


\section{Version revisitée}
\subsection{Introduction}
Après avoir abordé la Gravité Linéaire, dans sa version standard, nous allons à présent présenter une version quelques peu différente dans le but d'aboutir à une analogie encore plus grande entre la gravité et l'électromagnétisme. Dans un premier temps, l'attention sera portée sur quelques "imperfections" des approches de Huei et de Caroll. Dans un deuxième temps, le cadre théorique de la nouvelle approche sera présenté de telle sorte à révéler les différences avec l'approche standard et les hypothèses adoptées seront énumérées.

\subsection{Imperfections de la version standard de la Gravité Linéaire}
Malgré l'analogie frappante entre l'électromagnétisme et la Gravité linéaire, néanmoins, nous allons attirer l'attention sur certaines "insuffisances" dont la résolution va nous permettre de revisiter la version standard de façon à aboutir à une meilleure analogie gravitation-électromagnétisme.


\subsubsection{Restriction à la jauge harmonique de la définition du champ gravitoélectrique}
Rappelons que dans l'approche de Huei, la définition du champ gravitoélectrique est donnée par
\begin{eqnarray}
    &&\left\{
      \begin{array}{ll}
        g^{i}=c^{2}\;\mathcal{G}^{00i} \\
        i=1,2,3
      \end{array}
    \right. 
    \label{gooi insufisance}
\end{eqnarray}
D'après (\ref{bmunulambda}), la définition précédente se met sous la forme
\begin{eqnarray}
        g^{i}&=&\displaystyle\frac{c^{2}}{4}\left(\overline{h}^{\,00,i}-\overline{h}^{\,0i,0}+\eta^{00}\,\overline{h}^{\,i\alpha}_{\;\;\;\;\;,\alpha}
    -\eta^{i0}\,\overline{h}^{\,0\alpha}_{\;\;\;\;\;,\alpha}\right) \nonumber\\
    &=&\displaystyle\frac{c^{2}}{4}\left(\overline{h}^{\,00,i}-\overline{h}^{\,0i,0}+\overline{h}^{\,i\alpha}_{\;\;\;\;\;,\alpha}
    \right) \nonumber\\
            &=&\displaystyle\frac{c^{2}}{4}\left(\partial^{i}\overline{h}^{\,00}-\partial^{0}\overline{h}^{\,0i}+\partial_{0}\overline{h}^{\,i0}
            +\partial_{j}\overline{h}^{\,ij}\right)
\end{eqnarray}
pour devenir finalement
\begin{equation}\label{gooi autre ecriture}
g^{i}=\displaystyle\frac{c^{2}}{4}\left(\partial^{i}\overline{h}^{\,00}+\partial_{j}\overline{h}^{\,ij}\right)
\end{equation}

Compte tenu des définitions des potentiels scalaire (\ref{potentiel gravitation}) et vecteur (\ref{hoi}), l'expression (\ref{gooi autre ecriture}) ne se met sous la forme usuelle d'électromagnétisme
\begin{equation}\label{relation champ type electrique potentiels}
    \overrightarrow{g}=-\overrightarrow{\nabla}\phi_{g}-\frac{\partial \overrightarrow{A}_{g}}{\partial t}
\end{equation}
que dans le cas de la jauge harmonique. En effet, la condition de jauge (\ref{jauge harmonique nouvelle}) est équivalente aux quatre relations
\begin{eqnarray}
      \partial_{\mu}\overline{h}^{\,\mu0} &=& 0 \label{jauge harm bar 0} \\
      \partial_{\mu}\overline{h}^{\,\mu i} &=& 0\label{jauge harm bar i}
\end{eqnarray}
une composante "temporelle" définie pour $\beta=0$ et trois composantes "spatiales" définies pour $\beta=i$. D'après (\ref{jauge harm bar i}), nous déduisons que
\begin{equation}
    \partial_{j}\overline{h}^{\,ij}=-\partial_{0}\overline{h}^{\,i0}
\end{equation}
ce qui permet de mettre (\ref{gooi autre ecriture}) sous la forme
\begin{eqnarray*}
g^{i}&=&\displaystyle\frac{c^{2}}{4}\left(\partial^{i}\overline{h}^{\,00}-\partial_{0}\overline{h}^{\,0i}\right)\\
&=&\partial^{i}\left(c^{2}\,\overline{h}^{\,00}/4\right)-c\,\partial_{0}\left(c\,\overline{h}^{\,0i}/4\right)\\
&=&\partial^{i}\phi_{g}-c\,\partial_{0}A^{i}_{g}\\
&=&-\partial_{i}\phi_{g}-c\,\frac{\partial A^{i}_{g}}{\partial (ct)}
\end{eqnarray*}
ou encore finalement
\begin{equation}
g^{i}=-\partial_{i}\phi_{g}-\frac{\partial A^{i}_{g}}{\partial t}.
\end{equation}

Le fait que la relation usuelle (\ref{relation champ type electrique potentiels}) entre le champ gravitoélectrique et les potentiels scalaire $\phi_{g}$ et vecteur $\overrightarrow{A}_{g}$ soit vérifiée uniquement dans une jauge particulière, harmonique, est une circonstance insatisfaisante car en électromagnétisme, l'analogue de cette relation est valable quelque soit la jauge utilisée.

La relation (\ref{relation champ type electrique potentiels}) devrait être satisfaite de façon indépendante de la jauge adoptée de manière à aboutir à une meilleure analogie entre la gravitation et l'électromagnétisme.


\subsubsection{Force de type Lorentz avec un facteur indésirable}
Dans l'approche de Huei, la force à laquelle est soumise une particule d'épreuve, astreinte à se déplacer suivant des géodésiques, est de type Lorentz
\begin{equation}\label{eq lorentz grav force bis}
    m_{i}\frac{d^{2}\overrightarrow{x}}{dt^{2}}\approx m_{g}\bigg[\overrightarrow{g}+\mathbf{4}\,\Big(\overrightarrow{v}\times\overrightarrow{B_{g}}\Big)\bigg],
\end{equation}
néanmoins la partie magnétique est entachée d'un facteur 4 indésirable. Bien sûr, ce facteur n'est pas dû à une erreur de calcul, comme l'a bien souligné Wald \cite{Wald}, mais trouve ses origines dans la définition des composantes du potentiel vecteur en fonction des composantes de la perturbation de la métrique (\ref{facteur 4 origine}).

\subsubsection{Restriction au régime stationnaire}
Le point fort, si on peut s'exprimer ainsi, de l'approche de Huei est que les équations d'Einstein linéarisés se réduisent à des équations de type Maxwell. A partir de l'équation des géodésique d'une particule test, il est possible de montrer que la particule est soumise à une force de type Lorentz. En plus du facteur 4 indésirable, précédemment cité, l'équation (\ref{eq lorentz grav force bis}) a été obtenue dans le régime stationnaire. En effet, les termes de type $\partial_{0}h_{ij}$ et $\partial_{0}h_{i0}$, survenant lors du calcul des symboles de Christoffel dans (\ref{eq geod gravitomag}), n'ont pu être annulés qu'en supposant que les champs sont independents du temps.

Cette circonstance est également insatisfaisante car la force de Lorentz à laquelle est soumise une charge électrique ponctuelle n'est pas restreinte seulement au régime stationnaire.

Pour terminer, nous pouvons rajouter également le fait que Carroll est parvenu à régler les problèmes relatifs à la force de type Lorentz, par la redefinition des champs (\ref{h00 degre liberte}), (\ref{h0i degre liberte}) et (\ref{hij degre liberte}), mais sans parvenir à mettre les équations d'Einstein linéaires sous forme d'équations de type Maxwell.

\subsection{Cadre théorique et hypothèses adoptées}
Dans ce qui suit, il est question de revisiter la Gravité Linéaire de manière à surmonter les imperfections, citées précédemment, pour aboutir à une plus forte similarité entre la gravité et l'électromagnétisme.

\subsubsection{Métrique adoptée}
Le champ gravitationnel est si faible que la métrique utilisée est une perturbation linéaire de la métrique minkowskienne (\ref{plate plus perturbation}). Le point de vue adopté est l'étude, dans la limite du champ faible, de la propagation d'une perturbation sur une métrique de "fond" stationnaire.

Une conséquence immédiate de l'utilisation du même point de vue que la version standard de la Gravité Linéaire est que les expressions linéarisées des symboles de Christofel (\ref{christoffel}), du tenseur de Riemann (\ref{tenseur de riemann}), du tenseur de Ricci (\ref{tenseur ricci}), de la courbure scalaire (\ref{courbure scalaire}) et du tenseur d'Einstein (\ref{tenseur d'einstein linearise}) sont aussi valables dans cette nouvelle approche.

\subsubsection{Potentiels scalaire et vecteur}
Proposons une nouvelle définition du potentiel scalaire
\begin{equation}\label{potentiel scalaire bel bouda}
    \mathcal{A}_{g}^{0}=\frac{c}{2}\,h^{00}=\frac{\mathcal{\phi}_{g}}{c}
\end{equation}
et du potentiel vecteur
\begin{eqnarray}\label{potentiel vecteur bel bouda}
\left\{
      \begin{array}{ll}
        \mathcal{A}_{g}^{i}=c\,h^{0i} \\
        i=1,2,3
      \end{array}
         \right.
\end{eqnarray}
gravitationnels. En notation quadridimensionnelle, les définitions précédentes permettent de former le quadrivecteur potentiel
\begin{equation}
    \mathcal{A}_{g}^{\mu}=(\mathcal{\phi}_{g}/c,\overrightarrow{\mathcal{A}}_{g})=(\mathcal{A}_{g}^{0},\mathcal{A}_{g}^{1},\mathcal{A}_{g}^{2},\mathcal{A}_{g}^{3}).
\end{equation}

Attirons l'attention sur le fait que la définition (\ref{potentiel scalaire bel bouda}) est tout à fait compatible avec l'expression de la composante temporelle $g_{00}$ de la métrique (\ref{metrique 00 et phi}), obtenue à l'approximation newtonienne.


\subsubsection{Champs gravitationnels types électrique et magnétique}
De façon analogue à l'électromagnétisme, introduisons le tenseur antisymétrique\footnote{$\mathcal{F}^{\mu\nu}_{g}=-\mathcal{F}^{\nu\mu}_{g}$}
\begin{equation}\label{tenseur antisymetrique grav}
    \mathcal{F}^{\mu\nu}_{g}=\partial^{\mu}\mathcal{A}_{g}^{\nu}-\partial^{\nu}\mathcal{A}_{g}^{\mu}.
\end{equation}

De manière analogue aux définitions employées en électromagnétisme, les trois composantes du champ gravitationnel type électrique\footnote{Comparer avec la définition de Huei (\ref{gooi}) du champ gravitoélectrique} $\overrightarrow{\mathbf{E}}_{g}(\mathbf{E}_{g}^{1},\mathbf{E}_{g}^{2},\mathbf{E}_{g}^{3})$ sont données par\footnote{Du fait de l'antisymétrie de $\mathcal{F}^{\mu\nu}_{g}$ nous avons aussi $\mathbf{E}_{g}^{i}=c\,\mathcal{F}^{i0}_{g}$}
\begin{equation}\label{composantes champ grav electric}
    \mathbf{E}_{g}^{i}=-c\,\mathcal{F}^{0i}_{g}
\end{equation}
alors que les trois composantes du champ gravitationnel type magnétique\footnote{Comparer avec la définition de Huei (\ref{bg}) du champ gravitomagnétique} $\overrightarrow{\mathbf{B}}_{g}(\mathbf{B}_{g}^{1},\mathbf{B}_{g}^{2},\mathbf{B}_{g}^{3})$ sont données par
\begin{equation}\label{composantes champ grav magnetic}
    \mathbf{B}_{g}^{i}=-\epsilon^{ijk}\,(\mathcal{F}_{g})_{jk}/2
\end{equation}
où les $\epsilon^{ijk}$ représentent les composantes du tenseur complètement antisymétrique de Lévi-Civita\footnote{Dans l'espace à 3-dimensions, le symbole de Levi-Civita, complètement antisymétrique, est défini par \cite{Kunz}
\begin{eqnarray}
\epsilon^{ijk}=\epsilon_{ijk}=\left\{
                                      \begin{array}{ll}
                                        +1 \hspace{0.5cm}\text{si}\;(i,j,k)\; \text{est une permutation paire de}\; (1,2,3) , \\
                                        -1 \hspace{0.5cm}\text{si}\; (i,j,k)\; \text{est une permutation impaire de}\; (1,2,3), \\
                                        \;\;0 \hspace{0.6cm} \text{sinon},\nonumber
                                      \end{array}
                                    \right.
\end{eqnarray} tel que $1=\epsilon^{123}=-\epsilon^{213}=\epsilon^{231}=-\epsilon^{321}$ et $0=\epsilon^{iij}=\epsilon^{iji}=\epsilon^{jii}=\epsilon^{iii}$} avec la convention $\epsilon^{123}=+1$.

Contrairement à l'approche de Huei où la relation usuelle (\ref{relation champ type electrique potentiels}) est vérifiée uniquement dans la jauge harmonique, dans cette nouvelle approche, indépendamment de la jauge utilisée, nous vérifions automatiquement la relation
\begin{equation}\label{relation champ type electrique potentiels bel bouda}
    \overrightarrow{\mathbf{E}}_{g}=-\overrightarrow{\nabla}\mathcal{\phi}_{g}-\frac{\partial \overrightarrow{\mathcal{A}}_{g}}{\partial t}
\end{equation}
par l'adoption des définitions (\ref{tenseur antisymetrique grav}) et (\ref{composantes champ grav electric}). En effet, nous avons
\begin{eqnarray}
  \mathbf{E}_{g}^{1}/c=\mathcal{F}^{10}_{g}&=&\partial^{1}\mathcal{A}_{g}^{0}-\partial^{0}\mathcal{A}_{g}^{1}=-\partial_{1}(\mathcal{\phi}_{g}/c)
  -\frac{1}{c}\,\left(\frac{\partial \mathcal{A}_{g}^{1}}{\partial t}\right) \label{comp 1 champ gravit ele} \\
  \mathbf{E}_{g}^{2}/c=\mathcal{F}^{20}_{g}&=&\partial^{2}\mathcal{A}_{g}^{0}-\partial^{0}\mathcal{A}_{g}^{2}=-\partial_{2}(\mathcal{\phi}_{g}/c)
-\frac{1}{c}\,\left(\frac{\partial \mathcal{A}_{g}^{2}}{\partial t}\right) \label{comp 2 champ gravit ele}\\
  \mathbf{E}_{g}^{3}/c=\mathcal{F}^{30}_{g}&=&\partial^{3}\mathcal{A}_{g}^{0}-\partial^{0}\mathcal{A}_{g}^{3}=-\partial_{3}(\mathcal{\phi}_{g}/c)
-\frac{1}{c}\,\left(\frac{\partial \mathcal{A}_{g}^{3}}{\partial t}\right)\label{comp 3 champ gravit ele}.
\end{eqnarray}

D'autre part, le champ gravitationnel type magnétique dérive du potentiel vecteur $\overrightarrow{\mathcal{A}}_{g}$ conformément à la relation (\ref{champ gravit mag deriv pot vect}). En effet, d'après la définition (\ref{composantes champ grav magnetic}), on peut vérifier que les composantes du champ gravitomagnétique sont
\begin{eqnarray}
  \mathbf{B}_{g}^{1} &=& -\frac{1}{2}\,\epsilon^{1jk}\,(\mathcal{F}_{g})_{jk}=
  \left(\overrightarrow{\nabla}\times\overrightarrow{\mathcal{A}_{g}}\right)^{1}\label{comp 1 champ gravit magnetique}
  \end{eqnarray}
  \begin{eqnarray}
  \mathbf{B}_{g}^{2} &=& -\frac{1}{2}\,\epsilon^{2jk}\,(\mathcal{F}_{g})_{jk}=
  \left(\overrightarrow{\nabla}\times\overrightarrow{\mathcal{A}_{g}}\right)^{2}\label{comp 2 champ gravit magnetique}
  \end{eqnarray}
  \begin{eqnarray}
  \mathbf{B}_{g}^{3} &=& -\frac{1}{2}\,\epsilon^{3jk}\,(\mathcal{F}_{g})_{jk}=
  \left(\overrightarrow{\nabla}\times\overrightarrow{\mathcal{A}_{g}}\right)^{3},\label{comp 3 champ gravit magnetique}
\end{eqnarray}
ce qu'on peut écrire sous la forme
\begin{equation}\label{champ gravit mag deriv pot vect bel bouda}
\overrightarrow{\mathbf{B}_{g}}=\overrightarrow{\nabla}\times\overrightarrow{\mathcal{A}_{g}}.
\end{equation}

\subsubsection{Equations d'Einstein linéarisées}
Le tenseur symétrique de la perturbation $h_{\mu\nu}$ vérifie les mêmes équations d'Einstein linéarisées
\begin{eqnarray}\label{equation einstein linearise revisite}
  \frac{8\pi G}{c^{4}}\,T_{\mu\nu}&\approx&\frac{1}{2}\Big(\partial_{\sigma}\partial_{\mu}h_{\nu}^{\sigma}+\partial_{\nu}\partial_{\sigma}h_{\mu}^{\sigma}-
   \partial_{\nu}\partial_{\mu}h-\Box h_{\mu\nu}
 -\eta_{\mu\nu}\,\partial_{\alpha}\partial_{\beta}h^{\alpha\beta}+\eta_{\mu\nu}\,\Box h\Big),
\end{eqnarray}
que la version standard.

\subsubsection{Les conditions de jauges utilisées}
Rappelons que la jauge harmonique (\ref{jauge harmonique}) est équivalente aux quatre conditions
\begin{eqnarray}
   \beta=0\hspace{0.5cm}\longrightarrow\hspace{0.5cm} \partial_{\mu}h^{\mu}_{0}-\displaystyle\frac{1}{2}\,\partial_{0}h=0 \label{comp temp jauge harmonique}\\
   \beta=i\hspace{0.5cm}\longrightarrow\hspace{0.5cm}\partial_{\mu}h^{\mu}_{i}-\displaystyle\frac{1}{2}\,\partial_{i}h=0.\label{comp spatiale jauge harmonique}
\end{eqnarray}
Par opposition à Caroll, dans le but d'exploiter les degrés de liberté associés aux composantes $h^{00}$ et $h^{0i}$, utilisons les trois composantes "spatiales" de la jauge harmonique (\ref{comp spatiale jauge harmonique}) et remplaçons la composante "temporelle" par la condition alternative
\begin{equation}\label{trace spatiale}
    h^{i}_{\;i}=h^{1}_{\;1}+h^{2}_{\;2}+h^{3}_{\;3}=0.
\end{equation}

En se plaçant dans ce cas de figure, la trace de la perturbation
\begin{equation}\label{trace perturbation bel bouda}
    h=h_{00}-(h_{11}+h_{22}+h_{33})=h_{00},
\end{equation}
est réduite à la composante temporelle de la perturbation de la métrique. Il est ainsi possible de réécrire les définitions des potentiels scalaire (\ref{potentiel scalaire bel bouda}) et vecteur (\ref{potentiel vecteur bel bouda}) sous la forme unifiée suivante
\begin{equation}
    \mathcal{A}^{\mu}_{g}=c\left(h^{0\mu}-\frac{1}{2}\,\eta^{0\mu}h\right).
\end{equation}
En effet,
\begin{eqnarray*}
  \mathcal{A}^{0}_{g} &=& c\Big(h^{00}-\eta^{00}h^{00}/2\Big)=c\,h^{00}/2 \\
  \mathcal{A}^{i}_{g} &=& c\Big(h^{0i}-\eta^{oi}h^{00}/2\Big)=c\,h^{0i}.
\end{eqnarray*}


Pour résumer, au lieu d'adopter les quatre équations (\ref{comp temp jauge harmonique}) et (\ref{comp spatiale jauge harmonique}), nous retenons comme conditions de jauge les relations (\ref{comp spatiale jauge harmonique}) et (\ref{trace spatiale}).

\subsection{Equations pour la Gravitation de type Maxwell}
Dans ce qui se suit, nous allons montrer que les équations d'Einstein, dans le contexte de la version revisitée de la Gravité linéaire, se réduisent aux équations de type Maxwell.

\subsubsection{$1^{\text{ier}}$ Groupe d'Equations de type Maxwell}
Le premier groupe des équations type Maxwell est automatiquement vérifié du fait que le tenseur antisymétrique $\mathcal{F}^{\mu\nu}_{g}$ est définit par (\ref{tenseur antisymetrique grav}). En effet,
\begin{eqnarray}\label{groupe eq maxwell grav 1}
  \partial^{\sigma}\mathcal{F}_{g}^{\mu\nu}+\partial^{\nu}\mathcal{F}_{g}^{\sigma\mu}+\partial^{\mu}\mathcal{F}_{g}^{\nu\sigma} &=& \partial^{\sigma}\left(\partial^{\mu}\mathcal{A}_{g}^{\nu}-\partial^{\nu}\mathcal{A}_{g}^{\mu}\right)+\partial^{\nu}\left(\partial^{\sigma}\mathcal{A}_{g}^{\mu}
  -\partial^{\mu}\mathcal{A}_{g}^{\sigma}\right)
  +\partial^{\mu}\left(\partial^{\nu}\mathcal{A}_{g}^{\sigma}-\partial^{\sigma}\mathcal{A}_{g}^{\nu}\right)\nonumber\\
  &=& \partial^{\sigma}\partial^{\mu}\mathcal{A}_{g}^{\nu}-\partial^{\sigma}\partial^{\nu}\mathcal{A}_{g}^{\mu}
  +\partial^{\nu}\partial^{\sigma}\mathcal{A}_{g}^{\mu}-\partial^{\nu}\partial^{\mu}\mathcal{A}_{g}^{\sigma}
  +\partial^{\mu}\partial^{\nu}\mathcal{A}_{g}^{\sigma}-\partial^{\mu}\partial^{\sigma}\mathcal{A}_{g}^{\nu}\nonumber\\
  &=& 0.
\end{eqnarray}

Comme en électromagnétisme, il est possible de vérifier que les équations (\ref{groupe eq maxwell grav 1}), écrites sous forme covariante, sont équivalentes aux équations type Maxwell suivantes
\begin{eqnarray}
  \text{div\;} \overrightarrow{\mathbf{B}_{g}} &=& 0 \label{divergence bg}\\
  \text{rot\;} \overrightarrow{\mathbf{E}_{g}} &=& -\frac{\partial \overrightarrow{\mathbf{B}_{g}}}{\partial t}.\label{rotationnel eg}
\end{eqnarray}

\subsubsection{$2^{\text{ème}}$ Groupe d'Equations de type Maxwell}
Dans le contexte de l'approche revisitée de la Gravité Linéaire, nous allons montrer que dans le domaine linéaire, les équations d'Einstein dans le vide
\begin{equation}\label{tenseur d'einstein linearise vide}
    G_{\mu\nu} \approx \frac{1}{2}\Big(\partial_{\sigma}\partial_{\mu}h_{\nu}^{\sigma}+\partial_{\nu}\partial_{\sigma}h_{\mu}^{\sigma}-
   \partial_{\nu}\partial_{\mu}h-\Box h_{\mu\nu}
 -\eta_{\mu\nu}\,\partial_{\alpha}\partial_{\beta}h^{\alpha\beta}+\eta_{\mu\nu}\,\Box h\Big)=0,
\end{equation}
se réduisent à des équations de type Maxwell.

\begin{enumerate}
  \item $\mathbf{G_{00}=0}$\\

La composante temporelle du tenseur d'Einstein
\begin{eqnarray}\label{tenseur d'einstein linearise 00 bel bouda}
    G_{00} &\approx& \frac{1}{2}\Big(\partial_{\sigma}\partial_{0}h_{0}^{\sigma}+\partial_{0}\partial_{\sigma}h_{0}^{\sigma}-
   \partial_{0}\partial_{0}h-\eta^{\alpha\beta}\partial_{\alpha}\partial_{\beta} h_{00}
 -\eta_{00}\,\partial_{\alpha}\partial_{\beta}h^{\alpha\beta}+\eta_{00}\,\eta^{\alpha\beta}\partial_{\alpha}\partial_{\beta} h\Big)\nonumber\\
  &=& \frac{1}{2}\Big(\partial_{0}\partial_{0}h_{0}^{0}+\partial_{i}\partial_{0}h^{i}_{0}+\partial_{0}\partial_{0}h_{0}^{0}
 +\partial_{0}\partial_{i}h_{0}^{i}-
   \partial_{0}\partial_{0}h-\eta^{00}\partial_{0}\partial_{0} h_{00}-\eta^{ii}\partial_{i}\partial_{i} h_{00}\nonumber\\
 &&\hspace{1cm}-\partial_{0}\partial_{\beta}h^{0\beta}-\partial_{i}\partial_{\beta}h^{i\beta}+\eta^{00}\partial_{0}\partial_{0} h+\eta^{ii}\partial_{i}\partial_{i} h\Big)\nonumber\\
  &=& \frac{1}{2}\Big(2\,\partial_{0}\partial_{0}h^{0}_{0}+2\,\partial_{i}\partial_{0}h^{i}_{0}-
   2\,\partial_{0}\partial_{0}h_{00}+\partial_{i}\partial_{i} h_{00}\nonumber\\
 &&\hspace{1cm}-\partial_{0}\partial_{0}h^{00}-\partial_{0}\partial_{i}h^{0i}-\partial_{i}\partial_{0}h^{i0}-\partial_{i}\partial_{j}h^{ij}+\partial_{0}\partial_{0} h^{00}-\partial_{i}\partial_{i} h\Big)\nonumber
 \end{eqnarray}
 est donnée par l'expression
 \begin{eqnarray}
 G_{00} &\approx& \frac{1}{2}\Big(\partial_{i}\partial_{i} h_{00}-\partial_{i}\partial_{i} h-\partial_{i}\partial_{j}h^{ij}\Big),\label{G00 sans restriction}
\end{eqnarray}
obtenue sans la moindre restriction ni recours à une quelconque jauge particulière. En utilisant les composantes "spatiales" de la jauge harmonique (\ref{comp spatiale jauge harmonique}), nous avons
\begin{eqnarray}
  \partial_{j}h^{j}_{i} &=& -\partial_{0}h^{0}_{i}+\frac{1}{2}\,\partial_{i}h \nonumber\\
  \partial_{j}h^{ij} &=& -\left(\partial_{0}h^{0i}-\frac{1}{2}\,\partial^{i}h\right).\label{jauge harmonique lineaire i equiv}
\end{eqnarray}
Ainsi en remplaçant (\ref{jauge harmonique lineaire i equiv}) dans (\ref{G00 sans restriction}) nous obtenons la composante\footnote{$G_{00}=G^{00}=G^{0}_{\;0}$}
\begin{eqnarray}
  G^{00} &\approx& \frac{1}{2}\left[\partial_{i}\partial_{i} h^{00}-\partial_{i}\partial_{i} h+\partial_{i}\left(\partial_{0}h^{0i}-\frac{1}{2}\,\partial^{i}h\right)\right].\label{G00 jauge harm lin comp spatiale}
\end{eqnarray}

En ayant recours à la condition de nullité de la trace spatiale (\ref{trace spatiale}), nous avons alors d'une part
\begin{eqnarray}
  G^{00} &\approx& \frac{1}{2}\left[\partial_{i}\partial_{i} h^{00}-\partial_{i}\partial_{i} h^{00}+\partial_{i}\left(\partial_{0}h^{0i}-\frac{1}{2}\,\partial^{i}h^{00}\right)\right]\nonumber\\
   &=& \frac{1}{2}\partial_{i}\left(\partial_{0}h^{0i}-\frac{1}{2}\,\partial^{i}h^{00}\right).\nonumber
  \end{eqnarray}
D'autre part, compte tenu des définitions des potentiels scalaire (\ref{potentiel scalaire bel bouda}) et vecteur (\ref{potentiel vecteur bel bouda}), l'expression précédente se met sous la forme
\begin{eqnarray}
G^{00} &\approx& -\frac{1}{2c}\partial_{i}\bigg[\partial^{i}(c\,h^{00}/2)-\partial^{0}(c\,h^{0i})\bigg].\nonumber\\
             &=&  -\frac{1}{2c}\partial_{i}\bigg(\partial^{i}\mathcal{A}_{g}^{0}-\partial^{0}\mathcal{A}_{g}^{i}\bigg).
\end{eqnarray}
D'après la définition du tenseur antisymétrique (\ref{tenseur antisymetrique grav}), nous aboutissons finalement à la relation
\begin{eqnarray}\label{juin17}
  G^{00} &\approx& -\frac{1}{2c}\;\partial_{i}\mathcal{F}_{g}^{i0}.
\end{eqnarray}

De façon analogue à l'électromagnétisme, en utilisant la définition du champ gravitationnel type électrique (\ref{composantes champ grav electric}), il est possible de vérifier que
\begin{eqnarray*}
\partial_{i}\mathcal{F}_{g}^{i0} =\partial_{i}\mathbf{E}_{g}^{i}/c=\text{div\;} \overrightarrow{\mathbf{E}_{g}}/c.
\end{eqnarray*}

D'après la relation (\ref{juin17}), nous pouvons donc conclure que l'équation de type Maxwell, $\text{div\;} \overrightarrow{\mathbf{E}_{g}}=0$, découle de la composante $G^{00}$ de l'équation d'Einstein dans le vide
\begin{eqnarray}\label{resume 1}
  \left[\begin{tabular}{c}
         Equation d'Einstein dans le vide  \\
         $G^{00}=0$ \\
       \end{tabular}
  \right]\Longrightarrow\left[\partial_{i}\mathcal{F}_{g}^{i0}=0 \Longleftrightarrow  \text{div\;} \overrightarrow{\mathbf{E}_{g}}=0 \right].
\end{eqnarray}


  \item $\mathbf{G_{0i}=0}$\\

D'autre part, d'après (\ref{tenseur d'einstein linearise vide})
\begin{eqnarray}
    G_{0i} &\approx& \frac{1}{2}\Big(\partial_{\sigma}\partial_{0}h_{i}^{\sigma}+\partial_{i}\partial_{\sigma}h_{0}^{\sigma}-
   \partial_{i}\partial_{0}h-\Box h_{0i}\Big)\nonumber\\
           &=& \frac{1}{2}\Big(\partial_{0}\partial_{0}h_{i}^{0}+\partial_{j}\partial_{0}h^{j}_{i}+\partial_{i}\partial_{0}h_{0}^{0}
 +\partial_{i}\partial_{j}h_{0}^{j}-
   \partial_{i}\partial_{0}h-\partial^{0}\partial_{0} h_{0i}-\partial^{j}\partial_{j} h_{0i}\Big)\nonumber
\end{eqnarray}
nous obtenons la composante
\begin{eqnarray}
 G_{0i} &\approx& \frac{1}{2}\Big[-\partial^{j}\partial_{j} h_{0i}+\mathbf{\partial_{j}\partial_{0}h^{j}_{i}}+\partial_{i}\partial_{j}h_{0}^{j}+\left(\partial_{i}\partial_{0}h_{0}^{0}-
   \partial_{i}\partial_{0}h\right)\Big],\label{G0i sans restriction}
\end{eqnarray}
sans aucune restriction ni recours à une quelconque jauge particulière. En utilisant les composantes "spatiales" de la jauge harmonique (\ref{comp spatiale jauge harmonique}), 
la composante (\ref{G0i sans restriction}) devient
\begin{eqnarray}
   \hspace*{-1cm}G_{0i} &\approx& \frac{1}{2}\left[-\partial^{j}\partial_{j} h_{0i}+\mathbf{\partial_{0}\left(\frac{1}{2}\,\partial_{i}h-\partial_{0}h^{0}_{i}\right)}+\partial_{i}\partial_{j}h_{0}^{j}+\left(\partial_{i}\partial_{0}h_{0}^{0}-
   \partial_{i}\partial_{0}h\right)\right].
 \end{eqnarray}
 En élevant les indices, pour déterminer la composante contravariante correspondante, nous avons\footnote{$G_{0i}=-G^{0i}=-G^{i}_{\;0}=G_{\;i}^{0}$}
 \begin{eqnarray}
-G^{0i} &\approx& \frac{1}{2}\left[-\partial^{j}\partial_{j} \left(-h^{0i}\right)-\frac{1}{2}\,\partial_{0}\partial^{i}h+\partial_{0}\partial^{0}h^{0i}-\partial^{i}\partial_{j}h^{j0}-\left(\partial^{i}\partial_{0}h_{0}^{0}-
   \partial^{i}\partial_{0}h\right)\right]\nonumber\\
G^{0i} &\approx& -\frac{1}{2}\left[\partial^{j}\partial_{j}h^{0i}-\frac{1}{2}\,\partial_{0}\partial^{i}h+\partial_{0}\partial^{0}h^{0i}-
\partial^{i}\partial_{j}h^{j0}-\left(\partial^{i}\partial_{0}h_{0}^{0}-
   \partial^{i}\partial_{0}h\right)\right].\label{comp Goi avec trace spatiale nulle}
\end{eqnarray}

De plus, si on a recours à la condition de nullité de la trace spatiale (\ref{trace spatiale}), de telle sorte que $h=h_{0}^{0}=h^{00}$, alors la composante (\ref{comp Goi avec trace spatiale nulle}) devient
\begin{eqnarray}
G^{0i} &\approx& -\frac{1}{2}\left[\partial^{j}\partial_{j}h^{0i}-\frac{1}{2}\,\partial_{0}\partial^{i}h^{00}+\partial_{0}\partial^{0}h^{0i}-
\partial^{i}\partial_{j}h^{j0}-\left(\partial^{i}\partial_{0}h_{0}^{0}-
   \partial^{i}\partial_{0}h_{0}^{0}\right)\right]\nonumber\\
    &=& -\frac{1}{2}\left(\partial^{j}\partial_{j}h^{0i}-\frac{1}{2}\,\partial_{0}\partial^{i}h^{00}+\partial_{0}\partial^{0}h^{0i}-
\partial^{i}\partial_{j}h^{j0}\right).
\end{eqnarray}

Compte tenu des définitions du tenseur antisymétrique (\ref{tenseur antisymetrique grav}), des potentiels scalaire (\ref{potentiel scalaire bel bouda}) et vecteur (\ref{potentiel vecteur bel bouda}), l'expression précédente se met sous la forme
\begin{eqnarray}
 G^{0i} &\approx& -\frac{1}{2}\bigg[\partial_{0}\big(\partial^{0}h^{0i}-\partial^{i}h^{00}/2\big)+\partial_{j}\big(\partial^{j}h^{0i}-
                   \partial^{i}h^{j0}\big)\bigg]\nonumber\\
              &=&  -\frac{1}{2c}\bigg[\partial_{0}\big(\partial^{0}\mathcal{A}_{g}^{i}-\partial^{i}\mathcal{A}_{g}^{0}\big)
  +\partial_{j}\left(\partial^{j}\mathcal{A}_{g}^{i}-\partial^{i}\mathcal{A}_{g}^{j}\right)\bigg]\nonumber\\
  &=&   -\frac{1}{2c} \bigg(\partial_{0}\mathcal{F}_{g}^{0i}+\partial_{j}\mathcal{F}_{g}^{ji}\bigg),
\end{eqnarray}
ou encore finalement
\begin{eqnarray}\label{explicite maxwell 2 froupe 2}
 G^{0i}\approx -\frac{1}{2c}\partial_{\mu}\mathcal{F}_{g}^{\mu i}.
\end{eqnarray}

De façon analogue à l'électromagnétisme, en utilisant les définitions des champ gravitationnels types électrique (\ref{composantes champ grav electric}) et magnétique (\ref{composantes champ grav magnetic}), il est possible de vérifier que
$$\partial_{\mu}\mathcal{F}_{g}^{\mu i}=(\text{rot\;} \overrightarrow{\mathbf{B}_{g}})^{i}-\frac{1}{c^{2}}\,\frac{\partial (\overrightarrow{\mathbf{E}_{g}})^{i}}{\partial t}.$$

Donc, d'après la relation (\ref{explicite maxwell 2 froupe 2}), nous pouvons conclure que l'équation de type Maxwell, $\text{rot\;} \overrightarrow{\mathbf{B}_{g}}=\frac{1}{c^{2}}\,\frac{\partial \overrightarrow{\mathbf{E}_{g}}}{\partial t}$, découle de la composante $G^{0i}$ de l'équation d'Einstein dans le vide
\begin{eqnarray}\label{resume 2}
  \left[\begin{tabular}{c}
         Equation d'Einstein dans le vide  \\
         $G^{0i}=0$ \\
       \end{tabular}
  \right] \Longrightarrow \left[\partial_{\mu}\mathcal{F}_{g}^{\mu i}=0 \Longleftrightarrow   \overrightarrow{\nabla}\times\overrightarrow{\mathbf{B}_{g}}=\frac{1}{c^{2}}\,\frac{\partial \overrightarrow{\mathbf{E}_{g}}}{\partial t} \right]\nonumber\\
\end{eqnarray}

Les relations (\ref{juin17}) et (\ref{explicite maxwell 2 froupe 2}), écrites sous la forme unifiée
\begin{equation}\label{ecriture unifie gonu}
G^{0\nu}\approx -\frac{1}{2c}\partial_{\mu}\mathcal{F}_{g}^{\mu\nu},
\end{equation}
nous permettent finalement de conclure que le deuxième groupe d'équations type Maxwell, dans le vide, découle des composantes $G^{0\nu}$ des équations d'Einstein dans le vide
$$G^{0\nu}=0\hspace{0.5cm}\Longrightarrow\hspace{0.5cm} \partial_{\mu}\mathcal{F}_{g}^{\mu\nu}=0.$$

  \item $\mathbf{G_{ij}=0}$\\
Nous allons nous intéresser à des degrés de liberté supplémentaires par rapport à ceux utilisés en électromagnétisme. Nous allons voir que ces degrés de liberté, de sens physique inconnu encore, vérifient une équation de propagation libre.

D'après (\ref{tenseur d'einstein linearise vide}), nous avons

\begin{eqnarray}
    G_{ij} &\approx& \frac{1}{2}\Big[\left(\partial_{\sigma}\partial_{i}h_{j}^{\sigma}+\partial_{\sigma}\partial_{i}h_{j}^{\sigma}\right)
    +\partial_{j}\partial_{\sigma}h_{i}^{\sigma}-
   \partial_{j}\partial_{i}h-\Box h_{ij}-\eta_{ij}\,\partial_{\alpha}\partial_{\beta}h^{\alpha\beta}+\eta_{ij}\,\Box h\Big]\nonumber\\
          &=& \frac{1}{2}\Big[\left(\partial_{0}\partial_{i}h_{j}^{0}+\partial_{k}\partial_{i}h_{j}^{k}\right)
    +\left(\partial_{j}\partial_{0}h_{i}^{0}+\partial_{j}\partial_{k}h_{i}^{k}\right)-
   \partial_{j}\partial_{i}h-\Box h_{ij}-\eta_{ij}\,\partial_{\alpha}\partial_{\beta}h^{\alpha\beta}\nonumber\\
   &&\hspace{1cm}+\eta_{ij}\,\Box h\Big]\nonumber\\
 &=& \frac{1}{2}\Big[\left(\partial_{0}\partial_{i}h_{j}^{0}+\partial_{k}\partial_{i}h_{j}^{k}\right)
    +\left(\partial_{j}\partial_{0}h_{i}^{0}+\partial_{j}\partial_{k}h_{i}^{k}\right)-\partial_{j}\partial_{i}h\nonumber\\
   &&\hspace{5cm}-\Box h_{ij}+\eta_{ij}\,\left[\Box h-\eta^{\beta\rho}\,\partial_{\beta}\left(\partial_{\alpha}h^{\alpha}_{\rho}\right)\right]\Big\}\nonumber\\
   &=& \frac{1}{2}\Big[\partial_{i}\left(\partial_{0}h_{j}^{0}+\partial_{k}h_{j}^{k}\right)
    +\partial_{j}\left(\partial_{0}h_{i}^{0}+\partial_{k}h_{i}^{k}\right)-\partial_{j}\partial_{i}h\nonumber\\
   &&\hspace{5cm}-\Box h_{ij}+\eta_{ij}\,\left[\Box h-\eta^{\beta\rho}\,\partial_{\beta}\left(\partial_{\alpha}h^{\alpha}_{\rho}\right)\right]\Big\}. \label{Gij lineaire sans restriction }
\end{eqnarray}

En utilisant la jauge harmonique (\ref{jauge harmonique}), l'expression précédente se met sous la forme
\begin{eqnarray}
  G_{ij} &\approx& \frac{1}{2}\left\{\partial_{i}\left(\frac{1}{2}\,\partial_{j}h\right)
    +\partial_{j}\left(\frac{1}{2}\,\partial_{i}h\right)-\partial_{j}\partial_{i}h
   -\Box h_{ij}+\eta_{ij}\,\left[\Box h-\eta^{\beta\rho}\,\partial_{\beta}\left(\frac{1}{2}\,\partial_{\rho}h\right)\right]\right\}\nonumber\\
    &=& \frac{1}{2}\left\{
   -\Box h_{ij}+\eta_{ij}\,\left[\Box h-\frac{1}{2}\,\partial^{\rho}\partial_{\rho}h\right]\right\} \nonumber\\
    &=& \frac{1}{2}\left\{
   -\Box h_{ij}+\eta_{ij}\,\left[\Box h-\frac{1}{2}\,\Box h\right]\right\} \nonumber\\
   &=& -\frac{1}{2}\,\Box\left(
    h_{ij}-\frac{1}{2}\,\eta_{ij}\,h\right), \label{Gij lineaire harmonique}
\end{eqnarray}
ou encore
\begin{equation}
    G_{ij} \approx -\frac{1}{2}\;\Box\overline{h}_{\,ij},
\end{equation}
compte tenu de (\ref{changement de metrique}).

Dans la jauge harmonique, les degrés de liberté supplémentaire $\overline{h}_{ij}$ se propagent à la vitesse de la lumière

\begin{eqnarray}
  \left[\begin{tabular}{c}
         Equation d'Einstein dans le vide  \\
         $G^{ij}=0$ \\
       \end{tabular}
  \right]\Longrightarrow\left[\Box\left(
    h^{ij}-\frac{1}{2}\,\eta^{ij}\,h\right)=\Box\overline{h}^{\,ij}=0 \right].
\end{eqnarray}
\end{enumerate}

Finalement, pour résumer la situation nous pouvons proposer ce schéma récapitulatif

\begin{eqnarray}
 \hspace*{-1cm} \left[\underbrace{G^{\mu\nu}=0}_
  {\text{Equation d'Einstein}}\bigwedge\underbrace{\left\{
    \begin{array}{ll}
      \partial_{\mu}h^{\mu}_{0}-\displaystyle\frac{1}{2}\,\partial_{0}h=0, & \hbox{\tiny{non utilisée}} \\\\
      \partial_{\mu}h^{\mu}_{i}-\displaystyle\frac{1}{2}\,\partial_{i}h=0, & \hbox{\tiny{utilisée}}
    \end{array}
  \right.}_{\text{Jauge harmonique}}
  \bigwedge\underbrace{h^{1}_{1}+h^{2}_{2}+h^{3}_{3}=0}_{\text{restriction sur la trace}}\bigwedge
\left\{
  \begin{array}{ll}
    \begin{tabular}{l}
    $\mathcal{F}_{g}^{\mu\nu}=\partial^{\mu}\mathcal{A}_{g}^{\nu}-\partial^{\nu}\mathcal{A}_{g}^{\mu}$ \\\\
    $\mathcal{A}_{g}^{0}=\frac{c}{2}\,h^{00}$ \\\\
    $\mathcal{A}_{g}^{i}=c\,h^{0i}$ \\
  \end{tabular}
  \end{array}
\right.\right]\nonumber
  \end{eqnarray}
  \begin{eqnarray}
  \Downarrow\nonumber
  \end{eqnarray}
  \begin{eqnarray}
  \hspace*{-0.5cm}\left[\underbrace{\partial_{\mu}\mathcal{F}_{g}^{\mu\nu}=0\hspace{0.2cm}\bigwedge\hspace{0.2cm} \partial^{\sigma}\mathcal{F}_{g}^{\mu\nu}+\partial^{\nu}\mathcal{F}_{g}^{\sigma\mu}+\partial^{\mu}\mathcal{F}_{g}^{\nu\sigma} = 0}_{\text{Equations de Maxwell}}
  \hspace{0.2cm}\bigwedge\hspace{0.2cm}\underbrace{\Box\left(
    h^{ij}-\frac{1}{2}\,\eta^{ij}\,h\right)=\Box\overline{h}^{\,ij}=0}_{\text{propagation des degrés de liberté $h^{ij}$ dans la jauge harmonique}}\right]\nonumber
  \end{eqnarray}

\subsection{Force gravitationnelle de type Lorentz}
Dans ce qui suit, nous allons montrer qu'une particule d'épreuve, se déplaçant suivant des géodésiques, est soumise à une force gravitationnelle type Lorentz dans le domaine des champs et des vitesses faibles.

L'équation des géodésiques
\begin{eqnarray}
  \displaystyle\frac{d^{2}x^{\mu}}{d\tau^{2}} &=& -\Gamma^{\mu}_{\rho\sigma}\,\displaystyle\frac{dx^{\rho}}{d\tau}\,\displaystyle\frac{dx^{\sigma}}{d\tau} \nonumber\\
  &=& -\Gamma^{\mu}_{00}\,\displaystyle\frac{dx^{0}}{d\tau}\,\displaystyle\frac{dx^{0}}{d\tau}
  -2\,\Gamma^{\mu}_{0i}\,\displaystyle\frac{dx^{0}}{d\tau}\,\displaystyle\frac{dx^{i}}{d\tau}
  -\Gamma^{\mu}_{il}\,\displaystyle\frac{dx^{i}}{d\tau}\,\displaystyle\frac{dx^{l}}{d\tau} \nonumber
\end{eqnarray}
permet de déterminer les composantes spatiales
\begin{equation}\label{bel bouda acceleration relativiste}
    \displaystyle\frac{d^{2}x^{j}}{d\tau^{2}}=-c^{2}\,\Gamma^{j}_{00}\,\left(\displaystyle\frac{dt}{d\tau}\right)^{2}
    -2\,\Gamma^{j}_{0i}\,\left(\displaystyle\frac{dt}{d\tau}\right)\,c\,u^{i}-\Gamma^{\mu}_{il}\,u^{i}\,u^{l}.
\end{equation}
Rappelons que le quadrvecteur vitesse est par définition égale à la dérivée du quadrivecteur position par rapport au temps propre
\begin{equation}
    u^{\nu}=\frac{dx^{\nu}}{d\tau}=\frac{dx^{\mu}}{dt}\,\frac{dt}{d\tau}=(\gamma_{v}\,c,\gamma_{v}\,\overrightarrow{v})
\end{equation}
tel que $u^{0}=\gamma_{v}\,c$, $u^{i}=\gamma_{v}\,v^{i}$ et $\gamma_{v}=(1-\overrightarrow{v}^{2}/c^{2})^{-1/2}$.

Dans le cas linéaire, les symboles de Christoffel figurant dans (\ref{bel bouda acceleration relativiste}) sont donnés par
\begin{eqnarray}
  \Gamma^{j}_{00} &\approx& \frac{1}{2}\,\eta^{j\lambda} \left(\partial_{0}h_{0\lambda}+\partial_{0}h_{0\lambda}-\partial_{\lambda}h_{00}\right) =
   \partial_{0}h_{0}^{j}-\frac{1}{2}\,\partial^{j}h_{00}\label{bel bouda gama j00} \\
  \Gamma^{j}_{0i} &\approx& \frac{1}{2}\,\eta^{j\lambda} \left(\partial_{0}h_{i\lambda}+\partial_{i}h_{0\lambda}-\partial_{\lambda}h_{0i}\right) =
  \frac{1}{2}\,\left(\partial_{0}h_{i}^{j}+\partial_{i}h_{0}^{j}-\partial^{j}h_{0i}\right)\label{bel bouda gama j0i}\\
  \Gamma^{j}_{il} &\approx& \frac{1}{2}\,\eta^{j\lambda} \left(\partial_{i}h_{l\lambda}+\partial_{l}h_{i\lambda}-\partial_{\lambda}h_{il}\right) =
  \frac{1}{2}\,\left(\partial_{i}h_{l}^{j}+\partial_{l}h_{i}^{j}-\partial^{j}h_{il}\right),\label{bel bouda gama jil}
\end{eqnarray}
de sorte à avoir
\begin{eqnarray}
\displaystyle\frac{d^{2}x^{j}}{d\tau^{2}}&=&-c^{2}\,\left(\displaystyle\frac{dt}{d\tau}\right)^{2}(\partial_{0}h_{0}^{j}-\frac{1}{2}\,\partial^{j}h_{00})
    -c\,u^{i}\,\left(\displaystyle\frac{dt}{d\tau}\right)\left(\partial_{0}h_{i}^{j}+\partial_{i}h_{0}^{j}-\partial^{j}h_{0i}\right)\nonumber\\
    &&-\frac{1}{2}\,u^{i}\,u^{l}\,\left(\partial_{i}h_{l}^{j}+\partial_{l}h_{i}^{j}-\partial^{j}h_{il}\right) \nonumber\\ \nonumber\\
&=&-c^{2}\,\left(\displaystyle\frac{dt}{d\tau}\right)^{2}(\partial_{0}h_{0}^{j})+\frac{c^{2}}{2}
\,\left(\displaystyle\frac{dt}{d\tau}\right)^{2}(\partial^{j}h_{00})
    -c\,u^{i}\,\left(\displaystyle\frac{dt}{d\tau}\right)\left(\partial_{i}h_{0}^{j}-\partial^{j}h_{0i}\right)\nonumber\\
    &&-c\,u^{i}\,(\partial_{0}h_{i}^{j})\left(\displaystyle\frac{dt}{d\tau}\right)
    -\frac{1}{2}\,u^{i}\,u^{l}\,\left(\partial_{i}h_{l}^{j}+\partial_{l}h_{i}^{j}-\partial^{j}h_{il}\right),\label{acceleration complete}
\end{eqnarray}
ou encore finalement
\begin{eqnarray}
\hspace*{-1cm}\displaystyle\frac{d^{2}x^{j}}{d\tau^{2}}&=&c^{2}\Bigg[\frac{1}{2}\left(\displaystyle\frac{dt}{d\tau}\right)^{2}\partial^{j}h^{00}-\frac{1}{c}\,\left(\displaystyle\frac{dt}{d\tau}\right)^{2}\frac{\partial h^{0j}}{\partial t}\nonumber\\
&&-\frac{u_{i}}{c}\left(\displaystyle\frac{dt}{d\tau}\right)\left(\partial^{i}h^{0j}-\partial^{j}h^{0i}\right)
-\frac{u_{i}}{c^{2}}\,\displaystyle\frac{dt}{d\tau}\,\frac{\partial h^{ij}}{\partial t}
    -\frac{u_{i}u_{l}}{c^{2}}\left(\partial^{i}h^{jl}-\frac{1}{2}\,\partial^{j}h^{il}\right)\Bigg].\label{acceleration complete bis}
\end{eqnarray}

Pour des vitesses faibles\footnote{ Les champs faibles permettent d'accélérer les particules à des vitesses faibles} $v\ll c$, où
$\frac{dt}{d\tau}\rightarrow 1$ et $u^{i}\rightarrow v^{i}=dx^{i}/dt$, il est possible de négliger les termes proportionnels à $1/c^{2}$ dans (\ref{acceleration complete bis}), pour avoir\footnote{on néglige les termes quadratiques en vitesses $v^{2}/c^{2}$}
\begin{eqnarray}\label{bel bouda comp accel fontion h}
  \displaystyle\frac{d^{2}x^{j}}{dt^{2}} &\approx& c^{2}\bigg[\frac{1}{2}\left(\partial^{j}h^{00}\right)-\frac{1}{c}\,\frac{\partial h^{0j}}{\partial t}-\frac{v_{i}}{c}\left(\partial^{i}h^{0j}-\partial^{j}h^{0i}\right)\bigg].
\end{eqnarray}
En tenant compte des définitions des potentiels scalaire (\ref{potentiel scalaire bel bouda}) et vecteur (\ref{potentiel vecteur bel bouda}), les composantes d'accélérations (\ref{bel bouda comp accel fontion h}) prennent la forme
\begin{eqnarray}\label{bel bouda comp accel type lorentz}
  \displaystyle\frac{d^{2}x^{j}}{dt^{2}} &\approx& \left(c\,\partial^{j}\mathcal{A}_{g}^{0}-\frac{\partial \mathcal{A}_{g}^{j}}{\partial t}\right)-v_{i}\left(\partial^{i}\mathcal{A}_{g}^{j}-\partial^{j}\mathcal{A}_{g}^{i}\right).
\end{eqnarray}
Pour $j=1$
\begin{eqnarray*}
  \displaystyle\frac{d^{2}x^{1}}{dt^{2}} &\approx& \left(c\,\partial^{1}\mathcal{A}_{g}^{0}-\frac{\partial \mathcal{A}_{g}^{1}}{\partial t}\right)-v_{i}\left(\partial^{i}\mathcal{A}_{g}^{1}-\partial^{1}\mathcal{A}_{g}^{i}\right) \\
   &=& \left(-\partial_{1}\phi_{g}-\frac{\partial \mathcal{A}_{g}^{1}}{\partial t}\right)-v_{2}\left(\partial^{2}\mathcal{A}_{g}^{1}-\partial^{1}\mathcal{A}_{g}^{2}\right)
-v_{3}\left(\partial^{3}\mathcal{A}_{g}^{1}-\partial^{1}\mathcal{A}_{g}^{3}\right) \\
&=& \left(-\partial_{1}\phi_{g}-\frac{\partial \mathcal{A}_{g}^{1}}{\partial t}\right)+v^{2}\left(\partial_{1}\mathcal{A}_{g}^{2}-\partial_{2}\mathcal{A}_{g}^{1}\right)
-v^{3}\left(\partial_{3}\mathcal{A}_{g}^{1}-\partial_{1}\mathcal{A}_{g}^{3}\right)
\end{eqnarray*}
et compte tenu de la définition de la première composante du champ gravitoélectrique (\ref{comp 1 champ gravit ele}) et des composantes du champ gravitomagnétique (\ref{comp 2 champ gravit magnetique}) et (\ref{comp 3 champ gravit magnetique}), nous aboutissons finalement à la forme
\begin{equation}
    \displaystyle\frac{d^{2}x^{1}}{dt^{2}} \approx \mathbf{E}_{g}^{1}+\left(\overrightarrow{v}\times \overrightarrow{\mathbf{B}}_{g}\right)^{1}\label{force grav type lorentz comp 1}
\end{equation}

De même pour $j=2$ et $j=3$, on trouve
\begin{equation}
    \displaystyle\frac{d^{2}x^{2}}{dt^{2}} \approx \mathbf{E}_{g}^{2}+\left(\overrightarrow{v}\times \overrightarrow{\mathbf{B}}_{g}\right)^{2},\label{force grav type lorentz comp 2}
\end{equation}
et
\begin{equation}
    \displaystyle\frac{d^{2}x^{3}}{dt^{2}} \approx \mathbf{E}_{g}^{3}+\left(\overrightarrow{v}\times \overrightarrow{\mathbf{B}}_{g}\right)^{3}.\label{force grav type lorentz comp 3}
\end{equation}
Les équations (\ref{force grav type lorentz comp 1}), (\ref{force grav type lorentz comp 2}) et (\ref{force grav type lorentz comp 3}), résumées sous forme vectorielle
\begin{equation}
    \displaystyle\frac{d^{2}\overrightarrow{r}}{dt^{2}} \approx \overrightarrow{\mathbf{E}}_{g}+\left(\overrightarrow{v}\times \overrightarrow{\mathbf{B}}_{g}\right),\label{force grav type lorentz}
\end{equation}
montent clairement qu'une particule test, dans le cas des champs et vitesses faibles\footnote{à l'ordre $v/c$}, est soumise à une force gravitationnelle de type Lorentz
\begin{equation}
    m_{i}\,\displaystyle\frac{d^{2}\overrightarrow{r}}{dt^{2}} \approx m_{g}\bigg[\overrightarrow{\mathbf{E}}_{g}+\left(\overrightarrow{v}\times \overrightarrow{\mathbf{B}}_{g}\right)\bigg].\label{force grav type lorentz}
\end{equation}
Contrairement au modèle de Huei \cite{Huei} et Wald \cite{Wald} où un facteur 4 indésirable apparaît dans la partie magnétique de la force, il est clair que dans notre modèle, la force de type Lorentz (\ref{eq lorentz grav force bis}) ne souffre pas de cette imperfection.

\section{Discussion des résultats et conclusion}
Une analyse critique de la version standard de la Gravité Linéaire nous a permis de souligné quelques "insuffisances". La situation peut être résumée comme suit
\begin{enumerate}
  \item Dans l'approche de Huei, les équations d'Einstein se réduisent, dans le cadre de l'approximation du champ faible, à des équations de type Maxwell; néanmoins, la définition, usuelle en électromagnétisme, du champ gravitoélectrique en fonction des potentiels scalaire et vecteur n'est valable que dans le cas particulier de la jauge harmonique. De plus, bien que la force gravitationnelle agissant sur une particule test qui se déplace suivant des géodésiques est de type Lorentz, néanmoins, la partie magnétique de cette force est entachée d'un facteur 4 indésirable. En dernier lieu, la force type magnétique n'est obtenue qu'en se limitant au régime stationnaire.
  \item Dans l'approche de Carroll, les champs sont définis de telle sorte à résoudre le problème relatif au facteur 4 indésirable de la force gravitationnelle de type Lorentz, néanmoins, les équations d'Einstein ne prennent plus la forme d'équations de type Maxwell.
\end{enumerate}

Dans le but de remédier à tous ces problèmes, nous avons procédé par les étapes suivantes
\begin{enumerate}
  \item Identification des composantes $h^{00}$ et $h^{0i}$ de la métrique de perturbation, respectivement par les potentiels scalaire et vecteur
\begin{eqnarray*}
       \mathcal{A}_{g}^{0}&=&\frac{c}{2}\,h^{00}=\frac{\mathcal{\phi}_{g}}{c}\\
        \mathcal{A}_{g}^{i}&=&c\,h^{0i}.
\end{eqnarray*}
  \item Définition des champs gravitoélectrique et gravitomagnétique
\begin{eqnarray*}
    \mathbf{E}_{g}^{i}&=&-c\,\mathcal{F}^{0i}_{g}\\
    \mathbf{B}_{g}^{i}&=&-\epsilon^{ijk}\,(\mathcal{F}_{g})_{jk}/2
\end{eqnarray*}
par l'intermédiaire du tenseur antisymétrique $\mathcal{F}^{\mu\nu}_{g}=\partial^{\mu}\mathcal{A}_{g}^{\nu}-\partial^{\nu}\mathcal{A}_{g}^{\mu}$.
  \item Nous avons montré que ces définitions sont équivalentes à
\begin{eqnarray*}
\overrightarrow{\mathbf{E}_{g}}&=&-\overrightarrow{\nabla}\phi_{g}-\partial_{t}\overrightarrow{\mathcal{A}}_{g}\\
\overrightarrow{\mathbf{B}_{g}}&=&\overrightarrow{\nabla}\times\overrightarrow{\mathcal{A}_{g}},
\end{eqnarray*}
indépendamment de la jauge utilisée.
  \item Nous avons montré que les champs gravitationnels, dans le domaine linéaire, vérifient des équations de type Maxwell
\begin{enumerate}
  \item Le premier groupe des équations type Maxwell
\begin{eqnarray*}
\partial^{\sigma}\mathcal{F}_{g}^{\mu\nu}+\partial^{\nu}\mathcal{F}_{g}^{\sigma\mu}+\partial^{\mu}\mathcal{F}_{g}^{\nu\sigma} &=& 0
\hspace{0.2cm}\Longleftrightarrow\hspace{0.2cm}\left\{
  \begin{array}{ll}
    \text{div\;} \overrightarrow{\mathbf{B}_{g}} = 0 \\\\
    \text{rot\;} \overrightarrow{\mathbf{E}_{g}} = -\frac{\partial \overrightarrow{\mathbf{B}_{g}}}{\partial t}
  \end{array}
\right.
\end{eqnarray*}
est automatiquement vérifié compte tenu du caractère antisymétrique de $\mathcal{F}^{\mu\nu}_{g}$ et de la commutation des opérateurs dérivées.
  \item Pour retrouver le deuxième groupe des équations type Maxwell
\begin{eqnarray*}
    \partial_{\mu}\mathcal{F}_{g}^{\mu\nu}=0\hspace{0.3cm}\Longleftrightarrow\hspace{0.3cm}\left\{
                                                                               \begin{array}{ll}
                                                                                 \text{div\;} \overrightarrow{\mathbf{E}_{g}} =0\\\\
  \text{rot\;} \overrightarrow{\mathbf{B}_{g}} = \frac{1}{c^{2}}\,\frac{\partial \overrightarrow{\mathbf{E}_{g}}}{\partial t}
                                                                               \end{array}
                                                                             \right.
\end{eqnarray*}
nous avons adopté un point de vue quelque peu différent de l'approche standard. En effet, au lieu d'utiliser toutes les conditions de la jauge harmonique
\begin{eqnarray*}
 \partial_{\alpha}h^{\alpha}_{0}-\frac{1}{2}\,\partial_{0}h &=& 0 \\
 \partial_{\alpha}h^{\alpha}_{i}-\frac{1}{2}\,\partial_{i}h &=& 0
\end{eqnarray*}
nous avons exploité les trois composantes "spatiales" et nous avons substitué la composante "temporelle" par la condition alternative de trace spatiale nulle
\begin{equation*}
    h^{i}_{\;i}=0.
\end{equation*}
Dans ce cas, nous avons pu montrer la relation $G^{0\nu}\approx-\frac{1}{2c}\,\partial_{\mu}\mathcal{F}_{g}^{\mu\nu}$ qui nous permet de conclure que les équations d'Einstein dans le vide se réduisent au deuxième groupe d'équations type Maxwell
$$G^{0\nu}=0\hspace{0.5cm}\Longrightarrow\hspace{0.5cm} \partial_{\mu}\mathcal{F}_{g}^{\mu\nu}=0.$$
\item Avant de terminer, nous attirons l'attention sur un fait très intéressant. Une conséquence de la condition de trace spatiale nulle est que la trace de perturbation est réduite à sa seule composante temporelle $h=h^{00}$. Dans ce cas, la composante "temporelle" de la jauge harmonique, non utilisée jusqu'à présent, se met sous la forme
\begin{eqnarray}
0 &=&  \partial_{\alpha}h^{\alpha}_{0}-\frac{1}{2}\,\partial_{0}h \nonumber\\
  &=& \partial_{0}h^{00}+\partial_{i}h^{i0}-\frac{1}{2}\,\partial_{0}h^{00} = \frac{1}{2}\,\partial_{0}h^{00}+\partial_{i}h^{i0}  \nonumber\\
   &=& \frac{1}{c}\,\partial_{0}(c\,h^{00}/2)+\frac{1}{c}\,\partial_{i}(c\,h^{i0})  = \frac{1}{c}\left(\partial_{0}\mathcal{A}_{g}^{0}+\partial_{i}\mathcal{A}_{g}^{i}\right)  \nonumber\\
  &=& \frac{1}{c}\,\partial_{\mu}\mathcal{A}_{g}^{\mu}.  \label{equivalence harmonique1 et lorentz}
\end{eqnarray}
Cette dernière relation n'est autre que la \textbf{jauge de Lorentz} de l'électromagnétisme !

Après avoir retrouvé les équations de Maxwell, il est ainsi possible d'adopter la composante temporelle de la jauge harmonique (jauge de Lorentz), dans le but de découpler les équations de propagation des potentiels scalaire et vectoriel
\begin{eqnarray}\label{propagation potentiels quadrim}
    \Box \mathcal{A}_{g}^{\mu}=\mu_{0}\,J^{\mu}
    \hspace{0.5cm}\Longleftrightarrow\hspace{0.5cm}\left\{
                                   \begin{array}{ll}
                                     \Box \phi=\displaystyle\frac{\rho_{\text{\tiny{m}}}}{\varepsilon_{0}},\\
                                     \Box \overrightarrow{\mathcal{A}_{g}}=\mu_{0}\,\overrightarrow{j},
                                   \end{array}
                                 \right.
\end{eqnarray}
où $J^{\mu}=(c\,\rho_{\text{\tiny{m}}},\overrightarrow{j})$ , $\mathcal{A}_{g}^{\mu}=(\phi_{g}/c,\overrightarrow{\mathcal{A}_{g}})$ et $\overrightarrow{j}=\rho_{\text{\tiny{m}}}\overrightarrow{v}$.

Dans ce cas, plusieurs interprétations de la condition de trace spatiale nulle $h^{i}_{i}=0$ peuvent être avancée
\begin{itemize}
  \item Elle peut être vue comme l'adoption d'un système de coordonnées particulier dans lequel la métrique de perturbation prend une forme particulière. Il faut admettre qu'une telle interprétation serait très compromettante car elle enlèverait le caractère de généralité  à cette approche.
  \item Elle peut être vue comme la prépondérance de la composante temporelle $h^{00}$. Cette prépondérance peut s'expliquer par exemple par le fait d'avoir de façon individuelle $h^{00}\gg h^{11}$, $h^{00}\gg h^{22}$ et $h^{00}\gg h^{33}$ ce qui impliquerait que $h=h^{00}-(h^{11}+h^{22}+h^{33})\approx h^{00}$. Dans ce cas, quel serait le raison profonde de la petitesse des composantes $h^{ii}$ par rapport à $h^{00}$ ?
  \item L'interprétation la plus plausible, à notre avis, est que la condition de nullité de la trace spatiale est une condition nécessaire pour que la jauge harmonique englobe la jauge de Lorentz. Elle permet à la composante temporelle de la jauge harmonique de se réduire à la condition de jauge de Lorentz, afin de pouvoir découpler les équations de propagation des degrés de liberté $h^{00}$ et $h^{0i}$.
\end{itemize}

Il reste encore la question très intéressante des degrés de liberté supplémentaires $h^{ij}$. En effet, les $h^{0\mu}$ ont été utilisés pour définir $\mathcal{A}_{g}^{\mu}$, mais les $h^{ij}$ vérifient la condition $h^{11}+h^{22}+h^{33}=0$.  Nous pensons qu'ils ne vérifient pas seulement la condition de champ faible $h^{ij}\ll 1$, mais qu'ils sont négligeables devant les $h^{0\mu}$ de telle sorte que leurs effets ne sont pas détectés dans le domaine de cette application. On a vu que dans le cadre de la jauge harmonique, à l'extérieur de la source, les $\overline{h}^{ij}=h^{ij}-\eta^{ij}\,h/2$ se propagent à la vitesse de la lumière car ils vérifient des équations de propagation $\Box \overline{h}^{ij}=0$.
\end{enumerate}
\item Nous avons montré qu'une particule d'épreuve, astreinte à se déplacer suivant des géodésiques, est soumise à une force gravitationnelle de type Lorentz
\begin{equation}
    \displaystyle\frac{d^{2}\overrightarrow{r}}{dt^{2}} \approx \bigg[\overrightarrow{\mathbf{E}}_{g}+\left(\overrightarrow{v}\times \overrightarrow{\mathbf{B}}_{g}\right)\bigg].
\end{equation}
Ce résultat a été obtenu, pour des champs faibles et pour des vitesses d'ordre $v/c$, sans aucune restriction au régime stationnaire, de plus la partie magnétique ne souffre plus du facteur 4 indésirable. Bien que ce résultat ne puisse s'obtenir qu'en négligeant les termes en $v^{2}/c^{2}$, ce qui laisse à penser qu'il y a une restriction aux faibles vitesses, néanmoins, il faut signaler que la force de Lorentz, est par définition exprimée à l'ordre $v/c$ (la partie magnétique).
\end{enumerate}

\newpage
\pagestyle{fancy} \lhead{chapitre\;4}\rhead{Électromagnétisme}
\chapter{Électromagnétisme}

\section{Introduction}
Dans ce chapitre, nous allons développer une nouvelle approche pour décrire l'interaction électromagnétique. Notre démarche est motivée, d'une part, par la très grande similarité entre la gravité et l'électromagnétisme, obtenue au chapitre précédent, et d'autre part, par les résultats "surprenants" de C.C.Barros \cite{Barros1} pour décrire l'atome d'hydrogène.

Rappelons que dans le cadre d'une symétrie sphérique, l'auteur adopte une métrique similaire à celle de Schwarzschild, dans laquelle il incorpore le potentiel coulombien électron-ptoton et retrouve le spectre relativiste de Dirac pour l'atome d'Hydrogène.  Ces résultats nous incitent à émettre les hypothèses suivantes
\begin{itemize}
  \item L'extension du Principe d'Equivalence permettant d'absorber l'effet du champ électro-magnétique agissant sur une charge électrique dans la métrique de l'espace-temps de telle sorte à aboutir à une particule libre se déplaçant suivant des géodésiques.
   \item L'existence d'une version des équations d'Einstein pour l'interaction électromagnétique $G^{\mu\nu}=\chi_{e}\,T^{\mu\nu}$, où la constante de proportionnalité $\chi_{e}$ entre le tenseur d'Einstein et le tenseur énergie-impulsion sera déterminée ultérieurement.
\end{itemize}

Dans le cadre de la limite du champ faible, nous allons montrer qu'au premier ordre de la perturbation, les Equations de Maxwell découlent des Equations type Einstein pour l'électromagnétisme, alors que l'équation des géodésiques, dans le cas des faibles vitesses, permettent de reproduire l'équation de mouvement d'une charge électrique soumise à une force de Lorentz.

En poussant l'étude perturbative au deuxième ordre, des corrections des équations de Maxwell seront apportées et l'analyse qualitative des ordres supérieurs de la perturbation des équations type Einstein va révéler que l'électromagnétisme, dans son domaine actuel d'application, est essentiellement une interaction linéaire, contrairement à la gravité qui est essentiellement non linéaire.

Nous terminons par une discussion dans laquelle les résultats sont analysés.

\section{Cadre théorique et hypothèses adoptées}
Dans ce qui suit, il est question d'appliquer l'approche de la Gravité Linéaire revisitée à l'interaction électromagnétique. L'extension du Principe d'Equivalence sera envisagée de telle manière à absorber l'effet d'un champ éléctromagnétique agissant sur une charge électrique dans la métrique de l'espace-temps; autrement-dit effectuer une transformation de coordonnées pour passer à un référentiel dans lequel la charge électrique libre n'est soumise à aucune force et se déplace suivant des géodésiques. Cette éventuelle extension du Principe d'Equivalence pourrait aussi être envisagée pour les autres interactions non gravitationnelles en adoptant un point de vue selon lequel toute forme d'énergie est susceptible d'affecter la structure de l'espace-temps \cite{Cardone}.

\subsection{Métrique adoptée}
Contrairement à l'interaction gravitationnelle qui peut affecter considérablement les propriétés de l'espace-temps, nous supposons que l'influence du champ électromagnétique sur la géométrie de l'espace-temps est si faible\footnote{A notre connaissance, il n'y a pas de confirmation expérimentale de ce phénomène.} que la métrique utilisée est toujours une perturbation linéaire de la métrique minkowskienne (\ref{plate plus perturbation}). Cette approximation sera justifiée ultérieurement dans le domaine actuel d'application de l'électromagnétisme.


\subsection{Potentiels scalaire et vecteur}
De façon analogue aux relations (\ref{potentiel scalaire bel bouda}) et (\ref{potentiel vecteur bel bouda}) qui expriment le lien entre les potentiels scalaire et vecteur gravitationnels et la perturbation de la métrique, proposons en présence d'une particule test de masse $m$ et de charge $q$ les définitions suivantes du potentiel scalaire
\begin{equation}\label{potentiel scalaire bel bouda electrom}
    \mathcal{A}^{0}=\frac{mc}{2q}\,h^{00}=\frac{\mathcal{\phi}}{c}
\end{equation}
et du potentiel vecteur
\begin{eqnarray}\label{potentiel vecteur bel bouda electrom}
\left\{
      \begin{array}{ll}
        \mathcal{A}^{i}=\displaystyle\frac{mc}{q}\,h^{0i} \\
        i=1,2,3
      \end{array}
         \right.
\end{eqnarray}
électromagnétiques\footnote{En fait les définitions (\ref{potentiel scalaire bel bouda}) et (\ref{potentiel vecteur bel bouda}) sont plutôt données par $\mathcal{A}^{0}=(m_{i}c/2m_{g})h^{00}=\mathcal{\phi_{g}}/c$ et $\mathcal{A}^{k}=(m_{i}c/m_{g})h^{0k}$, où le rapport des masses inerte et grave est égale à l'unité $m_{i}/m_{g}=1$}. En notation quadridimensionnelle, les définitions précédentes permettent de former le quadrivecteur potentiel
\begin{equation}
    \mathcal{A}^{\mu}=(\mathcal{\phi}/c,\overrightarrow{\mathcal{A}})=(\mathcal{A}^{0},\mathcal{A}^{1},\mathcal{A}^{2},\mathcal{A}^{3}).
\end{equation}

Attirons l'attention sur le fait que la définition (\ref{potentiel scalaire bel bouda electrom}) est tout à fait compatible avec l'expression de la composante temporelle $g_{00}$ de la métrique (\ref{metrique ele 00 phi}), obtenue à l'approximation newtonienne.


\subsection{Champs électrique et magnétique}
Le champ électromagnétique est le tenseur antisymétrique d'ordre 2 suivant
\begin{equation}\label{tenseur antisymetrique grav electrom}
    \mathcal{F}^{\mu\nu}=\partial^{\mu}\mathcal{A}^{\nu}-\partial^{\nu}\mathcal{A}^{\mu}.
\end{equation}

Les trois composantes du champ électrique $\overrightarrow{\mathbf{E}}(\mathbf{E}^{1},\mathbf{E}^{2},\mathbf{E}^{3})$ sont données par 
\begin{equation}\label{composantes champ grav electric electrom}
    \mathbf{E}^{i}=-c\,\mathcal{F}^{0i}
\end{equation}
alors que les trois composantes du champ magnétique $\overrightarrow{\mathbf{B}}(\mathbf{B}^{1},\mathbf{B}^{2},\mathbf{B}^{3})$ sont données par
\begin{equation}\label{composantes champ grav magnetic electrom}
    \mathbf{B}^{i}=-\epsilon^{ijk}\,\mathcal{F}_{jk}/2
\end{equation}
où les $\epsilon^{ijk}$ représentent les composantes du tenseur complètement antisymétrique de Lévi-Civita avec la convention $\epsilon^{123}=+1$.

Il est aisé de vérifier que les définitions (\ref{tenseur antisymetrique grav electrom}), (\ref{composantes champ grav electric electrom}) et (\ref{composantes champ grav magnetic electrom}) permettent d'écrire les relations habituelles
\begin{equation}\label{relation champ type electrique potentiels bel bouda electrom}
    \overrightarrow{\mathbf{E}}=-\overrightarrow{\nabla}\mathcal{\phi}-\frac{\partial \overrightarrow{\mathcal{A}}}{\partial t}
\end{equation}
et
\begin{equation}\label{champ gravit mag deriv pot vect bel bouda electrom}
\overrightarrow{\mathbf{B}}=\overrightarrow{\nabla}\times\overrightarrow{\mathcal{A}},
\end{equation}
de façon indépendante de la jauge utilisée.

\subsection{Equations de type Einstein linéarisées}
Au premier ordre, le tenseur symétrique de la perturbation $h_{\mu\nu}$ vérifie les équations de type Einstein linéarisées
\begin{eqnarray}\label{equation einstein linearise revisite electrom}
  \chi_{e}\,T_{\mu\nu}&\approx&\frac{1}{2}\Big(\partial_{\sigma}\partial_{\mu}h_{\nu}^{\sigma}+\partial_{\nu}\partial_{\sigma}h_{\mu}^{\sigma}-
   \partial_{\nu}\partial_{\mu}h-\Box h_{\mu\nu}
 -\eta_{\mu\nu}\,\partial_{\alpha}\partial_{\beta}h^{\alpha\beta}+\eta_{\mu\nu}\,\Box h\Big),
\end{eqnarray}
pour l'électromagnétisme.

\subsection{Les conditions de jauges utilisées}
De façon analogue à la démarche adoptée dans la Gravité Linéaire revisitée, nous allons utiliser les trois composantes "spatiales" de la jauge harmonique (\ref{jauge harmonique})
\begin{eqnarray}\label{comp spa jaug harm electrom}
    \partial_{\mu}h^{\mu}_{i}-\displaystyle\frac{1}{2}\,\partial_{i}h=0,
\end{eqnarray}
et substituer la composante "temporelle" de cette jauge par la condition alternative
\begin{equation}\label{trace spatiale electrom}
    h^{i}_{\;i}=h^{1}_{\;1}+h^{2}_{\;2}+h^{3}_{\;3}=0.
\end{equation}

En se plaçant dans le cas de figure, la trace de la perturbation est
\begin{equation}\label{trace perturbation bel bouda electrom}
    h=h_{00}-(h_{11}+h_{22}+h_{33})=h_{00}.
\end{equation}
Il est ainsi possible de réécrire les définitions du potentiel scalaire (\ref{potentiel scalaire bel bouda electrom}) et vecteur (\ref{potentiel vecteur bel bouda electrom}) sous la forme unifiée suivante\footnote{la masse qui figure dans l'expression est la masse inertielle $m=m_{i}$.}
\begin{equation}\label{amu electro et jauge}
    \mathcal{A}^{\mu}=\frac{mc}{q}\left(h^{0\mu}-\frac{1}{2}\,\eta^{0\mu}h\right).
\end{equation}
En effet,
\begin{eqnarray*}
  \mathcal{A}^{0} &=& \frac{mc}{q}\Big(h^{00}-\eta^{00}h^{00}/2\Big)=\frac{mc}{2q}\,h^{00} \\
  \mathcal{A}^{i} &=& \frac{mc}{q}\Big(h^{0i}-\eta^{oi}h^{00}/2\Big)=\frac{mc}{q}\,h^{0i}.
\end{eqnarray*}


\section{Ordre 1 de la perturbation: Equations de Maxwell comme conséquence d'une version électromagnétique des équations d'Einstein}
Dans ce qui se suit, nous allons montrer qu'au premier ordre de la perturbation les équations de type Einstein de l'électromagnétisme se réduisent aux équations de Maxwell.

\subsection{$1^{\text{ier}}$ Groupe d'Equations de Maxwell}
Le premier groupe des équations de Maxwell est automatiquement vérifié du fait que le tenseur électromagnétique $\mathcal{F}^{\mu\nu}$ est défini par (\ref{tenseur antisymetrique grav electrom}). En effet,
\begin{eqnarray}
  \partial^{\sigma}\mathcal{F}^{\mu\nu}+\partial^{\nu}\mathcal{F}^{\sigma\mu}+\partial^{\mu}\mathcal{F}^{\nu\sigma} &=& \partial^{\sigma}\left(\partial^{\mu}\mathcal{A}^{\nu}-\partial^{\nu}\mathcal{A}^{\mu}\right)+\partial^{\nu}\left(\partial^{\sigma}\mathcal{A}^{\mu}
  -\partial^{\mu}\mathcal{A}^{\sigma}\right)
  +\partial^{\mu}\left(\partial^{\nu}\mathcal{A}^{\sigma}-\partial^{\sigma}\mathcal{A}^{\nu}\right)\nonumber\\
  &=& \partial^{\sigma}\partial^{\mu}\mathcal{A}^{\nu}-\partial^{\sigma}\partial^{\nu}\mathcal{A}^{\mu}
  +\partial^{\nu}\partial^{\sigma}\mathcal{A}^{\mu}-\partial^{\nu}\partial^{\mu}\mathcal{A}^{\sigma}
  +\partial^{\mu}\partial^{\nu}\mathcal{A}^{\sigma}-\partial^{\mu}\partial^{\sigma}\mathcal{A}^{\nu}\nonumber\\
  &=& 0.\label{groupe eq maxwell grav 1 electrom}
\end{eqnarray}

Il est possible de vérifier que les équations (\ref{groupe eq maxwell grav 1 electrom}), exprimées sous forme covariante, sont équivalentes aux équations de Maxwell suivantes
\begin{eqnarray}
  \text{div\;} \overrightarrow{\mathbf{B}} &=& 0 \label{divergence bg electrom}\\
  \text{rot\;} \overrightarrow{\mathbf{E}} &=& -\frac{\partial \overrightarrow{\mathbf{B}}}{\partial t}.\label{rotationnel eg electrom}
\end{eqnarray}

\subsection{$2^{\text{ème}}$ Groupe d'Equations de Maxwell}
Rappelons le deuxième groupe des équations de Maxwell, dans le vide,
\begin{eqnarray}\label{maxwell 2 groupe electrom}
    \partial_{\mu}\mathcal{F}^{\mu\nu}=0\hspace{0.5cm}\Longleftrightarrow\hspace{0.5cm}\left\{
                                                                               \begin{array}{ll}
                                                                                 0=\partial_{\mu}\mathcal{F}^{\mu0}=\partial_{0}\mathcal{F}^{00}
                                                                                 +\partial_{i}\mathcal{F}^{i0}=\partial_{i}\mathcal{F}^{i0} & \hbox{pour $\nu=0$}\\\\
                                                                                 0=\partial_{\mu}\mathcal{F}^{\mu i} & \hbox{pour $\nu=i=1,2,3$}
                                                                               \end{array}
                                                                             \right.
\end{eqnarray}
écrites sous forme covariante.

Il est possible de montrer que les équations (\ref{maxwell 2 groupe electrom}) sont équivalentes aux équations de Maxwell suivantes
\begin{eqnarray}
  \text{div\;} \overrightarrow{\mathbf{E}} &=& 0 \label{divergence eg electrom}\\
  \text{rot\;} \overrightarrow{\mathbf{B}} &=& \frac{1}{c^{2}}\,\frac{\partial \overrightarrow{\mathbf{E}}}{\partial t}.\label{rotationnel bg electrom}
\end{eqnarray}

Dans le contexte du premier ordre de la perturbation, nous allons montrer que les équations de Maxwell (\ref{maxwell 2 groupe electrom}) sont contenues dans une version électromagnétique d'équations Einstein dans le vide
\begin{equation}\label{tenseur d'einstein linearise vide electrom}
    G_{\mu\nu} \approx \frac{1}{2}\Big(\partial_{\sigma}\partial_{\mu}h_{\nu}^{\sigma}+\partial_{\nu}\partial_{\sigma}h_{\mu}^{\sigma}-
   \partial_{\nu}\partial_{\mu}h-\Box h_{\mu\nu}
 -\eta_{\mu\nu}\,\partial_{\alpha}\partial_{\beta}h^{\alpha\beta}+\eta_{\mu\nu}\,\Box h\Big)=0.
\end{equation}

\begin{enumerate}
  \item $\mathbf{G_{00}=0}$\\

La composante temporelle du tenseur d'Einstein,
\begin{eqnarray}\label{tenseur d'einstein linearise 00 bel bouda electrom}
    G_{00} &\approx& \frac{1}{2}\Big(\partial_{\sigma}\partial_{0}h_{0}^{\sigma}+\partial_{0}\partial_{\sigma}h_{0}^{\sigma}-
   \partial_{0}\partial_{0}h-\eta^{\alpha\beta}\partial_{\alpha}\partial_{\beta} h_{00}
 -\eta_{00}\,\partial_{\alpha}\partial_{\beta}h^{\alpha\beta}+\eta_{00}\,\eta^{\alpha\beta}\partial_{\alpha}\partial_{\beta} h\Big)\nonumber\\
  &=& \frac{1}{2}\Big(\partial_{0}\partial_{0}h_{0}^{0}+\partial_{i}\partial_{0}h^{i}_{0}+\partial_{0}\partial_{0}h_{0}^{0}
 +\partial_{0}\partial_{i}h_{0}^{i}-
   \partial_{0}\partial_{0}h-\eta^{00}\partial_{0}\partial_{0} h_{00}-\eta^{ii}\partial_{i}\partial_{i} h_{00}\nonumber\\
 &&\hspace{1cm}-\partial_{0}\partial_{\beta}h^{0\beta}-\partial_{i}\partial_{\beta}h^{i\beta}+\eta^{00}\partial_{0}\partial_{0} h+\eta^{ii}\partial_{i}\partial_{i} h\Big)\nonumber\\
  &=& \frac{1}{2}\Big(2\,\partial_{0}\partial_{0}h^{0}_{0}+2\,\partial_{i}\partial_{0}h^{i}_{0}-
   2\,\partial_{0}\partial_{0}h_{00}+\partial_{i}\partial_{i} h_{00}\nonumber\\
 &&\hspace{1cm}-\partial_{0}\partial_{0}h^{00}-\partial_{0}\partial_{i}h^{0i}-\partial_{i}\partial_{0}h^{i0}-\partial_{i}\partial_{j}h^{ij}+\partial_{0}\partial_{0} h^{00}-\partial_{i}\partial_{i} h\Big),\nonumber
 \end{eqnarray}
 est donnée par l'expression
 \begin{eqnarray}
 G_{00} &\approx& \frac{1}{2}\Big(\partial_{i}\partial_{i} h_{00}-\partial_{i}\partial_{i} h-\partial_{i}\partial_{j}h^{ij}\Big),\label{G00 sans restriction electrom}
\end{eqnarray}
qui est obtenue sans la moindre restriction ni recours à une quelconque jauge particulière.

En utilisant les composantes "spatiales" de la jauge harmonique (\ref{comp spa jaug harm electrom}), nous avons
\begin{eqnarray}
  \partial_{j}h^{j}_{i} &=& -\partial_{0}h^{0}_{i}+\frac{1}{2}\,\partial_{i}h \nonumber\\
  \partial_{j}h^{ij} &=& -\left(\partial_{0}h^{0i}-\frac{1}{2}\,\partial^{i}h\right).\label{jauge harmonique lineaire i equiv electrom}
\end{eqnarray}
Ainsi en remplaçant (\ref{jauge harmonique lineaire i equiv electrom}) dans (\ref{G00 sans restriction electrom}) nous obtenons la composante\footnote{$G_{00}=G^{00}=G^{0}_{\;0}$}
\begin{eqnarray}
  G^{00} &\approx& \frac{1}{2}\left[\partial_{i}\partial_{i} h^{00}-\partial_{i}\partial_{i} h+\partial_{i}\left(\partial_{0}h^{0i}-\frac{1}{2}\,\partial^{i}h\right)\right].\label{G00 jauge harm lin comp spatiale electrom}
\end{eqnarray}

En ayant recours à la condition de nullité de la trace spatiale (\ref{trace perturbation bel bouda electrom}), qui permet de réduire la trace de la perturbation à sa composante temporelle $h=h_{o}^{o}$, nous montrons que (\ref{G00 jauge harm lin comp spatiale electrom}) se met sous la forme
\begin{eqnarray}
  G^{00} &\approx& \frac{1}{2}\left[\partial_{i}\partial_{i} h^{00}-\partial_{i}\partial_{i} h^{00}+\partial_{i}\left(\partial_{0}h^{0i}-\frac{1}{2}\,\partial^{i}h^{00}\right)\right]\nonumber\\
         &=& \frac{1}{2}\partial_{i}\left(\partial_{0}h^{0i}-\frac{1}{2}\,\partial^{i}h^{00}\right).
  \end{eqnarray}

Compte tenu des définitions des potentiels scalaire (\ref{potentiel scalaire bel bouda electrom}) et vecteur (\ref{potentiel vecteur bel bouda electrom}), nous avons
\begin{eqnarray}
G^{00} &\approx& -\frac{q}{2mc}\partial_{i}\left[\partial^{i}\left(\frac{mc}{2q}\,h^{00}\right)-\partial^{0}\left(\frac{mc}{q}\,h^{0i}\right)\right]\nonumber\\
             &=& -\frac{q}{2mc}\partial_{i}\left(\partial^{i}\mathcal{A}^{0}-\partial^{0}\mathcal{A}^{i}\right).
\end{eqnarray}
La définition (\ref{tenseur antisymetrique grav electrom}) du champ électromagnétique, permet de mettre la relation précédente sous la forme
\begin{equation}\label{explicite maxwell 1 froupe 2 electrom}
G^{00}\approx -\frac{q}{2mc}\partial_{i}\mathcal{F}^{i0}.
\end{equation}

En utilisant la définition du champ électrique (\ref{composantes champ grav electric electrom}), il est possible de vérifier que
\begin{eqnarray*}
\partial_{i}\mathcal{F}^{i0} =\partial_{i}\mathbf{E}^{i}/c=\text{div\;} \overrightarrow{\mathbf{E}}/c.
\end{eqnarray*}

La relation (\ref{explicite maxwell 1 froupe 2 electrom}), nous permet de conclure que l'équation de Maxwell, $\text{div\;} \overrightarrow{\mathbf{E}}=0$, découle de la composante $G^{00}$ de l'équation type Einstein dans le vide pour décrire l'interaction électromagnétique
\begin{eqnarray}\label{resume 1 electrom}
  \left[\begin{tabular}{c}
         Version électromagnétique \\
         des équations d'Einstein dans le vide  \\
         $G^{00}=0$ \\
       \end{tabular}
  \right]\Longrightarrow\left[\partial_{i}\mathcal{F}^{i0}=0 \Longleftrightarrow  \text{div\;} \overrightarrow{\mathbf{E}}=0 \right].
\end{eqnarray}

  \item $\mathbf{G_{0i}=0}$\\

D'autre part, d'après (\ref{tenseur d'einstein linearise vide electrom})
\begin{eqnarray}
    G_{0i} &\approx& \frac{1}{2}\Big(\partial_{\sigma}\partial_{0}h_{i}^{\sigma}+\partial_{i}\partial_{\sigma}h_{0}^{\sigma}-
   \partial_{i}\partial_{0}h-\Box h_{0i}\Big)\nonumber\\
  &=& \frac{1}{2}\Big(\partial_{0}\partial_{0}h_{i}^{0}+\partial_{j}\partial_{0}h^{j}_{i}+\partial_{i}\partial_{0}h_{0}^{0}
 +\partial_{i}\partial_{j}h_{0}^{j}-
   \partial_{i}\partial_{0}h-\partial^{0}\partial_{0} h_{0i}-\partial^{j}\partial_{j} h_{0i}\Big)\nonumber
\end{eqnarray}
nous obtenons la composante
\begin{eqnarray}
 G_{0i} &\approx& \frac{1}{2}\Big[-\partial^{j}\partial_{j} h_{0i}+\mathbf{\partial_{j}\partial_{0}h^{j}_{i}}+\partial_{i}\partial_{j}h_{0}^{j}+\left(\partial_{i}\partial_{0}h_{0}^{0}-
   \partial_{i}\partial_{0}h\right)\Big],\label{G0i sans restriction electrom}
\end{eqnarray}
sans aucune restriction ni recours à une quelconque jauge particulière.

En utilisant les composantes "spatiales" de la jauge harmonique (\ref{comp spa jaug harm electrom}), la composante (\ref{G0i sans restriction electrom}) devient
\begin{eqnarray}
  \hspace*{-0.5cm} G_{0i} &\approx& \frac{1}{2}\left[-\partial^{j}\partial_{j} h_{0i}+\mathbf{\partial_{0}\left(\frac{1}{2}\,\partial_{i}h-\partial_{0}h^{0}_{i}\right)}+\partial_{i}\partial_{j}h_{0}^{j}+\left(\partial_{i}\partial_{0}h_{0}^{0}-
   \partial_{i}\partial_{0}h\right)\right].
 \end{eqnarray}
 En élevant les indices, pour déterminer la composante contravariante correspondante, nous avons\footnote{$G_{0i}=-G^{0i}=-G^{i}_{\;0}=G_{\;i}^{0}$}
 \begin{eqnarray}
-G^{0i} &\approx& \frac{1}{2}\left[-\partial^{j}\partial_{j} \left(-h^{0i}\right)-\frac{1}{2}\,\partial_{0}\partial^{i}h+\partial_{0}\partial^{0}h^{0i}-\partial^{i}\partial_{j}h^{j0}-\left(\partial^{i}\partial_{0}h_{0}^{0}-
   \partial^{i}\partial_{0}h\right)\right]\nonumber\\
G^{0i} &\approx& -\frac{1}{2}\left[\partial^{j}\partial_{j}h^{0i}-\frac{1}{2}\,\partial_{0}\partial^{i}h+\partial_{0}\partial^{0}h^{0i}-
\partial^{i}\partial_{j}h^{j0}-\left(\partial^{i}\partial_{0}h_{0}^{0}-
   \partial^{i}\partial_{0}h\right)\right].\label{comp Goi avec trace spatiale nulle electrom}
\end{eqnarray}

De plus, si on a recours à la condition de nullité de la trace spatiale (\ref{trace spatiale electrom}), de telle sorte que $h=h^{00}$, alors la composante (\ref{comp Goi avec trace spatiale nulle electrom}) devient
\begin{eqnarray}
G^{0i} &\approx& -\frac{1}{2}\left[\partial^{j}\partial_{j}h^{0i}-\frac{1}{2}\,\partial_{0}\partial^{i}h^{00}+\partial_{0}\partial^{0}h^{0i}-
\partial^{i}\partial_{j}h^{j0}-\left(\partial^{i}\partial_{0}h_{0}^{0}-
   \partial^{i}\partial_{0}h_{0}^{0}\right)\right]\nonumber\\
    &=& -\frac{1}{2}\left(\partial^{j}\partial_{j}h^{0i}-\frac{1}{2}\,\partial_{0}\partial^{i}h^{00}+\partial_{0}\partial^{0}h^{0i}-
\partial^{i}\partial_{j}h^{j0}\right).
\end{eqnarray}

Compte tenu des définitions (\ref{tenseur antisymetrique grav electrom}) du champ électromagnétique, des potentiels scalaire (\ref{potentiel scalaire bel bouda electrom}) et vecteur (\ref{potentiel vecteur bel bouda electrom}), l'expression précédente se met sous la forme
\begin{eqnarray}
  G^{0i} &\approx& -\frac{1}{2}\bigg[\partial_{0}\big(\partial^{0}h^{0i}-\partial^{i}h^{00}/2\big)+\partial_{j}\big(\partial^{j}h^{0i}-
\partial^{i}h^{j0}\big)\bigg]\nonumber\\
              &=& -\frac{q}{2mc}\bigg[\partial_{0}\big(\partial^{0}\mathcal{A}^{i}-\partial^{i}\mathcal{A}^{0}\big)
  +\partial_{j}\big(\partial^{j}\mathcal{A}^{i}-\partial^{i}\mathcal{A}^{j}\big)\bigg]\nonumber\\
              &=& -\frac{q}{2mc}\bigg(\partial_{0}\mathcal{F}^{0i}+\partial_{j}\mathcal{F}^{ji}\bigg),
\end{eqnarray}
ou encore finalement
\begin{eqnarray}\label{explicite maxwell 2 froupe 2 electrom}
    G^{0i} &\approx& -\frac{q}{2mc}\,\partial_{\mu}\mathcal{F}^{\mu i}.
\end{eqnarray}

En utilisant les définitions des champs électrique (\ref{relation champ type electrique potentiels bel bouda electrom}) et magnétique (\ref{champ gravit mag deriv pot vect bel bouda electrom}), il est possible de vérifier que
$$\partial_{\mu}\mathcal{F}^{\mu i}=(\text{rot\;} \overrightarrow{\mathbf{B}})^{i}-\frac{1}{c^{2}}\,\frac{\partial (\overrightarrow{\mathbf{E}})^{i}}{\partial t}.$$

D'après la relation (\ref{explicite maxwell 2 froupe 2 electrom}), nous sommes en mesure de conclure que l'équation de Maxwell, $\text{rot\;} \overrightarrow{\mathbf{B}}=\frac{1}{c^{2}}\,\frac{\partial \overrightarrow{\mathbf{E}}}{\partial t}$, découle de la composante $G^{0i}$ de l'équation type Einstein dans le vide pour décrire l'interaction électromagnétique
\begin{eqnarray}\label{resume 2 electrom}
 \hspace*{-0.5cm} \left[\begin{tabular}{c}
         Version électromagnétique \\
         des équations d'Einstein dans le vide  \\
         $G^{0i}=0$ \\
       \end{tabular}
  \right] \Longrightarrow \left[\partial_{\mu}\mathcal{F}^{\mu i}=0 \Longleftrightarrow   \overrightarrow{\nabla}\times\overrightarrow{\mathbf{B}}=\frac{1}{c^{2}}\,\frac{\partial \overrightarrow{\mathbf{E}}}{\partial t} \right]\nonumber\\
\end{eqnarray}

Les relations (\ref{explicite maxwell 1 froupe 2 electrom}) et (\ref{explicite maxwell 2 froupe 2 electrom}), écrites sous la forme unifiée
\begin{equation}\label{ecriture unifie gonu}
G^{0\nu}\approx -\frac{q}{2mc}\partial_{\mu}\mathcal{F}^{\mu\nu},
\end{equation}
nous permettent finalement de conclure que le deuxième groupe d'équations de Maxwell dans le vide découle des composantes $G^{0\nu}$ de la version électromagnétique des équations d'Einstein dans le vide
\begin{equation}\label{resume electrom maxwell}
G^{0\nu}=0\hspace{0.5cm}\Longrightarrow\hspace{0.5cm} \partial_{\mu}\mathcal{F}^{\mu\nu}=0.
\end{equation}

  \item $\mathbf{G_{ij}=0}$\\
Nous allons nous intéresser à des degrés de liberté supplémentaires par rapport à ceux utilisés en électromagnétisme standard (les $\mathcal{A}^{\mu}$). Nous allons voir que ces degrés de liberté, de sens physique inconnu jusqu'à présent, vérifient une équation de propagation libre. D'après (\ref{tenseur d'einstein linearise vide electrom}), nous avons

\begin{eqnarray}
    G_{ij} &\approx& \frac{1}{2}\Big[\left(\partial_{\sigma}\partial_{i}h_{j}^{\sigma}+\partial_{\sigma}\partial_{i}h_{j}^{\sigma}\right)
    +\partial_{j}\partial_{\sigma}h_{i}^{\sigma}-
   \partial_{j}\partial_{i}h-\Box h_{ij}-\eta_{ij}\,\partial_{\alpha}\partial_{\beta}h^{\alpha\beta}+\eta_{ij}\,\Box h\Big]\nonumber\\
  &=& \frac{1}{2}\Big[\left(\partial_{0}\partial_{i}h_{j}^{0}+\partial_{k}\partial_{i}h_{j}^{k}\right)
    +\left(\partial_{j}\partial_{0}h_{i}^{0}+\partial_{j}\partial_{k}h_{i}^{k}\right)-
   \partial_{j}\partial_{i}h-\Box h_{ij}-\eta_{ij}\,\partial_{\alpha}\partial_{\beta}h^{\alpha\beta}\nonumber\\
   &&\hspace{1cm}+\eta_{ij}\,\Box h\Big]\nonumber\\
 &=& \frac{1}{2}\Big[\left(\partial_{0}\partial_{i}h_{j}^{0}+\partial_{k}\partial_{i}h_{j}^{k}\right)
    +\left(\partial_{j}\partial_{0}h_{i}^{0}+\partial_{j}\partial_{k}h_{i}^{k}\right)-\partial_{j}\partial_{i}h\nonumber\\
   &&\hspace{5cm}-\Box h_{ij}+\eta_{ij}\,\left[\Box h-\eta^{\beta\rho}\,\partial_{\beta}\left(\partial_{\alpha}h^{\alpha}_{\rho}\right)\right]\Big\}\nonumber\\
   &=& \frac{1}{2}\Big[\partial_{i}\left(\partial_{0}h_{j}^{0}+\partial_{k}h_{j}^{k}\right)
    +\partial_{j}\left(\partial_{0}h_{i}^{0}+\partial_{k}h_{i}^{k}\right)-\partial_{j}\partial_{i}h\nonumber\\
   &&\hspace{5cm}-\Box h_{ij}+\eta_{ij}\,\left[\Box h-\eta^{\beta\rho}\,\partial_{\beta}\left(\partial_{\alpha}h^{\alpha}_{\rho}\right)\right]\Big\}. \label{Gij lineaire sans restriction electrom}
\end{eqnarray}

En utilisant la jauge harmonique (\ref{jauge harmonique}), l'expression précédente se met sous la forme
\begin{eqnarray}
  G_{ij} &\approx& \frac{1}{2}\left\{\partial_{i}\left(\frac{1}{2}\,\partial_{j}h\right)
    +\partial_{j}\left(\frac{1}{2}\,\partial_{i}h\right)-\partial_{j}\partial_{i}h
   -\Box h_{ij}+\eta_{ij}\,\left[\Box h-\eta^{\beta\rho}\,\partial_{\beta}\left(\frac{1}{2}\,\partial_{\rho}h\right)\right]\right\}\nonumber\\
    &=& \frac{1}{2}\left\{
   -\Box h_{ij}+\eta_{ij}\,\left[\Box h-\frac{1}{2}\,\partial^{\rho}\partial_{\rho}h\right]\right\} \nonumber\\
    &=& \frac{1}{2}\left\{
   -\Box h_{ij}+\eta_{ij}\,\left[\Box h-\frac{1}{2}\,\Box h\right]\right\} \nonumber\\
   &=& -\frac{q}{2mc}\,\Box\left[\frac{mc}{q}\left(
    h_{ij}-\frac{1}{2}\,\eta_{ij}\,h\right)\right] \label{Gij lineaire harmonique electrom}
\end{eqnarray}
ou encore
\begin{equation}
    G_{ij} \approx -\frac{q}{2mc}\;\Box\widehat{h}_{\,ij},
\end{equation}
tel que
\begin{equation}
    \widehat{h}_{\,ij}=\frac{mc}{q}\left(h_{ij}-\frac{1}{2}\,\eta_{ij}\,h\right).
\end{equation}

Dans la jauge harmonique et au premier ordre de la perturbation, les degrés de liberté supplémentaires $\widehat{h}_{ij}$ se propagent à la vitesse de la lumière
\begin{eqnarray}
  \hspace*{-1cm}\left[\begin{tabular}{c}
         Version électromagnétique \\
         des équations d'Einstein dans le vide  \\
         $G^{ij}=0$ \\
       \end{tabular}
  \right]\Longrightarrow\left[\frac{mc}{q}\Box\left(
    h_{ij}-\frac{1}{2}\,\eta_{ij}\,h\right)=\Box\widehat{h}^{\,ij}=0 \right].
\end{eqnarray}

\end{enumerate}

\subsection{Récapitulatif}

Finalement, pour résumer la situation nous pouvons proposer ce schéma récapitulatif

\begin{eqnarray}
  \hspace*{-1cm}\left[\underbrace{G^{\mu\nu}=0}_
  {\text{Equation type Einstein}}\bigwedge\underbrace{\left\{
    \begin{array}{ll}
      \partial_{\mu}h^{\mu}_{0}-\displaystyle\frac{1}{2}\,\partial_{0}h=0, & \hbox{\tiny{non utilisée}} \\\\
      \partial_{\mu}h^{\mu}_{i}-\displaystyle\frac{1}{2}\,\partial_{i}h=0, & \hbox{\tiny{utilisée}}
    \end{array}
  \right.}_{\text{Jauge harmonique}}
  \bigwedge\underbrace{h^{1}_{1}+h^{2}_{2}+h^{3}_{3}=0}_{\text{restriction sur la trace}}\bigwedge
\left\{
  \begin{array}{ll}
    \begin{tabular}{l}
    $\mathcal{F}^{\mu\nu}=\partial^{\mu}\mathcal{A}^{\nu}-\partial^{\nu}\mathcal{A}^{\mu}$ \\\\
    $\mathcal{A}^{0}=\frac{mc}{2q}\,h^{00}$ \\\\
    $\mathcal{A}^{i}=\frac{mc}{q}\,h^{0i}$ \\
  \end{tabular}
  \end{array}
\right.\right]\nonumber
  \end{eqnarray}
  \begin{eqnarray}
  \Downarrow\nonumber
  \end{eqnarray}
  \begin{eqnarray}
  \hspace*{-0.5cm}\left[\underbrace{\partial_{\mu}\mathcal{F}^{\mu\nu}=0\hspace{0.2cm}\bigwedge\hspace{0.2cm} \partial^{\sigma}\mathcal{F}^{\mu\nu}+\partial^{\nu}\mathcal{F}^{\sigma\mu}+\partial^{\mu}\mathcal{F}^{\nu\sigma} = 0}_{\text{Equations de Maxwell}}
  \hspace{0.2cm}\bigwedge\hspace{0.2cm}\underbrace{\frac{mc}{q}\Box\left(
    h_{ij}-\frac{1}{2}\,\eta_{ij}\,h\right)=\Box\widehat{h}^{\,ij}=0}_{\text{propagation des degrés de liberté $h^{ij}$ dans la jauge harmonique}}\right]\nonumber
  \end{eqnarray}

  \subsection{La jauge de Lorentz}
Une conséquence d'utilisation de la condition de trace spatiale nulle est que la trace de la perturbation est réduite à sa seule composante temporelle $h=h^{00}$. Dans ce cas, la composante "temporelle" de la jauge harmonique, non utilisée jusqu'à présent, se met sous la forme
\begin{eqnarray}
0 &=&  \partial_{\alpha}h^{\alpha}_{0}-\frac{1}{2}\,\partial_{0}h \nonumber\\
  &=& \partial_{0}h^{00}+\partial_{i}h^{i0}-\frac{1}{2}\,\partial_{0}h^{00} = \frac{1}{2}\,\partial_{0}h^{00}+\partial_{i}h^{i0}  \nonumber\\
   &=& \frac{q}{mc}\,\partial_{0}\left(\frac{mc}{2q}\,h^{00}\right)+\frac{q}{mc}\,\partial_{i}\left(\frac{mc}{q}\,h^{i0}\right)  = \frac{1}{c}\left(\partial_{0}\mathcal{A}^{0}+\partial_{i}\mathcal{A}^{i}\right)  \nonumber\\
  &=& \frac{q}{mc}\,\partial_{\mu}\mathcal{A}^{\mu},  \label{equivalence harmonique1 et lorentz electromagnetisme}
\end{eqnarray}
ce qui représente la \textbf{jauge de Lorentz}.

La condition de nullité de la trace spatiale (\ref{trace spatiale electrom}) peut ainsi être vue comme une condition nécessaire pour que la composante temporelle de la jauge harmonique se réduise à la jauge de Lorentz $\partial_{\mu}\mathcal{A}^{\mu}=0$; autrement dit, c'est une condition nécessaire pour que la jauge de Lorentz soit contenue dans la jauge harmonique.

Après avoir retrouvé les équations de Maxwell, il est ainsi possible d'adopter la composante temporelle de la jauge harmonique (jauge de Lorentz), dans le but de découpler les équations de propagation des potentiels scalaire et vectoriel
\begin{eqnarray}\label{propagation potentiels quadrim electromagnetisme}
    \Box \mathcal{A}^{\mu}=\mu_{0}\,J^{\mu}
    \hspace{0.5cm}\Longleftrightarrow\hspace{0.5cm}\left\{
                                   \begin{array}{ll}
                                     \Box \phi=\displaystyle\frac{\rho}{\varepsilon_{0}},\\
                                     \Box \overrightarrow{\mathcal{A}}=\mu_{0}\,\overrightarrow{j},
                                   \end{array}
                                 \right.
\end{eqnarray}
où $J^{\mu}=(c\,\rho,\overrightarrow{j})$ , $\mathcal{A}^{\mu}=(\phi/c,\overrightarrow{\mathcal{A}})$ et $\overrightarrow{j}=\rho\overrightarrow{v}$.

\subsection{Réflexions sur la dérivation des Equations de Maxwell à partir d'une version électromagnétique des Equations d'Einstein dans le cas intérieur à la source}
 Pour décrire l'électromagnétisme, nous avons postulé l'existence d'équations de type Einstein sous la forme
 \begin{equation}
    G^{\mu\nu}=\chi_{e}\,T^{\mu\nu}
\end{equation}
où $G^{\mu\nu}$ représente le tenseur d'Einstein décrivant les propriétés géométrique de l'espace-temps et $T^{\mu\nu}$ est un certain tenseur décrivant la présence des charges électriques et courants.

Nous supposons que la constante de proportionnalité $\chi_{e}$ de l'électromagnétisme est différente de la constante  $\chi=4\pi G/c^{4}$ de la gravitation car les deux interactions affectent de façon différente la structure de l'espace-temps.

\subsubsection{Les $\mathbf{h_{0\mu}}$}

On exploite une analogie entre de gravitation et l'électromagnétisme, en remplaçant la densité de masse $\rho_{m}$ par la densité de charge $\rho$. En considérant que la source est un fluide parfait nous établissons l'analogie suivante
\begin{center}
\begin{tabular}{|l|l|}
  \hline
  Gravitation & Eléctromagnétisme \\
\hline\hline
  $T^{00}\simeq \left(\frac{m_{g}}{m_{i}}\right)\rho_{m}c^{2}$ & $T^{00}\simeq \left(\frac{q}{m}\right)\rho\,c^{2}$ \\
\hline
  $T^{0j}\simeq \left(\frac{m_{g}}{m_{i}}\right)\rho_{m}c\,v^{j}$ & $T^{0j}\simeq \left(\frac{q}{m}\right)\rho\,c\,v^{j}$ \\
\hline
  $T^{kl}\simeq \left(\frac{m_{g}}{m_{i}}\right)\rho_{m}\,v^{k}v^{l}$ & $T^{kl}\simeq \left(\frac{q}{m}\right)\rho\,v^{k}v^{l}$ \\
\hline
\end{tabular}
\end{center}
avec la différence fondamentale que le rapport de la masse grave par la masse inerte est égale à l'unité $m_{g}/m_{i}=1$, alors que le rapport de la charge électrique par la masse inertielle d'une particule d'épreuve $q/m$ est une caractéristique intrinsèque de chaque particule.

En définissant, à l'approximation linéaire,
\begin{eqnarray}\label{t0mu jmu0}
    T^{0\nu}=\frac{q\,c}{m}\,J^{\nu}\hspace{0.2cm}\Leftrightarrow\hspace{0.2cm}\left\{
                                                                        \begin{array}{ll}
                                                                          T^{00}=\displaystyle\frac{q\,c}{m}(c\,\rho) \\\\
                                                                          T^{0i}=\displaystyle\frac{q\,c}{m}(\rho\,v^{i})
                                                                        \end{array}
                                                                      \right.
\end{eqnarray}
nous allons déterminer la constante de proportionnalité $\chi_{e}$ figurant dans les Equations type Einstein de l'électromagnétisme
\begin{eqnarray}
 G^{0\nu} &=& \chi_{e}\,T^{0\nu}\label{G0mu t mu0}.
\end{eqnarray}

Pour ce faire, nous avons d'après (\ref{resume electrom maxwell})

\begin{eqnarray}
 G^{0\nu} &=& \chi_{e}\,T^{0\nu}\nonumber\\
 -\frac{q}{2mc}\,\partial_{\mu}\mathcal{F}^{\mu\nu} &\approx& \chi_{e}\left(\frac{q\,c}{m}\,J^{\nu}\right)\nonumber\\
 \Rightarrow\hspace{0.5cm} \partial_{\mu}\mathcal{F}^{\mu\nu} &\approx&\underbrace{\left(-\frac{2\,m\,c}{q}\,\chi_{e}\,\frac{q\,c}{m}\right)}_{\mu_{0}}\,J^{\nu}.
\end{eqnarray}
Pour retrouver le deuxième groupe d'équations de Maxwell avec source, $\partial_{\mu}\mathcal{F}^{\mu\nu}=\mu_{0}J^{\nu}$, il suffit d'imposer que
\begin{eqnarray}\label{lambda}
  \chi_{e} = -\frac{\mu_{0}}{2c^{2}}= -\frac{(\mu_{0}\varepsilon_{0})}{2c^{2}\,\varepsilon_{0}}= -\frac{1}{2c^{4}\,\varepsilon_{0}}=-\left(\frac{4\pi}{2c^{4}}\right)\frac{1}{4\pi\varepsilon_{0}}\nonumber
\end{eqnarray}
ou encore
\begin{equation}\label{lambda}
    \chi_{e} =-\frac{2\pi K}{c^{4}}
\end{equation}
avec $K=(4\pi\varepsilon_{0})^{-1}$ est la constante figurant dans la loi de Coulomb.

Nous avons finalement l'équations type Einstein pour l'électromagnétisme\footnote{En utilisant les définitions des tenseurs de courbure et de Ricci, donnés dans \cite{Weinberg}, les équations d'Einstein seront définies avec un facteur (-1) global par rapport aux équations d'Einstein figurant dans cette thèse, où les définitions de Landau \cite{landau} ont été adoptées.}
\begin{equation}\label{version electrom einstein}
    G^{\mu\nu}=-\left(\frac{2\pi K}{c^{4}}\right)T^{\mu\nu}.
\end{equation}
Attirons l'attention sur le fait que, conformément à (\ref{equation einstein lineaire harmonique bis}) et (\ref{amu electro et jauge}), nous avons
 \begin{eqnarray}
   G^{0\nu} &\approx& -\frac{1}{2}\,\Box \left(h^{0\nu}-\frac{1}{2}\,\eta^{0\nu}\,h\right)=
   -\frac{q}{2mc}\,\Box \mathcal{A}^{\nu}.  \label{G0mu box amu}
 \end{eqnarray}
 Compte tenu de (\ref{t0mu jmu0}) et (\ref{lambda}), l'équation $G^{0\nu}=\chi_{e}T^{0\nu}$ conduit aux équations des potentiels découplés $\Box\mathcal{A}^{\nu}=\mu_{0}J^{\nu}$.

En comparant les coefficients figurant dans les équations d'Einstein de la gravité et de l'électro-magnétisme, nous avons
\begin{equation}
   \frac{\chi_{e}}{\chi}=-\displaystyle\frac{\left(\displaystyle\frac{2\pi K}{c^{4}}\right)}{\left(\displaystyle\frac{8\pi G}{c^{4}}\right)}=-\frac{K}{4G}
\end{equation}
de telle sorte que l'interaction électromagnétique est négligeable devant l'interaction gravitationnelle à grande échelle, et vice versa à l'échelle microscopique.

Le facteur 4 figurant dans ce rapport rappelle le facteur 4 indésirable qu'on a pu éliminer de la partie magnétique de la force gravitationnelle type Lorentz. La réapparition de ce facteur est tout à fait intrigante; au delà de l'aspect formel de l'irréductibilité de ce facteur 4, nous pensons que cette situation est peut-être révélatrice d'un fait fondamental à élucider si on veut comprendre la différence fondamentale entre la gravité et l'électromagnétisme.

\subsubsection{Les $\mathbf{h_{ij}}$}

Intéressons-nous à présent aux $h_{ij}$. D'après (\ref{Gij lineaire harmonique electrom}) nous avons
\begin{eqnarray}
  G^{ij} \approx -\frac{1}{2}\,\Box\overline{h}^{\,ij}&=&-\displaystyle\frac{2\pi K}{c^{4}}\;T^{ij} \\
   &=& -\displaystyle\frac{2\pi K}{c^{4}}\left(\frac{q}{m}\rho\,v^{i}v^{j}\right)
\end{eqnarray}
de telle sorte que $$\overline{h}_{ij}=\left(h_{ij}-\frac{1}{2}\,\eta_{ij}\,h\right),$$ ce qui conduit à l'équation de propagation suivante
\begin{eqnarray}\label{eq propag hih bar}
    \Box\overline{h}^{\,ij}&=&\frac{2\pi K}{c^{4}}\;\frac{2q}{m}\;\rho(\overrightarrow{x},t)\,v^{i}v^{j}\nonumber\\
                           &=&-4\pi\left(-\frac{Kq}{m\,c^{4}}\;\rho(\overrightarrow{x},t)\,v^{i}v^{j}\right)
\end{eqnarray}

D'une manière générale, une équation de propagation avec second membre
\begin{equation}\label{eq propag second membre}
    \Box \psi(\overrightarrow{x},t)=-4\pi\;f(\overrightarrow{x},t)
\end{equation}
admet une solution\footnote{Seule la solution retardée et retenue et la solution avancée est ignorée.}
$$\psi(\overrightarrow{x},t)=\int d^{3}x^{'}\;\frac{\left[f(\overrightarrow{x}^{'},t^{'})\right]_{ret}}{\mid\overrightarrow{x}-\overrightarrow{x}^{'}\mid}$$
où $\left[f(\overrightarrow{x^{'}},t^{'})\right]_{ret}=f\left(\overrightarrow{x}^{'},t^{'}-\frac{\mid\overrightarrow{x}-\overrightarrow{x}^{'}\mid}{c}\right)$.

Dans ce cas, l'équation de propagation (\ref{eq propag hih bar}) admet la solution
\begin{eqnarray}
 \overline{h}_{\,ij} &=& -\frac{Kq}{m\,c^{4}}\;\int d^{3}x^{'}\frac{\left[\rho(\overrightarrow{x}^{'},t^{'})\,v^{i}v^{j})\right]_{ret}}{\mid\overrightarrow{x}-\overrightarrow{x}^{'}\mid}
\end{eqnarray}
avec la restriction sur la trace $h=h_{00}$.

Il faut savoir que nous avons négligé les termes quadratiques $v^{i}v^{j}\ll c^{2}$ pour pouvoir retrouver une force de Lorentz. Dans le cadre de cette approximation l'équation de propagation (\ref{eq propag hih bar}) devient plutôt
$$\Box\overline{h}^{\,ij}\approx 0.$$

Les solutions d'une telle équation de propagation sont des ondes planes qui se propagent à l'infini à la vitesse $c$.

En supposant que les $h_{ij}$ soient très négligeables devant les degrés de liberté de l'électro-magnétisme de telle sorte que $h_{00}\gg h_{0i}\gg h_{ij}$. Les composantes $h_{ij}$ représentent des degrés de liberté "supplémentaires", par rapport au quadri-potentiel électromagnétique $\mathcal{A}^{\mu}$, et sont très négligeables dans le cadre de cette application mais qui jouerait probablement un rôle à une échelle plus petite.

Au premier ordre de la perturbation, les degrés de liberté $h^{ij}$ sont complètement découplés des degrés de liberté effectif de l'électromagnétisme standard $h^{00}$ et $h^{0i}$. Nous allons voir qu'au second ordre de la perturbation, ces $h^{ij}$ vont être couplés aux degrés de liberté effectifs et ne peuvent plus être négligés.

Pour terminer, nous dirons que les $h^{ij}$ sont plutôt des degrés de liberté qui commencent à se révéler à partir du second ordre de la perturbation; c'est pour cette raison qu'en électromagnétisme standard, développée au premier ordre de la perturbation, ils ne se sont pas encore révélés car complètement découplés de $\mathcal{A}^{\mu}$.

\subsection{Justification de l'approximation lineaire}
Les équations de Maxwell ont été déterminées à partir d'une nouvelle version d'équations d'Einstein au premier ordre de la perturbation. Nous allons présenter quelques arguments pour montrer que les termes d'ordres supérieurs sont complètement négligeables dans les domaines actuels d'application de l'électromagnétisme. Pour ce faire, reprenons les définitions (\ref{composantes champ grav electric electrom}) et (\ref{composantes champ grav magnetic electrom}) des champs électrique et magnétique qui prennent la forme
\begin{equation}\label{electric field}
    \mathbf{E}^{i}=-c\,\mathcal{F}^{0i}=-\frac{mc^{2}}{q}\left(\partial^{0}h^{0i}-\frac{1}{2}\,\partial^{i}h^{00}\right)
\end{equation}
et
\begin{equation}\label{magnetic field}
    \mathbf{B}^{i}=-\frac{1}{2}\,\epsilon^{ijk}\mathcal{F}_{jk}=-\frac{mc^{2}}{2q}\,\epsilon^{ijk}\left(\partial_{j}h_{0k}-\partial_{k}h_{0j}\right),
\end{equation}
compte tenu de (\ref{tenseur antisymetrique grav electrom}) et (\ref{amu electro et jauge}).

Pour éviter les effets quantiques, nous n'allons pas choisir un système subatomique comme particule test. Au lieu de cela, considérons dans le domaine des hautes tensions une particule de poussière de diamètre $10^{-6}$ m et de masse de l'ordre de  $m\approx 10^{-14}$ Kg. Cette particule peut être, ou bien à l'intérieur d'un précipitateur électrostatique où peut régner un champ électrique $\mathbf{E}$ intense, ou bien à l'intérieur de l'entrefer d'une machine électrique tournante où peut régner un champ magnétique $\mathbf{B}$ intense. Pour de telles particules, il est bien connu \cite{parker} que la charge électrique de saturation est de l'ordre de $q\approx 10^{-18}$ C. Même pour des valeurs des champs $\mathbf{E}\approx 10^{8}$ V.m$^{-1}$ et $\mathbf{B}\approx 10$ T, rarement atteintes, les quantités $\partial_{\nu}h^{0\mu}$ figurant dans (\ref{electric field}) et (\ref{magnetic field}) ne prennent que des valeurs très petites, de l'ordre de  $q\mathbf{E}^{i}/mc^{2}\approx 10^{-13}$ m$^{-1}$ à $q\mathbf{B}^{i}/mc\approx 10^{-11}$ m$^{-1}$.

D'autre part, il est bien connu que la charge maximale $q$ que porte une particule de diamètre $\Phi$ est proportionnelle à $\Phi^{2}$ \cite{parker, mizuno}, alors que sa masse est proportionnelle à $\Phi^{3}$. Il s'en suit que le rapport $q/m$ est une fonction décroissante puisqu'il est proportionnel à $\Phi^{-1}$. Autrement dit, si nous choisissons une particule test avec une plus grande charge électrique, sa masse sera assez grande pour obtenir des valeurs encore plus petites de $\partial_{\nu}h^{0\mu}$. Pour avoir des valeurs plus importantes de $\partial_{\nu}h^{0\mu}$, il est nécessaire d'utiliser une particule test avec une plus petite masse. Pour avoir des valeurs significatives de $\partial_{\nu}h^{0\mu}$, la masse doit être si petite que les effets quantiques ne peuvent désormais plus être négligées. Par exemple, dans le cas où la particule test est un proton, avec les mêmes valeurs précédentes pour $\mathbf{E}$ et $\mathbf{B}$, nous obtenons des valeurs non négligeables, de l'ordre de $qE^{i}/mc^{2}\approx 0.1$ m$^{-1}$ à $qB^{i}/mc\approx 1$ m$^{-1}$.

Pour des systèmes non quantiques, le fait que les dérivées $\partial_{\nu}h^{0\mu}$ tendent vers zéro implique que les $h^{0\mu}$ prennent des valeurs quasiment constantes. En exigeant la convergence asymptotique de la métrique de l'espace-temps vers la métrique plate minkowskienne, très loin de la source, ces constantes prennent ainsi des valeurs très proches de zéro. Donc, en chaque point de l'espace-temps, les relations (\ref{electric field}) and (\ref{magnetic field}) montrent clairement que même pour des champs électromagnétiques intenses, nous avons toujours $h^{0\mu}\ll 1$. Ceci justifie l'approximation linéaire adoptée dans notre approche et indique que les termes d'ordres supérieurs sont tout à fait négligeables.

Il est claire que pour des champs électriques et magnétiques faibles, l'approximation linéaire est pleinement justifiée. En se plaçant de le cas de figure le plus défavorable, i.e. pour les champs les plus intenses possibles, compte tenu des avancées technologiques actuelles, nous avons montré que l'approximation linéaire reste valable. Si la technologie se développe pour accéder à des champs encore plus intenses, dans ce cas, les termes non linéaires joueraient un rôle important et ne peuvent désormais plus être négligés; les équations de Maxwell ne décriraient plus correctement les champs et il faudrait plutôt faire appel à la version électromagnétique des équations d'Einstein (\ref{version electrom einstein}).

\section{Force de Lorentz}
Dans ce qui suit, nous allons montrer qu'une particule d'épreuve, se déplaçant suivant des géodésiques, est soumise à une force de Lorentz dans le domaine des champs et des vitesses faibles.

L'équations des géodésiques
\begin{eqnarray}
  \displaystyle\frac{d^{2}x^{\mu}}{d\tau^{2}}
  &=& -\Gamma^{\mu}_{00}\,\displaystyle\frac{dx^{0}}{d\tau}\,\displaystyle\frac{dx^{0}}{d\tau}
  -2\,\Gamma^{\mu}_{0i}\,\displaystyle\frac{dx^{0}}{d\tau}\,\displaystyle\frac{dx^{i}}{d\tau}
  -\Gamma^{\mu}_{il}\,\displaystyle\frac{dx^{i}}{d\tau}\,\displaystyle\frac{dx^{l}}{d\tau} \nonumber
\end{eqnarray}
permet de déterminer les composantes spatiales
\begin{equation}\label{bel bouda acceleration relativiste electrom}
    \displaystyle\frac{d^{2}x^{j}}{d\tau^{2}}=-c^{2}\,\Gamma^{j}_{00}\,\left(\displaystyle\frac{dt}{d\tau}\right)^{2}
    -2\,\Gamma^{j}_{0i}\,\left(\displaystyle\frac{dt}{d\tau}\right)\,c\,u^{i}-\Gamma^{\mu}_{il}\,u^{i}\,u^{l},
\end{equation}
en fonction du quadrvecteur vitesse
\begin{equation}
    u^{\nu}=\frac{dx^{\nu}}{d\tau}=\frac{dx^{\mu}}{dt}\,\frac{dt}{d\tau}=(\gamma_{v}\,c,\gamma_{v}\,\overrightarrow{v})
\end{equation}
tel que $u^{0}=\gamma_{v}\,c$, $u^{i}=\gamma_{v}\,v^{i}$ et $\gamma_{v}=(1-\overrightarrow{v}^{2}/c^{2})^{-1/2}$.

Dans le cas linéaire, les symboles de Christoffel
\begin{equation}\label{christoffel bel bouda electrom}
     \Gamma^{\rho}_{\mu\nu} \approx \frac{1}{2}\,\eta^{\rho\lambda} \left(\partial_{\mu}h_{\nu\lambda}+\partial_{\nu}h_{\lambda\mu}-\partial_{\lambda}h_{\mu\nu}\right).
\end{equation}
figurant dans (\ref{bel bouda acceleration relativiste electrom}) sont donnés par
\begin{eqnarray}
  \Gamma^{j}_{00} &\approx& \frac{1}{2}\,\eta^{j\lambda} \left(\partial_{0}h_{0\lambda}+\partial_{0}h_{0\lambda}-\partial_{\lambda}h_{00}\right) \approx
   \partial_{0}h_{0}^{j}-\frac{1}{2}\,\partial^{j}h_{00}\label{bel bouda gama j00 electrom} \\
  \Gamma^{j}_{0i} &\approx& \frac{1}{2}\,\eta^{j\lambda} \left(\partial_{0}h_{i\lambda}+\partial_{i}h_{0\lambda}-\partial_{\lambda}h_{0i}\right) \approx
  \frac{1}{2}\,\left(\partial_{0}h_{i}^{j}+\partial_{i}h_{0}^{j}-\partial^{j}h_{0i}\right)\label{bel bouda gama j0i electrom}\\
  \Gamma^{j}_{il} &\approx& \frac{1}{2}\,\eta^{j\lambda} \left(\partial_{i}h_{l\lambda}+\partial_{l}h_{i\lambda}-\partial_{\lambda}h_{il}\right) \approx
  \frac{1}{2}\,\left(\partial_{i}h_{l}^{j}+\partial_{l}h_{i}^{j}-\partial^{j}h_{il}\right),\label{bel bouda gama jil electrom}
\end{eqnarray}
de sorte à avoir
\begin{eqnarray}
\displaystyle\frac{d^{2}x^{j}}{d\tau^{2}}&=&-c^{2}\,\left(\displaystyle\frac{dt}{d\tau}\right)^{2}(\partial_{0}h_{0}^{j}-\frac{1}{2}\,\partial^{j}h_{00})
    -c\,u^{i}\,\left(\displaystyle\frac{dt}{d\tau}\right)\left(\partial_{0}h_{i}^{j}+\partial_{i}h_{0}^{j}-\partial^{j}h_{0i}\right)\nonumber\\
    &&-\frac{1}{2}\,u^{i}\,u^{l}\,\left(\partial_{i}h_{l}^{j}+\partial_{l}h_{i}^{j}-\partial^{j}h_{il}\right) \nonumber\\ \nonumber\\
&=&-c^{2}\,\left(\displaystyle\frac{dt}{d\tau}\right)^{2}(\partial_{0}h_{0}^{j})+\frac{c^{2}}{2}
\,\left(\displaystyle\frac{dt}{d\tau}\right)^{2}(\partial^{j}h_{00})
    -c\,u^{i}\,\left(\displaystyle\frac{dt}{d\tau}\right)\left(\partial_{i}h_{0}^{j}-\partial^{j}h_{0i}\right)\nonumber\\
    &&-c\,u^{i}\,(\partial_{0}h_{i}^{j})\left(\displaystyle\frac{dt}{d\tau}\right)
    -\frac{1}{2}\,u^{i}\,u^{l}\,\left(\partial_{i}h_{l}^{j}+\partial_{l}h_{i}^{j}-\partial^{j}h_{il}\right)\label{acceleration complete electrom}
\end{eqnarray}
ou encore finalement
\begin{eqnarray}
\displaystyle\frac{d^{2}x^{j}}{d\tau^{2}}&=&c^{2}\Bigg[\frac{1}{2}\left(\displaystyle\frac{dt}{d\tau}\right)^{2}\partial^{j}h^{00}-\frac{1}{c}\,\left(\displaystyle\frac{dt}{d\tau}\right)^{2}\frac{\partial h^{0j}}{\partial t}\nonumber\\
&&-\frac{u_{i}}{c}\left(\displaystyle\frac{dt}{d\tau}\right)\left(\partial^{i}h^{0j}-\partial^{j}h^{0i}\right)
-\frac{u_{i}}{c^{2}}\,\displaystyle\frac{dt}{d\tau}\,\frac{\partial h^{ij}}{\partial t}
    -\frac{u_{i}u_{l}}{c^{2}}\left(\partial^{i}h^{jl}-\frac{1}{2}\,\partial^{j}h^{il}\right)\Bigg].\label{acceleration complete bis electrom}
\end{eqnarray}

Pour des vitesses faibles $v\ll c$, où
$\frac{dt}{d\tau}\rightarrow 1$ et $u^{i}\rightarrow v^{i}=dx^{i}/dt$, il est possible de négliger les termes proportionnels à $1/c^{2}$ dans (\ref{acceleration complete bis electrom}), pour avoir\footnote{on néglige les termes quadratiques en vitesses $v^{2}/c^{2}$}
\begin{eqnarray}\label{bel bouda comp accel fontion h electrom}
  \displaystyle\frac{d^{2}x^{j}}{dt^{2}} &\approx& c^{2}\bigg[\frac{1}{2}\left(\partial^{j}h^{00}\right)-\frac{1}{c}\,\frac{\partial h^{0j}}{\partial t}-\frac{v_{i}}{c}\left(\partial^{i}h^{0j}-\partial^{j}h^{0i}\right)\bigg].
\end{eqnarray}
En tenant compte des définitions des potentiels scalaire (\ref{potentiel scalaire bel bouda electrom}) et vecteur (\ref{potentiel vecteur bel bouda electrom}), les composantes d'accélérations (\ref{bel bouda comp accel fontion h electrom}) prennent la forme
\begin{eqnarray}\label{bel bouda comp accel type lorentz electrom}
  \displaystyle\frac{d^{2}x^{j}}{dt^{2}} &\approx& \frac{q}{m}\Bigg[ \left(c\,\partial^{j}\mathcal{A}^{0}-\frac{\partial \mathcal{A}^{j}}{\partial t}\right)-v_{i}\left(\partial^{i}\mathcal{A}^{j}-\partial^{j}\mathcal{A}^{i}\right)\Bigg].
\end{eqnarray}
Pour $j=1$
\begin{eqnarray*}
  \displaystyle\frac{d^{2}x^{1}}{dt^{2}} &\approx& \frac{q}{m}\Bigg[\left(c\,\partial^{1}\mathcal{A}^{0}-\frac{\partial \mathcal{A}^{1}}{\partial t}\right)-v_{i}\left(\partial^{i}\mathcal{A}^{1}-\partial^{1}\mathcal{A}^{i}\right)\Bigg] \\
   &\approx& \frac{q}{m}\Bigg[\left(-\partial_{1}\phi-\frac{\partial \mathcal{A}^{1}}{\partial t}\right)-v_{2}\left(\partial^{2}\mathcal{A}^{1}-\partial^{1}\mathcal{A}^{2}\right)
-v_{3}\left(\partial^{3}\mathcal{A}^{1}-\partial^{1}\mathcal{A}^{3}\right)\Bigg] \\
&\approx& \frac{q}{m}\Bigg[\left(-\partial_{1}\phi-\frac{\partial \mathcal{A}^{1}}{\partial t}\right)+v^{2}\left(\partial_{1}\mathcal{A}^{2}-\partial_{2}\mathcal{A}^{1}\right)
-v^{3}\left(\partial_{3}\mathcal{A}^{1}-\partial_{1}\mathcal{A}^{3}\right)\Bigg],
\end{eqnarray*}
et compte tenu des définitions des champs électrique (\ref{relation champ type electrique potentiels bel bouda electrom}) et magnétique (\ref{champ gravit mag deriv pot vect bel bouda electrom}), nous aboutissons finalement à
\begin{equation}
    \displaystyle\frac{d^{2}x^{1}}{dt^{2}} \approx \frac{q}{m}\bigg[\mathbf{E}^{1}+\left(\overrightarrow{v}\times \overrightarrow{\mathbf{B}}\right)^{1}\bigg].\label{force grav type lorentz comp 1 electrom}
\end{equation}
De même pour $j=2$ et $j=3$, nous aboutissons à
\begin{equation}
    \displaystyle\frac{d^{2}x^{2}}{dt^{2}} \approx \frac{q}{m}\bigg[\mathbf{E}^{2}+\left(\overrightarrow{v}\times \overrightarrow{\mathbf{B}}\right)^{2}\bigg],\label{force grav type lorentz comp 2 electrom}
\end{equation}
et
\begin{equation}
    \displaystyle\frac{d^{2}x^{3}}{dt^{2}} \approx \frac{q}{m}\bigg[\mathbf{E}^{3}+\left(\overrightarrow{v}\times \overrightarrow{\mathbf{B}}\right)^{3}\bigg].\label{force grav type lorentz comp 3 electrom}
\end{equation}
Les équations (\ref{force grav type lorentz comp 1 electrom}), (\ref{force grav type lorentz comp 2 electrom}) et (\ref{force grav type lorentz comp 3 electrom}), résumées sous forme vectorielle
\begin{equation}
    \displaystyle\frac{d^{2}\overrightarrow{r}}{dt^{2}} \approx \frac{q}{m}\bigg[\overrightarrow{\mathbf{E}}+\left(\overrightarrow{v}\times \overrightarrow{\mathbf{B}}\right)\bigg],\label{force grav type lorentz electrom}
\end{equation}
montent clairement qu'une particule test, dans le cas des champs et vitesses faibles\footnote{à l'ordre $v/c$}, est soumise à la force de Lorentz
\begin{equation}
    m\,\displaystyle\frac{d^{2}\overrightarrow{r}}{dt^{2}} \approx q\bigg[\overrightarrow{\mathbf{E}}+\left(\overrightarrow{v}\times \overrightarrow{\mathbf{B}}\right)\bigg].\label{force grav type lorentz electrom}
\end{equation}


\section{Ordre 2 de la perturbation: Corrections des Equations de Maxwell}

\subsection{Introduction}
Après avoir vu que dans le contexte de l'interaction électromagnétique, il était possible de montrer que les équations d'Einstein linéarisées, à l'ordre 1 de la perturbation, se réduisent aux équations de Maxwell, une étude perturbative à l'ordre 2 sera entreprise, et ce dans le but d'apporter des corrections aux équations de Maxwell et à la force de Lorentz. On espère ainsi rendre compte des phénomènes non linéaires d'auto-interaction de la charge électrique avec son propre champ électromagnétique.

Dans ce contexte, on pose que la métrique de l'espace-temps est donnée par
\begin{eqnarray}\label{plate plus perturbation ordre 2}
    g_{\mu\nu}=\eta_{\mu\nu}+h_{\mu\nu},\hspace{1cm}\mid h_{\mu\nu}\mid \ll 1.
\end{eqnarray}
tel que $\eta_{\mu\nu}=(1,-1,-1,-1)$, alors qu'à l'ordre 2 de la perturbation nous posons
\begin{eqnarray}\label{metrique inversebouda}
  g^{\mu\nu} &\approx& \eta^{\mu\nu}-h^{\mu\nu}+f^{\mu\nu},
\end{eqnarray}
où $f^{\mu\nu}$ est par hypothèse un terme de seconde ordre.

Dans un premier temps, déterminons l'expression de $f^{\mu\nu}$. Pour ce faire imposons aux composantes covariantes et contravariantes du tenseur métrique de satisfaire la relation d'orhogonalité
\begin{eqnarray}
  g_{\mu\nu}\,g^{\nu\rho} &=& \delta_{\mu}^{\rho}\nonumber\\
  &\approx&\left(\eta_{\mu\nu}+h_{\mu\nu}\right)\left(\eta^{\nu\rho}-h^{\nu\rho}+f^{\nu\rho}\right) \nonumber\\
  &=& \delta_{\mu}^{\rho}-\eta_{\mu\nu}\,h^{\nu\rho}+\eta_{\mu\nu}\,f^{\nu\rho}+\eta^{\nu\rho}\,h_{\mu\nu}-h_{\mu\nu}\,h^{\nu\rho}+\mathcal{O}(h^{3}).\nonumber
\end{eqnarray}
Une identification ordre par ordre permet de montrer
\begin{enumerate}
  \item \textbf{A l'ordre 0}
  \begin{equation}
  \eta_{\mu\nu}\,\eta^{\nu\rho}=\delta_{\mu}^{\rho}.
  \end{equation}

  \item \textbf{A l'ordre 1}
\begin{eqnarray}
  \eta^{\nu\rho}\,h_{\mu\nu}-\eta_{\mu\nu}\,h^{\nu\rho}&=&0\nonumber\\
  \eta_{\mu\nu}\,h^{\nu\rho}&=&\eta^{\nu\rho}\,h_{\mu\nu}\nonumber\\
  \eta^{\lambda\mu}\,\eta_{\mu\nu}\,h^{\nu\rho}&=&\eta^{\lambda\mu}\,\eta^{\nu\rho}\,h_{\mu\nu}\nonumber\\
  \delta^{\lambda}_{\nu}\,h^{\nu\rho}&=&\eta^{\lambda\mu}\,\eta^{\nu\rho}\,h_{\mu\nu}\nonumber\\
  h^{\lambda\rho}&=&\eta^{\lambda\mu}\,\eta^{\nu\rho}\,h_{\mu\nu}
\end{eqnarray}
  \item \textbf{A l'ordre 2}
\begin{eqnarray}
\eta_{\mu\nu}\,f^{\nu\rho}-h_{\mu\nu}\,h^{\nu\rho}&=&0\nonumber\\
\eta_{\mu\nu}\,f^{\nu\rho}&=&h_{\mu\nu}\,h^{\nu\rho}\nonumber\\
\eta^{\lambda\mu}\,\eta_{\mu\nu}\,f^{\nu\rho}&=&\eta^{\lambda\mu}\,h_{\mu\nu}\,h^{\nu\rho}\nonumber\\
\delta^{\lambda}_{\nu}\,f^{\nu\rho}&=&\eta^{\lambda\mu}\,h_{\mu\nu}\,h^{\nu\rho}\nonumber\\
f^{\lambda\rho}&=&\eta^{\lambda\mu}\,h_{\mu\nu}\,h^{\nu\rho}=h^{\lambda}_{\;\;\nu}\,h^{\nu\rho}
\end{eqnarray}
\end{enumerate}

Dans un deuxième temps, assurons-nous qu'il est possible d'élever ou d'abaisser les indices, à l'ordre 2, en utilisant simplement $\eta_{\mu\nu}$ et $\eta^{\mu\nu}$. Pour ce faire, en exigeant la satisfaction de la transformation entre composantes covariantes et contravariantes du tenseur métrique $g_{\mu\nu} = g_{\mu\sigma}\,g_{\rho\nu}\,g^{\sigma\rho}$ à l'ordre 2
\begin{eqnarray}\label{ordre 1 plus ordre 2 contrav}
            \hspace*{-1.5cm}g_{\mu\nu} &\approx& \left(\eta_{\mu\sigma}+h_{\mu\sigma}\right)\left(\eta_{\rho\nu}+h_{\rho\nu}\right)g^{\sigma\rho} \nonumber\\
  \hspace*{-1.5cm}g_{\mu\nu}&\approx& \left[\eta_{\mu\sigma}\,\eta_{\rho\nu}+\eta_{\mu\sigma}\,h_{\rho\nu}+\eta_{\rho\nu}\,h_{\mu\sigma}+h_{\mu\sigma}\,h_{\rho\nu}+\mathcal{O}(h^{3})\right]g^{\sigma\rho} \nonumber\\
  \hspace*{-1.5cm}g_{\mu\nu}&\approx& \left[\eta_{\mu\sigma}\,\eta_{\rho\nu}+\eta_{\mu\sigma}\,h_{\rho\nu}+\eta_{\rho\nu}\,h_{\mu\sigma}+h_{\mu\sigma}\,h_{\rho\nu}+\mathcal{O}(h^{3})\right]
  \big(\eta^{\sigma\rho}-h^{\sigma\rho}+\eta_{\alpha\beta}\,h^{\sigma\alpha}\,h^{\beta\rho}\big)\nonumber
\end{eqnarray}
nous aboutissons finalement à la relation
\begin{eqnarray}
  \hspace*{-1.5cm}\eta_{\mu\nu}+h_{\mu\nu}&\approx& \eta_{\mu\sigma}\,\eta_{\rho\nu}\,\eta^{\sigma\rho}+\Big(-\eta_{\mu\sigma}\,\eta_{\rho\nu}\,h^{\sigma\rho}
  +\eta_{\mu\sigma}\,\eta^{\sigma\rho}\,h_{\rho\nu}+\eta_{\rho\nu}\,\eta^{\sigma\rho}\,h_{\mu\sigma}\Big) \nonumber\\
  &&+\Big(\eta_{\mu\sigma}\,\eta_{\rho\nu}\,\eta_{\alpha\beta}\,h^{\sigma\alpha}\,h^{\beta\rho}
  -\eta_{\mu\sigma}\,h_{\rho\nu}\,h^{\sigma\rho}\nonumber\\
  &&-\eta_{\rho\nu}\,h_{\mu\sigma}\,h^{\sigma\rho}+\eta^{\sigma\rho}\,h_{\mu\sigma}\,h_{\rho\nu}+\mathcal{O}(h^{3})\Big).
\end{eqnarray}
Par identification ordre par ordre nous avons
\begin{enumerate}
  \item \textbf{A l'ordre 0}
\begin{equation}
    \eta_{\mu\nu}=\eta_{\mu\sigma}\,\eta_{\rho\nu}\,\eta^{\sigma\rho}.
\end{equation}
  \item \textbf{A l'ordre 1}
\begin{eqnarray*}
  h_{\mu\nu} &=& -\eta_{\mu\sigma}\,\eta_{\rho\nu}\,h^{\sigma\rho}
  +\underbrace{\eta_{\mu\sigma}\,\eta^{\sigma\rho}}_{\delta_{\mu}^{\rho}}\,h_{\rho\nu}
  +\underbrace{\eta_{\rho\nu}\,\eta^{\sigma\rho}}_{\delta_{\nu}^{\sigma}}\,h_{\mu\sigma}\nonumber \\
   &=& -\eta_{\mu\sigma}\,\eta_{\rho\nu}\,h^{\sigma\rho}+h_{\mu\nu}+h_{\mu\nu} \\
\end{eqnarray*}
il suffit que \begin{equation}\eta_{\mu\sigma}\,\eta_{\rho\nu}\,h^{\sigma\rho}=h_{\mu\nu}.\end{equation}
  \item \textbf{A l'ordre 2}
\begin{eqnarray*}
  0 &=& \eta_{\mu\sigma}\,\eta_{\rho\nu}\,\eta_{\alpha\beta}\,h^{\sigma\alpha}\,h^{\beta\rho}
  -\eta_{\mu\sigma}\,h_{\rho\nu}\,h^{\sigma\rho}-\eta_{\rho\nu}\,h_{\mu\sigma}\,h^{\sigma\rho}+\eta^{\sigma\rho}\,h_{\mu\sigma}\,h_{\rho\nu}.\nonumber
\end{eqnarray*}
Sachant que
 \begin{eqnarray*}
          \eta_{\mu\sigma}\,\eta_{\rho\nu}\,\eta_{\alpha\beta}\,h^{\sigma\alpha}\,h^{\beta\rho} &=& \eta_{\mu\sigma}\,\eta_{\rho\nu}\,h^{\sigma}_{\hspace{0.1cm}\beta}\,h^{\beta\rho} \nonumber\\
          &=& \eta_{\mu\sigma}\,h^{\sigma}_{\hspace{0.1cm}\beta}\,h^{\beta}_{\hspace{0.1cm}\nu} \nonumber\\
           &=&  h_{\mu\beta}\,h^{\beta}_{\hspace{0.1cm}\nu}
\end{eqnarray*}
nous obtenons ainsi la relation
\begin{eqnarray*}
  0 &=& h_{\mu\beta}\,h^{\beta}_{\nu}
  -h_{\rho\nu}\,h^{\rho}_{\mu}-h_{\mu\sigma}\,h^{\sigma}_{\nu}+h^{\rho}_{\mu}\,h^{}_{\rho\nu}=0\nonumber
\end{eqnarray*}
\end{enumerate}

Résumons la situation à l'ordre 2 de la perturbation. Les composantes covariantes et contravariantes sont respectivement données par\footnote{En posant $g_{\mu\nu}=\eta_{\mu\nu}+\epsilon\,\mathbf{h}_{\mu\nu}$ où $\epsilon\ll1$ dans ce cas il est clair que $g^{\mu\nu} \approx \eta^{\mu\nu}-\epsilon\,\mathbf{h}^{\mu\nu}+\epsilon^{2}\,\eta^{\mu\lambda}\,\mathbf{h}_{\lambda\rho}\,\mathbf{h}^{\rho\nu}$ est du second ordre en $\epsilon$.}
\begin{eqnarray}
    g_{\mu\nu}&=&\eta_{\mu\nu}+h_{\mu\nu},\hspace{1cm}\mid h_{\mu\nu}\mid \ll 1.\\
    g^{\mu\nu} &\approx& \eta^{\mu\nu}-h^{\mu\nu}+f^{\mu\nu}
\end{eqnarray}
tel que
\begin{eqnarray}
  h^{\lambda\rho}&=&\eta^{\lambda\mu}\,\eta^{\nu\rho}\,h_{\mu\nu} \\
  f^{\lambda\rho}&=&\eta^{\lambda\mu}\,h_{\mu\nu}\,h^{\nu\rho}=h^{\lambda}_{\;\;\nu}\,h^{\nu\rho}.
\end{eqnarray}

\subsection{Equations de type Einstein à l'ordre 2 de la perturbation}
Les effets des ordres supérieurs à l'ordre deux de la perturbation sont négligés, on considère un champ de tenseurs symétrique $h_{\mu\nu}$ qui se propage dans un espace-temps plat de Minkowski.

\subsubsection{Symboles de Christoffel}

Par définition $$\Gamma^{\rho}_{\mu\nu} = \frac{1}{2}\,g^{\rho\lambda} \left(\partial_{\mu}g_{\nu\lambda}+\partial_{\nu}g_{\lambda\mu}-\partial_{\lambda}g_{\mu\nu}\right).$$
A l'ordre 2, compte tenu de (\ref{plate plus perturbation ordre 2}) et et (\ref{metrique inversebouda}), l'expression précédente devient
\begin{eqnarray}
  \Gamma^{\rho\;(2)}_{\mu\nu} &=& \frac{1}{2}\,\left(\eta^{\rho\lambda}-h^{\rho\lambda}+h^{\rho}_{\;\;\sigma}\,h^{\sigma\lambda}\right) \Big[\partial_{\mu}\left(\eta_{\nu\lambda}+h_{\nu\lambda}\right)+\partial_{\nu}\left(\eta_{\lambda\mu}+h_{\lambda\mu}\right)
  -\partial_{\lambda}\left(\eta_{\mu\nu}+h_{\mu\nu}\right)\Big]\nonumber\\
    &=& \frac{1}{2}\,\eta^{\rho\lambda} \left(\partial_{\mu}h_{\nu\lambda}+\partial_{\nu}h_{\lambda\mu}-\partial_{\lambda}h_{\mu\nu}\right)
    -\frac{1}{2}\,h^{\rho\lambda} \left(\partial_{\mu}h_{\nu\lambda}+\partial_{\nu}h_{\lambda\mu}-\partial_{\lambda}h_{\mu\nu}\right).
\end{eqnarray}
Or à l'odre 1, d'après (\ref{christoffel}), nous avons
\begin{equation}\label{christoffel ordre 1}
     \Gamma^{\rho\;(1)}_{\mu\nu} = \frac{1}{2}\,\eta^{\rho\lambda} \left(\partial_{\mu}h_{\nu\lambda}+\partial_{\nu}h_{\lambda\mu}-\partial_{\lambda}h_{\mu\nu}\right).
\end{equation}
Ainsi, les symboles de Christoffel à l'ordre 2 sont donnés finalement par
\begin{eqnarray}
     \Gamma^{\rho\;(2)}_{\mu\nu} &=& \Gamma^{\rho\;(1)}_{\mu\nu}-\frac{1}{2}\,h^{\rho\lambda} \left(\partial_{\mu}h_{\nu\lambda}+\partial_{\nu}h_{\lambda\mu}-\partial_{\lambda}h_{\mu\nu}\right).\label{christoffel ordre 2bouda}
\end{eqnarray}

\subsubsection{Tenseur de Riemann}
Le tenseur de Riemann est défini comme suit
\begin{eqnarray}
  R_{\mu\nu\rho\sigma} &=& g_{\mu\lambda}\,R^{\lambda}_{\;\;\nu\rho\sigma} = g_{\mu\lambda}\left(\partial_{\rho}\Gamma^{\lambda}_{\nu\sigma}-\partial_{\sigma}\Gamma^{\lambda}_{\nu\rho}
                       +\Gamma^{\lambda}_{n\rho}\,\Gamma^{n}_{\nu\sigma}-\Gamma^{\lambda}_{n\sigma}\,\Gamma^{n}_{\nu\rho}\right).
\end{eqnarray}

A l'ordre 2, compte tenu de (\ref{plate plus perturbation ordre 2}) et (\ref{christoffel ordre 2bouda}), l'expression précédente devient
\begin{eqnarray}
  \hspace*{-3cm}R_{\mu\nu\rho\sigma}^{\;(2)} &=& \left(\eta_{\mu\lambda}+h_{\mu\lambda}\right)\Bigg\{\partial_{\rho}\Big[\Gamma^{\lambda\;(1)}_{\nu\sigma}
  -\frac{1}{2}\,h^{\lambda\alpha}\left(\partial_{\nu}h_{\sigma\alpha}+\partial_{\sigma}h_{\alpha\nu}-\partial_{\alpha}h_{\nu\sigma}\right)\Big]\nonumber\\
  &&-\partial_{\sigma}\Big[\Gamma^{\lambda\;(1)}_{\nu\rho}-\frac{1}{2}\,h^{\lambda\alpha} \left(\partial_{\nu}h_{\rho\alpha}+\partial_{\rho}h_{\alpha\nu}-\partial_{\alpha}h_{\nu\rho}\right)\Big]
  +\Gamma^{\lambda\;(1)}_{n\rho}\,\Gamma^{n\;(1)}_{\nu\sigma}-\Gamma^{\lambda\;(1)}_{n\sigma}\,\Gamma^{n\;(1)}_{\nu\rho}\Bigg\}\nonumber
  \end{eqnarray}
  \begin{eqnarray}
  \hspace*{-3cm}R_{\mu\nu\rho\sigma}^{\;(2)}&=&\eta_{\mu\lambda}\left(\partial_{\rho}\Gamma_{\nu\sigma}^{\lambda\;(1)}
  -\partial_{\sigma}\Gamma_{\nu\rho}^{\lambda\;(1)}\right)\nonumber\\
  &&-\frac{1}{2}\,\eta_{\mu\lambda}\,h^{\lambda\alpha}
  \left(\partial_{\rho}\partial_{\nu}h_{\sigma\alpha}+\partial_{\rho}\partial_{\sigma}h_{\alpha\nu}
  -\partial_{\rho}\partial_{\alpha}h_{\nu\sigma}\right)\nonumber\\ \nonumber\\
  &&-\frac{1}{2}\,\eta_{\mu\lambda}\,\partial_{\rho}h^{\lambda\alpha}
  \left(\partial_{\nu}h_{\sigma\alpha}+\partial_{\sigma}h_{\alpha\nu}
  -\partial_{\alpha}h_{\nu\sigma}\right)\nonumber\\ \nonumber\\
  &&+\frac{1}{2}\,\eta_{\mu\lambda}\,h^{\lambda\alpha}
  \left(\partial_{\sigma}\partial_{\nu}h_{\rho\alpha}
  +\partial_{\sigma}\partial_{\rho}h_{\alpha\nu}-\partial_{\sigma}\partial_{\alpha}h_{\nu\rho}\right)\nonumber\\ \nonumber\\
  &&+\frac{1}{2}\,\eta_{\mu\lambda}\,\partial_{\sigma}h^{\lambda\alpha}
  \left(\partial_{\nu}h_{\rho\alpha}
  +\partial_{\rho}h_{\alpha\nu}-\partial_{\alpha}h_{\nu\rho}\right)
  +h_{\mu\lambda}\left(\partial_{\rho}\Gamma_{\nu\sigma}^{\lambda\;(1)}-\partial_{\sigma}\Gamma_{\nu\rho}^{\lambda\;(1)}\right)\nonumber\\ \nonumber\\
  &&+\eta_{\mu\lambda}\Gamma^{\lambda\;(1)}_{n\rho}\,\Gamma^{n\;(1)}_{\nu\sigma}
  -\eta_{\mu\lambda}\Gamma^{\lambda\;(1)}_{n\sigma}\,\Gamma^{n\;(1)}_{\nu\rho}.\label{tenseur riemann ordre2 1}
  \end{eqnarray}
Or à l'ordre 1, nous avons d'une part
  \begin{equation}\label{tenseur riemann ordre 1}
    R_{\mu\nu\rho\sigma}^{\;(1)}=\eta_{\mu\lambda}\left(\partial_{\rho}\Gamma_{\nu\sigma}^{\lambda\;(1)}-\partial_{\sigma}\Gamma_{\nu\rho}^{\lambda\;(1)}\right)
  \end{equation}
  et d'autre part
  \begin{eqnarray*}
  h_{\mu\lambda}\left(\partial_{\rho}\Gamma_{\nu\sigma}^{\lambda\;(1)}-\partial_{\sigma}\Gamma_{\nu\rho}^{\lambda\;(1)}\right)&=&\frac{1}{2}\,\eta^{\lambda\alpha}\,
  h_{\mu\lambda}\Bigg\{\partial_{\rho}\Big(\partial_{\nu}h_{\sigma\alpha}+\partial_{\sigma}h_{\alpha\nu}-\partial_{\alpha}h_{\nu\sigma}\Big)\\
  &&-\partial_{\sigma}\Big(\partial_{\nu}h_{\rho\alpha}+\partial_{\rho}h_{\alpha\nu}-\partial_{\alpha}h_{\nu\rho}\Big)\Bigg\}\\
  &=&\frac{1}{2}\,h_{\mu}^{\alpha}\Big(\partial_{\rho}\partial_{\nu}h_{\sigma\alpha}-\partial_{\rho}\partial_{\alpha}h_{\nu\sigma}
  -\partial_{\sigma}\partial_{\nu}h_{\rho\alpha}+\partial_{\sigma}\partial_{\alpha}h_{\nu\rho}\Big).
  \end{eqnarray*}
  Dans ce cas (\ref{tenseur riemann ordre2 1}) devient
  \begin{eqnarray}
  \hspace*{-2cm}R_{\mu\nu\rho\sigma}^{\;(2)} &=& R_{\mu\nu\rho\sigma}^{\;(1)}\nonumber\\ \nonumber\\
  &&+\frac{1}{2}\,h^{\alpha}_{\mu}
  \Big(\partial_{\sigma}\partial_{\nu}h_{\rho\alpha}
  +\partial_{\sigma}\partial_{\rho}h_{\alpha\nu}-\partial_{\sigma}\partial_{\alpha}h_{\nu\rho}-\partial_{\rho}\partial_{\nu}h_{\sigma\alpha}
  -\partial_{\rho}\partial_{\sigma}h_{\alpha\nu}+\partial_{\rho}\partial_{\alpha}h_{\nu\sigma}\Big)\nonumber\\ \nonumber\\
  &&+\frac{1}{2}\,\partial_{\sigma}h^{\alpha}_{\mu}\Big(\partial_{\nu}h_{\rho\alpha}
  +\partial_{\rho}h_{\alpha\nu}-\partial_{\alpha}h_{\nu\rho}\Big)
  -\frac{1}{2}\,\partial_{\rho}h^{\alpha}_{\mu}\Big(\partial_{\nu}h_{\sigma\alpha}
  +\partial_{\sigma}h_{\alpha\nu}-\partial_{\alpha}h_{\nu\sigma}\Big)\nonumber\\ \nonumber\\
  &&+\frac{1}{4}\,\underbrace{\eta_{\mu\lambda}\,\eta^{\lambda\alpha}}_{\delta_{\mu}^{\alpha}}\,\eta^{n\beta}
  \Big(\partial_{n}h_{\rho\alpha}+\partial_{\rho}h_{\alpha n}-\partial_{\alpha}h_{n\rho}\Big)
  \Big(\partial_{\nu}h_{\sigma\beta}+\partial_{\sigma}h_{\beta\nu}-\partial_{\beta}h_{\nu\sigma}\Big)\nonumber\\
  &&-\frac{1}{4}\,\underbrace{\eta_{\mu\lambda}\,\eta^{\lambda\alpha}}_{\delta_{\mu}^{\alpha}}\,\eta^{n\beta}
  \Big(\partial_{n}h_{\sigma\alpha}+\partial_{\sigma}h_{\alpha n}-\partial_{\alpha}h_{n\sigma}\Big)
  \Big(\partial_{\nu}h_{\rho\beta}+\partial_{\rho}h_{\beta\nu}-\partial_{\beta}h_{\nu\rho}\Big)\nonumber \\
  &&+\frac{1}{2}\,h_{\mu}^{\alpha}\Big(\partial_{\rho}\partial_{\nu}h_{\sigma\alpha}-\partial_{\rho}\partial_{\alpha}h_{\nu\sigma}
  -\partial_{\sigma}\partial_{\nu}h_{\rho\alpha}+\partial_{\sigma}\partial_{\alpha}h_{\nu\rho}\Big)\nonumber
  \end{eqnarray}
  \begin{eqnarray}
  \hspace*{-2cm}R_{\mu\nu\rho\sigma}^{\;(2)} &=&
  R_{\mu\nu\rho\sigma}^{\;(1)}\nonumber\\ \nonumber\\
  &&+\frac{1}{2}\,h^{\alpha}_{\mu}
  \Big(\partial_{\sigma}\partial_{\nu}h_{\rho\alpha}
  -\partial_{\sigma}\partial_{\alpha}h_{\nu\rho}-\partial_{\rho}\partial_{\nu}h_{\sigma\alpha}
  +\partial_{\rho}\partial_{\alpha}h_{\nu\sigma}\Big)\nonumber\\ \nonumber\\
  &&+\frac{1}{2}\,\partial_{\sigma}h^{\alpha}_{\mu}\Big(\partial_{\nu}h_{\rho\alpha}
  +\partial_{\rho}h_{\alpha\nu}-\partial_{\alpha}h_{\nu\rho}\Big)
  -\frac{1}{2}\,\partial_{\rho}h^{\alpha}_{\mu}\Big(\partial_{\nu}h_{\sigma\alpha}
  +\partial_{\sigma}h_{\alpha\nu}-\partial_{\alpha}h_{\nu\sigma}\Big)\nonumber\\ \nonumber\\
  &&+\frac{1}{4}\Big(\partial_{n}h_{\rho\mu}+\partial_{\rho}h_{\mu n}-\partial_{\mu}h_{n\rho}\Big)
  \Big(\partial_{\nu}h^{n}_{\sigma}+\partial_{\sigma}h^{n}_{\nu}-\partial^{n}h_{\nu\sigma}\Big)\nonumber\\ \nonumber\\
  &&-\frac{1}{4}\Big(\partial_{n}h_{\sigma\mu}+\partial_{\sigma}h_{\mu n}-\partial_{\mu}h_{n\sigma}\Big)
  \Big(\partial_{\nu}h^{n}_{\rho}+\partial_{\rho}h^{n}_{\nu}-\partial^{n}h_{\nu\rho}\Big)\nonumber\\ \nonumber\\
  &&+\frac{1}{2}\,h_{\mu}^{\alpha}\Big(\partial_{\rho}\partial_{\nu}h_{\sigma\alpha}-\partial_{\rho}\partial_{\alpha}h_{\nu\sigma}
  -\partial_{\sigma}\partial_{\nu}h_{\rho\alpha}+\partial_{\sigma}\partial_{\alpha}h_{\nu\rho}\Big).
\end{eqnarray}
Le cinquième et sixième terme de l'expression précédente se mettent respectivement sous la forme
\begin{eqnarray}
  &&\Big(\partial_{n}h_{\rho\mu}+\partial_{\rho}h_{\mu n}-\partial_{\mu}h_{n\rho}\Big)
  \Big(\partial_{\nu}h^{n}_{\sigma}+\partial_{\sigma}h^{n}_{\nu}-\partial^{n}h_{\nu\sigma}\Big) \nonumber\\
  &&\hspace{1.5cm}=\Big(\partial^{n}h_{\rho\mu}+\partial_{\rho}h^{n}_{\mu}-\partial_{\mu}h^{n}_{\rho}\Big)
  \Big(\partial_{\nu}h_{\sigma n}+\partial_{\sigma}h_{\nu n}-\partial_{n}h_{\nu\sigma}\Big)\\ \nonumber\\
  &&\Big(\partial_{n}h_{\sigma\mu}+\partial_{\sigma}h_{\mu n}-\partial_{\mu}h_{n\sigma}\Big)
  \Big(\partial_{\nu}h^{n}_{\rho}+\partial_{\rho}h^{n}_{\nu}-\partial^{n}h_{\nu\rho}\Big) \nonumber\\
  &&\hspace{1.5cm}=\Big(\partial^{n}h_{\sigma\mu}+\partial_{\sigma}h^{n}_{\mu}-\partial_{\mu}h^{n}_{\sigma}\Big)
  \Big(\partial_{\nu}h_{\rho n}+\partial_{\rho}h_{\nu n}-\partial_{n}h_{\nu\rho}\Big).
\end{eqnarray}
Donc
  \begin{eqnarray}
  \hspace*{-2cm}R_{\mu\nu\rho\sigma}^{\;(2)} &=&
  R_{\mu\nu\rho\sigma}^{\;(1)}\nonumber\\ \nonumber\\
  &&+\frac{1}{2}\,\partial_{\sigma}h^{\alpha}_{\mu}\Big(\partial_{\nu}h_{\rho\alpha}
  +\partial_{\rho}h_{\alpha\nu}-\partial_{\alpha}h_{\nu\rho}\Big)
  -\frac{1}{2}\,\partial_{\rho}h^{\alpha}_{\mu}\Big(\partial_{\nu}h_{\sigma\alpha}
  +\partial_{\sigma}h_{\alpha\nu}-\partial_{\alpha}h_{\nu\sigma}\Big)\nonumber\\ \nonumber\\
  &&+\frac{1}{4}
  \Big(\partial^{\alpha}h_{\rho\mu}+\partial_{\rho}h^{\alpha}_{\mu}-\partial_{\mu}h^{\alpha}_{\rho}\Big)
  \Big(\partial_{\nu}h_{\sigma\alpha}+\partial_{\sigma}h_{\nu\alpha}-\partial_{\alpha}h_{\nu\sigma}\Big)\nonumber\\ \nonumber\\
  &&-\frac{1}{4}
  \Big(\partial^{\alpha}h_{\sigma\mu}+\partial_{\sigma}h^{\alpha}_{\mu}-\partial_{\mu}h^{\alpha}_{\sigma}\Big)
  \Big(\partial_{\nu}h_{\rho\alpha}+\partial_{\rho}h_{\nu\alpha}-\partial_{\alpha}h_{\nu\rho}\Big)\nonumber
  \end{eqnarray}
  \begin{eqnarray}
  \hspace*{-2cm}R_{\mu\nu\rho\sigma}^{\;(2)} &=&
  R_{\mu\nu\rho\sigma}^{\;(1)}\nonumber\\ \nonumber\\
  &&+\Big(\partial_{\nu}h_{\rho\alpha}+\partial_{\rho}h_{\alpha\nu}-\partial_{\alpha}h_{\nu\rho}\Big)
  \Bigg[\frac{1}{2}\,\partial_{\sigma}h^{\alpha}_{\mu}
  -\frac{1}{4}\Big(\partial^{\alpha}h_{\sigma\mu}+\partial_{\sigma}h^{\alpha}_{\mu}-\partial_{\mu}h^{\alpha}_{\sigma}\Big)\Bigg]\nonumber\\ \nonumber\\
  &&-\Big(\partial_{\nu}h_{\sigma\alpha}+\partial_{\sigma}h_{\alpha\nu}-\partial_{\alpha}h_{\nu\sigma}\Big)
  \Bigg[\frac{1}{2}\,\partial_{\rho}h^{\alpha}_{\mu}
  -\frac{1}{4}\Big(\partial^{\alpha}h_{\rho\mu}+\partial_{\rho}h^{\alpha}_{\mu}-\partial_{\mu}h^{\alpha}_{\rho}\Big)\Bigg].\nonumber
  \end{eqnarray}
  Nous obtenons l'expression finale du tenseur de Riemann, à l'ordre 2 de la perturbation,
  \begin{eqnarray}
  \hspace*{-2cm}R_{\mu\nu\rho\sigma}^{\;(2)} &=&
  R_{\mu\nu\rho\sigma}^{\;(1)}\nonumber\\ \nonumber\\
  &&+\frac{1}{4}\Big(\partial_{\nu}h_{\rho\alpha}+\partial_{\rho}h_{\alpha\nu}-\partial_{\alpha}h_{\nu\rho}\Big)
  \Big(\partial_{\mu}h^{\alpha}_{\sigma}-\partial^{\alpha}h_{\sigma\mu}+\partial_{\sigma}h^{\alpha}_{\mu}\Big)\nonumber\\ \nonumber\\
  &&-\frac{1}{4}\Big(\partial_{\nu}h_{\sigma\alpha}+\partial_{\sigma}h_{\alpha\nu}-\partial_{\alpha}h_{\nu\sigma}\Big)
  \Big(\partial_{\mu}h^{\alpha}_{\rho}-\partial^{\alpha}h_{\rho\mu}+\partial_{\rho}h^{\alpha}_{\mu}\Big),\label{tenseur riemann 2 electro}
\end{eqnarray}
ou encore, en remplaçant $R_{\mu\nu\rho\sigma}^{\;(1)}$ par son expression (\ref{tenseur de riemann})
\begin{eqnarray}
\hspace*{-2cm}R_{\mu\nu\rho\sigma}^{\;(2)} &=&
  \hspace{0.2cm}\frac{1}{2}\Big(\partial_{\rho}\partial_{\nu}h_{\sigma\mu}
  -\partial_{\rho}\partial_{\mu}h_{\nu\sigma}-\partial_{\sigma}\partial_{\nu}h_{\rho\mu}
  +\partial_{\sigma}\partial_{\mu}h_{\nu\rho}\Big)\nonumber\\ \nonumber\\
  &&+\frac{1}{4}\Big(\partial_{\nu}h_{\rho\alpha}+\partial_{\rho}h_{\alpha\nu}-\partial_{\alpha}h_{\nu\rho}\Big)
  \Big(\partial_{\mu}h^{\alpha}_{\sigma}+\partial_{\sigma}h^{\alpha}_{\mu}-\partial^{\alpha}h_{\sigma\mu}\Big)\nonumber\\ \nonumber\\
  &&-\frac{1}{4}\Big(\partial_{\nu}h_{\sigma\alpha}+\partial_{\sigma}h_{\alpha\nu}-\partial_{\alpha}h_{\nu\sigma}\Big)
  \Big(\partial_{\mu}h^{\alpha}_{\rho}+\partial_{\rho}h^{\alpha}_{\mu}-\partial^{\alpha}h_{\rho\mu}\Big).\label{tenseur riemann 2 bis electro}
\end{eqnarray}
\subsubsection{Tenseur de Ricci}
Par définition
\begin{eqnarray}
  R_{\nu\sigma} = g^{\mu\rho}\,R_{\mu\nu\rho\sigma}.
 \end{eqnarray}
 A l'ordre 2, compte tenu des expressions de la métrique inverse (\ref{metrique inversebouda}) et du tenseur de Riemann (\ref{tenseur riemann 2 electro}), nous avons
  \begin{eqnarray}
 \hspace*{-4cm}R_{\nu\sigma}^{\;(2)}&=& \Big(\eta^{\mu\rho}-h^{\mu\rho}+h^{\mu}_{\hspace{0.1cm}\alpha}\,h^{\alpha\rho}\Big)R_{\mu\nu\rho\sigma}^{\;(2)}\nonumber
 \end{eqnarray}
 \begin{eqnarray}
 \hspace*{-4cm}R_{\nu\sigma}^{\;(2)}&=&\Big(\eta^{\mu\rho}-h^{\mu\rho}+h^{\mu}_{\hspace{0.1cm}\alpha}\,h^{\alpha\rho}\Big)
 \Bigg[R_{\mu\nu\rho\sigma}^{\;(1)}\nonumber\\
  &&+\frac{1}{4}\Big(\partial_{\nu}h_{\rho\alpha}+\partial_{\rho}h_{\alpha\nu}-\partial_{\alpha}h_{\nu\rho}\Big)
  \Big(\partial_{\mu}h^{\alpha}_{\sigma}-\partial^{\alpha}h_{\sigma\mu}+\partial_{\sigma}h^{\alpha}_{\mu}\Big)\nonumber\\ \nonumber\\
  &&-\frac{1}{4}\Big(\partial_{\nu}h_{\sigma\alpha}+\partial_{\sigma}h_{\alpha\nu}-\partial_{\alpha}h_{\nu\sigma}\Big)
  \Big(\partial_{\mu}h^{\alpha}_{\rho}-\partial^{\alpha}h_{\rho\mu}+\partial_{\rho}h^{\alpha}_{\mu}\Big)\Bigg]\nonumber
\end{eqnarray}
\begin{eqnarray}
\hspace*{-4cm}R_{\nu\sigma}^{\;(2)}&=&\eta^{\mu\rho}\,R_{\mu\nu\rho\sigma}^{\;(1)}\nonumber\\ \nonumber\\
  &&+\frac{1}{4}\eta^{\mu\rho}\Big(\partial_{\nu}h_{\rho\alpha}+\partial_{\rho}h_{\alpha\nu}-\partial_{\alpha}h_{\nu\rho}\Big)
  \Big(\partial_{\mu}h^{\alpha}_{\sigma}-\partial^{\alpha}h_{\sigma\mu}+\partial_{\sigma}h^{\alpha}_{\mu}\Big)\nonumber\\ \nonumber\\
  &&-\frac{1}{4}\eta^{\mu\rho}\Big(\partial_{\nu}h_{\sigma\alpha}+\partial_{\sigma}h_{\alpha\nu}-\partial_{\alpha}h_{\nu\sigma}\Big)
  \Big(\partial_{\mu}h^{\alpha}_{\rho}-\partial^{\alpha}h_{\rho\mu}+\partial_{\rho}h^{\alpha}_{\mu}\Big)\nonumber\\ \nonumber\\
  &&-h^{\mu\rho}\,R_{\mu\nu\rho\sigma}^{\;(1)} \nonumber
\end{eqnarray}
Or d'après l'expression (\ref{ricci tensor 1 electrom}) du tenseur de Ricci, au premier ordre, nous avons
 \begin{eqnarray}
\hspace*{-4cm}R_{\nu\sigma}^{\;(2)}&=&R_{\nu\sigma}^{\;(1)}\nonumber\\ \nonumber\\
&&+\frac{1}{4}\eta^{\mu\rho}\Big(\partial_{\nu}h_{\rho\alpha}+\partial_{\rho}h_{\alpha\nu}-\partial_{\alpha}h_{\nu\rho}\Big)
  \Big(\partial_{\mu}h^{\alpha}_{\sigma}-\partial^{\alpha}h_{\sigma\mu}+\partial_{\sigma}h^{\alpha}_{\mu}\Big)\nonumber\\ \nonumber\\
  &&-\frac{1}{4}\eta^{\mu\rho}\Big(\partial_{\nu}h_{\sigma\alpha}+\partial_{\sigma}h_{\alpha\nu}-\partial_{\alpha}h_{\nu\sigma}\Big)
  \Big(\partial_{\mu}h^{\alpha}_{\rho}-\partial^{\alpha}h_{\rho\mu}+\partial_{\rho}h^{\alpha}_{\mu}\Big)\nonumber\\ \nonumber\\
  &&-h^{\mu\rho}\,R_{\mu\nu\rho\sigma}^{\;(1)} \nonumber
 \end{eqnarray}
\begin{eqnarray}
\hspace*{-4cm}R_{\nu\sigma}^{\;(2)}&=&R_{\nu\sigma}^{\;(1)}\nonumber\\ \nonumber\\
&&+\frac{1}{4}\eta^{\mu\rho}\Big(\partial_{\nu}h_{\rho\alpha}+\partial_{\rho}h_{\alpha\nu}-\partial_{\alpha}h_{\nu\rho}\Big)
  \Big(\partial_{\mu}h^{\alpha}_{\sigma}-\partial^{\alpha}h_{\sigma\mu}+\partial_{\sigma}h^{\alpha}_{\mu}\Big)\nonumber\\ \nonumber\\
  &&-\frac{1}{4}\eta^{\mu\rho}\Big(\partial_{\nu}h_{\sigma\alpha}+\partial_{\sigma}h_{\alpha\nu}-\partial_{\alpha}h_{\nu\sigma}\Big)
  \Big(\partial_{\mu}h^{\alpha}_{\rho}-\partial^{\alpha}h_{\rho\mu}+\partial_{\rho}h^{\alpha}_{\mu}\Big)\nonumber\\ \nonumber\\
  &&-\frac{1}{2}\,h^{\mu\rho}\Big(\partial_{\rho}\partial_{\nu}h_{\sigma\mu}
  -\partial_{\rho}\partial_{\mu}h_{\nu\sigma}-\partial_{\sigma}\partial_{\nu}h_{\rho\mu}
  +\partial_{\sigma}\partial_{\mu}h_{\nu\rho}\Big)
  \end{eqnarray}
  tel que

$h=\eta^{\alpha\beta}\,h_{\alpha\beta}=h^{\beta}_{\beta}$ (trace de la perturbation à l'ordre 1).

$\Box=\eta^{\alpha\beta}\,\partial_{\alpha}\partial_{\beta}=\frac{1}{c^{2}}\partial_{t}^{2}-\overrightarrow{\nabla}^{2}$ (d'Alembertien).

  En exprimant le tenseur de Ricci à l'ordre 1 par son expression explicite, il vient alors
  \begin{eqnarray}
  R_{\nu\sigma}^{\;(2)}&=&\frac{1}{2}\Big(\partial_{\rho}\partial_{\nu}h^{\rho}_{\sigma}
  -\Box h_{\nu\sigma}-\partial_{\sigma}\partial_{\nu}h
  +\partial_{\sigma}\partial_{\mu}h^{\mu}_{\nu}\Big)\nonumber\\ \nonumber\\
  &&+\frac{1}{4}\eta^{\mu\rho}\Big(\partial_{\nu}h_{\rho\alpha}+\partial_{\rho}h_{\alpha\nu}-\partial_{\alpha}h_{\nu\rho}\Big)
  \Big(\partial_{\mu}h^{\alpha}_{\sigma}-\partial^{\alpha}h_{\sigma\mu}+\partial_{\sigma}h^{\alpha}_{\mu}\Big)\nonumber\\ \nonumber\\
  &&-\frac{1}{4}\eta^{\mu\rho}\Big(\partial_{\nu}h_{\sigma\alpha}+\partial_{\sigma}h_{\alpha\nu}-\partial_{\alpha}h_{\nu\sigma}\Big)
  \Big(\partial_{\mu}h^{\alpha}_{\rho}-\partial^{\alpha}h_{\rho\mu}+\partial_{\rho}h^{\alpha}_{\mu}\Big)\nonumber\\ \nonumber\\
  &&-\frac{1}{2}\,h^{\mu\rho}\Big(\partial_{\rho}\partial_{\nu}h_{\sigma\mu}
  -\partial_{\rho}\partial_{\mu}h_{\nu\sigma}-\partial_{\sigma}\partial_{\nu}h_{\rho\mu}
  +\partial_{\sigma}\partial_{\mu}h_{\nu\rho}\Big).\label{tenseur ricci 2 electro 2}
\end{eqnarray}

Dans le but de simplifier un peu plus l'expression précédente, procédons par étapes. Pour ce faire, posons
\begin{eqnarray}
   B &=&\frac{1}{4}\Big(\partial_{\nu}h^{\mu}_{\alpha}+\partial^{\mu}h_{\alpha\nu}-\partial_{\alpha}h^{\mu}_{\nu}\Big)
  \Big(\underline{\partial_{\mu}h^{\alpha}_{\sigma}}-\underline{\partial^{\alpha}h_{\sigma\mu}}+\partial_{\sigma}h^{\alpha}_{\mu}\Big)\nonumber
   \end{eqnarray}
  \begin{eqnarray}
  B&=&+\frac{1}{4}\Big(\partial_{\nu}h^{\mu}_{\alpha}+\partial^{\mu}h_{\alpha\nu}-\partial_{\alpha}h^{\mu}_{\nu}\Big)\partial_{\mu}h^{\alpha}_{\sigma}
  -\frac{1}{4}\Big(\partial_{\nu}h^{\mu}_{\alpha}+\partial^{\mu}h_{\alpha\nu}-\partial_{\alpha}h^{\mu}_{\nu}\Big)\partial^{\alpha}h_{\sigma\mu}\nonumber\\
  &&+\frac{1}{4}\Big(\partial_{\nu}h^{\mu}_{\alpha}+\partial^{\mu}h_{\alpha\nu}-\partial_{\alpha}h^{\mu}_{\nu}\Big)\partial_{\sigma}h^{\alpha}_{\mu}\nonumber
  \end{eqnarray}
  En modifiant l'indice muet $\mu$ à $\alpha$, dans le deuxième terme, nous avons
  \begin{eqnarray}
  \hspace*{-3cm}B&=&+\frac{1}{4}\Big(\partial_{\nu}h^{\mu}_{\alpha}+\partial^{\mu}h_{\alpha\nu}-\partial_{\alpha}h^{\mu}_{\nu}\Big)\partial_{\mu}h^{\alpha}_{\sigma}
  -\frac{1}{4}\Big(\partial_{\nu}h^{\alpha}_{\mu}+\partial^{\alpha}h_{\mu\nu}-\partial_{\mu}h^{\alpha}_{\nu}\Big)\partial^{\mu}h_{\sigma\alpha}\nonumber\\
  &&+\frac{1}{4}\Big(\partial_{\nu}h^{\mu}_{\alpha}+\partial^{\mu}h_{\alpha\nu}-\partial_{\alpha}h^{\mu}_{\nu}\Big)\partial_{\sigma}h^{\alpha}_{\mu}\nonumber
  \end{eqnarray}
  \begin{eqnarray}
  \hspace*{-3cm}B&=&\frac{1}{4}\Big(\underline{\partial_{\nu}h^{\mu}_{\alpha}}+
  \partial^{\mu}h_{\alpha\nu}-\partial_{\alpha}h^{\mu}_{\nu}\Big)\partial_{\mu}h^{\alpha}_{\sigma}
  -\frac{1}{4}\Big(\underline{\partial_{\nu}h^{\mu}_{\alpha}}+
  \partial_{\alpha}h^{\mu}_{\nu}-\partial^{\mu}h_{\nu\alpha}\Big)\partial_{\mu}h^{\alpha}_{\sigma}\nonumber\\
  &&+\frac{1}{4}\Big(\partial_{\nu}h^{\mu}_{\alpha}+\partial^{\mu}h_{\alpha\nu}
  -\partial_{\alpha}h^{\mu}_{\nu}\Big)\partial_{\sigma}h^{\alpha}_{\mu}\nonumber
  \end{eqnarray}
  ou encore
  \begin{eqnarray}
  B &=&\frac{1}{2}\Big(\partial^{\mu}h_{\alpha\nu}-\partial_{\alpha}h^{\mu}_{\nu}\Big)\partial_{\mu}h^{\alpha}_{\sigma}
  +\frac{1}{4}\Big(\partial_{\nu}h^{\mu}_{\alpha}+\partial^{\mu}h_{\alpha\nu}-\partial_{\alpha}h^{\mu}_{\nu}\Big)\partial_{\sigma}h^{\alpha}_{\mu}.\label{B}
\end{eqnarray}

En remplaçant (\ref{B}) dans (\ref{tenseur ricci 2 electro 2})
\begin{eqnarray}
  \hspace*{-3cm}R_{\nu\sigma}^{\;(2)}&=&\frac{1}{2}\Big(\partial_{\rho}\partial_{\nu}h^{\rho}_{\sigma}
  -\Box h_{\nu\sigma}-\partial_{\sigma}\partial_{\nu}h
  +\partial_{\sigma}\partial_{\mu}h^{\mu}_{\nu}\Big)\nonumber\\ \nonumber\\
  &&-\frac{1}{2}\,h^{\mu\rho}\Big(\partial_{\rho}\partial_{\nu}h_{\sigma\mu}
  -\partial_{\rho}\partial_{\mu}h_{\nu\sigma}-\partial_{\sigma}\partial_{\nu}h_{\rho\mu}
  +\partial_{\sigma}\partial_{\mu}h_{\nu\rho}\Big)\nonumber\\ \nonumber\\
  &&+\frac{1}{2}\Big(\partial^{\mu}h_{\alpha\nu}-\partial_{\alpha}h^{\mu}_{\nu}\Big)\partial_{\mu}h^{\alpha}_{\sigma}
  +\frac{1}{4}\Big(\partial_{\nu}h^{\mu}_{\alpha}+\partial^{\mu}h_{\alpha\nu}-\partial_{\alpha}h^{\mu}_{\nu}\Big)\partial_{\sigma}h^{\alpha}_{\mu}\nonumber\\ \nonumber\\
  &&-\frac{1}{2}\Big(\partial_{\nu}h_{\sigma\alpha}+\partial_{\sigma}h_{\alpha\nu}-\partial_{\alpha}h_{\nu\sigma}\Big)
  \Big(\partial^{\rho}h^{\alpha}_{\rho}-\frac{1}{2}\,\partial^{\alpha}h\Big)
\end{eqnarray}
nous obtenons finalement l'expression du tenseur de Ricci à l'ordre 2 de perturbation
\begin{eqnarray}\label{tenseur ricci ordre 2}
  \hspace*{-3cm}R_{\nu\sigma}^{\;(2)}&=&\frac{1}{2}\Big(\partial_{\rho}\partial_{\nu}h^{\rho}_{\sigma}
  -\Box h_{\nu\sigma}-\partial_{\sigma}\partial_{\nu}h
  +\partial_{\sigma}\partial_{\mu}h^{\mu}_{\nu}\Big)\nonumber\\ \nonumber\\
  &&-\frac{1}{2}\,h^{\mu\rho}\Big(\partial_{\rho}\partial_{\nu}h_{\sigma\mu}
  -\partial_{\rho}\partial_{\mu}h_{\nu\sigma}-\partial_{\sigma}\partial_{\nu}h_{\rho\mu}
  +\partial_{\sigma}\partial_{\mu}h_{\nu\rho}\Big)\nonumber\\ \nonumber\\
  &&+\frac{1}{2}\Big(\underline{\partial^{\mu}h_{\alpha\nu}-\partial_{\alpha}h^{\mu}_{\nu}}\Big)\partial_{\mu}h^{\alpha}_{\sigma}
  +\frac{1}{4}\Big(\partial_{\nu}h^{\mu}_{\alpha}+\underline{\partial^{\mu}h_{\alpha\nu}-\partial_{\alpha}h^{\mu}_{\nu}}\Big)\partial_{\sigma}h^{\alpha}_{\mu}\nonumber\\ \nonumber\\
  &&-\frac{1}{2}\Big(\partial_{\nu}h^{\alpha}_{\sigma}+\partial_{\sigma}h^{\alpha}_{\nu}-\partial^{\alpha}h_{\nu\sigma}\Big)
  \Big(\partial_{\rho}h^{\rho}_{\alpha}-\frac{1}{2}\,\partial_{\alpha}h\Big)
\end{eqnarray}
qui peut se mettre encore sous la forme
\begin{eqnarray}\label{tenseur ricci ordre 2 bis}
  \hspace*{-3cm}R_{\nu\sigma}^{\;(2)}&=&\frac{1}{2}\Big(\partial_{\rho}\partial_{\nu}h^{\rho}_{\sigma}
  -\Box h_{\nu\sigma}-\partial_{\sigma}\partial_{\nu}h
  +\partial_{\sigma}\partial_{\mu}h^{\mu}_{\nu}\Big)\nonumber\\ \nonumber\\
  &&-\frac{1}{2}\,h^{\mu\rho}\Big(\partial_{\rho}\partial_{\nu}h_{\sigma\mu}
  -\partial_{\rho}\partial_{\mu}h_{\nu\sigma}-\partial_{\sigma}\partial_{\nu}h_{\rho\mu}
  +\partial_{\sigma}\partial_{\mu}h_{\nu\rho}\Big)\nonumber\\ \nonumber\\
  &&+\frac{1}{2}\Big(\partial^{\mu}h_{\alpha\nu}-\partial_{\alpha}h^{\mu}_{\nu}\Big)\Big(\partial_{\mu}h^{\alpha}_{\sigma}
  +\frac{1}{2}\,\partial_{\sigma}h^{\alpha}_{\mu}\Big)+\frac{1}{4}\,\left(\partial_{\nu}h_{\alpha}^{\mu}\right)\partial_{\sigma}h_{\mu}^{\alpha}\nonumber\\ \nonumber\\
  &&-\frac{1}{2}\Big(\partial_{\nu}h^{\alpha}_{\sigma}+\partial_{\sigma}h^{\alpha}_{\nu}-\partial^{\alpha}h_{\nu\sigma}\Big)
  \Big(\partial_{\rho}h^{\rho}_{\alpha}-\frac{1}{2}\,\partial_{\alpha}h\Big).
\end{eqnarray}
\subsubsection{Courbure scalaire}
Par définition $$R=g^{\nu\sigma}\,R_{\nu\sigma}.$$
A l'ordre 2, compte tenu des expressions de la métrique inverse (\ref{metrique inversebouda}) et du tenseur de Ricci (\ref{tenseur ricci ordre 2}), nous avons
\begin{eqnarray}
\hspace*{-3cm}R^{\;(2)} &=& \Big(\eta^{\nu\sigma}-h^{\nu\sigma}+h^{\nu}_{\hspace{0.1cm}\alpha}\,h^{\alpha\sigma}\Big)R_{\nu\sigma}^{\;(2)}\nonumber\\
 R^{\;(2)} &=&\Big(\eta^{\nu\sigma}-h^{\nu\sigma}+h^{\nu}_{\hspace{0.1cm}\alpha}\,h^{\alpha\sigma}\Big)
  \Bigg[R_{\nu\sigma}^{\;(1)}\nonumber\\ \nonumber\\
  &&-\frac{1}{2}\,h^{\mu\rho}\Big(\partial_{\rho}\partial_{\nu}h_{\sigma\mu}
  -\partial_{\rho}\partial_{\mu}h_{\nu\sigma}-\partial_{\sigma}\partial_{\nu}h_{\rho\mu}
  +\partial_{\sigma}\partial_{\mu}h_{\nu\rho}\Big)\nonumber\\ \nonumber\\
  &&+\frac{1}{2}\Big(\partial^{\mu}h_{\alpha\nu}-\partial_{\alpha}h^{\mu}_{\nu}\Big)\Big(\partial_{\mu}h^{\alpha}_{\sigma}
  +\frac{1}{2}\,\partial_{\sigma}h_{\mu}^{\alpha}\Big)
  +\frac{1}{4}\,\left(\partial_{\nu}h^{\mu}_{\alpha}\right)\partial_{\sigma}h^{\alpha}_{\mu}\nonumber\\ \nonumber\\
  &&-\frac{1}{2}\Big(\partial_{\nu}h^{\alpha}_{\sigma}+\partial_{\sigma}h^{\alpha}_{\nu}-\partial^{\alpha}h_{\nu\sigma}\Big)
  \Big(\partial_{\mu}h^{\mu}_{\alpha}-\frac{1}{2}\,\partial_{\alpha}h\Big)\Bigg]\nonumber\\ \nonumber\\
R^{\;(2)}  &=&\eta^{\nu\sigma}\,R_{\nu\sigma}^{\;(1)}-\frac{1}{2}\,h^{\mu\rho}\Big(\partial_{\rho}\partial_{\nu}h_{\sigma\mu}
  -\partial_{\rho}\partial_{\mu}h_{\nu\sigma}-\partial_{\sigma}\partial_{\nu}h_{\rho\mu}
  +\partial_{\sigma}\partial_{\mu}h_{\nu\rho}\Big)\nonumber\\ \nonumber\\
  &&+\frac{1}{2}\Big(\partial^{\mu}h_{\alpha\nu}-\partial_{\alpha}h^{\mu}_{\nu}\Big)\Big(\partial_{\mu}h^{\alpha\nu}
  +\frac{1}{2}\,\partial^{\nu}h_{\mu}^{\alpha}\Big)+\frac{1}{4}\,\left(\partial^{\sigma}h^{\mu}_{\alpha}\right)\partial_{\sigma}h^{\alpha}_{\mu}\nonumber\\
  \nonumber\\
  &&-\frac{1}{2}\Big(\underbrace{\partial^{\sigma}h^{\alpha}_{\sigma}+\partial^{\nu}h^{\alpha}_{\nu}}_{2\,\partial^{\mu}h^{\alpha}_{\mu}}
  -\partial^{\alpha}h\Big)
  \Big(\partial_{\mu}h^{\mu}_{\alpha}-\frac{1}{2}\,\partial_{\alpha}h\Big)-h^{\nu\sigma}\,R_{\nu\sigma}^{\;(1)}.\label{courbure scalaire 2 electrom 1}
\end{eqnarray}

Dans le but de procéder par étapes, posons d'une part,
\begin{eqnarray}
  C &=& \frac{1}{2}\Big(\partial^{\mu}h_{\alpha\nu}-\partial_{\alpha}h^{\mu}_{\nu}\Big)\Big(\partial_{\mu}h^{\alpha\nu}
  +\frac{1}{2}\,\partial^{\nu}h_{\mu}^{\alpha}\Big)+\frac{1}{4}\,\left(\partial^{\sigma}h^{\mu}_{\alpha}\right)\partial_{\sigma}h^{\alpha}_{\mu}\nonumber\\
   C &=& \frac{1}{2}\left(\partial^{\mu}h_{\alpha\nu}\right)\partial_{\mu}h^{\alpha\nu}
    +\frac{1}{4}\left(\partial^{\mu}h_{\alpha\nu}\right)\partial^{\nu}h_{\mu}^{\alpha}
    -\frac{1}{2}\left(\partial_{\alpha}h^{\mu}_{\nu}\right)\partial_{\mu}h^{\alpha\nu}\nonumber\\
    &&-\frac{1}{4}\underbrace{\left(\partial_{\alpha}h^{\mu}_{\nu}\right)\partial^{\nu}h_{\mu}^{\alpha}}_{\left(\partial^{\mu}h_{\alpha\nu}\right)\partial^{\nu}h_{\mu}^{\alpha}}
    +\frac{1}{4}\,\underbrace{\left(\partial^{\sigma}h^{\mu}_{\alpha}\right)\partial_{\sigma}h^{\alpha}_{\mu}}
    _{\left(\partial^{\mu}h_{\alpha\nu}\right)\partial_{\mu}h^{\alpha\nu}}\nonumber\\ \nonumber\\
    C&=& \frac{3}{4}\left(\partial^{\mu}h_{\alpha\nu}\right)\partial_{\mu}h^{\alpha\nu}
    -\frac{1}{2}\left(\partial_{\alpha}h^{\mu}_{\nu}\right)\partial_{\mu}h^{\alpha\nu}.\label{C}
\end{eqnarray}
Dans ce cas, en tenant compte de (\ref{C}), la courbure scalaire (\ref{courbure scalaire 2 electrom 1}) devient
\begin{eqnarray}
\hspace*{-3cm}R^{\;(2)}
  &=&R^{\;(1)}-\frac{1}{2}\,h^{\mu\rho}\Big(\partial_{\rho}\partial_{\nu}h^{\nu}_{\mu}
  -\partial_{\rho}\partial_{\mu}h-\Box h_{\rho\mu}
  +\partial_{\sigma}\partial_{\mu}h^{\sigma}_{\rho}\Big)\nonumber\\ \nonumber\\
  &&+\frac{3}{4}\left(\partial^{\mu}h_{\alpha\nu}\right)\partial_{\mu}h^{)\alpha\nu}
    -\frac{1}{2}\left(\partial_{\alpha}h^{\mu}_{\nu}\right)\partial_{\mu}h^{\alpha\nu}
  -\Big(\partial^{\mu}h^{\alpha}_{\mu}-\frac{1}{2}\,\partial^{\alpha}h\Big)
  \Big(\partial_{\mu}h^{\mu}_{\alpha}-\frac{1}{2}\,\partial_{\alpha}h\Big)\nonumber\\ \nonumber\\
  &&-\frac{1}{2}\,h^{\nu\sigma}\,\Big(\partial_{\rho}\partial_{\nu}h_{\sigma}^{\rho}-\Box h_{\nu\sigma}-\partial_{\sigma}\partial_{\nu}h+\partial_{\sigma}\partial_{\mu}h_{\nu}^{\mu}\Big).\label{courbure scalaire 2 electrom 2}
\end{eqnarray}
D'autre part, posons
\begin{eqnarray}
  D &=&-\frac{1}{2}\,h^{\mu\rho}\Big(\partial_{\rho}\partial_{\nu}h^{\nu}_{\mu}
  -\partial_{\rho}\partial_{\mu}h-\Box h_{\rho\mu}
  +\partial_{\sigma}\partial_{\mu}h^{\sigma}_{\rho}\Big)\nonumber\\
  &&-\frac{1}{2}\,h^{\nu\sigma}\,\Big(\partial_{\rho}\partial_{\nu}h_{\sigma}^{\rho}-\Box h_{\nu\sigma}-\partial_{\sigma}\partial_{\nu}h+\partial_{\sigma}\partial_{\mu}h_{\nu}^{\mu}\Big)\nonumber\\
 D &=&-\frac{1}{2}\,h^{\nu\sigma}\Big(\partial_{\sigma}\partial_{\rho}h^{\rho}_{\nu}
  -\partial_{\sigma}\partial_{\nu}h-\Box h_{\sigma\nu}
  +\partial_{\rho}\partial_{\nu}h^{\rho}_{\sigma}\Big)\nonumber\\
  &&-\frac{1}{2}\,h^{\nu\sigma}\,\Big(\partial_{\rho}\partial_{\nu}h_{\sigma}^{\rho}-\Box h_{\nu\sigma}-\partial_{\sigma}\partial_{\nu}h+\partial_{\sigma}\partial_{\mu}h_{\nu}^{\mu}\Big)\nonumber\\ \nonumber\\
  D&=&h^{\nu\sigma}
  \Big(\Box h_{\nu\sigma}-\partial_{\rho}\partial_{\nu}h^{\rho}_{\sigma}-\partial_{\sigma}\partial_{\rho}h_{\nu}^{\rho}
  +\partial_{\sigma}\partial_{\nu}h\Big).\label{D}
\end{eqnarray}
Finalement, compte tenu de (\ref{D}), l'expression (\ref{courbure scalaire 2 electrom 2}) de la courbure scalaire, à l'odre 2 de la perturbation, devient
\begin{eqnarray}
\hspace*{-3cm}R^{\;(2)}
  &=&R^{\;(1)}+\frac{3}{4}\left(\partial^{\mu}h_{\alpha\nu}\right)\underline{\partial_{\mu}h^{\alpha\nu}}
  -\frac{1}{2}\left(\partial_{\alpha}h^{\mu}_{\nu}\right)\underline{\partial_{\mu}h^{\alpha\nu}}\nonumber\\ \nonumber\\
    &&-\Big(\partial^{\mu}h^{\alpha}_{\mu}-\frac{1}{2}\,\partial^{\alpha}h\Big)
  \Big(\partial_{\mu}h^{\mu}_{\alpha}-\frac{1}{2}\,\partial_{\alpha}h\Big)\nonumber\\ \nonumber\\
  &&+\,h^{\nu\sigma}
  \Big(\Box h_{\nu\sigma}-\partial_{\rho}\partial_{\nu}h^{\rho}_{\sigma}-\partial_{\sigma}\partial_{\rho}h_{\nu}^{\rho}
  +\partial_{\sigma}\partial_{\nu}h\Big),
\end{eqnarray}
ou encore
\begin{eqnarray}\label{courbure scalaire ordre 2}
\hspace*{-3cm}R^{\;(2)}
  &=&\Big(\partial_{\rho}\partial_{\nu}h^{\rho\nu}-\Box h\Big)+\frac{1}{2}\,\partial_{\mu}h^{\alpha\nu}\left[\frac{3}{2}\left(\partial^{\mu}h_{\alpha\nu}\right)
    -\left(\partial_{\alpha}h^{\mu}_{\nu}\right)\right]\nonumber\\
  &&-\Big(\partial^{\mu}h^{\alpha}_{\mu}-\frac{1}{2}\,\partial^{\alpha}h\Big)
  \Big(\partial_{\mu}h^{\mu}_{\alpha}-\frac{1}{2}\,\partial_{\alpha}h\Big)\nonumber \nonumber\\
  &&+\,h^{\nu\sigma}
  \Big(\Box h_{\nu\sigma}-\partial_{\rho}\partial_{\nu}h^{\rho}_{\sigma}-\partial_{\sigma}\partial_{\rho}h_{\nu}^{\rho}
  +\partial_{\sigma}\partial_{\nu}h\Big).
\end{eqnarray}
\subsubsection{Tenseur d'Einstein à l'ordre 2 de la perturbation}
Par définition
\begin{equation}
    G_{\nu\sigma} = R_{\nu\sigma}-\frac{1}{2}\,g_{\nu\sigma}\,R.
\end{equation}

A l'ordre 2, compte tenu de (\ref{plate plus perturbation ordre 2}), (\ref{tenseur ricci ordre 2}) et (\ref{courbure scalaire ordre 2}), nous avons
\begin{eqnarray}
\hspace*{-3cm}G_{\nu\sigma}^{\;(2)} &=& R_{\nu\sigma}^{\;(2)}-\frac{1}{2}\Big(\eta_{\nu\sigma}+h_{\nu\sigma}\Big)R^{\;(2)}\nonumber\\ \nonumber\\
G_{\nu\sigma}^{\;(2)}&=&\Bigg\{R^{\;(1)}_{\nu\sigma}-\frac{1}{2}\,h^{\mu\rho}\Big(\partial_{\rho}\partial_{\nu}h_{\sigma\mu}
  -\partial_{\rho}\partial_{\mu}h_{\nu\sigma}-\partial_{\sigma}\partial_{\nu}h_{\rho\mu}
  +\partial_{\sigma}\partial_{\mu}h_{\nu\rho}\Big)\nonumber\\ \nonumber\\
  &&+\frac{1}{2}\Big(\partial^{\mu}h_{\alpha\nu}-\partial_{\alpha}h^{\mu}_{\nu}\Big)\Big(\partial_{\mu}h^{\alpha}_{\sigma}
  +\frac{1}{2}\,\partial_{\sigma}h^{\alpha}_{\mu}\Big)+\frac{1}{4}\,\left(\partial_{\nu}h_{\alpha}^{\mu}\right)\partial_{\sigma}h_{\mu}^{\alpha}\nonumber\\ \nonumber\\
  &&-\frac{1}{2}\Big(\partial_{\nu}h^{\alpha}_{\sigma}+\partial_{\sigma}h^{\alpha}_{\nu}-\partial^{\alpha}h_{\nu\sigma}\Big)
  \Big(\partial_{\rho}h^{\rho}_{\alpha}-\frac{1}{2}\,\partial_{\alpha}h\Big)\Bigg\}\nonumber\\ \nonumber\\
  &&-\frac{1}{2}\,\eta_{\nu\sigma}\Bigg\{R^{\;(1)}
 +\frac{1}{2}\,\partial_{\mu}h^{\alpha\beta}\left[\frac{3}{2}\left(\partial^{\mu}h_{\alpha\beta}\right)
    -\left(\partial_{\alpha}h^{\mu}_{\beta}\right)\right]\nonumber\\
  &&-\Big(\partial^{\mu}h^{\alpha}_{\mu}-\frac{1}{2}\,\partial^{\alpha}h\Big)
  \Big(\partial_{\mu}h^{\mu}_{\alpha}-\frac{1}{2}\,\partial_{\alpha}h\Big)\nonumber\\ \nonumber\\
  &&+h^{\alpha\beta}
  \Big(\Box h_{\alpha\beta}-\partial_{\rho}\partial_{\alpha}h^{\rho}_{\beta}-\partial_{\beta}\partial_{\rho}h_{\alpha}^{\rho}
  +\partial_{\beta}\partial_{\alpha}h\Big)\Bigg\}-\frac{1}{2}\,h_{\nu\sigma}\,\Big(\partial_{\rho}\partial_{\alpha}h^{\rho\alpha}-\Box h\Big) \nonumber\\
\end{eqnarray}

\begin{eqnarray}
\hspace*{-3cm}G_{\nu\sigma}^{\;(2)}&=&\left(R^{\;(1)}_{\nu\sigma}-\frac{1}{2}\,\eta_{\nu\sigma}\,R^{\;(1)}\right)\nonumber\\ \nonumber\\
  &&-\Bigg[\frac{1}{2}\,h^{\mu\rho}\Big(\partial_{\rho}\partial_{\nu}h_{\sigma\mu}
  -\partial_{\rho}\partial_{\mu}h_{\nu\sigma}-\partial_{\sigma}\partial_{\nu}h_{\rho\mu}
  +\partial_{\sigma}\partial_{\mu}h_{\nu\rho}\Big)\nonumber\\ \nonumber\\
  &&+\frac{1}{2}\,\eta_{\nu\sigma}\,h_{\alpha\beta}
  \Big(\Box h^{\alpha\beta}-\partial_{\rho}\partial^{\alpha}h^{\rho\beta}-\partial^{\beta}\partial_{\rho}h^{\rho\alpha}
  +\partial^{\beta}\partial^{\alpha}h\Big)\nonumber\\ \nonumber\\
  &&+\frac{1}{2}\,h_{\nu\sigma}\,\Big(\partial_{\rho}\partial_{\alpha}h^{\rho\alpha}-\Box h\Big)\Bigg]+\frac{1}{2}\Big(\partial^{\mu}h_{\alpha\nu}-\partial_{\alpha}h^{\mu}_{\nu}\Big)\Big(\partial_{\mu}h^{\alpha}_{\sigma}
  +\frac{1}{2}\,\partial_{\sigma}h^{\alpha}_{\mu}\Big)\nonumber\\
  &&+\frac{1}{2}\,\eta_{\nu\sigma}\Big(\partial^{\mu}h^{\alpha}_{\mu}-\frac{1}{2}\,\partial^{\alpha}h\Big)
  \Big(\partial_{\mu}h^{\mu}_{\alpha}-\frac{1}{2}\,\partial_{\alpha}h\Big)\nonumber\\ \nonumber\\
  &&+\frac{1}{4}\,\left(\partial_{\nu}h_{\alpha}^{\mu}\right)\partial_{\sigma}h_{\mu}^{\alpha}
  +\frac{1}{4}\,\eta_{\nu\sigma}\left(\partial_{\mu}h^{\alpha\beta}\right)\left[\left(\partial_{\alpha}h^{\mu}_{\beta}\right)
  -\frac{3}{2}\left(\partial^{\mu}h_{\alpha\beta}\right)
    \right]\nonumber\\ \nonumber\\
    &&-\frac{1}{2}\Big(\partial_{\nu}h^{\alpha}_{\sigma}+\partial_{\sigma}h^{\alpha}_{\nu}-\partial^{\alpha}h_{\nu\sigma}\Big)
  \Big(\partial_{\mu}h^{\mu}_{\alpha}-\frac{1}{2}\,\partial_{\alpha}h\Big).
\end{eqnarray}
Or le tenseur d'Einstein, à l'ordre 1, est donné par $G_{\nu\sigma}^{\;(1)}=R^{\;(1)}_{\nu\sigma}-\frac{1}{2}\,\eta_{\nu\sigma}\,R^{\;(1)}$, ce qui permet finalement d'exprimer le tenseur d'Einstein à l'ordre 2 sous la forme
\begin{eqnarray}\label{tenseur einstein ordre 2 electrom}
\hspace*{-3cm}G_{\nu\sigma}^{\;(2)}&=&G_{\nu\sigma}^{\;(1)}-\Bigg[\frac{1}{2}\,h^{\mu\rho}\Big(\partial_{\rho}\partial_{\nu}h_{\sigma\mu}
  -\partial_{\rho}\partial_{\mu}h_{\nu\sigma}-\partial_{\sigma}\partial_{\nu}h_{\rho\mu}
  +\partial_{\sigma}\partial_{\mu}h_{\nu\rho}\Big)\nonumber\\ \nonumber\\
  &&+\frac{1}{2}\,\eta_{\nu\sigma}\,h_{\alpha\beta}
  \Big(\Box h^{\alpha\beta}-\partial_{\rho}\partial^{\alpha}h^{\rho\beta}-\partial^{\beta}\partial_{\rho}h^{\rho\alpha}
  +\partial^{\beta}\partial^{\alpha}h\Big)\nonumber\\ \nonumber\\
  &&+\frac{1}{2}\,h_{\nu\sigma}\,\Big(\partial_{\rho}\partial_{\alpha}h^{\rho\alpha}-\Box h\Big)\Bigg]+\frac{1}{2}\Big(\partial^{\mu}h_{\alpha\nu}-\partial_{\alpha}h^{\mu}_{\nu}\Big)\Big(\partial_{\mu}h^{\alpha}_{\sigma}
  +\frac{1}{2}\,\partial_{\sigma}h^{\alpha}_{\mu}\Big)\nonumber\\
  &&+\frac{1}{2}\,\eta_{\nu\sigma}\Big(\partial^{\mu}h^{\alpha}_{\mu}-\frac{1}{2}\,\partial^{\alpha}h\Big)
  \Big(\partial_{\mu}h^{\mu}_{\alpha}-\frac{1}{2}\,\partial_{\alpha}h\Big)\nonumber\\ \nonumber\\
  &&+\frac{1}{4}\,\left(\partial_{\nu}h_{\alpha}^{\mu}\right)\partial_{\sigma}h_{\mu}^{\alpha}
  +\frac{1}{4}\,\eta_{\nu\sigma}\left(\partial_{\mu}h^{\alpha\beta}\right)\left[\left(\partial_{\alpha}h^{\mu}_{\beta}\right)
  -\frac{3}{2}\left(\partial^{\mu}h_{\alpha\beta}\right)
    \right]\nonumber\\ \nonumber\\
    &&-\frac{1}{2}\Big(\partial_{\nu}h^{\alpha}_{\sigma}+\partial_{\sigma}h^{\alpha}_{\nu}-\partial^{\alpha}h_{\nu\sigma}\Big)
  \Big(\partial_{\mu}h^{\mu}_{\alpha}-\frac{1}{2}\,\partial_{\alpha}h\Big),
\end{eqnarray}
ou encore
\begin{eqnarray}\label{tenseur einstein ordre 2 bis electrom}
\hspace*{-3cm}G_{\nu\sigma}^{\;(2)}&=&\frac{1}{2}\Big(\partial_{\rho}\partial_{\nu}h^{\rho}_{\sigma}
  -\Box h_{\nu\sigma}-\partial_{\sigma}\partial_{\nu}h
  +\partial_{\sigma}\partial_{\mu}h^{\mu}_{\nu}-\eta_{\nu\sigma}\,\partial_{\rho}\partial_{\alpha}h^{\rho\alpha}+\eta_{\nu\sigma}\Box h\Big)\nonumber\\ \nonumber\\
  &&-\Bigg[\frac{1}{2}\,h^{\mu\rho}\Big(\partial_{\rho}\partial_{\nu}h_{\sigma\mu}
  -\partial_{\rho}\partial_{\mu}h_{\nu\sigma}-\partial_{\sigma}\partial_{\nu}h_{\rho\mu}
  +\partial_{\sigma}\partial_{\mu}h_{\nu\rho}\Big)\nonumber\\ \nonumber\\
  &&+\frac{1}{2}\,\eta_{\nu\sigma}\,h_{\alpha\beta}
  \Big(\Box h^{\alpha\beta}-\partial_{\rho}\partial^{\alpha}h^{\rho\beta}-\partial^{\beta}\partial_{\rho}h^{\rho\alpha}
  +\partial^{\beta}\partial^{\alpha}h\Big)\nonumber\\ \nonumber\\
  &&+\frac{1}{2}\,h_{\nu\sigma}\,\Big(\partial_{\rho}\partial_{\alpha}h^{\rho\alpha}-\Box h\Big)\Bigg]+\frac{1}{2}\Big(\partial^{\mu}h_{\alpha\nu}-\partial_{\alpha}h^{\mu}_{\nu}\Big)\Big(\partial_{\mu}h^{\alpha}_{\sigma}
  +\frac{1}{2}\,\partial_{\sigma}h^{\alpha}_{\mu}\Big)\nonumber\\
  &&+\frac{1}{2}\,\eta_{\nu\sigma}\Big(\partial^{\mu}h^{\alpha}_{\mu}-\frac{1}{2}\,\partial^{\alpha}h\Big)
  \Big(\partial_{\mu}h^{\mu}_{\alpha}-\frac{1}{2}\,\partial_{\alpha}h\Big)\nonumber\\ \nonumber\\
  &&+\frac{1}{4}\,\left(\partial_{\nu}h_{\alpha}^{\mu}\right)\partial_{\sigma}h_{\mu}^{\alpha}
  +\frac{1}{4}\,\eta_{\nu\sigma}\left(\partial_{\mu}h^{\alpha\beta}\right)\left[\left(\partial_{\alpha}h^{\mu}_{\beta}\right)
  -\frac{3}{2}\left(\partial^{\mu}h_{\alpha\beta}\right)
    \right]\nonumber\\ \nonumber\\
    &&-\frac{1}{2}\Big(\partial_{\nu}h^{\alpha}_{\sigma}+\partial_{\sigma}h^{\alpha}_{\nu}-\partial^{\alpha}h_{\nu\sigma}\Big)
  \Big(\partial_{\mu}h^{\mu}_{\alpha}-\frac{1}{2}\,\partial_{\alpha}h\Big).
\end{eqnarray}

\subsection{Corrections des équations de Maxwell à l'ordre 2 de la perturbation}
\subsubsection{Description de la procédure adoptée}
Nous avons montré, qu'au premier ordre de la perturbation les équations de type Einstein dans le vide pour l'interaction électromagnétiques se réduisent au deuxième groupe d'équations de Maxwell. En effet, d'après (\ref{resume electrom maxwell}) nous avons
\begin{equation}\label{resume electrom maxwell ordre 1}
G_{0\nu}^{(1)}=-\frac{q}{2mc}\,\partial^{\mu}\mathcal{F}_{\mu\nu}=0\hspace{0.5cm}\Longrightarrow\hspace{0.5cm} \partial^{\mu}\mathcal{F}_{\mu\nu}=0.
\end{equation}

Dans ce qui suit, nous allons réécrire les équations de type Einstein dans le vide
$$G_{0\nu}^{(2)}=-\frac{q}{2mc}\,\partial^{\mu}\mathcal{F}_{\mu\nu}+\text{{\small Corrections d'ordre 2}}=0$$
en ayant recours aux conditions les conditions de nullité de la trace spatiale (\ref{trace spatiale electrom}) et les composantes spatiales" de la jauge harmonique (\ref{comp spa jaug harm electrom}), dans le but d'apporter des corrections, à l'ordre 2 de la perturbation, aux équations de Maxwell
$$\partial^{\mu}\mathcal{F}_{\mu\nu}+\text{{\small Corrections d'ordre 2}}=0.$$

\subsubsection{Corrections du deuxième groupe d'équations de Maxwell à l'ordre 2 de la perturbation}
A l'ordre 2 de la perturbation et conformément à (\ref{tenseur einstein ordre 2 electrom}), les équations d'Einstein dans le vide $$G_{\nu\sigma}^{\;(2)}=0,$$ données explicitement par
\begin{eqnarray}
\hspace*{-3cm}0&=&G_{\nu\sigma}^{\;(1)}-\Bigg[\frac{1}{2}\,h^{\mu\rho}\Big(\partial_{\rho}\partial_{\nu}h_{\sigma\mu}
  -\partial_{\rho}\partial_{\mu}h_{\nu\sigma}-\partial_{\sigma}\partial_{\nu}h_{\rho\mu}
  +\partial_{\sigma}\partial_{\mu}h_{\nu\rho}\Big)\nonumber\\ \nonumber\\
  &&+\frac{1}{2}\,\eta_{\nu\sigma}\,h_{\alpha\beta}
  \Big(\Box h^{\alpha\beta}-\partial_{\rho}\partial^{\alpha}h^{\rho\beta}-\partial^{\beta}\partial_{\rho}h^{\rho\alpha}
  +\partial^{\beta}\partial^{\alpha}h\Big)+\frac{1}{2}\,h_{\nu\sigma}\,\Big(\partial_{\rho}\partial_{\alpha}h^{\rho\alpha}-\Box h\Big)\Bigg]\nonumber\\
  &&+\frac{1}{2}\Big(\partial^{\mu}h_{\alpha\nu}-\partial_{\alpha}h^{\mu}_{\nu}\Big)\Big(\partial_{\mu}h^{\alpha}_{\sigma}
  +\frac{1}{2}\,\partial_{\sigma}h^{\alpha}_{\mu}\Big)+\frac{1}{2}\,\eta_{\nu\sigma}\Big(\partial^{\mu}h^{\alpha}_{\mu}-\frac{1}{2}\,\partial^{\alpha}h\Big)
  \Big(\partial_{\mu}h^{\mu}_{\alpha}-\frac{1}{2}\,\partial_{\alpha}h\Big)\nonumber\\ \nonumber\\
  &&+\frac{1}{4}\,\left(\partial_{\nu}h_{\alpha}^{\mu}\right)\partial_{\sigma}h_{\mu}^{\alpha}
  +\frac{1}{4}\,\eta_{\nu\sigma}\left(\partial_{\mu}h^{\alpha\beta}\right)\left[\left(\partial_{\alpha}h^{\mu}_{\beta}\right)
  -\frac{3}{2}\left(\partial^{\mu}h_{\alpha\beta}\right)
    \right]\nonumber\\ \nonumber\\
    &&-\frac{1}{2}\Big(\partial_{\nu}h^{\alpha}_{\sigma}+\partial_{\sigma}h^{\alpha}_{\nu}-\partial^{\alpha}h_{\nu\sigma}\Big)
  \Big(\partial_{\mu}h^{\mu}_{\alpha}-\frac{1}{2}\,\partial_{\alpha}h\Big),\label{eq einstein ordre 2 bisbouda electrom}
\end{eqnarray}
sont équivalentes aux équations
\begin{eqnarray}
  G_{0\sigma}^{\;(2)}&=&0 \label{einstein 2 0gama}\\
  G_{ij}^{\;(2)} &=& 0 \label{einstein 2 ij}.
\end{eqnarray}

Nous allons nous restreindre à l'équation (\ref{einstein 2 0gama}) qu'on va écrire sous la forme
\begin{equation}\label{dat bis}
    G_{0\sigma}^{\;(1)}+\mathbf{W}_{0\sigma}^{\;(2)}=0,
\end{equation}
où
\begin{eqnarray}
\hspace*{-3cm}\mathbf{W}_{0\sigma}^{\;(2)}&=&-\Bigg[\frac{1}{2}\,h^{\mu\rho}\Big(\partial_{\rho}\partial_{0}h_{\sigma\mu}
  -\partial_{\rho}\partial_{\mu}h_{0\sigma}-\partial_{\sigma}\partial_{0}h_{\rho\mu}
  +\partial_{\sigma}\partial_{\mu}h_{0\rho}\Big)\nonumber\\ \nonumber\\
  &&+\frac{1}{2}\,\eta_{0\sigma}\,h_{\alpha\beta}
  \Big(\Box h^{\alpha\beta}-\partial_{\rho}\partial^{\alpha}h^{\rho\beta}-\partial^{\beta}\partial_{\rho}h^{\rho\alpha}
  +\partial^{\beta}\partial^{\alpha}h\Big)+\frac{1}{2}\,h_{0\sigma}\,\Big(\partial_{\rho}\partial_{\alpha}h^{\rho\alpha}-\Box h\Big)\Bigg]\nonumber\\
  &&+\frac{1}{2}\Big(\partial^{\mu}h_{\alpha0}-\partial_{\alpha}h^{\mu}_{0}\Big)\Big(\partial_{\mu}h^{\alpha}_{\sigma}
  +\frac{1}{2}\,\partial_{\sigma}h^{\alpha}_{\mu}\Big)+\frac{1}{2}\,\eta_{0\sigma}\Big(\partial^{\mu}h^{\alpha}_{\mu}-\frac{1}{2}\,\partial^{\alpha}h\Big)
  \Big(\partial_{\mu}h^{\mu}_{\alpha}-\frac{1}{2}\,\partial_{\alpha}h\Big)\nonumber\\ \nonumber\\
  &&+\frac{1}{4}\,\left(\partial_{0}h_{\alpha}^{\mu}\right)\partial_{\sigma}h_{\mu}^{\alpha}
  +\frac{1}{4}\,\eta_{0\sigma}\left(\partial_{\mu}h^{\alpha\beta}\right)\left[\left(\partial_{\alpha}h^{\mu}_{\beta}\right)
  -\frac{3}{2}\left(\partial^{\mu}h_{\alpha\beta}\right)
    \right]\nonumber\\ \nonumber\\
    &&-\frac{1}{2}\Big(\partial_{0}h^{\alpha}_{\sigma}+\partial_{\sigma}h^{\alpha}_{0}-\partial^{\alpha}h_{0\sigma}\Big)
  \Big(\partial_{0}h^{\mu}_{\alpha}-\frac{1}{2}\,\partial_{\alpha}h\Big),\label{dat bis bis}
\end{eqnarray}
représente les termes d'ordre 2 du tenseur d'Einstein $G_{0\sigma}^{\;(2)}$.

En adoptant, à la fois, les conditions de nullité de la trace spatiale (\ref{trace spatiale electrom}) et les composantes spatiales" de la jauge harmonique (\ref{comp spa jaug harm electrom}), le premier terme de l'équation (\ref{dat bis}) se met sous la forme
\begin{equation}\label{dat 1}
    G_{0\sigma}^{\;(1)}=-\frac{q}{2mc}\,\partial^{\mu}\mathcal{F}_{\mu\sigma}.
\end{equation}
alors que, conformément à (\ref{moutarde 1 index}), le deuxième terme (\ref{dat bis bis}) se met sous la forme
\begin{eqnarray}
\hspace*{-3cm}\mathbf{W}_{0\sigma}^{\;(2)}&=&-\frac{1}{2}\,h^{ij}\Big[\partial_{j}\left(\partial_{0}h_{\sigma i}
  -\partial_{i}h_{0\sigma}\right)-\partial_{\sigma}\left(\partial_{0}h_{ji}
  +\partial_{i}h_{0j}\right)\Big]\nonumber\\
  &&-\frac{1}{2}\,h^{0i}\Big[\partial_{\sigma}\left(\partial_{i}h_{00}/2+\partial_{j}h^{j}_{\;i}\right)
  -\partial_{0}\left(\partial_{i}h_{0\sigma}+\partial_{0}h_{\sigma i}\right)\Big]\nonumber\\
  &&-\frac{1}{2}\,\eta_{0\sigma}\bigg\{h_{00}
  \Big[\Box h^{00}-2\,\partial^{0}\left(\partial_{\rho}h^{\rho 0}\right)
  +\partial^{0}\partial^{0}h^{00}\Big]+h_{ij}\,\Box h^{ij}\nonumber\\
  &&+2\,h_{0i}\Big[\Box h^{0i}+\partial^{0}\left(\partial^{i}h^{00}/2\right)-\partial^{i}\left(\partial_{\rho}h^{\rho 0}\right)\Big]\bigg\}\nonumber\\
  &&-\frac{1}{2}\bigg[\Big(\partial^{i}h_{j0}-\partial_{j}h^{i}_{0}\Big)\Big(\partial_{i}h^{j}_{\sigma}
  +\frac{1}{2}\,\partial_{\sigma}h^{j}_{i}\Big)-\Big(\partial_{i}h_{\;0}^{0}/2+\partial_{j}h^{j}_{i}\Big)\Big(\partial_{0}h^{i}_{\sigma}
  -\partial^{i}h^{0}_{\sigma}\Big)\bigg]\nonumber\\
  &&-\frac{1}{2}\,\eta_{0\sigma}\Big(\partial^{0}h^{0}_{0}/2+\partial^{i}h^{0}_{i}\Big)^{2}\nonumber\\
  &&-\frac{1}{4}\bigg[\left(\partial_{0}h_{0}^{0}\right)\partial_{\sigma}h_{0}^{0}
  +\left(\partial_{0}h_{j}^{i}\right)\partial_{\sigma}h_{i}^{j}+2\left(\partial_{i}h_{0}^{0}/2-\partial_{j}h^{j}_{i}\right)\partial_{\sigma}h_{0}^{i}\bigg]\nonumber\\
  &&+\frac{1}{4}\,\eta_{0\sigma}\Bigg\{-\frac{1}{2}\left(\partial_{0}h^{00}\right)^{2}
  +\left(\partial_{0}h^{ks}\right)\left[\left(\partial_{k}h^{0}_{s}\right)-\frac{3}{2}\left(\partial^{0}h_{ks}\right)\right]
  +\left(\partial_{0}h^{k0}\right)\big[2\,\partial_{i}h^{i}_{k}\big]\nonumber\\
  &&+\left(\partial_{i}h^{00}\right)\left[-\partial^{i}h^{00}-\partial_{j}h^{ji}\right]
  +\left(\partial_{i}h^{ks}\right)\left[\left(\partial_{k}h^{i}_{s}\right)-\frac{3}{2}\left(\partial^{i}h_{ks}\right)\right]\nonumber\\
  &&+\left(\partial_{i}h^{k0}\right)\bigg[\left(\partial_{k}h^{i}_{0}\right)+\left(\partial_{0}h^{i}_{k}\right)
  -3\left(\partial^{i}h_{0k}\right)\bigg]\Bigg\}\nonumber\\
  &&+\frac{1}{2}\Big(\partial_{0}h^{0}_{\sigma}+\partial_{\sigma}h^{0}_{0}-\partial^{0}h_{0\sigma}\Big)
  \Big(\partial_{0}h^{0}_{0}+\partial_{j}h^{j}_{0}-\frac{1}{2}\,\partial_{0}h^{00}\Big).\label{moutarde 1}
 \end{eqnarray}

D'après (\ref{dat bis}), (\ref{dat 1}) et (\ref{moutarde 1}), nous aboutissons finalement à une équation du type
\begin{eqnarray}\label{maxwell ordre 2}
\left(\frac{q}{m}\right)\frac{1}{2c}\,\partial^{\mu}\mathcal{F}_{\mu\sigma}+\left(\frac{q}{m}\right)^{2}f^{(2)}=0
\end{eqnarray}
où $f^{(2)}$ représente les termes correctifs du second ordre en $\mathcal{A}^{\mu}$, ses dérivées et de champs analogues définis par les $h^{ij}$.

\section{Ordres supérieurs de la perturbation}
De façon analogue à ce qui a été fait à l'ordre 2 de la perturbation, il est possible de développer le tenseur d'Einstein à des ordres supérieurs pour obtenir
\begin{eqnarray}
G_{0\sigma}=-\left(\frac{q}{m}\right)\frac{1}{2c}\,\partial^{\mu}\mathcal{F}_{\mu\sigma}-\left(\frac{q}{m}\right)^{2}f^{(2)}-\left(\frac{q}{m}\right)^{3}f^{(3)}-\cdots
\end{eqnarray}\label{maxwell ordre n}
où $f^{(n)}\hspace{0.2cm}(n=2,3,\cdots)$ représentent les termes correctifs d'ordres $n$ en $\mathcal{A}^{\mu}$, ses dérivées et de champs analogues définis par les $h^{ij}$. Les équations d'Einstein $G_{0\sigma}=0$, impliquent
\begin{eqnarray}\label{maxwell ordre n bis}
\frac{1}{2c}\,\partial^{\mu}\mathcal{F}_{\mu\sigma}+\left(\frac{q}{m}\right)f^{(2)}+\left(\frac{q}{m}\right)^{2}f^{(3)}+\cdots=0.
\end{eqnarray}
Les termes linéaires de l'équations (\ref{maxwell ordre n bis}) représentent le deuxième groupe d'équations de Maxwell, alors que les termes non linéaires sont représentés par les termes d'ordres supérieurs $(q/m)^{n-1}f^{(n)}$.

\section{Discussion des résultats}
Au terme de cette étude, nous allons faire quelques remarques et critiques relatives aux résultats obtenus.

\subsection{La force de Lorentz}
   La première remarque concerne la loi de Lorentz, obtenue à partir des équations de géodésiques sans la restriction au régime stationnaire, mais pour des vitesses faibles (à l'ordre $v/c$). Bien que la force de Lorentz est, par définition, exprimée à l'ordre $v/c$, néanmoins pour des vitesses arbitraires, il est facile de voir que l'équation de mouvement d'une charge électrique soumise à la force de Lorentz ne pourrait jamais être écrite sous forme de l'équations d'une géodésique, que si les composantes de la métrique ont une dépendance explicite de la vitesse (Géométrie de Finsler). Cette circonstance pose des problèmes, à la fois, conceptuels et d'interprétation physique: c'est comme si la métrique est doté d'une mémoire qui lui permet d'être imprégnée des propriétés dynamiques de la particule; chaque fois qu'une particule traverse une région de l'espace-temps, elle provoque une sorte de deformation plastique de la métrique. 
   Cette circonstance constitue un sérieux obstacle au souhait de généraliser le Principe d'Equivalence.

    Cependant, la force de Lorentz n'est pas fondamentale \cite{gralla, Jackson} du fait qu'elle ne permet pas de rendre compte des effets d'interaction de la charge électrique avec son propre champ électromagnétique (self-force) ni du rayonnement électromagnétique. Peut-être que la loi complète décrivant la force électromagnétique est sous forme d'une géodésique pour des vitesses arbitraires. La question demeure toujours en suspend et nécessite plus d'investigations car la possibilité d'étendre le Principe d'Equivalence est peut-être reliée à elle. Ce qui est sûr est le fait de pouvoir montrer que les équations de Maxwell sont contenues dans les équations de type Einstein même si la possibilité de l'extension du Principe d'Equivalence n'est pas tout à fait élucidée.

\subsection{Termes non linéaires}
 La deuxième remarque concerne les termes non-linéaires figurant dans (\ref{maxwell ordre n bis}) et la courbure de l'espace-temps.

Pour la gravité, le rapport $q/m$ est remplacé par $m_{g}/m_{i}$, de sorte que l'analogue de l'équation (\ref{maxwell ordre n bis}) pour l'interaction gravitationnelle
est
\begin{eqnarray}\label{grav maxwell ordre n bis}
\frac{1}{2c}\,\partial^{\mu}\mathcal{F}_{\mu\sigma}+\left(\frac{m_{g}}{m_{i}}\right)f^{(2)}+\left(\frac{m_{g}}{m_{i}}\right)^{2}f^{(3)}+\cdots=0.
\end{eqnarray}
L'égalité entre la masse grave et inerte d'un même corps fait que leur rapport possède une valeur constante pour n'importe quelle particule\footnote{Moyenant un choix d'unités adéquat,i lest possible de se ramener au cas où $m_{g}/m_{i}=1$} de sorte que (\ref{grav maxwell ordre n bis}) devient
\begin{eqnarray}\label{grav maxwell ordre n bis}
\frac{1}{2c}\,\partial^{\mu}\mathcal{F}_{\mu\sigma}+f^{(2)}+f^{(3)}+\cdots=0.
\end{eqnarray}
Il est clair que les termes d'ordre supérieurs $f^{(n)}$ demeurent toujours présents, ce qui fait de la gravité une interaction à caractère essentiellement non linéaire.

Dans le cas de l'électromagnétisme, le rapport $q/m$ ne prend pas la même valeur pour des particules différentes. En particulier, en choisissant une particule d'épreuve de telle sorte que $q\rightarrow 0$ et $m_{i}\neq 0$, tous les termes d'ordres supérieurs dans (\ref{maxwell ordre n bis}) disparaissent. Dans ce cas (\ref{maxwell ordre n bis}) se réduit aux équations de Maxwell, $\partial^{\mu}\mathcal{F}_{\mu\sigma}=0$, et l'espace-temps devient minkowskien. Cette caractéristique peut aussi être vérifiée dans le cas à symétrie sphérique où la solution de Schwarzschild est utilisée. Pour ce faire, réécrivons l'expression (\ref{xi}) pour une particule test de charge $q$ soumise à l'action d'une autre charge fixe $Q$
\begin{equation}\label{xi Q}
    \xi_{Q}(r)=1-\frac{2\,K}{c^{2}}\frac{Q}{r}\frac{q}{m_{i}}.
\end{equation}
Si $q\rightarrow 0$ et $m_{i}\neq 0$, l'équation (\ref{xi Q}) indique que $\xi_{Q}(r)\rightarrow 1$ et l'espace-temps est minkowskien. Nous déduisons donc que la présence de la charge $Q$ n'est pas une condition suffisante pour affecter la géométrie de l'espace-temps; ce n'est qu'à travers une interaction entre la source et la particule test que les termes d'ordres supérieurs sont révélés de telle sorte à affecter la métrique de l'espace-temps. Cette caractéristique constitue l'une des distinctions fondamentales entre les interactions gravitationnelle et électromagnétique. Ceci ouvre la voie pour un nouveau concept de champ permettant de tenir compte des propriétés dynamiques de la particule test.

Nous avons montré que l'effet des ces termes d'ordres supérieurs est négligeable même dans le cas des champs intenses, néanmoins ces effets deviennent significatifs dans le domaine subatomique où les masses des particules test sont très petites et les effets quantiques sont dominants. Ceci expliquerait peut-être pourquoi l'approche de Barros permet de reproduire correctement le spectre de l'atome d'hydrogène.

\subsection{La Relativité Restreinte Déformée}
La dépendance intrigante des champs vis-à-vis des propriétés de la particule test n'est pas si étrange. En effet, dans la Relativité Restreinte Déformée (DSR), pour garantir l'invariance de la longueur de Planck, la loi de transformation des coordonnées \cite{Kimberly} doit dépendre, à la fois, de l'impulsion et de l'énergie de la particule d'épreuve, ce qui induit une dépendance du champ électromagnétique, dans le domaine des hautes énergies, des propriétés de la particule test \cite{Harikumar}.

\subsection{Lien entre la Jauge harmonique et la jauge de Lorentz}
La jauge harmonique (\ref{jauge harmonique}) est équivalente aux quatre conditions (\ref{comp temp jauge harmonique}) et (\ref{comp spatiale jauge harmonique}). Dans le contexte de la Gravité Linéaire revistée, nous avons gardé les composantes "spatiales" de la jauge harmonique et substitué la composante "temporelle" par une condition alternative annulant la trace spatiale de la perturbation de la métrique (\ref{trace spatiale}). Dans ce cas, nous avons pu montré que les équations d'Einstein se réduisent, au premier ordre de la perturbation, à des équations de type Maxwell.

De plus, nous avons montré que la condition de trace spatiale nulle n'est qu'une condition nécessaire pour que la composante "temporelle" (\ref{comp spatiale jauge harmonique}) se réduise à la condition de Lorentz $\partial_{\mu}\mathcal{A}^{\mu}_{g}=0$. Dans ce cas, pour avoir en plus des équations de propagation découplées pour les potentiels $\phi_{g}$ et $\overrightarrow{\mathcal{A}}_{g}$ il ne reste qu'à adopter la composante temporelle de la jauge harmonique.


\subsection{Problème d'invariance de jauge des termes supérieurs}
L'un des problèmes inhérents à l'approche perturbative des équations d'Einstein est que les corrections perturbatives, au delà du premier ordre, ne satisfont plus l'exigence d'invariance de jauge \cite{Carroll1}. En effet, il est facile de se convaincre que le tenseur de Riemann, exprimé à l'ordre 2 de la perturbation, n'est pas invariant sous la transformation de jauge (\ref{jauge}); autrement dit
la variation $$\delta R_{\mu\nu\rho\sigma}=R_{\mu\nu\rho\sigma}^{(2)\,'}(x^{\beta})-R^{(2)}_{\mu\nu\rho\sigma}(x^{\beta})$$ n'est pas nulle, contrairement à la variation (\ref{invariance de jauge riemann}) calculée au premier ordre.

Ce problème d'invariance de jauge pose des problèmes d'interprétation physique des corrections perturbatives, à partir de l'ordre 2, car elles ne sont pas indépendantes du référentiel de coordonnées (c'est comme si elles sont calculées pour un référentiel particulier).

Nous pensons que ce problème trouve son origine dans le fait que la transformation de jauge (\ref{jauge}) est exprimée exclusivement au premier ordre de la perturbation. Il va falloir penser à généraliser cette transformation de jauge aux ordre supérieurs de la perturbation.

\subsection{Absorption de l'action du champs électromagnétique agissant sur une charge électrique dans la métrique de l'espace-temps}
En vertu du nouveau Principe d'Equivalence, étendu à l'interaction électro-magnétique, il est toujours possible d'effectuer une transformation de coordonnées pour se ramener à un référentiel où l'effet du champ électromagnétique sur une charge électrique est annulé. Dans ce qui suit, nous allons effectuer un calcul dans lequel on va absorber dans le métrique l'action du champ électromagnétique sur une charge électrique, de façon à voir la charge comme une particule libre astreinte à se déplacer suivant des géodésiques.

\subsubsection{Approximation des champs et vitesses faibles}

Le lagrangien d'une particule de masse $m$ et de charge électrique $q$, soumise à un champ éléctromagnétique $\mathcal{A}^{\mu}(\phi/c,\overrightarrow{\mathcal{A}})$, est donné par
\begin{eqnarray}\label{lagrangien electrom}
  L &=& -m\,c^{2}\,\sqrt{1-\frac{\overrightarrow{v}^{2}}{c^{2}}}+q\,\overrightarrow{\mathcal{A}}.\overrightarrow{v}-q\,\phi.
\end{eqnarray}
Son action est
\begin{equation}\label{action electromagn}
    S=\displaystyle\int\left(-m\,c^{2}\,\sqrt{1-\frac{\overrightarrow{v}^{2}}{c^{2}}}+q\,\overrightarrow{\mathcal{A}}.\overrightarrow{v}-q\,\phi\right)\,dt.
\end{equation}

L'objectif que nous nous fixons est d'exprimer (\ref{action electromagn}) sous la forme d'une action pour une particule libre
\begin{equation}\label{action libre}
    S=\displaystyle-m\,c\,\int ds
\end{equation}
où $ds$ représente l'intervalle infinitésimal
\begin{eqnarray}
    ds^{2}&=&g_{\mu\nu}\,dx^{\mu}\,dx^{\nu}=g_{00}\,\left(dx^{0}\right)^{2}+2\,g_{0i}\,dx^{0}\,dx^{i}+g_{ij}\,dx^{i}\,dx^{j}.\label{carre intervalle explicite}
\end{eqnarray}
En effet, (\ref{action electromagn}) peut se mettre sous la forme 
\begin{equation}
    S=-m\,c\,\displaystyle\int\left(c\,\sqrt{1-\frac{\overrightarrow{v}^{2}}{c^{2}}}-\frac{q}{mc}\,\overrightarrow{\mathcal{A}}.\overrightarrow{v}
    +\frac{q}{mc}\,\phi\right)\,dt,
\end{equation}
de sorte qu'au lieu de considérer la chargé électrique comme soumise à l'action d'un champ électromagnétique extérieur, il est possible de la voir comme une particule libre astreinte à suivre des géodésiques définies par l'intervalle
\begin{eqnarray}
  ds &=& \left(c\,\sqrt{1-\frac{\overrightarrow{v}^{2}}{c^{2}}}-\frac{q}{mc}\,\overrightarrow{\mathcal{A}}.\overrightarrow{v},
    +\frac{q}{mc}\,\phi\right)\,dt \label{intervalle infinitesimal}
\end{eqnarray}
autrement dit, l'action du champ électromagnétique sur la particule chargée est "absorbée" dans la métrique.

Dans le cadre de l'approximation des champs et vitesses faibles, nous avons
\begin{eqnarray}
  ds &=& \left[c\,\left(1-\frac{1}{2}\,\frac{v^{2}}{c^{2}}+\mathcal{O}(v^{4}/c^{4})\right)-\frac{q}{mc}\,\overrightarrow{\mathcal{A}}.\overrightarrow{v}
    +\frac{q}{mc}\,\phi\right]\,dt\nonumber\\
   ds &\approx& \left(c-\frac{v^{2}}{2c}+\frac{\lambda}{c}\right)\,dt
\end{eqnarray}
où $\lambda=\frac{q}{m}\,\left(\phi-\overrightarrow{\mathcal{A}}.\overrightarrow{v}\right)$.\\
\noindent En élevant au carré l'expression précédente du carré d'intervalle, tout en négligeant les termes supérieurs à l'ordre $v^{2}/c^{2}$
\begin{eqnarray}
  ds^{2} &\approx& \left(c-\frac{v^{2}}{2\,c}+\frac{\lambda}{c}\right)^{2}\,dt^{2} \nonumber\\
   &\approx& \left(c^{2}+2\,\lambda
  \right)\,dt^{2}-v^{2}\,dt^{2} \nonumber\\
  &\approx& \left(c^{2}+2\,\frac{q}{m}\,\phi
  \right)\,dt^{2}-2\,\frac{q}{m}\,\overrightarrow{\mathcal{A}}.\overrightarrow{v}\,dt^{2}-dx^{2}-dy^{2}-dz^{2} \nonumber\\
  &\approx& \left(1+2\,\frac{q\,\phi}{mc^{2}}
  \right)\,(c\,dt)^{2}-2\,\frac{q}{m}\,\left(\mathcal{A}_{x}\,\frac{dx}{dt}+\mathcal{A}_{y}\,\frac{dy}{dt}+\mathcal{A}_{z}\,\frac{dz}{dt}\right)\,dt^{2}-dx^{2}-dy^{2}-dz^{2} \nonumber\\
  &\approx& \left(1+2\,\frac{q\,\phi}{mc^{2}}
  \right)\,(c\,dt)^{2}+2\,\frac{q}{mc}\,\mathcal{A}_{i}\,dx^{i}\left(c\,dt\right)-\left(dx^{1}\right)^{2}-\left(dx^{2}\right)^{2}-\left(dx^{3}\right)^{2}, \nonumber
 \end{eqnarray}
  nous aboutissons à l'expression\footnote{noter que $\mathcal{A}^{1}=\mathcal{A}_{x}$ donc $\mathcal{A}_{1}=-\mathcal{A}_{x}$, $\cdots$}
 \begin{eqnarray}
ds^{2}  &\approx& \left(1+2\,\frac{q\,\phi}{mc^{2}}
  \right)\,(dx^{0})^{2}+2\,\frac{q}{mc}\,\mathcal{A}_{i}\,dx^{i}\,dx^{0}-\left(dx^{1}\right)^{2}-\left(dx^{2}\right)^{2}-\left(dx^{3}\right)^{2} \label{carre intervalle electrom}
\end{eqnarray}
L'identification membre à membre de (\ref{carre intervalle explicite}) et (\ref{carre intervalle electrom}), permet de retrouver les expressions des composantes de la métrique en fonction des potentiels électromagnétiques
\begin{eqnarray}
  g_{00} &=& 1+2\,\frac{q\,\phi}{mc^{2}} \\
  g_{0i} &=& \frac{q\,\mathcal{A}_{i}}{mc} \\
  g_{11} &=& -1 \\
  g_{22} &=& -1 \\
  g_{33} &=& -1.
\end{eqnarray}
En considérant la métrique comme une perturbation de la métrique de Minkowski $$g_{\mu\nu}=\eta_{\mu\nu}+h_{\mu\nu}$$
nous déduisons finalement les expressions des perturbations
\begin{eqnarray}
  h_{00} &=& 2\,\frac{q\,\phi}{mc^{2}}\ll 1 \\
  h_{0i} &=& \frac{q\,\mathcal{A}_{i}}{mc}\ll 1 \\ \nonumber\\
  h_{ij} &=& 0.
\end{eqnarray}
Il faut remarquer que ces équations sont conformes aux définitions (\ref{potentiel scalaire bel bouda electrom}), (\ref{potentiel vecteur bel bouda electrom}) et que $h^{i}_{i}=0$.

\subsubsection{Approximation des champs faibles pour des vitesses quelconques}
Dans ce qui suit nous allons reprendre la même démarche précédente, à la seule différence qu'aucune restriction sur la vitesse ne sera introduite; seule l'approximation du champ faible est utilisée.

Reprenons l'expression de l'intervalle infinitésimal  (\ref{intervalle infinitesimal}), que nous écrivons sous forme condensée
  \begin{eqnarray}
  ds &=& c\,\sqrt{1-\frac{\overrightarrow{v}^{2}}{c^{2}}}\,dt-\frac{q}{mc}\,\mathcal{A}_{\mu}dx^{\mu}\label{condense}
\end{eqnarray}
où $dx^{\mu}=(dx^{0},dx^{1},dx^{2},dx^{3})$ et
\begin{eqnarray}
   \mathcal{A}_{\mu}\,dx^{\mu}&=&\eta_{\mu\nu}\,\mathcal{A}^{\mu}\,dx^{\nu}\nonumber\\
                    &=&\eta_{00}\,\mathcal{A}^{0}\,dx^{0}+\eta_{ij}\,\mathcal{A}^{i}\,dx^{j}\nonumber\\
                    &=&\frac{\phi}{c}\,(c\,dt)-\overrightarrow{\mathcal{A}}.\overrightarrow{v}\;dt\nonumber\\
                    &=&\phi\,dt-\overrightarrow{\mathcal{A}}.\overrightarrow{v}\;dt.
\end{eqnarray}
Isolons le terme contenant la racine carrée dans (\ref{condense})
 \begin{eqnarray}
  c\,\sqrt{1-\frac{\overrightarrow{v}^{2}}{c^{2}}}\;dt  &=& ds+\frac{q}{mc}\,\mathcal{A}_{\mu}\,dx^{\mu}.\nonumber
\end{eqnarray}
et élevons au carré (la première fois) pour obtenir
\begin{eqnarray}\label{terme cont ds}
  &&\hspace{2.6cm}c^{2}\,\left(1-\frac{\overrightarrow{v}^{2}}{c^{2}}\right)\;dt^{2} = ds^{2}+\frac{q^{2}}{m^{2}c^{2}}\,\mathcal{A}_{\mu}\,\mathcal{A}_{\nu}\,dx^{\mu}\,dx^{\nu}+2\,\frac{q}{mc}\,\mathcal{A}_{\mu}\,dx^{\mu}\,ds\nonumber\\
  &&\;c^{2}\,dt^{2}-(dx^{1})^{2}-(dx^{2})^{2}-(dx^{3})^{2}  = g_{\mu\nu}\,dx^{\mu}\,dx^{\nu}+\frac{q^{2}}{m^{2}c^{2}}\,\mathcal{A}_{\mu}\,\mathcal{A}_{\nu}\,dx^{\mu}\,dx^{\nu}
  +2\,\frac{q}{mc}\,\mathcal{A}_{\mu}\,dx^{\mu}\,ds\nonumber\\
  &&\hspace{4cm}\eta_{\mu\nu}\,dx^{\mu}\,dx^{\nu}= g_{\mu\nu}\,dx^{\mu}\,dx^{\nu}+\frac{q^{2}}{m^{2}c^{2}}\,\mathcal{A}_{\mu}\,\mathcal{A}_{\nu}\,dx^{\mu}\,dx^{\nu}
  +2\,\frac{q}{mc}\,\mathcal{A}_{\mu}\,dx^{\mu}\,ds.\nonumber\\
\end{eqnarray}
Isolons à présent le terme contenant $ds$ dans (\ref{terme cont ds})
\begin{eqnarray}
  2\,\frac{q}{mc}\,\mathcal{A}_{\mu}\,dx^{\mu}\,ds&=& \eta_{\mu\nu}\,dx^{\mu}\,dx^{\nu}-g_{\mu\nu}\,dx^{\mu}\,dx^{\nu}-\frac{q^{2}}{m^{2}c^{2}}\,\mathcal{A}_{\mu}\,\mathcal{A}_{\nu}\,dx^{\mu}\,dx^{\nu}\nonumber\\
  &=& \left(\eta_{\mu\nu}-g_{\mu\nu}-\frac{q^{2}}{m^{2}c^{2}}\,\mathcal{A}_{\mu}\,\mathcal{A}_{\nu}\right)\,dx^{\mu}\,dx^{\nu}\nonumber
\end{eqnarray}
et élevons au carré (la deuxième fois) pour avoir
\begin{eqnarray}\label{second membre}
  4\,\frac{q^{2}}{m^{2}c^{2}}\,\mathcal{A}_{\mu}\mathcal{A}_{\nu}dx^{\mu}dx^{\nu}ds^{2}
  = \left(\eta_{\mu\nu}-g_{\mu\nu}-\frac{q^{2}}{m^{2}\,c^{2}}\,\mathcal{A}_{\mu}\mathcal{A}_{\nu}\right)
  \left(\eta_{\rho\sigma}-g_{\rho\sigma}-\frac{q^{2}}{m^{2}c^{2}}\,\mathcal{A}_{\rho}\mathcal{A}_{\sigma}\right)\,dx^{\mu}dx^{\nu}dx^{\rho}dx^{\sigma}.\nonumber\\
  \end{eqnarray}
  En tenant compte de (\ref{intervalle infinitesimal}) tout en développant le second membre de (\ref{second membre}), nous obtenons
  \begin{eqnarray}
  \hspace*{-0.8cm}&&4\,\frac{q^{2}}{m^{2}c^{2}}\,\mathcal{A}_{\mu}\,\mathcal{A}_{\nu}\,\,g_{\rho\sigma}\,dx^{\mu}\,dx^{\nu}\,dx^{\rho}\,dx^{\sigma}
  =\Big(\eta_{\rho\sigma}\,\eta_{\rho\sigma}+
  g_{\rho\sigma}\,g_{\rho\sigma}+\frac{q^{4}}{m^{4}c^{4}}\,\mathcal{A}_{\mu}\,\mathcal{A}_{\nu}\,\mathcal{A}_{\rho}\,\mathcal{A}_{\sigma}
  -2\,g_{\mu\nu}\eta_{\rho\sigma}\nonumber\\
  \hspace*{-0.8cm}&&\hspace{6.5cm}-2\,\frac{q^{2}}{m^{2}c^{2}}\,\mathcal{A}_{\mu}\,\mathcal{A}_{\nu}\,\eta_{\rho\sigma}
  +2\,\frac{q^{2}}{m^{2}c^{2}}\,\mathcal{A}_{\mu}\,\mathcal{A}_{\nu}\,g_{\rho\sigma}\Big)
  \,dx^{\mu}\,dx^{\nu}\,dx^{\rho}\,dx^{\sigma}.\nonumber
  \end{eqnarray}
  En regroupant les termes
  \begin{eqnarray}
  \left[\frac{q^{4}}{m^{4}c^{4}}\,\mathcal{A}_{\mu}\mathcal{A}_{\nu}\mathcal{A}_{\rho}\mathcal{A}_{\sigma}
  -2\,\frac{q^{2}}{m^{2}c^{2}}\,\mathcal{A}_{\mu}\mathcal{A}_{\nu}\left(g_{\rho\sigma}+\eta_{\rho\sigma}\right)+\left(g_{\rho\sigma}g_{\rho\sigma}
  +\eta_{\rho\sigma}\eta_{\rho\sigma}-2\,g_{\mu\nu}\eta_{\rho\sigma}\right)\right]\,dx^{\mu}dx^{\nu}dx^{\rho}dx^{\sigma}=0\nonumber
  \end{eqnarray}
  et après simplification, nous aboutissons à l'expression
  \begin{eqnarray}\label{kk}
 \left[\frac{q^{4}}{m^{4}c^{4}}\,\mathcal{A}_{\mu}\,\mathcal{A}_{\nu}\,\mathcal{A}_{\rho}\,\mathcal{A}_{\sigma}
  -2\,\frac{q^{2}}{m^{2}c^{2}}\,\mathcal{A}_{\mu}\,\mathcal{A}_{\nu}\,\left(g_{\rho\sigma}+\eta_{\rho\sigma}\right)+\left(g_{\mu\nu}-\eta_{\mu\nu}\right)
  \left(g_{\rho\sigma}-\eta_{\rho\sigma}\right)\right]\,dx^{\mu}\,dx^{\nu}\,dx^{\rho}\,dx^{\sigma}=0.\nonumber\\
\end{eqnarray}
Le développement explicite des termes de (\ref{kk}) permet d'aboutir finalement à l'expression
\begin{eqnarray}
 0 &=&(dx^{0})^{4}\;\left\{\frac{q^{4}}{m^{4}c^{4}}\,\left(\mathcal{A}_{0}\right)^{4}
 -2\,\frac{q^{2}}{m^{2}c^{2}}\,\left(\mathcal{A}_{0}\right)^{2}\big(g_{00}+1\big)+\big(g_{00}-1\big)^{2}\right\}\nonumber\\
   &&+(dx^{0})^{3}dx^{i}\;\left\{\frac{q^{4}}{m^{4}c^{4}}\,\left(\mathcal{A}_{0}\right)^{3}\mathcal{A}_{i}
   -\,\frac{2\,q^{2}}{m^{2}c^{2}}\,\Big[2\left(\mathcal{A}_{0}\right)^{2}g_{0i}
   +2\,\mathcal{A}_{0}\,\mathcal{A}_{i}\big(g_{00}+1\big)\Big]+4\,g_{00}\,g_{i0}\right\}\nonumber\\
   &&+(dx^{0})^{2}dx^{i}dx^{j}\;\bigg\{\frac{6\,q^{4}}{m^{4}c^{4}}\,\left(\mathcal{A}_{0}\right)^{2}\mathcal{A}_{i}\mathcal{A}_{j}-\,\frac{2\,q^{2}}{m^{2}c^{2}}\,
   \Big[\left(\mathcal{A}_{0}\right)^{2}\big(g_{ij}+\eta_{ij}\big)
   +4\,\mathcal{A}_{0}\,\mathcal{A}_{i}(g_{0i})\nonumber\\
   &&\hspace{3cm}+\mathcal{A}_{i}\mathcal{A}_{j}\big(g_{00}+1\big)\Big]+\Big[2\big({g_{ij}-\eta_{ij}}\big)\big(g_{00}-1\big)+4\,g_{i0}\,g_{0j}\Big]\bigg\}\nonumber\\
   &&+dx^{0}dx^{i}dx^{j}dx^{k}\;\bigg\{\frac{4\,q^{4}}{m^{4}c^{4}}\,\left(\mathcal{A}_{0} \mathcal{A}_{i}\mathcal{A}_{j}\mathcal{A}_{k}\right)-\frac{2\,q^{2}}{m^{2}c^{2}}\,\Big[2\,\mathcal{A}_{0}\mathcal{A}_{i}\big(g_{jk}+\eta_{jk}\big)+
   2\,\mathcal{A}_{i}\mathcal{A}_{j}\,g_{0k}\Big]\nonumber\\
   &&\hspace{3cm}+3\,g_{0i}\,g_{jk}\bigg\}\nonumber\\
   &&+dx^{i}dx^{j}dx^{k}dx^{l}\;\bigg\{\frac{q^{4}}{m^{4}c^{4}}\,\left(\mathcal{A}_{i}\mathcal{A}_{j}\mathcal{A}_{k}\mathcal{A}_{l}\right)
   -\frac{2\,q^{2}}{m^{2}c^{2}}\,\mathcal{A}_{i}\mathcal{A}_{j}\big(g_{kl}+\eta_{kl}\big)\nonumber\\
   &&\hspace{3cm}+\big(g_{ij}-\eta_{ij}\big)\big(g_{kl}-\eta_{kl}\big)\bigg\}.\nonumber\\
\end{eqnarray}
Dans ce qui suit nous allons exploiter le fait que les $dx^{\mu}dx^{\nu}dx^{\rho}dx^{\sigma}$ soient indépendants, ce qui implique que les coefficients de proportionnalité soient identiquement nuls.

\textbf{a. Pour $\mu=\nu=\rho=\sigma=0$}

 \begin{eqnarray}
    \frac{q^{4}}{m^{4}c^{4}}\,\left(\mathcal{A}_{0}\right)^{4}
 -2\,\frac{q^{2}}{m^{2}c^{2}}\,\left(\mathcal{A}_{0}\right)^{2}\big(g_{00}+1\big)+\big(g_{00}-1\big)^{2} &=& 0
  \end{eqnarray}
  Pour déterminer $\mathcal{A}_{0}=\mathcal{A}_{0}(g_{00})$, calculons le discriminant $\triangle = \frac{16\,q^{4}}{m^{4}c^{4}}\,g_{00}$, pour avoir
  \begin{eqnarray}
    (\mathcal{A}_{0})^{2} &=& \frac{m^{2}c^{2}}{q^{2}}\Big(1+g_{00}\pm 2\,\sqrt{g_{00}}\Big)
  \end{eqnarray}
  Le choix de la solution est conditionné à l'exigence de la convergence asymptotique de la métrique à celle de Minkowski très loin des charges, autrement dit, $$g_{00}\longrightarrow 1\hspace{0.2cm}\Longrightarrow\hspace{0.2cm}\mathcal{A}_{0}\longrightarrow 0$$
  ainsi
  \begin{equation}\label{a0carre}
    (\mathcal{A}_{0})^{2} = \frac{m^{2}c^{2}}{q^{2}}\Big(1+g_{00}-2\,\sqrt{g_{00}}\Big).
  \end{equation}
  A l'approximation du champ faible $g_{00}\approx 1+h_{00}$ où $h_{00}\ll 1$, nous avons
  \begin{eqnarray*}
    (A_{0})^{2} &\approx& \frac{m^{2}c^{2}}{q^{2}}\Big(1+(1+h_{00})-2\,\sqrt{1+h_{00}}\Big) \\
    &\approx& \frac{m^{2}c^{2}}{q^{2}}\left\{2+h_{00}-2\left[1+\frac{h_{00}}{2}-\frac{1}{8}\,(h_{00})^{2}+\cdots\right]\right\} \\
    &\approx& \frac{m^{2}c^{2}}{4\,q^{2}}\,(h_{00})^{2}\\
    \Rightarrow\hspace{0.5cm} \mathcal{A}_{0} &\approx& \frac{mc}{2\,q}\,h_{00} \\
    \Leftrightarrow\hspace{0.5cm}h_{00} &\approx& \frac{2\,q}{mc}\,\mathcal{A}_{0}\simeq\frac{2\,q}{mc}\left(\frac{\phi}{c}\right)
  \end{eqnarray*}
  Finalement
  \begin{equation}\label{h00}
    h_{00}\approx2\,\frac{q\,\phi}{mc^{2}}.
  \end{equation}
  Soulignos que ce résultat a été obtenu dans le cadre de l'approximation du champ faible et aucune approximation sur la vitesse n'a été adoptée (valable à des vitesses élevées).

 \textbf{ b. Pour $\mu=\nu=\rho=\sigma=i$}
  \begin{eqnarray}
    \frac{q^{4}}{m^{4}c^{4}}\,\left(\mathcal{A}_{i}\right)^{4}-2\,\frac{q^{2}}{m^{2}c^{2}}\,\left(\mathcal{A}_{i}\right)^{2}\big(g_{ii}-1\big)+\big(g_{ii}+1\big)^{2} &=& 0
  \end{eqnarray}
  Pour déterminer $\mathcal{A}_{i}=\mathcal{A}_{i}(g_{ii})$, le calcul du discriminant $\triangle = -\frac{16\,q^{4}}{m^{4}c^{4}}\,g_{ii} > 0$, permet d'avoir
  \begin{eqnarray}
    (\mathcal{A}_{i})^{2} = \frac{m^{2}c^{2}}{q^{2}}\Big(-1+g_{ii}\pm 2\,\sqrt{-g_{ii}}\Big)
  \end{eqnarray}
  Pour choisir la solution exigeons que $$g_{ii}\longrightarrow -1\hspace{0.2cm}\Longrightarrow\hspace{0.2cm}\mathcal{A}_{i}\longrightarrow 0$$
  de sorte que
  \begin{equation}\label{a0carre}
    (\mathcal{A}_{i})^{2} = \frac{m^{2}c^{2}}{q^{2}}\Big(-1+g_{ii}+2\,\sqrt{-g_{ii}}\Big)
  \end{equation}
  A l'approximation du champ faible, $g_{ii}\approx-1-h_{ii}$ où $h_{ii}\ll 1$, nous avons
  \begin{eqnarray*}
    (\mathcal{A}_{i})^{2} &\approx& \frac{m^{2}c^{2}}{q^{2}}\Big[-1+(-1-h_{ii})+2\,\sqrt{-(-1-h_{ii})}\Big] \\
    &\approx& \frac{m^{2}c^{2}}{q^{2}}\left\{-2-h_{ii}+2\left[1+\frac{h_{ii}}{2}+\frac{1}{8}\,(h_{ii})^{2}+\cdots\right]\right\} \\
    &\approx& \frac{m^{2}c^{2}}{4\,q^{2}}\,(h_{ii})^{2}.
  \end{eqnarray*}
  ce qui conduit finalement à
  \begin{equation}
    \mathcal{A}_{i}\approx \frac{mc}{2\,q}\,h_{ii}.
  \end{equation}
  Ce résultat semble en contradiction avec l'exigence d'annuler la trace spatiale $h_{11}+h_{22}+h_{33}$.

  Il est très instructif de reprendre la même démarche précédente pour l'interaction gravitaionnelle. Dans ce cas, les mêmes problèmes précédents persistent.

\subsubsection{Conclusion}
\begin{itemize}
    \item Finalement nous nous rendons compte qu'il y a une incompatibilité entre la force de Lorentz et l'équation des géodésiques pour des vitesses quelconques (précisément aux vitesses élevées). Le dilemme est que la force de Lorentz est par définition une force écrite à l'ordre $v/c$, auquel nous nous sommes restreints.
    \item Il semble que la méthode proposée, sensée être valable à n'importe quelle vitesse, comporte des insuffisances qui conduisent à des contradictions même pour le cas de la gravitation.
    \item Les deux approximations du champ faible et des faibles vitesses sont étroitement liées et ne peuvent pas être envisagées séparément. Dans ce cas, les champs faibles agissent sur la particule test de sorte qu'elle ne puisse être accélérée pour atteindre des vitesses importantes.
    \item Pour pouvoir s'intéresser à des vitesses élevées, l'approximation du champ faible ne sera plus valable. Il faut envisager les équations d'Einstein sans le moindre recours à une quelconque approximation en considérant aussi le cas intérieur à la source du champ.
  \end{itemize}

\newpage

\pagestyle{fancy} \lhead{chapitre\;5}\rhead{Conclusion générale}
\chapter{Conclusion générale}

\noindent Cette Thèse s'inscrit dans l'effort d'unification des interactions fondamentales. Dans le but de fournir une description géométrique unifiée des interactions gravitationnelle et électromagnétique, nous nous sommes appuyés sur deux contributions antérieures
\begin{itemize}
  \item La formulation de P. Huei \cite{Huei} de la Gravité Linéaire, dans laquelle les équations d'Einstein linéarisées se réduisent à des équations de Type Maxwell. Une analyse de cette approche a révélé quelques imperfections, notamment la restriction de l'étude, à la fois, au régime stationnaire et à la jauge harmonique ainsi que l'apparition d'un facteur 4 indésirable au niveau de la partie magnétique de la force de Type Lorentz.
  \item L'approche de C.C. Barros \cite{Barros1, Barros2, Barros3} où l'atome d'hydrogène est décrit de façon tout à fait inédite. En effet, au lieu d'introduire l'interaction coulombienne à laquelle est soumis l'électron par la démarche habituelle qui consiste à appliquer une transformation, dite de couplage minimal, sur l'équation de Dirac libre, l'auteur adopte plutôt une démarche qui consiste à choisir une métrique similaire à celle de Schwarzschild dans laquelle il incorpore l'interaction électrostatique "proton-électron" de façon analogue à ce qui se fait en RG pour la gravité. Dans le cadre de l'approximation du champ faible (l'énergie potentielle considérée est négligeable devant l'énergie au repos de l'électron), le potentiel d'ineraction reprend sa place habituelle dans l'équation d'onde, conduisant ainsi au même spectre d'énergie relativiste prévu par la Théorie de Dirac.
\end{itemize}
\vspace{0.5cm}
\noindent L'analyse de l'approche de Barros nous a incité à admettre l'existence d'une version électro-magnétique des équations d'Einstein d'une part, et d'autre part, à explorer les possibilités d'étendre le Principe d'Equivalence à l'interaction électromagnétique. En s'intéressant au cadre de validité du principe d'Equivalence, nous avons remarqué que la notion de localité, telle qu'elle est exploitée mathématiquement dans le formalisme de la RG, n'est rigoureusement exacte qu'en la réduisant à un point. Ainsi se présente la possibilité d'étendre le Postulat d'Equivalence à toutes les interactions fondamentales, car en un point donné, il est possible d'annuler l'action de n'importe quel champ moyennant une transformation de coordonnées adéquate.

\vspace{0.5cm}
\noindent Ce rapprochement entre l'électromagnétisme et la gravité nous incite à réexaminer la Gravité Linéaire, qui permet aux équations d'Einstein de prendre la même forme que celles de Maxwell. Néanmoins, certaines imperfections nous empêchent d'avoir une analogie parfaite. Ces insuffisances se résument comme suit:


\begin{enumerate}
  \item Dans l'approche de Huei, les équations d'Einstein se réduisent, dans le cadre de l'approximation du champ faible, à des équations de type Maxwell, mais avec une définition du champ gravitoélectrique valable uniquement dans le cas particulier de la jauge harmonique. De plus, la partie magnétique de la force gravitationnelle agissant sur une particule test, astreinte à se déplace suivant des géodésiques, est entachée d'un facteur 4 indésirable. En dernier lieu, la force type magnétique n'est obtenue qu'en se limitant au régime stationnaire.
  \item Dans l'approche de Carroll, les champs sont définis de telle sorte à résoudre le problème relatif au facteur 4 indésirable de la force gravitationnelle de type Lorentz, néanmoins, les équations d'Einstein ne prennent plus la forme d'équations de type Maxwell.
\end{enumerate}
Dans le but de remédier à tous ces problèmes, nous avons procédé par les étapes suivantes
\begin{enumerate}
  \item Identification des composantes $h^{00}$ et $h^{0i}$ de la métrique de la perturbation, respectivement par les potentiels scalaire et vecteur.
  \item Définition des champs gravitoélectrique $\mathbf{E}_{g}^{i}$ et gravitomagnétique $\mathbf{B}_{g}^{i}$, par l'intermédiaire du tenseur antisymétrique $\mathcal{F}^{\mu\nu}_{g}=\partial^{\mu}\mathcal{A}_{g}^{\nu}-\partial^{\nu}\mathcal{A}_{g}^{\mu}$. Nous avons montré que les définitions adoptées sont équivalentes à
\begin{eqnarray*}
\overrightarrow{\mathbf{E}_{g}}&=&-\overrightarrow{\nabla}\phi_{g}-\partial_{t}\overrightarrow{\mathcal{A}}_{g}\\
\overrightarrow{\mathbf{B}_{g}}&=&\overrightarrow{\nabla}\times\overrightarrow{\mathcal{A}_{g}},
\end{eqnarray*}
indépendamment de la jauge utilisée.
  \item Nous avons montré que les champs gravitationnels, dans le domaine linéaire, vérifient des équations de type Maxwell
\begin{enumerate}
  \item Le premier groupe des équations type Maxwell
\begin{eqnarray*}
\partial^{\sigma}\mathcal{F}_{g}^{\mu\nu}+\partial^{\nu}\mathcal{F}_{g}^{\sigma\mu}+\partial^{\mu}\mathcal{F}_{g}^{\nu\sigma} &=& 0
\hspace{0.2cm}\Longleftrightarrow\hspace{0.2cm}\left\{
  \begin{array}{ll}
    \text{div\;} \overrightarrow{\mathbf{B}_{g}} = 0 \\\\
    \text{rot\;} \overrightarrow{\mathbf{E}_{g}} = -\frac{\partial \overrightarrow{\mathbf{B}_{g}}}{\partial t}
  \end{array}
\right.
\end{eqnarray*}
est automatiquement vérifié compte tenu du caractère antisymétrique de $\mathcal{F}^{\mu\nu}_{g}$ et de la commutation des opérateurs dérivées.
  \item Pour retrouver le deuxième groupe des équations type Maxwell
\begin{eqnarray*}
    \partial_{\mu}\mathcal{F}_{g}^{\mu\nu}=0\hspace{0.3cm}\Longleftrightarrow\hspace{0.3cm}\left\{
                                                                               \begin{array}{ll}
                                                                                 \text{div\;} \overrightarrow{\mathbf{E}_{g}} =0\\\\
  \text{rot\;} \overrightarrow{\mathbf{B}_{g}} = \frac{1}{c^{2}}\,\frac{\partial \overrightarrow{\mathbf{E}_{g}}}{\partial t}
                                                                               \end{array}
                                                                             \right.
\end{eqnarray*}
nous avons adopté un point de vue quelque peu différent de l'approche standard. En effet, au lieu d'utiliser toutes les conditions de la jauge harmonique,
nous avons exploité les trois composantes "spatiales" et nous avons substitué la composante "temporelle" par la condition alternative de trace spatiale nulle
\begin{equation*}
    h^{i}_{\;i}=0.
\end{equation*}
Dans ce cas, nous avons pu montrer la relation $G^{0\nu}\approx-\frac{1}{2c}\,\partial_{\mu}\mathcal{F}_{g}^{\mu\nu}$ qui nous permet de conclure que les équations d'Einstein dans le vide se réduisent au deuxième groupe d'équations type Maxwell
$$G^{0\nu}=0\hspace{0.5cm}\Longrightarrow\hspace{0.5cm} \partial_{\mu}\mathcal{F}_{g}^{\mu\nu}=0.$$
\item L'interprétation la plus plausible, à notre sens, de la condition de nullité de la trace spatiale $h^{i}_{\;i}=0$, ou de façon équivalente $h=h^{00}$, est que cette condition est nécessaire pour que la jauge de Lorentz, $\partial_{\mu}\mathcal{A}_{g}^{\mu}=0$, soit contenue dans la composante "temporelle" de la jauge harmonique. 

Dans ce cas, après avoir retrouvé les équations de Maxwell, il est ainsi possible d'adopter la composante temporelle de la jauge harmonique (jauge de Lorentz), dans le but de découpler les équations de propagation des potentiels scalaire et vectoriel $\Box \mathcal{A}_{g}^{\mu}=\mu_{0}\,J^{\mu}$.

Nous avons également attiré l'attention sur la possibilité d'avancer d'autres interprét-ations potentielles de la condition de trace nulle $h^{i}_{i}=0$
\begin{itemize}
  \item Elle peut être vue comme l'adoption d'un système de coordonnées particulier dans lequel la perturbation de la métrique prend une forme particulière. Une telle interprétation enlèverait le caractère de généralité à cette approche car elle serait valable uniquement dans un référentiel particulier.
  \item Elle peut être vue comme la prépondérance de la composante temporelle $h^{00}$ sur les composantes spatiales $h^{ii}$. Cette prépondérance peut s'expliquer par exemple par le fait d'avoir de façon individuelle $h^{00}\gg h^{11}$, $h^{00}\gg h^{22}$ et $h^{00}\gg h^{33}$, ce qui impliquerait que $h=h^{00}-(h^{11}+h^{22}+h^{33})\approx h^{00}$. Dans ce cas, quel serait le raison profonde de la petitesse des composantes $h^{ii}$ par rapport à $h^{00}$ ?
\end{itemize}

Pour ce qui est des degrés de liberté supplémentaires $h^{ij}$, nous pensons qu'ils ne vérifient pas seulement la condition de champ faible $h^{ij}\ll 1$, mais qu'ils sont négligeables devant les $h^{0\mu}$ (utilisés pour définir $\mathcal{A}_{g}^{\mu}$) de telle sorte que leurs effets ne soient pas détectés dans le domaine lineéaire. On a vu que dans le cadre de la jauge harmonique, les $\overline{h}^{ij}=h^{ij}-\eta^{ij}\,h/2$, se propagent à la vitesse de la lumière à l'extérieur de la source.
\end{enumerate}

\item Nous avons montré qu'une particule d'épreuve, astreinte à se déplacer suivant des géodésiques, est soumise à une force gravitationnelle de type Lorentz
\begin{equation}
    \displaystyle\frac{d^{2}\overrightarrow{r}}{dt^{2}} \approx \bigg[\overrightarrow{\mathbf{E}}_{g}+\left(\overrightarrow{v}\times \overrightarrow{\mathbf{B}}_{g}\right)\bigg].\nonumber
\end{equation}
Ce résultat a été obtenu, pour des champs faibles et pour des vitesses d'ordre $v/c$, sans aucune restriction au régime stationnaire et sans le facteur 4 indésirable dans la partie magnétique qui apparaît dans la version standard. 
\end{enumerate}

\vspace{0.5cm}
\noindent Nous avons ensuite appliqué l'approche de la Gravité Linéaire revisitée à l'interaction électro-magnétique. Nous avons montré que le champ électromagnétique est décrit par des équations de Type Einstein qui se réduisent, dans le cas linéaire aux équations de Maxwell. De plus, en poussant l'etude perturbative jusqu'au deuxième ordre, des corrections aux équations de Maxwell ont pu être apportées. Finalement, nous avons montré que les termes d'ordres supérieurs sont négligeables dans le domaine usuel d'application de l'électromagnétisme et nous avons montré que l'équation des géodésiques se réduit, dans le cas des vitesses faible, à l'équation de mouvement d'une charge électrique soumise à la force de Lorentz.

Nous avons terminé notre étude par quelques discussions et critiques que nous pouvons résumer comme suit:
\begin{enumerate}
  \item La force de Lorentz a été obtenue sans restriction au régime stationnaire mais en négligeant les termes supérieurs à $v^{2}/c^{2}$. Nous avons attiré l'attention sur le fait que la restriction aux faibles vitesses, bien qu'elle constitue un point faible de cette nouvelle approche, est suffisante pour déterminer la partie magnétique proportionnelle à $v/c$. Pour des vitesses arbitraires il est nécessaire de faire dépendre la métrique explicitement de la vitesse de la particule test, ce qui conduit à utiliser une géométrie de Finsler au lieu de celle de Riemann. Nous avons aussi souligné le fait que la force de Lorentz ne permet pas de tenir compte des phénomènes de self-interaction et de rayonnement de la charge électrique qui sont incompatibles avec le Principe d'Equivalence.
  \item L'étude qualitative des termes non linéaires des équations type Einstein nous a permis de souligner une différence fondamentale entre la gravité et l'électromagnétisme. En effet, alors que pour la gravité l'égalité entre la masse grave et inerte d'un même corps fait que les termes d'ordre supérieurs demeurent toujours présents et ne peuvent être négligés, ce qui fait de la gravité une interaction à caractère essentiellement non linéaire, dans le cas de l'électromagnétisme, le rapport $q/m$ ne prend pas la même valeur pour des particules différentes. En particulier, en choisissant une particule d'épreuve de telle sorte que $q\rightarrow 0$ et $m_{i}\neq 0$, tous les termes d'ordres supérieurs disparaissent pour donner lieu aux équations de Maxwell, $\partial^{\mu}\mathcal{F}_{\mu\sigma}=0$, et l'espace-temps devient minkowskien. Contrairement à la gravité, la présence d'une charge $Q$ n'est pas une condition suffisante pour affecter la géométrie de l'espace-temps; ce n'est qu'à travers une interaction entre la source et la particule test que les termes d'ordres supérieurs sont révélés de telle sorte à affecter la métrique de l'espace-temps. Cette caractéristique constitue l'une des distinctions fondamentales entre les interactions gravitationnelle et électromagnétique. Ceci ouvre la voie à une nouvelle conception du champ permettant de tenir compte des propriétés dynamiques de la particule test. En électromagnétisme, nous avons aussi montré que l'effet des termes d'ordres supérieurs est négligeable même dans le cas des champs intenses, mais que leurs effets deviennent significatifs dans le domaine subatomique où les masses des particules sont très petites et les effets quantiques dominants. Ceci pourrait expliquer pourquoi l'approche de Barros permet de reproduire correctement le spectre de l'atome d'hydrogène.
  \item Nous avons attiré l'attention sur le fait que la dépendance intrigante des champs vis-à-vis des propriétés de la particule test est déjà signalée en Relativité Restreinte Déformée (DSR) où la loi de transformation des coordonnées \cite{Kimberly} doit dépendre, à la fois, de l'impulsion et l'énergie de la particule d'épreuve. Ceci induit une dépendance du champ électromagnétique, dans le domaine des hautes énergies, des propriétés de la particule test \cite{Harikumar}.
\item Dans le contexte de la Gravité Linéaire revistée, nous avons pu montré que les équations d'Einstein se réduisent, au premier ordre de la perturbation, aux équations de Maxwell. De plus, nous avons montré que la condition de trace spatiale nulle n'est qu'une condition nécessaire pour que la composante "temporelle" se réduise à la condition de Lorentz $\partial_{\mu}\mathcal{A}^{\mu}=0$. Dans ce cas, pour avoir en plus des équations de propagation découplées pour les potentiels $\phi$ et $\overrightarrow{\mathcal{A}}$ il ne reste qu'à adopter la composante temporelle de la jauge harmonique.
\item Nous avons signalé le problème de non respect du principe d'invariance de jauge pour les termes d'ordres supérieurs \cite{Carroll1}. Ce problème pose des problèmes d'interprétation physique des corrections perturbatives, à partir de l'ordre 2, car elles ne sont pas indépendantes du référentiel de coordonnées. Nous pensons que ce problème trouve son origine dans le fait que la transformation de jauge est exprimée exclusivement au premier ordre de la perturbation et qu'il va falloir penser à généraliser cette transformation aux ordre supérieurs de la perturbation.
\item En essayant d'absorber l'effet du champs électromagnétique agissant sur une charge électrique dans la métrique de l'espace-temps, nous nous sommes rendu compte que cela n'est possible qu'en se limitant aux faibles vitesses. Cette circonstance pourrait trouver son origine dans le fait que les deux approximations des champs faibles et de vitesses faibles ne peuvent pas être envisagées de façon indépendantes, ce qui nous oblige à adopter le point de vue selon lequel: il faut se placer dans le cas où les champs sont suffisamment faibles pour que les particules test ne peuvent pas être accélérées à des vitesses importantes. De même, l'absorption de l'interaction de la charge électrique avec son propre champ (self-interaction), n'est pas également évidente.
\end{enumerate}

\vspace{0.5cm}
\noindent Malgré la grande similitude dans la description des interactions gravitationnelle et électromagnétique, rendue possible grâce à cette nouvelle approche, il n'en demeure pas que des différences fondamentales sont signalées:
          \begin{enumerate}
            \item En premier lieu, rappelons que contrairement au champ électromagnétique, généré par deux sortes de charges électriques, le champ de gravitation n'est généré que par une seule catégorie de charge: la masse (ou bien l'énergie). Il faut savoir que le champ gravitationnel "ressent" toujours la présence des masses alors que le champ électromagnétique n'interagit pas avec les corps neutres; cette circonstance fait que les effets magnétiques d'ordres $v^{2}/c^{2}$ sont facilement détectables en se plaçant dans des cas où le champ électrique est nul, alors que les effets gravitomagnétiques, toujours du même ordre, sont plus difficiles à détecter car il n'est pas possible d'annuler l'effet du champ gravoélectrique \cite{tarantola}.
            \item Le rayonnement électromagnétique est dû à la variation du moment quarupolaire alors que le rayonnement électromagnétique est de nature dipolaire \cite{Carroll1}.
            \item L'effet du champ de gravitation sur les propriétés de l'espace-temps est plus important que celui du champ électromagnétique.
            \item Contrairement à la gravité, la présence d'une charge électrique (source du champ) n'est pas une condition suffisante pour affecter la métrique de l'espace-temps; c'est à travers une interaction du champ électromagnétique et d'une charge électrique test que la structure de l'espace-temps est affectée.
             \item Les propriétés de l'espace-temps sont indépendantes des particules d'épreuve qui subissent l'action du champs gravitationnel; autrement dit, il n'y a qu'une seule scène pour toutes les masses, alors que les propriétés du champ électromagnétique sont liées aux propriétés des charges électriques avec lequel elles interagissent. 
             \item L'égalité des masses pesante et inerte fait que les termes d'ordres supérieurs sont toujours présents pour la gravitation ce qui fait d'elle une interaction à caractère non linéaire, i.e. décrite par une équation non linéaire. Par contre, pour l'électromagnétisme, les termes d'ordres supérieurs ne sont nuls que s'il n'y a pas de charge électrique test dont l'interaction avec le champ va conduire à affecter la structure de l'espace-temps. Il faut noter que même pour des champs considérés comme très élevés actuellement, ces termes supérieurs demeurent négligeables par rapports aux termes du premier ordre, ce qui fait de l'électromagnétisme une interaction à caractère essentiellement linéaire, décrite par les équations de Maxwell, alors que l'effet de ces termes ne peut plus être négligé dans le cas des particules subatomiques où les effets quantiques sont dominants.
          \end{enumerate}

\vspace{0.5cm}
\noindent Au terme de cette investigation où une nouvelle reinterpretation des équations d'Einstein a été proposée pour étudier d'autres interactions fondamentales que la gravité; les équations de Maxwell pouvaient être déduites à partir d'une nouvelle version des équations d'Einstein et l'effet du champ électromagnétique, de manière similaire à la gravité, peut être décrit directement à partir de la métrique de l'espace-temps sans à avoir à l'introduire comme champ extérieur. Néanmoins, plusieurs questions restent en suspens et nécessitent plus d'investigations. En guise de perspectives, plusieurs pistes peuvent être envisagées
   \begin{enumerate}
      \item Essayer de s'intéresser aux effets quantiques afin de comprendre les résultats de Barros.
      \item Etudier les phénomènes de rayonnement et de self-interaction électromagnétiques par le biais des équations non linéaires et essayer d'exploiter l'analogie entre l'électromagnétisme et la gravité pour étudier les ondes gravitationnelles.
      \item A long terme, s'intéresser aux possibilités d'interprétation géométrique des interactions faible et nucléaire forte (déjà jumelées avec l'interaction électromagnétique).
    \end{enumerate} 




\chapter*{Appendices}
\appendix
\addcontentsline{toc}{chapter}{\hspace{5cm}Appendices}
\newpage
\pagestyle{fancy} \lhead{Appendices}\rhead{Transformation de jauge des composantes du tenseur de perturbation}
\chapter{Transformation de jauge des composantes du tenseur de perturbation}

Dans cette annexe, nous allons voir comment se transforment les composantes $\phi$, $\omega_{i}$, $\psi$ et les $S_{ij}$ du tenseur de perturbation via la transformation de jauge (\ref{jauge ondes gravit}).
\section{Transformation du $\phi$}\label{transf phi ondes grav index titre}
D'après (\ref{jauge ondes gravit}) et (\ref{h00 degre liberte}), nous déduisons
\begin{eqnarray*}
  h_{00} &\longrightarrow& h_{00}-\partial_{0}\xi_{0}-\partial_{0}\xi_{0} \\
  \left(2\,\phi/c^{2}\right) &\longrightarrow& \left(2\,\phi/c^{2}\right)-2\,\partial_{0}\xi_{0} 
\end{eqnarray*}
la transformation du potentiel scalaire
\begin{equation}\label{transf phi ondes grav index}
    \phi \longrightarrow \phi-c^{2}\,\partial_{0}\xi^{0}.
\end{equation}

\section{Transformation des $\omega_{i}$}\label{transf omegai ondes grav index titre}
D'après (\ref{jauge ondes gravit}) et (\ref{h0i degre liberte}), nous déduisons
\begin{eqnarray*}
  h_{0i} &\longrightarrow& h_{0i}-\partial_{0}\xi_{i}-\partial_{i}\xi_{0} \\
  \left(-\omega_{i}/c\right) &\longrightarrow& \left(-\omega_{i}/c\right)-\left(\partial_{0}\xi_{i}+\partial_{i}\xi_{0}\right)
\end{eqnarray*}
la transformation des composantes du potentiel vecteur
\begin{equation}\label{transf omegai ondes grav index}
    \omega_{i} \longrightarrow \omega_{i}+c\left(\partial_{0}\xi_{i}+\partial_{i}\xi_{0}\right).
\end{equation}

\section{Transformation du $\psi$}\label{transf psi ondes grav index titre}
D'après (\ref{jauge ondes gravit}) et (\ref{psi degre liberte})
\begin{eqnarray*}
  h_{ij} &\longrightarrow& h_{ij}-\left(\partial_{i}\xi_{j}+\partial_{j}\xi_{i}\right) \\
  \delta^{ij}\,h_{ij} &\longrightarrow& \delta^{ij}\,h_{ij}-\delta^{ij}\left(\partial_{i}\xi_{j}+\partial_{j}\xi_{i}\right) \\
  6\,\psi &\longrightarrow& 6\,\psi+\eta^{ij}\left(\partial_{i}\xi_{j}+\partial_{j}\xi_{i}\right)\\
  6\,\psi &\longrightarrow& 6\,\psi+2\,\partial^{i}\xi_{i}
\end{eqnarray*}
nous retrouvons finalement la transformation de la trace spatiale de perturbation
\begin{equation}\label{transf psi ondes grav index}
    \psi \longrightarrow \psi+\frac{1}{3}\,\partial_{i}\xi^{i}.
\end{equation}

\section{Transformation des $S_{ij}$} \label{transf sij ondes grav index titre}
D'après (\ref{sij degre liberte}) et (\ref{psi degre liberte}) il vient que
\begin{equation}\label{sij annexe}
S_{ij}=-\frac{1}{2}\,h_{ij}+\delta_{ij}\,\psi.
\end{equation}

Conformément à (\ref{jauge ondes gravit}), nous avons d'une part
\begin{eqnarray}
  -\frac{1}{2}\,h_{ij} &\longrightarrow& -\frac{1}{2}\,h_{ij}+\frac{1}{2}\,\left(\partial_{i}\xi_{j}+\partial_{j}\xi_{i}\right),
\end{eqnarray}
et d'autre part
\begin{eqnarray}
  \delta_{ij}\,\psi &\longrightarrow& \delta_{ij}\left(\psi+\frac{1}{3}\,\partial_{k}\xi^{k}\right)\nonumber\\
  \delta_{ij}\,\psi &\longrightarrow& \delta_{ij}\,\psi+\frac{1}{3}\,\partial_{k}\xi^{k}\,\delta_{ij},
\end{eqnarray}
alors, nous déduisons que
\begin{eqnarray}
 \left(-\frac{1}{2}\,h_{ij}+\delta_{ij}\,\psi\right)
 &\longrightarrow& \left[-\frac{1}{2}\,h_{ij}+\frac{1}{2}\,\left(\partial_{i}\xi_{j}+\partial_{j}\xi_{i}\right)\right]+\bigg[\delta_{ij}\,\psi+\frac{1}{3}\,\partial_{k}\xi^{k}\,\delta_{ij}\bigg],\nonumber\\
 \left(-\frac{1}{2}\,h_{ij}+\delta_{ij}\,\psi\right)
 &\longrightarrow& \left(-\frac{1}{2}\,h_{ij}+\delta_{ij}\,\psi\right)+\frac{1}{2}\,\left(\partial_{i}\xi_{j}+\partial_{j}\xi_{i}\right)+\frac{1}{3}\,\partial_{k}\xi^{k}\,\delta_{ij},
\end{eqnarray}
En tenant compte de (\ref{sij annexe}), nous aboutissons finalement à la transformation
\begin{equation}\label{transf sij ondes grav index}
    S_{ij} \longrightarrow  S_{ij}+\frac{1}{2}\,\left(\partial_{i}\xi_{j}+\partial_{j}\xi_{i}\right)+\frac{1}{3}\,\partial_{k}\xi^{k}\,\delta_{ij},
\end{equation}
de la partie à trace nulle de la perturbation.

\newpage
\pagestyle{fancy} \lhead{Appendices}\rhead{Transformation des composantes du tenseur de perturbation}
\chapter{Définition de le jauge transverse}

En pratique, fixer une jauge revient à déterminer les quatre composantes de $\xi^{\mu}$ figurant dans la transformation
\begin{equation}\label{transf coord infinit ondes gravit index}
    x^{'\,\mu}=x^{\mu}+\xi^{\mu},
\end{equation}
autrement dit, imposer quatre relations pour déterminer $\xi^{0}$ et $\xi^{i}=(\xi^{1},\xi^{2},\xi^{3})$.

\section{Fixation de $\xi^{i}=(\xi^{1},\xi^{2},\xi^{3})$} \label{fixation xi i}
Le but est d'effectuer la transformation de coordonnées (\ref{transf coord infinit ondes gravit index}) de telle sorte à imposer aux $S_{ij}$ d'être spacialement transverses
\begin{equation}\label{transverse spatial sij appedice}
    \partial^{i}S_{ij}=0,
\end{equation}
dans le nouveau système de coordonnées.

D'après (\ref{transf sij ondes grav index}) nous déduisons que
\begin{eqnarray*}
  S_{ij} &\longrightarrow&  S_{ij}+\frac{1}{2}\,\left(\partial_{i}\xi_{j}+\partial_{j}\xi_{i}\right)+\frac{1}{3}\,\partial_{k}\xi^{k}\,\delta_{ij} \\
 \Longrightarrow\hspace{0.2cm} \partial^{i}S_{ij} &\longrightarrow&  \partial^{i}S_{ij}+\frac{1}{2}\,\partial^{i}\left(\partial_{i}\xi_{j}+\partial_{j}\xi_{i}\right)+\frac{1}{3}\,\partial^{i}\partial_{k}\xi^{k}\,\delta_{ij}
\end{eqnarray*}
Imposons maintenant que la condition (\ref{transverse spatial sij appedice}) soit satisfaite dans le nouveau système de coordonnées, $\partial^{i}S^{\,'}_{ij}=0$, de telle sorte que
\begin{eqnarray*}
  0 &=& \partial^{i}S_{ij}+\frac{1}{2}\,\partial^{i}\left(\partial_{i}\xi_{j}+\partial_{j}\xi_{i}\right)+\frac{1}{3}\,\partial^{i}\partial_{k}\xi^{k}\,\delta_{ij} \\
 0 &=& \partial^{i}S_{ij}-\frac{1}{2}\,\overrightarrow{\nabla}^{2}\xi_{j}+\frac{1}{2}\,\partial_{j}\partial^{i}\xi_{i}
 -\frac{1}{3}\underbrace{\left[\partial^{i}\,(-\delta_{ij})\right]}_{\partial_{j}}\partial_{k}\xi^{k} \\
 0 &=& \partial^{i}S_{ij}-\frac{1}{2}\,\overrightarrow{\nabla}^{2}\xi_{j}+\frac{1}{2}\,\partial_{j}\partial_{i}\xi^{i}
 -\frac{1}{3}\,\partial_{j}\partial_{i}\xi^{i}\\
 0 &=& \partial^{i}S_{ij}-\frac{1}{2}\,\overrightarrow{\nabla}^{2}\xi_{j}+\frac{1}{6}\,\partial_{j}\partial_{i}\xi^{i}.
\end{eqnarray*}
Nous déduisons finalement qu'il suffit d'imposer
\begin{equation}
 \overrightarrow{\nabla}^{2}\xi_{j}-\frac{1}{3}\,\partial_{j}\left(\partial_{i}\xi^{i}\right)=2\,\partial^{i}S_{ij},
\end{equation}
pour pouvoir déterminer les trois $\xi^{j}$, à condition d'imposer des conditions aux limites appropriées.

\section{Fixation de $\xi^{0}$}\label{fixation xi 0}
A présent, déterminons comment fixer $\xi^{0}$ pour rendre le vecteur de perturbation transverse
\begin{equation}\label{transv omegai ondes grav appedice}
    \partial^{i}\omega_{i}=0,
\end{equation}
dans le nouveau système de coordonnées.

D'après (\ref{transf omegai ondes grav index}) nous déduisons que
\begin{eqnarray*}
                               \omega_{i} &\longrightarrow& \omega_{i}+c\left(\partial_{0}\xi_{i}+\partial_{i}\xi_{0}\right)\\
  \Longrightarrow\hspace{0.2cm}\partial^{i}\omega_{i} &\longrightarrow& \partial^{i}\omega_{i}+c\left(\partial_{0}\partial^{i}\xi_{i}+\partial_{i}\partial^{i}\xi^{0}\right)\\
  \Longrightarrow\hspace{0.2cm}\partial^{i}\omega_{i} &\longrightarrow&
  \partial^{i}\omega_{i}+c\,\partial_{0}\partial^{i}\xi_{i}-c\,\overrightarrow{\nabla}^{2}\xi^{0}
\end{eqnarray*}
Dans le but de satisfaire (\ref{transv omegai ondes grav appedice}) dans le nouveau système de coordonnées, $\partial^{i}\omega^{\,'}_{i}=0$, imposons que
\begin{eqnarray*}
  \partial^{i}\omega_{i}+c\,\partial_{0}\partial^{i}\xi_{i}-c\,\overrightarrow{\nabla}^{2}\xi^{0}&=&0,
\end{eqnarray*}
ce qui conduit finalement à la relation
\begin{equation}
    \overrightarrow{\nabla}^{2}\xi^{0}=\frac{1}{c}\,\partial^{i}\omega_{i}+\partial_{0}\left(\partial^{i}\xi_{i}\right).
\end{equation}

\newpage
\pagestyle{fancy} \lhead{Appendices}\rhead{Termes correctifs à l'ordre 2 du tenseur d'Einstein}
\chapter{Termes correctifs à l'ordre 2 du tenseur d'Einstein}

\section{Introduction}
A l'ordre 2 de perturbation, le tenseur d'Einstein a été mis sous la forme
\begin{equation}\label{dat bis appedix}
    G_{0\sigma}^{\;(1)}+\mathbf{W}_{0\sigma}^{\;(2)}=0.
\end{equation}
Dans ce qui sui nous allons appliquer les conditions de nullité de la trace spatiale et les composantes "spatiales" de la jauge harmonique
\begin{eqnarray}\label{2 conditions }
         \left\{
  \begin{array}{ll}
    \partial_{0}h^{0}_{i}+\partial_{j}h^{j}_{i}=\partial_{i}h/2\\\\
    h=h^{00}=h_{00}=h^{0}_{\;0}
  \end{array}
\right.
\end{eqnarray}
ou de façon équivalente
\begin{equation}\label{cond unifiee}
    \partial_{0}h^{0}_{i}+\partial_{j}h^{j}_{i}=\partial_{i}h^{00}/2,
\end{equation}
aux termes $\mathbf{W}_{0\sigma}^{\;(2)}$  d'ordre 2 du tenseur d'Einstein (\ref{dat bis appedix}) dans le but d'apporter des corrections aux équations de Maxwell $\partial^{\mu}\mathcal{F}_{\mu\sigma}=0$.

\section{Définition des termes de $\mathbf{W}_{0\sigma}^{\;(2)}$}
Pour procéder par étapes, réécrivons les huit termes constituant $\mathbf{W}_{0\sigma}^{\;(2)}$ de (\ref{dat bis bis}), sous forme
\begin{eqnarray}
\mathbf{W}_{0\sigma}^{\;(2)}&=&\mathbf{W}_{1}+\mathbf{W}_{2}+\mathbf{W}_{3}
+\mathbf{W}_{4}+\mathbf{W}_{5}+\mathbf{W}_{6}+\mathbf{W}_{7}+\mathbf{W}_{8}\label{dat bis bis plusieurs termes appedix}
\end{eqnarray}
où
\begin{equation}\label{w1}
   \mathbf{W}_{1}=-\frac{1}{2}\,h^{\mu\rho}\Big(\partial_{\rho}\partial_{0}h_{\sigma\mu}
  -\partial_{\rho}\partial_{\mu}h_{0\sigma}-\partial_{\sigma}\partial_{0}h_{\rho\mu}
  +\partial_{\sigma}\partial_{\mu}h_{0\rho}\Big),
\end{equation}
\begin{equation}\label{w2}
    \mathbf{W}_{2}=-\frac{1}{2}\,\eta_{0\sigma}\,h_{\alpha\beta}
  \Big(\Box h^{\alpha\beta}-\partial_{\rho}\partial^{\alpha}h^{\rho\beta}-\partial^{\beta}\partial_{\rho}h^{\rho\alpha}
  +\partial^{\beta}\partial^{\alpha}h\Big),
\end{equation}
\begin{equation}\label{w3}
    \mathbf{W}_{3}=-\frac{1}{2}\,h_{0\sigma}\,\Big(\partial_{\rho}\partial_{\alpha}h^{\rho\alpha}-\Box h\Big),
\end{equation}
\begin{equation}\label{w4}
    \mathbf{W}_{4}=-\frac{1}{2}\Big(\partial^{\mu}h_{\alpha0}-\partial_{\alpha}h^{\mu}_{0}\Big)\Big(\partial_{\mu}h^{\alpha}_{\sigma}
  +\frac{1}{2}\,\partial_{\sigma}h^{\alpha}_{\mu}\Big),
\end{equation}
\begin{equation}\label{w5}
    \mathbf{W}_{5}=-\frac{1}{2}\,\eta_{0\sigma}\Big(\partial^{\mu}h^{\alpha}_{\mu}-\frac{1}{2}\,\partial^{\alpha}h\Big)
  \Big(\partial_{\mu}h^{\mu}_{\alpha}-\frac{1}{2}\,\partial_{\alpha}h\Big),
\end{equation}
\begin{equation}\label{w6}
    \mathbf{W}_{6}=-\frac{1}{4}\,\left(\partial_{0}h_{\alpha}^{\mu}\right)\partial_{\sigma}h_{\mu}^{\alpha},
\end{equation}
\begin{equation}\label{w7}
    \mathbf{W}_{7}=\frac{1}{4}\,\eta_{0\sigma}\left(\partial_{\mu}h^{\alpha\beta}\right)\left[\left(\partial_{\alpha}h^{\mu}_{\beta}\right)
  -\frac{3}{2}\left(\partial^{\mu}h_{\alpha\beta}\right)
    \right],
\end{equation}
\begin{equation}\label{w8}
    \mathbf{W}_{8}=\frac{1}{2}\Big(\partial_{0}h^{\alpha}_{\sigma}+\partial_{\sigma}h^{\alpha}_{0}-\partial^{\alpha}h_{0\sigma}\Big)
  \Big(\partial_{\mu}h^{\mu}_{\alpha}-\frac{1}{2}\,\partial_{\alpha}h\Big).
\end{equation}

\section{Application des conditions utilisées dans l'approche revisitée de la Gravité Linéaire pour les termes de $\mathbf{W}_{0\sigma}^{\;(2)}$}
En adoptant les conditions (\ref{2 conditions }), le premier terme de l'équation (\ref{dat bis appedix}) se met sous la forme
\begin{equation}\label{dat 1 appedix}
    G_{0\sigma}^{\;(1)}=\frac{q}{2mc}\,\partial^{\mu}\mathcal{F}_{\mu\sigma}.
\end{equation}
Nous allons appliquer les deux conditions précédentes pour simplifier chaque un des huit termes (\ref{w1}) à (\ref{w8}).
\subsection{$\mathbf{W}_{1}$}
Le premier terme est donné par
\begin{equation*}
   \mathbf{W}_{1}=-\frac{1}{2}\,h^{\mu\rho}\Big(\partial_{\rho}\partial_{0}h_{\sigma\mu}
  -\partial_{\rho}\partial_{\mu}h_{0\sigma}-\partial_{\sigma}\partial_{0}h_{\rho\mu}
  +\partial_{\sigma}\partial_{\mu}h_{0\rho}\Big).
\end{equation*}
En effectuant une sommation sur les indices muets $\mu$ et $\rho$, nous avons
\begin{eqnarray*}
  \mathbf{W}_{1}&=&-\frac{1}{2}\,h^{00}\Big(\partial_{0}\partial_{0}h_{\sigma0}
  -\partial_{0}\partial_{0}h_{0\sigma}-\partial_{\sigma}\partial_{0}h_{00}
  +\partial_{\sigma}\partial_{0}h_{00}\Big)\\
  &&-\frac{1}{2}\,h^{ij}\Big(\partial_{j}\partial_{0}h_{\sigma i}
  -\partial_{j}\partial_{i}h_{0\sigma}-\partial_{\sigma}\partial_{0}h_{ji}
  +\partial_{\sigma}\partial_{i}h_{0j}\Big)\\
  &&-\frac{1}{2}\,h^{0i}\bigg[\Big(\partial_{i}\partial_{0}h_{\sigma0}
  -\partial_{i}\partial_{0}h_{0\sigma}-\partial_{\sigma}\partial_{0}h_{i0}
  +\partial_{\sigma}\partial_{0}h_{0i}\Big)\\
  &&\hspace{1.1cm}+\Big(\partial_{0}\partial_{0}h_{\sigma i}
  -\partial_{0}\partial_{i}h_{0\sigma}-\partial_{\sigma}\partial_{0}h_{0i}
  +\partial_{\sigma}\partial_{i}h_{00}\Big)\bigg]
\end{eqnarray*}
Le premier terme (quand $(\rho,\mu)=(0,0)$) est nul, alors que le troisième terme (quand $(\rho,\mu)=(0,i)$ et $(\rho,\mu)=(i,0)$) comporte quelques simplifications
\begin{eqnarray*}
  \mathbf{W}_{1}&=&-\frac{1}{2}\,h^{ij}\Big(\partial_{j}\partial_{0}h_{\sigma i}
  -\partial_{j}\partial_{i}h_{0\sigma}-\partial_{\sigma}\partial_{0}h_{ji}
  +\partial_{\sigma}\partial_{i}h_{0j}\Big)\\
  &&-\frac{1}{2}\,h^{0i}\Big(\partial_{\sigma}\partial_{i}h_{00}-\partial_{\sigma}\partial_{0}h_{0i}
  -\partial_{i}\partial_{0}h_{0\sigma}+\partial_{0}\partial_{0}h_{\sigma i}\Big)
\end{eqnarray*}
de telle sorte à aboutir à
\begin{eqnarray*}
  \mathbf{W}_{1}&=&-\frac{1}{2}\,h^{ij}\Big[\partial_{j}\left(\partial_{0}h_{\sigma i}
  -\partial_{i}h_{0\sigma}\right)-\partial_{\sigma}\left(\partial_{0}h_{ji}
  +\partial_{i}h_{0j}\right)\Big]\\
  &&-\frac{1}{2}\,h^{0i}\Big[\partial_{\sigma}\left(\underline{\partial_{i}h_{00}-\partial_{0}h_{0i}}\right)
  -\partial_{0}\left(\partial_{i}h_{0\sigma}+\partial_{0}h_{\sigma i}\right)\Big].
\end{eqnarray*}
Or d'après (\ref{cond unifiee}) nous aboutissons finalement à l'expression
\begin{eqnarray}
  \mathbf{W}_{1}&=&-\frac{1}{2}\,h^{ij}\Big[\partial_{j}\left(\partial_{0}h_{\sigma i}
  -\partial_{i}h_{0\sigma}\right)-\partial_{\sigma}\left(\partial_{0}h_{ji}
  +\partial_{i}h_{0j}\right)\Big]\nonumber\\
  &&-\frac{1}{2}\,h^{0i}\Big[\partial_{\sigma}\left(\partial_{i}h_{00}/2+\partial_{j}h^{j}_{\;i}\right)
  -\partial_{0}\left(\partial_{i}h_{0\sigma}+\partial_{0}h_{\sigma i}\right)\Big].\label{w1 condition}
\end{eqnarray}

\subsection{$\mathbf{W}_{2}$}
Le deuxième terme est donné par
\begin{equation*}
    \mathbf{W}_{2}=-\frac{1}{2}\,\eta_{0\sigma}\,h_{\alpha\beta}
  \Big(\Box h^{\alpha\beta}-\partial_{\rho}\partial^{\alpha}h^{\rho\beta}-\partial^{\beta}\partial_{\rho}h^{\rho\alpha}
  +\partial^{\beta}\partial^{\alpha}h\Big).
\end{equation*}
En effectuant une sommation sur les indices muets $\alpha$ et $\beta$, nous avons
\begin{eqnarray*}
  \mathbf{W}_{2}&=&-\frac{1}{2}\,\eta_{0\sigma}\bigg[h_{00}
  \Big(\Box h^{00}-\partial_{\rho}\partial^{0}h^{\rho 0}-\partial^{0}\partial_{\rho}h^{\rho 0}
  +\partial^{0}\partial^{0}h\Big)\\
  &&\hspace{1.2cm}+h_{ij}
  \Big(\Box h^{ij}-\partial_{\rho}\partial^{i}h^{\rho j}-\partial^{j}\partial_{\rho}h^{\rho i}
  +\partial^{j}\partial^{i}h\Big)\\
  &&\hspace{1.2cm}+h_{0i}
  \Big(\Box h^{0i}-\partial_{\rho}\partial^{0}h^{\rho i}-\partial^{i}\partial_{\rho}h^{\rho 0}
  +\partial^{i}\partial^{0}h\Big)\\
  &&\hspace{1.2cm}+h_{i0}
  \Big(\Box h^{i0}-\partial_{\rho}\partial^{i}h^{\rho 0}-\partial^{0}\partial_{\rho}h^{\rho i}
  +\partial^{0}\partial^{i}h\Big)\bigg]
\end{eqnarray*}
\begin{eqnarray*}
  \mathbf{W}_{2}&=&-\frac{1}{2}\,\eta_{0\sigma}\bigg\{h_{00}
  \Big[\Box h^{00}-2\,\partial^{0}\left(\partial_{\rho}h^{\rho 0}\right)
  +\partial^{0}\partial^{0}h\Big]\\
  &&\hspace{1.3cm}+h_{ij}
  \Big[\Box h^{ij}-\partial^{i}\left(\partial_{\rho}h^{\rho j}\right)-\partial^{j}\left(\partial_{\rho}h^{\rho i}\right)
  +\partial^{j}\partial^{i}h\Big]\\
  &&\hspace{0.99cm}+2\,h_{0i}
  \Big[\Box h^{0i}-\partial^{0}\left(\partial_{\rho}h^{\rho i}\right)-\partial^{i}\left(\partial_{\rho}h^{\rho 0}\right)
  +\partial^{i}\partial^{0}h\Big]\bigg\}
\end{eqnarray*}
Or en utilisant les conditions (\ref{cond unifiee}) nous avons
\begin{eqnarray*}
  \mathbf{W}_{2}&=&-\frac{1}{2}\,\eta_{0\sigma}\bigg\{h_{00}
  \Big[\Box h^{00}-2\,\partial^{0}\left(\partial_{\rho}h^{\rho 0}\right)
  +\partial^{0}\partial^{0}h^{00}\Big]\\
  &&\hspace{1.3cm}+h_{ij}
  \Big[\Box h^{ij}-\partial^{i}\left(\partial^{j}h^{00}/2\right)-\partial^{j}\left(\partial^{i}h^{00}/2\right)
  +\partial^{j}\partial^{i}h^{00}\Big]\\
  &&\hspace{0.99cm}+2\,h_{0i}
  \Big[\Box h^{0i}-\partial^{0}\left(\partial^{i}h^{00}/2\right)-\partial^{i}\left(\partial_{\rho}h^{\rho 0}\right)
  +\partial^{i}\partial^{0}h^{00}\Big]\bigg\}
\end{eqnarray*}
ou encore finalement
\begin{eqnarray}
  \mathbf{W}_{2}&=&-\frac{1}{2}\,\eta_{0\sigma}\bigg\{h_{00}
  \Big[\Box h^{00}-2\,\partial^{0}\left(\partial_{\rho}h^{\rho 0}\right)
  +\partial^{0}\partial^{0}h^{00}\Big]+h_{ij}\,\Box h^{ij}\nonumber\\
  &&\hspace{3cm}+2\,h_{0i}
  \Big[\Box h^{0i}+\partial^{0}\left(\partial^{i}h^{00}/2\right)-\partial^{i}\left(\partial_{\rho}h^{\rho 0}\right)\Big]\bigg\}.\label{w2 condition}
\end{eqnarray}

Attirons l'attention sur le fait que si on utilise de plus la composante temporelle de la jauge harmonique $\partial_{\rho}h^{\rho 0}=\partial^{0}h/2=\partial^{0}h^{00}/2$, dans ce cas (\ref{w2 condition}) se met sous la forme
\begin{eqnarray*}
  \mathbf{W}_{2}&=&-\frac{1}{2}\,\eta_{0\sigma}\bigg\{h_{00}
  \Big[\Box h^{00}-2\,\partial^{0}\left(\partial^{0}h^{00}/2\right)
  +\partial^{0}\partial^{0}h^{00}\Big]+h_{ij}\,\Box h^{ij}\nonumber\\
  &&\hspace{3cm}+2\,h_{0i}
  \Big[\Box h^{0i}+\partial^{0}\left(\partial^{i}h^{00}/2\right)-\partial^{i}\left(\partial^{0}h^{00}/2\right)\Big]\bigg\},
\end{eqnarray*}
ou finalement
\begin{eqnarray}
  \mathbf{W}_{2}&=&-\frac{1}{2}\,\eta_{0\sigma}\bigg(h_{00}\,\Box h^{00}+h_{ij}\,\Box h^{ij}+2\,h_{0i}
  \,\Box h^{0i}\bigg).\label{w2 condition totale}
\end{eqnarray}

\subsection{$\mathbf{W}_{3}$}
Le troisième terme est donné par
\begin{equation*}
    \mathbf{W}_{3}=-\frac{1}{2}\,h_{0\sigma}\,\Big(\partial_{\rho}\partial_{\alpha}h^{\rho\alpha}-\Box h\Big).
\end{equation*}
En effectuant une sommation sur les indices muets $\rho$ et $\alpha$, nous avons
\begin{eqnarray*}
  \mathbf{W}_{3}&=&-\frac{1}{2}\,h_{0\sigma}\bigg[\Big(\partial_{0}\partial_{0}h^{00}+2\,\partial_{0}\partial_{i}h^{0i}+\partial_{i}\partial_{j}h^{ij}\Big)-\Box h\bigg]
\end{eqnarray*}
La combinaison du premier et quatrième terme, d'une part, et du deuxième et troisième terme, d'autre part, permet d'avoir
\begin{eqnarray*}
  \mathbf{W}_{3}&=&-\frac{1}{2}\,h_{0\sigma}\bigg[\left(\partial_{0}\partial_{0}h^{00}-\Box h\right)+2\,\partial_{i}\left(\partial_{0}h^{0i}+\partial_{j}h^{ij}\right)\bigg]
\end{eqnarray*}
En utilisant les conditions (\ref{cond unifiee}) l'expression précédente devient
\begin{eqnarray*}
  \mathbf{W}_{3}&=&-\frac{1}{2}\,h_{0\sigma}\bigg[\overrightarrow{\nabla}^{2}h^{00}+2\,\partial_{i}\left(\partial^{i}h^{00}/2\right)\bigg]=
  -\frac{1}{2}\,h_{0\sigma}\bigg[\overrightarrow{\nabla}^{2}h^{00}+\underbrace{\partial_{i}(\eta^{ij}\partial_{j}}_{-\delta^{ij}\partial_{i}\partial_{j}})h^{00}\bigg]
\end{eqnarray*}
pour donner lieu finalement à
\begin{eqnarray}
  \mathbf{W}_{3}&=&0.\label{w3 condition}
\end{eqnarray}

\subsection{$\mathbf{W}_{4}$}
Le quatrième terme est donné par
\begin{equation*}
    \mathbf{W}_{4}=-\frac{1}{2}\Big(\partial^{\mu}h_{\alpha0}-\partial_{\alpha}h^{\mu}_{0}\Big)\Big(\partial_{\mu}h^{\alpha}_{\sigma}
  +\frac{1}{2}\,\partial_{\sigma}h^{\alpha}_{\mu}\Big).
\end{equation*}
En effectuant une sommation sur les indices muets $\mu$ et $\alpha$, nous avons
\begin{eqnarray*}
  \mathbf{W}_{4}&=&-\frac{1}{2}\bigg[\Big(\partial^{0}h_{00}-\partial_{0}h^{0}_{0}\Big)\Big(\partial_{0}h^{0}_{\sigma}
  +\frac{1}{2}\,\partial_{\sigma}h^{0}_{0}\Big)
  +\Big(\partial^{i}h_{j0}-\partial_{j}h^{i}_{0}\Big)\Big(\partial_{i}h^{j}_{\sigma}
  +\frac{1}{2}\,\partial_{\sigma}h^{j}_{i}\Big)\\
  &&\hspace{0.5cm}+\Big(\underline{\partial^{0}h_{i0}-\partial_{i}h^{0}_{0}}\Big)\Big(\partial_{0}h^{i}_{\sigma}
  +\frac{1}{2}\,\partial_{\sigma}h^{i}_{0}\Big)
  +\Big(\underline{\partial^{i}h_{00}-\partial_{0}h^{i}_{0}}\Big)\Big(\partial_{i}h^{0}_{\sigma}
  +\frac{1}{2}\,\partial_{\sigma}h^{0}_{i}\Big)\bigg]
\end{eqnarray*}
Le premier terme s'annule, alors que le troisième et quatrième terme comportent un terme en facteur
\begin{eqnarray*}
  \mathbf{W}_{4}&=&-\frac{1}{2}\bigg\{
  \Big(\partial^{i}h_{j0}-\partial_{j}h^{i}_{0}\Big)\Big(\partial_{i}h^{j}_{\sigma}
  +\frac{1}{2}\,\partial_{\sigma}h^{j}_{i}\Big)\\
  &&\hspace{0.5cm}+\Big(\partial^{0}h_{i0}-\partial_{i}h^{0}_{0}\Big)\bigg[\Big(\partial_{0}h^{i}_{\sigma}
  +\frac{1}{2}\,\partial_{\sigma}h^{i}_{0}\Big)-\Big(\partial^{i}h^{0}_{\sigma}
  +\frac{1}{2}\,\partial_{\sigma}h^{0i}\Big)\bigg]\bigg\}
\end{eqnarray*}
ce qui permet d'avoir
\begin{eqnarray*}
  \mathbf{W}_{4}&=&-\frac{1}{2}\bigg[
  \Big(\partial^{i}h_{j0}-\partial_{j}h^{i}_{0}\Big)\Big(\partial_{i}h^{j}_{\sigma}
  +\frac{1}{2}\,\partial_{\sigma}h^{j}_{i}\Big)+\Big(\underline{\partial^{0}h_{i0}}-\partial_{i}h^{0}_{0}\Big)\Big(\partial_{0}h^{i}_{\sigma}
  -\partial^{i}h^{0}_{\sigma}\Big)\bigg].
\end{eqnarray*}
En utilisant les conditions (\ref{cond unifiee}) l'expression précédente devient finalement
\begin{eqnarray}
  \hspace*{-0.9cm}\mathbf{W}_{4}&=&-\frac{1}{2}\bigg[
  \Big(\partial^{i}h_{j0}-\partial_{j}h^{i}_{0}\Big)\Big(\partial_{i}h^{j}_{\sigma}
  +\frac{1}{2}\,\partial_{\sigma}h^{j}_{i}\Big)-\Big(\partial_{i}h_{\;0}^{0}/2+\partial_{j}h^{j}_{i}\Big)\Big(\partial_{0}h^{i}_{\sigma}
  -\partial^{i}h^{0}_{\sigma}\Big)\bigg].\label{w4 condition}
\end{eqnarray}

\subsection{$\mathbf{W}_{5}$}
Le cinquième terme est donné par
\begin{equation*}
    \mathbf{W}_{5}=-\frac{1}{2}\,\eta_{0\sigma}\Big(\partial^{\mu}h^{\alpha}_{\mu}-\frac{1}{2}\,\partial^{\alpha}h\Big)
  \Big(\partial_{\mu}h^{\mu}_{\alpha}-\frac{1}{2}\,\partial_{\alpha}h\Big).
\end{equation*}
En effectuant une sommation sur l'indice muet $\mu$
\begin{eqnarray*}
  \mathbf{W}_{5}&=&-\frac{1}{2}\,\eta_{0\sigma}\bigg[\Big(\partial^{0}h^{\alpha}_{0}+\partial^{i}h^{\alpha}_{i}-\frac{1}{2}\,\partial^{\alpha}h\Big)
  \Big(\partial_{0}h^{0}_{\alpha}+\partial_{j}h^{j}_{\alpha}-\frac{1}{2}\,\partial_{\alpha}h\Big)\bigg]
\end{eqnarray*}
et ensuite sur l'indice muet $\alpha$
\begin{eqnarray*}
  \mathbf{W}_{5}&=&-\frac{1}{2}\,\eta_{0\sigma}\bigg[\Big(\underline{\partial^{0}h^{0}_{0}}+\partial^{i}h^{0}_{i}-\underline{\partial^{0}h/2}\Big)
  \Big(\underline{\partial_{0}h^{0}_{0}}+\partial_{j}h^{j}_{0}-\underline{\partial_{0}h/2}\Big)\\
  &&\hspace{1.2cm}+\Big(\partial^{0}h^{k}_{0}+\underline{\partial^{i}h^{k}_{i}-\partial^{k}h/2}\Big)
  \Big(\partial_{0}h^{0}_{k}+\underline{\partial_{j}h^{j}_{k}-\partial_{k}h/2}\Big)\bigg].
\end{eqnarray*}
En utilisant les conditions (\ref{cond unifiee}) l'expression précédente devient
\begin{eqnarray*}
  \mathbf{W}_{5}&=&-\frac{1}{2}\,\eta_{0\sigma}\bigg[\Big(\partial^{0}h^{0}_{0}/2+\partial^{i}h^{0}_{i}\Big)
  \Big(\partial_{0}h^{0}_{0}/2+\partial_{j}h^{j}_{0}\Big)+\Big(\partial^{0}h^{k}_{0}-\partial^{0}h^{k}_{0}\Big)
  \Big(\partial_{0}h^{0}_{k}-\partial_{0}h^{0}_{k}\Big)\bigg].
\end{eqnarray*}
ou encore finalement
\begin{eqnarray}
  \hspace*{-0.9cm}\mathbf{W}_{5}&=&-\frac{1}{2}\,\eta_{0\sigma}\bigg[\Big(\partial^{0}h^{0}_{0}/2+\partial^{i}h^{0}_{i}\Big)
  \Big(\partial_{0}h^{0}_{0}/2+\partial_{j}h^{j}_{0}\Big)\bigg]=-\frac{1}{2}\,\eta_{0\sigma}\Big(\partial^{0}h^{0}_{0}/2+\partial^{i}h^{0}_{i}\Big)^{2}.\label{w5 condition}
\end{eqnarray}

Attirons l'attention sur le fait que si on utilise de plus la composante temporelle de la jauge harmonique
$\partial_{0}h^{00}+\partial_{i}h^{i0}=\partial^{0}h^{00}/2$, ou de façon équivalente $\partial_{i}h^{i0}=-\partial^{0}h^{00}/2$, l'expression (\ref{w5 condition}) s'annule
\begin{eqnarray}
\mathbf{W}_{5}&=&0.\label{w5 condition totale}
\end{eqnarray}

\subsection{$\mathbf{W}_{6}$}
Le sixième terme est donné par
\begin{equation*}
    \mathbf{W}_{6}=-\frac{1}{4}\,\left(\partial_{0}h_{\alpha}^{\mu}\right)\partial_{\sigma}h_{\mu}^{\alpha}.
\end{equation*}
En effectuant une sommation sue les indices muets $\mu$ et $\alpha$, nous avons
\begin{eqnarray*}
  \mathbf{W}_{6}&=&-\frac{1}{4}\bigg[\left(\partial_{0}h_{0}^{0}\right)\partial_{\sigma}h_{0}^{0}
  +\left(\partial_{0}h_{j}^{i}\right)\partial_{\sigma}h_{i}^{j}+\left(\partial_{0}h_{i}^{0}\right)\partial_{\sigma}h_{0}^{i}
  +\underbrace{(\partial_{0}h_{0}^{i})\partial_{\sigma}h_{i}^{0}}_{\left(\partial_{0}h_{i}^{0}\right)\partial_{\sigma}h_{0}^{i}}\bigg]
\end{eqnarray*}
Le troisième et quatrième terme se combinent, pour donner
\begin{eqnarray*}
  \mathbf{W}_{6}&=&-\frac{1}{4}\bigg[\left(\partial_{0}h_{0}^{0}\right)\partial_{\sigma}h_{0}^{0}
  +\left(\partial_{0}h_{j}^{i}\right)\partial_{\sigma}h_{i}^{j}+2\left(\partial_{0}h_{i}^{0}\right)\partial_{\sigma}h_{0}^{i}\bigg].
\end{eqnarray*}
En utilisant les conditions (\ref{cond unifiee}) l'expression précédente devient finalement
\begin{eqnarray}
  \mathbf{W}_{6}&=&-\frac{1}{4}\bigg[\left(\partial_{0}h_{0}^{0}\right)\partial_{\sigma}h_{0}^{0}
  +\left(\partial_{0}h_{j}^{i}\right)\partial_{\sigma}h_{i}^{j}+2\left(\partial_{i}h_{0}^{0}/2-\partial_{j}h^{j}_{i}\right)\partial_{\sigma}h_{0}^{i}\bigg].\label{w6 condition}
\end{eqnarray}

Dans le cas où on utilise la composante temporelle de la jauge harmonique
$\partial_{i}h^{i0}=-\partial^{0}h^{00}/2$, l'expression (\ref{w6 condition}) devient finalement
\begin{eqnarray}
  \mathbf{W}_{6}&=&-\frac{1}{4}\bigg[\left(-2\,\partial_{i}h^{i}_{0}\right)\partial_{\sigma}h_{0}^{0}
  +\left(\partial_{0}h_{j}^{i}\right)\partial_{\sigma}h_{i}^{j}+\left(\partial_{i}h_{0}^{0}-2\,\partial_{j}h^{j}_{i}\right)\partial_{\sigma}h_{0}^{i}\bigg].\label{w6 condition totale}
\end{eqnarray}

\subsection{$\mathbf{W}_{7}$}
Le septième terme est donné par
\begin{equation*}
    \mathbf{W}_{7}=\frac{1}{4}\,\eta_{0\sigma}\left(\partial_{\mu}h^{\alpha\beta}\right)\left[\left(\partial_{\alpha}h^{\mu}_{\beta}\right)
  -\frac{3}{2}\left(\partial^{\mu}h_{\alpha\beta}\right)\right].
\end{equation*}
La sommation sur l'indice muet $\mu$ permet d'avoir
\begin{eqnarray*}
  \mathbf{W}_{7}&=&\frac{1}{4}\,\eta_{0\sigma}\Bigg\{\left(\partial_{0}h^{\alpha\beta}\right)\left[\left(\partial_{\alpha}h^{0}_{\beta}\right)
  -\frac{3}{2}\left(\partial^{0}h_{\alpha\beta}\right)\right]+\left(\partial_{i}h^{\alpha\beta}\right)\left[\left(\partial_{\alpha}h^{i}_{\beta}\right)
  -\frac{3}{2}\left(\partial^{i}h_{\alpha\beta}\right)\right]\Bigg\}
\end{eqnarray*}
alors que la sommation sur les indices muets $\alpha$ et $\beta$ donne
\begin{eqnarray*}
  \mathbf{W}_{7}&=&\frac{1}{4}\,\eta_{0\sigma}\Bigg\{
  \left(\partial_{0}h^{00}\right)\left[\left(\partial_{0}h^{0}_{0}\right)-\frac{3}{2}\left(\partial^{0}h_{00}\right)\right]
  +\left(\partial_{0}h^{ks}\right)\left[\left(\partial_{k}h^{0}_{s}\right)-\frac{3}{2}\left(\partial^{0}h_{ks}\right)\right]\\
  &&\hspace{0.9cm}+\left(\underline{\partial_{0}h^{k0}}\right)\left[\left(\partial_{k}h^{0}_{0}\right)-\frac{3}{2}\left(\partial^{0}h_{k0}\right)\right]
  +\left(\underline{\partial_{0}h^{0k}}\right)\left[\left(\partial_{0}h^{0}_{k}\right)-\frac{3}{2}\left(\partial^{0}h_{0k}\right)\right]\Bigg\}\\
  &&+\frac{1}{4}\,\eta_{0\sigma}\Bigg\{
  \left(\partial_{i}h^{00}\right)\left[\left(\partial_{0}h^{i}_{0}\right)-\frac{3}{2}\left(\partial^{i}h_{00}\right)\right]
  +\left(\partial_{i}h^{ks}\right)\left[\left(\partial_{k}h^{i}_{s}\right)-\frac{3}{2}\left(\partial^{i}h_{ks}\right)\right]\\
  &&\hspace{1.25cm}+\left(\underline{\partial_{i}h^{k0}}\right)\left[\left(\partial_{k}h^{i}_{0}\right)-\frac{3}{2}\left(\partial^{i}h_{k0}\right)\right]
  +\left(\underline{\partial_{i}h^{0k}}\right)\left[\left(\partial_{0}h^{i}_{k}\right)-\frac{3}{2}\left(\partial^{i}h_{0k}\right)\right]\Bigg\}
\end{eqnarray*}

\begin{eqnarray*}
  \mathbf{W}_{7}&=&\frac{1}{4}\,\eta_{0\sigma}\Bigg\{-\frac{1}{2}\left(\partial_{0}h^{00}\right)^{2}
  +\left(\partial_{0}h^{ks}\right)\left[\left(\partial_{k}h^{0}_{s}\right)-\frac{3}{2}\left(\partial^{0}h_{ks}\right)\right]\\
  &&+\left(\partial_{0}h^{k0}\right)\bigg[\big(\partial_{k}h^{0}_{0}\big)+\underline{\big(\partial_{0}h^{0}_{k}\big)
  -3\left(\partial^{0}h_{k0}\right)}\bigg]\Bigg\}\\
  &&+\frac{1}{4}\,\eta_{0\sigma}\Bigg\{
  \left(\partial_{i}h^{00}\right)\left[(\partial_{0}h^{i}_{0})-\frac{3}{2}\left(\partial^{i}h_{00}\right)\right]
  +\left(\partial_{i}h^{ks}\right)\left[\left(\partial_{k}h^{i}_{s}\right)-\frac{3}{2}\left(\partial^{i}h_{ks}\right)\right]\\
  &&+\left(\partial_{i}h^{k0}\right)\bigg[\left(\partial_{k}h^{i}_{0}\right)+\left(\partial_{0}h^{i}_{k}\right)
  -3\left(\partial^{i}h_{0k}\right)\bigg]\Bigg\}.
\end{eqnarray*}
ou encore
\begin{eqnarray*}
  \mathbf{W}_{7}&=&\frac{1}{4}\,\eta_{0\sigma}\Bigg\{-\frac{1}{2}\left(\partial_{0}h^{00}\right)^{2}
  +\left(\partial_{0}h^{ks}\right)\left[\left(\partial_{k}h^{0}_{s}\right)-\frac{3}{2}\left(\partial^{0}h_{ks}\right)\right]\\
  &&+\left(\partial_{0}h^{k0}\right)\bigg[\big(\partial_{k}h^{0}_{0}\big)-2\left(\underline{\partial^{0}h_{k0}}\right)\bigg]\Bigg\}\\
  &&+\frac{1}{4}\,\eta_{0\sigma}\Bigg\{
  \left(\partial_{i}h^{00}\right)\left[(\underline{\partial_{0}h^{i}_{0}})-\frac{3}{2}\left(\partial^{i}h_{00}\right)\right]
  +\left(\partial_{i}h^{ks}\right)\left[\left(\partial_{k}h^{i}_{s}\right)-\frac{3}{2}\left(\partial^{i}h_{ks}\right)\right]\\
  &&+\left(\partial_{i}h^{k0}\right)\bigg[\left(\partial_{k}h^{i}_{0}\right)+\left(\partial_{0}h^{i}_{k}\right)
  -3\left(\partial^{i}h_{0k}\right)\bigg]\Bigg\}.
\end{eqnarray*}
En utilisant les conditions (\ref{cond unifiee}) l'expression précédente devient finalement
\begin{eqnarray}
  \mathbf{W}_{7}&=&\frac{1}{4}\,\eta_{0\sigma}\Bigg\{-\frac{1}{2}\left(\partial_{0}h^{00}\right)^{2}
  +\left(\partial_{0}h^{ks}\right)\left[\left(\partial_{k}h^{0}_{s}\right)-\frac{3}{2}\left(\partial^{0}h_{ks}\right)\right]
  +\left(\partial_{0}h^{k0}\right)\big[2\,\partial_{i}h^{i}_{k}\big]\Bigg\}\nonumber\\
  &&+\frac{1}{4}\,\eta_{0\sigma}\Bigg\{
  \left(\partial_{i}h^{00}\right)\left[-\partial^{i}h^{00}-\partial_{j}h^{ji}\right]
  +\left(\partial_{i}h^{ks}\right)\left[\left(\partial_{k}h^{i}_{s}\right)-\frac{3}{2}\left(\partial^{i}h_{ks}\right)\right]\nonumber\\
  &&+\left(\partial_{i}h^{k0}\right)\bigg[\left(\partial_{k}h^{i}_{0}\right)+\left(\partial_{0}h^{i}_{k}\right)
  -3\left(\partial^{i}h_{0k}\right)\bigg]\Bigg\}.\label{w7 condition}
\end{eqnarray}

\subsection{$\mathbf{W}_{8}$}
Le huitième terme est donné par
\begin{equation*}
    \mathbf{W}_{8}=\frac{1}{2}\Big(\partial_{0}h^{\alpha}_{\sigma}+\partial_{\sigma}h^{\alpha}_{0}-\partial^{\alpha}h_{0\sigma}\Big)
  \Big(\partial_{\mu}h^{\mu}_{\alpha}-\frac{1}{2}\,\partial_{\alpha}h\Big).
\end{equation*}
En effectuant une sommation sur l'indice muet $\alpha$
\begin{eqnarray*}
  \mathbf{W}_{8}&=&\frac{1}{2}\Big(\partial_{0}h^{0}_{\sigma}+\partial_{\sigma}h^{0}_{0}-\partial^{0}h_{0\sigma}\Big)
  \Big(\partial_{\mu}h^{\mu}_{0}-\frac{1}{2}\,\partial_{0}h\Big)\\
  &&+\frac{1}{2}\Big(\partial_{0}h^{i}_{\sigma}+\partial_{\sigma}h^{i}_{0}-\partial^{i}h_{0\sigma}\Big)
  \Big(\partial_{\mu}h^{\mu}_{i}-\frac{1}{2}\,\partial_{i}h\Big)
\end{eqnarray*}
ensuite sur l'indice muet $\mu$, nous avons
\begin{eqnarray*}
  \mathbf{W}_{8}&=&\frac{1}{2}\Big(\partial_{0}h^{0}_{\sigma}+\partial_{\sigma}h^{0}_{0}-\partial^{0}h_{0\sigma}\Big)
  \Big(\partial_{0}h^{0}_{0}+\partial_{j}h^{j}_{0}-\frac{1}{2}\,\partial_{0}h\Big)\\
  &&+\frac{1}{2}\Big(\partial_{0}h^{i}_{\sigma}+\partial_{\sigma}h^{i}_{0}-\partial^{i}h_{0\sigma}\Big)
  \Big(\underline{\partial_{0}h^{0}_{i}+\partial_{j}h^{j}_{i}}-\frac{1}{2}\,\partial_{i}h\Big).
\end{eqnarray*}
En utilisant les conditions (\ref{cond unifiee}) l'expression précédente devient finalement
\begin{eqnarray*}
  \mathbf{W}_{8}&=&\frac{1}{2}\Big(\partial_{0}h^{0}_{\sigma}+\partial_{\sigma}h^{0}_{0}-\partial^{0}h_{0\sigma}\Big)
  \Big(\underline{\partial_{0}h^{0}_{0}+\partial_{j}h^{j}_{0}}-\frac{1}{2}\,\partial_{0}h^{00}\Big).\label{w8 condition}
\end{eqnarray*}

Dans le cas où on utilise la composante temporelle de la jauge harmonique
$\partial_{j}h^{j0}=-\partial^{0}h^{00}/2$, l'expression (\ref{w8 condition}) s'annule
\begin{equation}\label{w8 condition totale}
    \mathbf{W}_{8}=0.
\end{equation}

\subsection{$\mathbf{W}_{0\sigma}^{\;(2)}$}
En utilisant les conditions (\ref{cond unifiee}) les termes correctifs à l'ordre 2 du tenseur d'Einstein $\mathbf{W}_{0\sigma}^{\;(2)}$, d'après (\ref{w1 condition}), (\ref{w2 condition}), (\ref{w3 condition}), (\ref{w4 condition}), (\ref{w5 condition}), (\ref{w6 condition}), (\ref{w7 condition}) et (\ref{w8 condition}) sont donnés finalement par
\begin{eqnarray}
\hspace*{-3cm}\mathbf{W}_{0\sigma}^{\;(2)}&=&-\frac{1}{2}\,h^{ij}\Big[\partial_{j}\left(\partial_{0}h_{\sigma i}
  -\partial_{i}h_{0\sigma}\right)-\partial_{\sigma}\left(\partial_{0}h_{ji}
  +\partial_{i}h_{0j}\right)\Big]\nonumber\\
  &&-\frac{1}{2}\,h^{0i}\Big[\partial_{\sigma}\left(\partial_{i}h_{00}/2+\partial_{j}h^{j}_{\;i}\right)
  -\partial_{0}\left(\partial_{i}h_{0\sigma}+\partial_{0}h_{\sigma i}\right)\Big]\nonumber\\
  &&-\frac{1}{2}\,\eta_{0\sigma}\bigg\{h_{00}
  \Big[\Box h^{00}-2\,\partial^{0}\left(\partial_{\rho}h^{\rho 0}\right)
  +\partial^{0}\partial^{0}h^{00}\Big]+h_{ij}\,\Box h^{ij}\nonumber\\
  &&+2\,h_{0i}\Big[\Box h^{0i}+\partial^{0}\left(\partial^{i}h^{00}/2\right)-\partial^{i}\left(\partial_{\rho}h^{\rho 0}\right)\Big]\bigg\}\nonumber\\
  &&-\frac{1}{2}\bigg[\Big(\partial^{i}h_{j0}-\partial_{j}h^{i}_{0}\Big)\Big(\partial_{i}h^{j}_{\sigma}
  +\frac{1}{2}\,\partial_{\sigma}h^{j}_{i}\Big)-\Big(\partial_{i}h_{\;0}^{0}/2+\partial_{j}h^{j}_{i}\Big)\Big(\partial_{0}h^{i}_{\sigma}
  -\partial^{i}h^{0}_{\sigma}\Big)\bigg]\nonumber\\
  &&-\frac{1}{2}\,\eta_{0\sigma}\Big(\partial^{0}h^{0}_{0}/2+\partial^{i}h^{0}_{i}\Big)^{2}\nonumber\\
  &&-\frac{1}{4}\bigg[\left(\partial_{0}h_{0}^{0}\right)\partial_{\sigma}h_{0}^{0}
  +\left(\partial_{0}h_{j}^{i}\right)\partial_{\sigma}h_{i}^{j}+2\left(\partial_{i}h_{0}^{0}/2-\partial_{j}h^{j}_{i}\right)\partial_{\sigma}h_{0}^{i}\bigg]\nonumber\\
  &&+\frac{1}{4}\,\eta_{0\sigma}\Bigg\{-\frac{1}{2}\left(\partial_{0}h^{00}\right)^{2}
  +\left(\partial_{0}h^{ks}\right)\left[\left(\partial_{k}h^{0}_{s}\right)-\frac{3}{2}\left(\partial^{0}h_{ks}\right)\right]
  +\left(\partial_{0}h^{k0}\right)\big[2\,\partial_{i}h^{i}_{k}\big]\nonumber\\
  &&+\left(\partial_{i}h^{00}\right)\left[-\partial^{i}h^{00}-\partial_{j}h^{ji}\right]
  +\left(\partial_{i}h^{ks}\right)\left[\left(\partial_{k}h^{i}_{s}\right)-\frac{3}{2}\left(\partial^{i}h_{ks}\right)\right]\nonumber\\
  &&+\left(\partial_{i}h^{k0}\right)\bigg[\left(\partial_{k}h^{i}_{0}\right)+\left(\partial_{0}h^{i}_{k}\right)
  -3\left(\partial^{i}h_{0k}\right)\bigg]\Bigg\}\nonumber\\
  &&+\frac{1}{2}\Big(\partial_{0}h^{0}_{\sigma}+\partial_{\sigma}h^{0}_{0}-\partial^{0}h_{0\sigma}\Big)
  \Big(\partial_{0}h^{0}_{0}+\partial_{j}h^{j}_{0}-\frac{1}{2}\,\partial_{0}h^{00}\Big).\label{moutarde 1 index}
\end{eqnarray}

Si de plus, nous utilisons la composante temporelle de la jauge harmonique
$\partial_{j}h^{j0}=-\partial^{0}h^{00}/2$, l'expression précédente devient
\begin{eqnarray}
\hspace*{-3cm}\mathbf{W}_{0\sigma}^{\;(2)}&=&-\frac{1}{2}\,h^{ij}\Big[\partial_{j}\left(\partial_{0}h_{\sigma i}
  -\partial_{i}h_{0\sigma}\right)-\partial_{\sigma}\left(\partial_{0}h_{ji}
  +\partial_{i}h_{0j}\right)\Big]\nonumber\\
  &&-\frac{1}{2}\,h^{0i}\Big[\partial_{\sigma}\left(\partial_{i}h_{00}/2+\partial_{j}h^{j}_{\;i}\right)
  -\partial_{0}\left(\partial_{i}h_{0\sigma}+\partial_{0}h_{\sigma i}\right)\Big]\nonumber\\
  &&-\frac{1}{2}\,\eta_{0\sigma}\bigg(h_{00}\,\Box h^{00}+h_{ij}\,\Box h^{ij}+2\,h_{0i}
  \,\Box h^{0i}\bigg)\nonumber\\
  &&-\frac{1}{2}\bigg[\Big(\partial^{i}h_{j0}-\partial_{j}h^{i}_{0}\Big)\Big(\partial_{i}h^{j}_{\sigma}
  +\frac{1}{2}\,\partial_{\sigma}h^{j}_{i}\Big)-\Big(\partial_{i}h_{\;0}^{0}/2+\partial_{j}h^{j}_{i}\Big)\Big(\partial_{0}h^{i}_{\sigma}
  -\partial^{i}h^{0}_{\sigma}\Big)\bigg]\nonumber\\
  &&-\frac{1}{4}\bigg[\left(-2\,\partial_{i}h^{i}_{0}\right)\partial_{\sigma}h_{0}^{0}
  +\left(\partial_{0}h_{j}^{i}\right)\partial_{\sigma}h_{i}^{j}+\left(\partial_{i}h_{0}^{0}-2\,\partial_{j}h^{j}_{i}\right)\partial_{\sigma}h_{0}^{i}\bigg]\nonumber\\
  &&+\frac{1}{4}\,\eta_{0\sigma}\Bigg\{-\frac{1}{2}\left(\partial_{0}h^{00}\right)^{2}
  +\left(\partial_{0}h^{ks}\right)\left[\left(\partial_{k}h^{0}_{s}\right)-\frac{3}{2}\left(\partial^{0}h_{ks}\right)\right]
  +\left(\partial_{0}h^{k0}\right)\big[2\,\partial_{i}h^{i}_{k}\big]\nonumber\\
  &&+\left(\partial_{i}h^{00}\right)\left[-\partial^{i}h^{00}-\partial_{j}h^{ji}\right]
  +\left(\partial_{i}h^{ks}\right)\left[\left(\partial_{k}h^{i}_{s}\right)-\frac{3}{2}\left(\partial^{i}h_{ks}\right)\right]\nonumber\\
  &&+\left(\partial_{i}h^{k0}\right)\bigg[\left(\partial_{k}h^{i}_{0}\right)+\left(\partial_{0}h^{i}_{k}\right)
  -3\left(\partial^{i}h_{0k}\right)\bigg]\Bigg\}.\label{moutarde 2 index}
\end{eqnarray}

\newpage
\pagestyle{fancy}\lhead{Bibliographie}\rhead{\;}

\end{document}